\documentclass[natbib209]{emulateapj}
\usepackage{graphicx}
\usepackage{rotating}
\usepackage[usenames]{color}
\usepackage{hyperref}

\shortauthors{Oh et al.}
\shorttitle{High-resolution mass models of dwarf galaxies from LITTLE THINGS}

\slugcomment{Accepted for publication on AJ}

\begin{document}

\newcommand{\C}{\ensuremath{\mathfrak{C}}}
\newcommand{\VNFW}{\ensuremath{V_{200}}}
\newcommand{\Vh}{\ensuremath{V_h}}
\newcommand{\kms}{\ensuremath{\mathrm{km}\,\mathrm{s}^{-1}}}
\newcommand{\cm}{\ensuremath{\mathrm{cm}^{-2}}}
\newcommand{\kmsnospace}{\ensuremath{\mathrm{km}\,\mathrm{s}^{-1}}}
\newcommand{\barkms}{\ensuremath{\mathrm{km}\,\mathrm{s}^{-1}\,\mathrm{kpc}^{-1}}}
\newcommand{\kmsMpc}{\ensuremath{\mathrm{km}\,\mathrm{s}^{-1}\,\mathrm{Mpc}^{-1}}}
\newcommand{\etal}{et al.}
\newcommand{\LCDM}{$\Lambda$CDM}
\newcommand{\ML}{\ensuremath{\Upsilon_{\star}}}
\newcommand{\MLfix}{\ensuremath{\Upsilon_{\star}^{fix}}}
\newcommand{\MLfree}{\ensuremath{\Upsilon_{\star}^{\rm free}}}
\newcommand{\MLsps}{\ensuremath{\Upsilon_{\star}^{3.6}}}
\newcommand{\MLspl}{\ensuremath{\Upsilon_{\star}^{4.5}}}
\newcommand{\MLk}{\ensuremath{\Upsilon_{\star}^{K}}}
\newcommand{\MLb}{\ensuremath{\Upsilon_{\star}^{B}}}
\newcommand{\MLmax}{\ensuremath{\Upsilon_{max}}}
\newcommand{\Lsun}{\ensuremath{\rm{{L}_{\odot}}}}
\newcommand{\Msun}{\ensuremath{\rm{{M}_{\odot}}}}
\newcommand{\mass}{\ensuremath{\rm{{\cal M}}}}
\newcommand{\magsq}{\ensuremath{\mathrm{mag}\,\mathrm{arcsec}^{-2}}}
\newcommand{\Lsundens}{\ensuremath{\rm L_{\odot}\,\mathrm{pc}^{-2}}}
\newcommand{\surfdens}{\ensuremath{\rm M_{\odot}\,\mathrm{pc}^{-2}}}
\newcommand{\cubedens}{\ensuremath{M_{\odot}\,\mathrm{pc}^{-3}}}

\long\def\Ignore#1{\relax}
\long\def\Comment#1{{\footnotesize #1}}

\title{High-resolution mass models of dwarf galaxies from LITTLE THINGS}

\author{Se-Heon Oh\altaffilmark{1,2}, Deidre A. Hunter\altaffilmark{3}, Elias
Brinks\altaffilmark{4}, Bruce G. Elmegreen\altaffilmark{5}, Andreas
Schruba\altaffilmark{6,7}, Fabian Walter\altaffilmark{8}, Michael P.
Rupen\altaffilmark{9}, Lisa M. Young\altaffilmark{10}, Caroline E.
Simpson\altaffilmark{11}, Megan Johnson\altaffilmark{12}, Kimberly A.
Herrmann\altaffilmark{13}, Dana Ficut-Vicas\altaffilmark{4}, Phil
Cigan\altaffilmark{10}, Volker Heesen\altaffilmark{14}, Trisha
Ashley\altaffilmark{15}, and Hong-Xin Zhang\altaffilmark{16}}

\altaffiltext{1}{The International Centre for Radio Astronomy Research (ICRAR),
The University of Western Australia,
35 Stirling Highway, Crawley, Western Australia 6009, Australia}
\altaffiltext{2}{ARC Centre of Excellence for All-sky Astrophysics (CAASTRO)}

\altaffiltext{3}{Lowell Observatory, 1400 West Mars Hill Road, Flagstaff,
Arizona 86001 USA}

\altaffiltext{4}{Centre for Astrophysics Research, University of Hertfordshire,
College Lane, Hatfield, AL10 9AB
United Kingdom}

\altaffiltext{5}{IBM T. J. Watson Research Center, PO Box 218, Yorktown Heights,
New York 10598 USA}

\altaffiltext{6}{Max-Planck-Institut f\"ur extraterrestrische Physik,
Giessenbachstrasse 1, 85748 Garching, Germany}

\altaffiltext{7}{Cahill Center for Astronomy and Astrophysics, California
Institute of Technology, 1200 East California Blvd, Pasadena, CA 91125, USA}

\altaffiltext{8}{Max-Planck-Institut f\"ur Astronomie, K\"onigstuhl 17, 69117
Heidelberg Germany}

\altaffiltext{9}{National Radio Astronomy Observatory, 1003 Lopezville Road,
Socorro, NM 87801 USA}

\altaffiltext{10}{Physics Department, New Mexico Institute of Mining and
Technology, Socorro, New Mexico 87801
USA}

\altaffiltext{11}{Department of Physics, Florida International University, CP
204, 11200 SW 8th St, Miami, Florida 33199 USA}

\altaffiltext{12}{Australia Telescope National Facility, CSIRO Astronomy and
Space Science, PO Box 76, Epping NSW 1710, Australia}

\altaffiltext{13}{Department of Physics, Pennsylvania State
University Mont Alto, Science Technology Building, Mont Alto, PA 17237 USA}

\altaffiltext{14}{School of Physics and Astronomy, University of
Southampton, Southampton SO17 1BJ UK}

\altaffiltext{15}{Bay Area Environmental Research Institute, Petaluma, CA 94952 USA}

\altaffiltext{16}{Peking University, Astronomy Department, No.
5 Yiheyuan Road, Haidian District,
Beijing, P. R. China 100871}

\email{se-heon.oh@icrar.org}

\begin{abstract}
We present high-resolution rotation curves and mass models of 26 dwarf galaxies
from `Local Irregulars That Trace Luminosity Extremes, The H{\sc i} Nearby
Galaxy Survey' (LITTLE THINGS). LITTLE THINGS is a high-resolution
($\sim$6\arcsec\ angular; $<$2.6 \kms\ velocity resolution) Very Large Array
(VLA) H{\sc i} survey for nearby dwarf galaxies in the local volume within 11
Mpc. The high-resolution H{\sc i} observations enable us to derive reliable
rotation curves of the sample galaxies in a homogeneous and consistent manner.
The rotation curves are then combined with {\it Spitzer} archival 3.6$\mu$m and
ancillary optical {\it U}, {\it B}, and {\it V} images to construct mass models
of the galaxies. This high quality multi-wavelength dataset significantly
reduces observational uncertainties and thus allows us to examine the mass
distribution in the galaxies in detail. We decompose the rotation curves in
terms of the dynamical contributions by baryons and dark matter halos, and
compare the latter with those of dwarf galaxies from The H{\sc i} Nearby Galaxy
Survey (THINGS) as well as \LCDM\ Smoothed Particle Hydrodynamic (SPH)
simulations in which the effect of baryonic feedback processes is included.
Being generally consistent with THINGS and simulated dwarf galaxies, most of the
LITTLE THINGS sample galaxies show a linear increase of the rotation curve in
their inner regions, which gives shallower logarithmic inner slopes $\alpha$ of
their dark matter density profiles. The mean value of the slopes of the 26
LITTLE THINGS dwarf galaxies is $\alpha =-0.32 \pm 0.24$ which is in accordance
with the previous results found for low surface brightness galaxies ($\alpha
=-0.2 \pm 0.2$) as well as the seven THINGS dwarf galaxies ($\alpha =-0.29 \pm
0.07$). However, this significantly deviates from the cusp-like dark matter
distribution predicted by dark-matter-only \LCDM\ simulations. Instead our
results are more in line with the shallower slopes found in the \LCDM\ SPH
simulations of dwarf galaxies in which the effect of baryonic feedback processes
is included. In addition, we discuss the central dark matter distribution of
DDO 210 whose stellar mass is relatively low in our sample to examine the
scenario of inefficient supernova feedback in low mass dwarf galaxies predicted
from recent \LCDM\ SPH simulations of dwarf galaxies where central cusps still remain.
\end{abstract}

\section{Introduction} \label{INTRO}
Dark matter the existence of which is indirectly invoked by its gravitational
effect in individual galaxies as well as in galaxy clusters, dominates, together with dark energy,
the energy budget in the Universe (\citeauthor{Zwicky_1937}
\citeyear{Zwicky_1937}; \citeauthor{van_den_Bergh_1961} \citeyear{van_den_Bergh_1961};
\citeauthor{1970ApJ...159..379R} \citeyear{1970ApJ...159..379R};
\citeauthor{Bosma_1978} \citeyear{Bosma_1978}; \citeauthor{Peebles_1982}
\citeyear{Peebles_1982}; \citeauthor{Riess_1998} \citeyear{Riess_1998};
\citeauthor{Perlmutter_1999} \citeyear{Perlmutter_1999};
\citeauthor{Spergel_2003} \citeyear{Spergel_2003}).
In particular, the role of dark matter (DM) is critical not only in forming and evolving
galaxies in the early Universe but also in shaping the large-scale structure in
the Universe through cosmic time (\citeauthor{Blumenthal_1984}
\citeyear{Blumenthal_1984}; \citeauthor{Colless_2001}
\citeyear{Colless_2001};  \citeauthor{Padmanabhan_2007}
\citeyear{Padmanabhan_2007}; \citeauthor{Jones_2009} \citeyear{Jones_2009};
\citeauthor{2011ApJS..192...18K} \citeyear{2011ApJS..192...18K}
etc.). The cosmological importance of
DM has driven efforts to explore the physical nature of DM 
particles and to attempt their direct detection 
(\citeauthor{Moore_2001} \citeyear{Moore_2001};
\citeauthor{Gaitskell_2004} \citeyear{Gaitskell_2004};
\citeauthor{2012EPJC...72.1971A} \citeyear{2012EPJC...72.1971A}; 
\citeauthor{2014PhRvL.112i1303A} \citeyear{2014PhRvL.112i1303A}, and references
therein). Of many candidates for DM particles, Cold Dark Matter
(CDM) has been envisaged as one of the most successful models in that numerical
simulations based on a paradigm combining CDM with the cosmological constant,
$\Lambda$ (so-called \LCDM). This describes well the large-scale structure in the Universe
traced by surveys such as the SDSS\footnote{The Sloan Digital Sky Survey
(\url{http://www.sdss.org/})} (\citeauthor{2000AJ....120.1579Y}
\citeyear{2000AJ....120.1579Y}; \citeauthor{2004A&A...418....7D} \citeyear{2004A&A...418....7D}), 2dFGRS\footnote{The
2dF Galaxy Redshift Survey (\url{http://www2.aao.gov.au/2dfgrs/})}
\citep{2001MNRAS.328.1039C}, 6dFGS\footnote{The 6dF Galaxy Survey
(\url{http://oldweb.aao.gov.au/local/www/6df/})} \citep{2004MNRAS.355..747J},
and CMB power spectrum observations (\citeauthor{Primack_2003}
\citeyear{Primack_2003}; \citeauthor{Spergel_2003} \citeyear{Spergel_2003},
\citeyear{Spergel_2007}; \citeauthor{2014A&A...566A..54P}
\citeyear{2014A&A...566A..54P}). 

However, despite the success of cosmological \LCDM\ simulations in producing the
large-scale structure of the Universe, distinct differences between the
simulations and observations have
been found in the DM distribution near the centre of individual
galaxies. The simulations have consistently predicted cusp-like DM 
distributions towards the centers of galaxies,
being described by a power law, $\rho$$\sim$$R^{\alpha}$ where $R$ is the galaxy
radius and $\alpha$$\sim$$-1.0$ (\citeauthor{Moore_1994}
\citeyear{Moore_1994}; \citeauthor{NFW_1996} \citeyear{NFW_1996};
\citeauthor{NFW_1997} \citeyear{NFW_1997}; \citeauthor{Moore_1999a}
\citeyear{Moore_1999a}; \citeauthor{Ghigna_2000} \citeyear{Ghigna_2000};
\citeauthor{Klypin_2001} \citeyear{Klypin_2001}; \citeauthor{Power_2002}
\citeyear{Power_2002}; \citeauthor{Stoehr_2003} \citeyear{Stoehr_2003};
\citeauthor{Navarro_2004} \citeyear{Navarro_2004}; \citeauthor{Diemand_2008}
\citeyear{Diemand_2008}; \citeauthor{2009MNRAS.398L..21S}
\citeyear{2009MNRAS.398L..21S}; \citeauthor{Navarro_2010} \citeyear{Navarro_2010};
\citeauthor{2013ApJ...767..146I} \citeyear{2013ApJ...767..146I} etc.). In contrast, inferred DM distributions in nearby dwarf galaxies in the local Universe have shown a
linear velocity increase towards their centers, giving rise to a sizable density-core
($\rho\sim$$R^{\alpha}$ where $\alpha$$\sim$$0.0$) (\citeauthor{Moore_1994}
\citeyear{Moore_1994}; \citeauthor{deBlok_1996}
\citeyear{deBlok_1996}; \citeauthor{deBlok_1997} \citeyear{deBlok_1997};
\citeauthor{deBlok_2001} \citeyear{deBlok_2001}; \citeauthor{deBlok_2002}
\citeyear{deBlok_2002}; \citeauthor{Weldrake_2003} \citeyear{Weldrake_2003};
\citeauthor{Spekkens_2005} \citeyear{Spekkens_2005};
\citeauthor{Kuzio_de_Naray_2006} \citeyear{Kuzio_de_Naray_2006};
\citeauthor{Kuzio_de_Naray_2008} \citeyear{Kuzio_de_Naray_2008};
\citeauthor{Oh_2008} \citeyear{Oh_2008}; \citeauthor{Oh_2011a}
\citeyear{Oh_2011a}; \citeauthor{Oh_2011b} \citeyear{Oh_2011b}; see, however,
\citeauthor{2014ApJ...789...63A} \citeyear{2014ApJ...789...63A} for a discussion
on gas kinematics which produces shallower density profiles than those from
stellar kinematics).
This clear discrepancy of the central DM
distribution in galaxies between \LCDM\ simulations and observations, the
so-called `cusp/core' problem has been one aspect of the small-scale crisis in
\LCDM\ cosmology which is likely connected to the `missing satellites' problem
\citep{2013ApJ...765...22B} and `too-big-to-fail' problem
(dense satellites; see \citeauthor{2011MNRAS.415L..40B}
\citeyear{2011MNRAS.415L..40B}, \citeyear{2012MNRAS.422.1203B}).  
Given that DM combined with 
the \LCDM\ paradigm is successful in explaining both the large-scale structure 
in the Universe as well as galaxy formation and evolution, there are good
reasons to explore ways to resolve the `cusp/core' problem. We refer to 
\cite{deBlok_2010} and \cite{Pontzen_2014} for the latest review of the `cusp/core' problem.

Late-type dwarf galaxies in the local Universe, with a simple dynamical structure 
(no bulge and spiral components)  have been used for addressing the central DM 
distribution in galaxies since their dynamics are usually dominated by DM,
enabling us to derive more accurately the DM distribution near their centers. These
diffuse dwarf galaxies have provided significant observational constraints on the central DM
distribution in galaxies. Over the past decade, several high-resolution neutral hydrogen (H{\sc i})
surveys of galaxies in the local Universe ($<$11 Mpc) using radio
interferometers, among others THINGS\footnote{The H{\sc i} Nearby Galaxy Survey (\url{http://www.mpia-hd.mpg.de/THINGS/Overview.html})} (\citeauthor{Walter_2008}
\citeyear{Walter_2008}), VLA-ANGST\footnote{Very Large Array \-- ACS Nearby Galaxy
Survey Treasury (\url{https://science.nrao.edu/science/surveys/vla-angst})} (\citeauthor{Ott_2012} \citeyear{Ott_2012}),
LITTLE THINGS\footnote{Local Irregulars That Trace Luminosity Extremes, The
H{\sc i} Nearby Galaxy Survey (\url{https://science.nrao.edu/science/surveys/littlethings})} (\citeauthor{Hunter_2012}
\citeyear{Hunter_2012}), FIGGS\footnote{Faint Irregular Galaxies GMRT Survey}
(\citeauthor{Begum_2008} \citeyear{Begum_2008}), SHIELD\footnote{The Survey of
HI in Extremely Low-mass Dwarfs (\url{http://www.macalester.edu/~jcannon/shield.html})} (\citeauthor{Cannon_2011}
\citeyear{Cannon_2011}), LVHIS\footnote{The Local Volume HI Survey (\url{http://www.atnf.csiro.au/research/LVHIS/})}
(\citeauthor{Koribalski_2010} \citeyear{Koribalski_2010}), have allowed us
to derive more reliable H{\sc i} rotation curves
of galaxies and examine their central mass distributions within 1 kpc where the
predictions of \LCDM\ simulations are most distinctive.
For example, high-resolution (0.1--0.2 kpc) DM density profiles of seven
dwarf galaxies from THINGS, complemented with the `{\it Spitzer} Infrared
Nearby Galaxies Survey' (SINGS; \citeauthor{Kennicutt_2003}
\citeyear{Kennicutt_2003}), were derived by \citeauthor{Oh_2008}
(\citeyear{Oh_2008}; \citeyear{Oh_2011a}). From this, they found that the mean value of the inner density
slopes, $\alpha$, of the seven dwarf galaxies is $-0.29\pm0.07$ which is in good
agreement with the value of $-0.2\pm0.2$ derived earlier from a larger number of Low Surface
Brightness (LSB) galaxies (\citeauthor{deBlok_2002} \citeyear{deBlok_2002}). In the past,
it has been argued that observational systematic effects such as beam smearing, center 
offsets and non-circular motions could have affected the
derived central DM distributions of galaxies as these observational biases
tend to flatten the derived inner DM density profiles,
hiding the central cusps. However, because of the high resolution and quality of the
above mentioned observations, these potential biases were significantly 
reduced. As a result, observational evidence for the core-like distribution of DM 
near the centers of dIrr galaxies is particularly strong.

In order to explain this behaviour, baryonic feedback processes have been proposed 
as a means for removing the central cusps expected
from \LCDM\ DM-only simulations. More specifically, it is expected that DM and baryons in
galaxies can be substantially redistributed by frequent explosions of supernovae
(SNe) (\citeauthor{Larson_1974}
\citeyear{Larson_1974}; \citeauthor{Navarro_1996} \citeyear{Navarro_1996};
\citeauthor{Dekel_2003} \citeyear{Dekel_2003}; \citeauthor{Mo_Mao_2004}
\citeyear{Mo_Mao_2004}; \citeauthor{Mashchenko_2008}
\citeyear{Mashchenko_2008}; \citeauthor{de_Souza_2011}
\citeyear{de_Souza_2011}; \citeauthor{Brook_2011} \citeyear{Brook_2011};
\citeauthor{Pontzen_2012} \citeyear{Pontzen_2012};
\citeauthor{2014MNRAS.437..415D} \citeyear{2014MNRAS.437..415D};
\citeauthor{2014ApJ...793...46O} \citeyear{2014ApJ...793...46O} etc.).
However, due to numerical difficulties in simulating
multi-phase gas physics as well as the lack of understanding of the detailed
baryonic physics in galaxies, taking the baryonic feedback into account in
hydrodynamical simulations of dwarf galaxies was considered difficult and limited to
simulations of high-redshift galaxies (e.g., \citeauthor{Mashchenko_2006}
\citeyear{Mashchenko_2006}).

\cite{Governato_2010} were the first to perform high-resolution cosmological
N--body+Smoothed Particle Hydrodynamic (SPH) simulations of dwarf galaxies which
include the effect of detailed baryonic feedback processes, in particular
physically motivated gas outflows driven by SN explosions. From this, they found
that the photometric and kinematic properties of the simulated dwarf galaxies
are in close, qualitative agreement with those of observed nearby dwarf galaxies.
More quantitatively, \cite{Oh_2011b} performed  an analysis of the baryonic
and DM mass distributions of the simulated dwarf galaxies, in exactly the same way 
as the THINGS dwarf galaxies were analysed, and showed that their derived rotation 
curves and the corresponding DM density profiles are consistent with those of the
THINGS dwarf galaxies and show a linear increase of velocity in the inner
region inherent of shallow DM density profiles. This suggests that
repeated gas outflows driven by SN explosions even without a burst of star
formation are able to play a fundamental role in removing the central cusps and
inducing flatter DM density slopes near the centers of dwarf galaxies.
This is in contrast to clusters of galaxies where galaxy interactions are more
likely to be the dominant mechanisms for the removal of central cusps rather
than star formation activities (\citeauthor{Richtler_2011}
\citeyear{Richtler_2011}). See \cite{Governato_2010} and
\citeauthor{Pontzen_2012} (\citeyear{Pontzen_2012}; \citeyear{Pontzen_2014})
for detailed discussions regarding the effect of SN explosions on the central cusps.

THINGS was only able to probe a small number (i.e., seven) of dwarf galaxies. It
is therefore essential to extend the investigation to a larger number of dwarf
galaxies in order to obtain a statistically robust observational sample to which
simulations can be compared to. 
Data on a larger sample of dIrr galaxies has now been provided 
by the latest H{\sc i} survey of nearby galaxies, LITTLE THINGS (\citeauthor{Hunter_2012}
\citeyear{Hunter_2012}). 
LITTLE THINGS is a high-resolution ($\sim$6$\arcsec$
angular; $\leq$ 2.6 \kms\ velocity resolution) H{\sc i} survey for 41 nearby
($<$ 11 Mpc) gas-rich dwarf galaxies undertaken with the NRAO\footnote{NRAO is a
facility of the National Science Foundation operated under cooperative agreement
by Associated Universities, Inc. These data were taken during the upgrade of the
VLA to the Expanded VLA or EVLA. In this paper we refer to the instrument as the
VLA, the retrofitted antennas as EVLA antennas, and non-retrofitted antennas as
VLA antennas. This emphasizes the hybrid nature of the instrument and
distinguishes it from the far more powerful Jansky VLA or JVLA it has become
since 2012.} Very Large Array (VLA)
in the northern sky. The H{\sc i} observations are complemented with other wavelength data, such as H$\alpha$,
optical {\it U}, {\it B}, {\it V}, and near infrared (\citeauthor{Hunter_2006}
\citeyear{Hunter_2006}), archival {\it Spitzer} infrared and {\it GALEX}
ultraviolet images as well as follow-up observations with ALMA and {\it
Herschel}. These high-quality multi-wavelength
data sets significantly reduce the observational uncertainties
inherent in low resolution data which may result in hiding the central cusps. Of the 41
galaxies, we select a sample of 26 dwarf galaxies (three of them are also in
THINGS) which show a regular rotation pattern in their velocity fields. In this
paper, we extract (1) bulk and non-circular motions of
the sample galaxies, (2) derive rotation curves, (3) decompose the derived rotation
curves in terms of the contributions by baryons and DM halos, and (4) address
the central DM distribution by making a direct comparison between the derived DM
distributions of the galaxies to those of SPH$+$N--body simulations of dwarf
galaxies.

The structure of this paper is as follows. The data used for deriving the mass
models of our sample galaxies are described in Section~\ref{DATA}. In
Section~\ref{MASS_MODELS}, we present the rotation curves, the mass models of
baryons, and the DM mass modeling of the galaxies. 
In Section~\ref{DARK_MATTER_DISTRIBUTION}, we discuss the central DM 
distributions of the sample galaxies by comparing them with those of dwarf galaxies
from both THINGS and simulations. We then discuss the effect of SN feedback
on the central cusp in Section~\ref{SN_FEEDBACK}, followed by the discussion of
the effect of beam smearing on the central DM distribution of galaxies
in Section~\ref{BEAM_SMEARING_EFFECT}. Lastly, we summarise the main results of this
paper and conclusions in Section~\ref{CONCLUSIONS}.

\section{The data} \label{DATA}
We use high-resolution H{\sc i} data of 26 nearby ($<$11 Mpc) dwarf galaxies
from LITTLE THINGS to address the central DM distribution of the galaxies. The
sample galaxies show a regular rotation pattern in their 2-dimensional (2D) H{\sc i} velocity
fields (see Appendix), allowing us to derive reliable rotation curves
which include the contributions to their kinematics of both their DM halo, i.e.,
non--baryonic and baryonic matter.
Considering the distances of the sample galaxies, the linear
resolutions of the LITTLE THINGS H{\sc i} data ($\sim$6\arcsec) range from $\sim$26 to 200 pc with an
average of 100 pc which is sufficient to resolve the inner 1 kpc region of the
galaxies in so far unmatched detail. This enables us to examine the central DM 
distributions of the galaxies in detail. In addition, observational systematic
effects inherent in low-resolution data (e.g., beam smearing, kinematic center
offset and non-circular motions) are significantly reduced in the
high-resolution H{\sc i} data, which allows us to derive more accurate underlying
kinematics of the sample galaxies. 

Although the total kinematics of late-type dwarf galaxies is dominated by DM
 (\citeauthor{Prada_2002} \citeyear{Prada_2002}), it is nonetheless important to separate the contribution
by baryons from the total rotation curve. This is achieved by using {\it Spitzer} archival {\it IRAC} 3.6$\mu$m and
ancillary optical color information (\citeauthor{Hunter_2006}
\citeyear{Hunter_2006}). {\it Spitzer IRAC} 3.6$\mu$m images are much less affected by dust than maps
at shorter wavelengths and trace the old stellar populations that occupy the dominant fraction of
the stellar mass in galaxies (\citeauthor{Walter_2007} \citeyear{Walter_2007}). Although there is some
contamination by PAH emission extending into the 3.6$\mu$m band (\citeauthor{2014ApJ...788..144M} \citeyear{2014ApJ...788..144M}) 
this is reduced in dIrr galaxies which, by virtue of having low heavy element abundances, have a correspondingly
lower dust content.

According to the simulations by
\cite{Governato_2012}, the degree of baryonic feedback is largely dependent on
the amount of stars, so past star formation activity in the sense that for the
same total mass budget (baryons + DM), those galaxies with a higher past star
formation activity will have had more significant outflows.
Therefore a reliable measurement of the
stellar mass in a galaxy is essential to investigating
the effect of baryonic feedback on the central cusp. The basic observational
properties of the sample galaxies are listed in Table~\ref{LT_TR_properties}.
We refer to \cite{Hunter_2012} for a complete description of the H{\sc i} observations
and data reduction. 

\begin{sidewaystable*}[htbp]
\centering
\caption{Properties and tilted-ring parameters of the LITTLE THINGS sample galaxies}
\label{LT_TR_properties}
\scriptsize
\scalebox{0.9}{
\begin{tabular}{@{}lccrrrrrccc}
\hline
\hline
\tablewidth{0pt}
Name & $\alpha$ (2000.0) & $\delta$ (2000.0) & $D$ & $V_{sys}$ & $\langle \rm P.A. \rangle$ & $\langle \rm i \rangle$ & $M_{\rm V}$ & $12+{\rm log}(O/H)$ & $\rm {log(SFR^{H\alpha}_{D})}$ & $\rm {log(SFR^{FUV}_{D})}$ \\
  & (h m s) & ($^{\circ}$ $'$ \arcsec) & (Mpc) & (\kms) & ($^{\circ}$) & ($^{\circ}$) & (mag) &  & ($\rm M_{\odot}\,yr^{-1}\,kpc^{-2}$) & ($\rm M_{\odot}\,yr^{-1}\,kpc^{-2}$) \\
 & (1) & (2) & (3) & (4) & (5) & (6) & (7) & (8) & (9) & (10) \\
\hline 
CVnIdwA     & 12 38 39.2 & 	$+$32 45 41.0  & 3.6  & 306.2 $\pm$ 1.3 &  48.4 $\pm$ 13.9 & 66.5 $\pm$ 5.2 & $-12.4$	& 7.3 $\pm$ 0.06	& $-2.58 \pm 0.01$	& $-2.48 \pm 0.01$ \\
DDO 43    	& 07 28 17.7 & 	$+$40 46 08.3  & 7.8  & 355.4 $\pm$ 3.6	&  294.1 $\pm$ 0.1 & 40.6 $\pm$ 0.1 & $-15.1$	& 8.3 $\pm$ 0.09	& $-1.78 \pm 0.01$	& $-1.55 \pm 0.01$ \\ 
DDO 46  	& 07 41 26.3 & 	$+$40 06 37.5  & 6.1  & 360.8 $\pm$	1.3 &  274.1 $\pm$ 5.0 & 27.9 $\pm$ 0.1 & $-14.7$	& 8.1 $\pm$ 0.10	& $-2.89 \pm 0.01$	& $-2.46 \pm 0.01$  \\ 
DDO 47 		& 07 41 55.3 & 	$+$16 48 07.1  & 5.2  & 270.7 $\pm$ 1.3	&  311.6 $\pm$ 11.9 & 45.5 $\pm$ 9.0 & $-15.5$	& 7.8 $\pm$ 0.20	& $-2.70 \pm 0.01$	& $-2.40 \pm 0.01$  \\
DDO 50 	    & 08 19 03.7 & 	$+$70 43 24.6  & 3.4  & 156.5 $\pm$ 1.2	&  175.7 $\pm$ 10.1 & 49.7 $\pm$ 6.0 & $-16.6$	& 7.7 $\pm$ 0.14	& $-1.67 \pm 0.01$	& $-1.55 \pm 0.01$  \\
DDO 52      & 08 28 28.4 & 	$+$41 51 26.5  & 10.3 & 399.0 $\pm$ 2.3	&  8.2 $\pm$ 5.3 & 43.0 $\pm$ 0.0 & $-15.4$		& 7.7 $\pm$ $...$ 	& $-3.20 \pm 0.01$ 	& $-2.43 \pm 0.01$   \\
DDO 53 	    & 08 34 06.4 & 	$+$66 10 47.9  & 3.6  & 18.6  $\pm$ 0.7	&  131.6 $\pm$ 0.1 & 27.0 $\pm$ 0.0 & $-13.8$	& 7.6 $\pm$ 0.11	& $-2.42 \pm 0.01$	& $-2.41 \pm 0.01$  \\
DDO 70 		& 10 00 00.9 & 	$+$05 20 12.9  & 1.3  & 303.5 $\pm$	1.2 &  44.5 $\pm$ 13.5 & 50.0 $\pm$ 0.0 & $-14.1$	& 7.5 $\pm$ 0.06	& $-2.85 \pm 0.01$	& $-2.16 \pm 0.00$  \\
DDO 87 	    & 10 49 34.9 & 	$+$65 31 47.9  & 7.7  & 340.0 $\pm$	1.0 &  235.1 $\pm$ 3.9 & 55.5 $\pm$ 4.8 & $-15.0$	& 7.8 $\pm$ 0.04	& $-1.36 \pm 0.01$	& $-1.00 \pm 0.01$  \\
DDO 101     & 11 55 39.1 &  $+$31 31 9.9   & 6.4  & 589.4 $\pm$	1.0 &  287.4 $\pm$ 2.8 & 51.0 $\pm$ 7.1 & $-15.0$	& 8.7 $\pm$ 0.03	& $-2.85 \pm 0.01$	& $-2.81 \pm 0.01$ \\
DDO 126     & 12 27 06.6 &  $+$37 08 15.9  & 4.9  & 219.4 $\pm$	1.8 &  138.0 $\pm$ 3.6 & 65.0 $\pm$ 0.0 & $-14.9$	& 7.8 $\pm$ $...$ 	& $-2.37 \pm 0.01$ 	& $-2.10 \pm 0.01$   \\
DDO 133     & 12 32 55.2 &  $+$31 32 19.1  & 3.5  & 330.7 $\pm$	1.0 &  359.6 $\pm$ 8.4 & 43.4 $\pm$ 0.1 & $-14.8$	& 8.2 $\pm$ 0.09    & $-2.88 \pm 0.01$	& $-2.62 \pm 0.01$ \\
DDO 154     & 12 54 05.7 &  $+$27 09 09.9  & 3.7  & 372.0 $\pm$	1.3 &  226.3 $\pm$ 3.1 & 68.2 $\pm$ 3.1 & $-14.2$	& 7.5 $\pm$ 0.09	& $-2.50 \pm 0.01$	& $-1.93 \pm 0.01$ \\
DDO 168     & 13 14 27.3 &  $+$45 55 37.3  & 4.3  & 192.6 $\pm$	1.2 &  275.5 $\pm$ 5.8 & 46.5 $\pm$ 0.1 & $-15.7$	& 8.3 $\pm$ 0.07	& $-2.27 \pm 0.01$	& $-2.04 \pm 0.01$ \\
DDO 210     & 20 46 51.6 &  $-$12 50 57.7  & 0.9  & $-$139.5 $\pm$ 1.0 &  65.0 $\pm$ 0.0 & 66.7 $\pm$ 0.1 & $-10.9$	& 7.2 $\pm$ $...$   & $...$ 			& $-2.71 \pm 0.06$   \\
DDO 216     & 23 28 34.7 &  $+$14 44 56.2  & 1.1  & $-$186.0 $\pm$ 1.1 &  133.6 $\pm$ 7.5 & 63.7 $\pm$ 4.6 & $-13.7$	& 7.9 $\pm$ 0.15	& $-4.10 \pm 0.07$	& $-3.21 \pm 0.01$ \\
F564-V3     & 09 02 54.0 &  $+$20 04 26.3  & 8.7  & 481.0 $\pm$	2.0 &  12.5 $\pm$ 0.0 & 56.5 $\pm$ 9.9 & $-14.0$	& 7.6 $\pm$ $...$ 	& $...$  			& $-2.79 \pm 0.02$   \\
IC 10		& 00 20 18.9 &  $+$59 17 49.9  & 0.7  & $-$348.0 $\pm$ 2.9 &  55.7 $\pm$ 10.1 & 47.0 $\pm$ 15.6 & $-16.3$		& 8.2 $\pm$ 0.12	& $-1.11 \pm 0.01$	& $...$ \\
IC 1613     & 01 04 49.6 &  $+$02 08 14.1  & 0.7  & $-$232.2 $\pm$ 2.2 &  73.7 $\pm$ 0.0 & 48.0 $\pm$ 0.0 & $-14.6$	& 7.6 $\pm$ 0.05	        & $-2.56 \pm 0.01$	& $-1.99 \pm 0.01$ \\
NGC 1569    & 04 30 46.2 &  $+$64 51 10.3  & 3.4  & $-$85.4 $\pm$ 5.6 &  122.5 $\pm$ 1.5 & 69.1 $\pm$ 0.1 & $-18.2$		& 8.2 $\pm$ 0.05	& $+0.19 \pm 0.01$	& $-0.01 \pm 0.01$ \\
NGC 2366    & 07 28 53.4 &  $+$69 12 49.6  & 3.4  & 103.5 $\pm$ 1.4  &  38.7 $\pm$ 4.2 & 63.0 $\pm$ 0.8 & $-16.8$	& 7.9 $\pm$ 0.01	& $-1.67 \pm 0.01$	& $-1.66 \pm 0.01$ \\
NGC 3738    & 11 35 46.9 &  $+$54 31 44.8  & 4.9  & 235.8 $\pm$ 1.2 	&  292.2 $\pm$ 5.1 & 22.6 $\pm$ 0.1 & $-17.1$	& 8.4 $\pm$ 0.01	& $-1.66 \pm 0.01$	& $-1.53 \pm 0.01$ \\
UGC 8508    & 13 30 44.9 &  $+$54 54 32.4  & 2.6  & 60.5 $\pm$ 1.1 	&  126.1 $\pm$ 3.2 & 82.5 $\pm$ 0.1 & $-13.6$	& 7.9 $\pm$ 0.20	& $-2.03 \pm 0.01$	& $...$ \\
WLM     	& 00 01 59.9 &  $-$15 27 57.2  & 1.0 & $-$122.3 $\pm$ 1.2 &  174.5 $\pm$ 2.8 & 74.0 $\pm$ 0.1 & $-14.4$	& 7.8 $\pm$ 0.06	        & $-2.77 \pm 0.01$	& $-2.05 \pm 0.01$ \\
Haro 29     & 12 26 18.4 &  $+$48 29 40.4  & 5.9  & 279.3 $\pm$ 2.2	&  214.5 $\pm$ 2.8 & 61.2 $\pm$ 4.0 & $-14.6$	& 7.9 $\pm$ 0.07	& $-0.77 \pm 0.01$	& $-1.07 \pm 0.01$ \\
Haro 36     & 12 46 56.6 &  $+$51 36 47.3  & 9.3  & 499.3 $\pm$ 3.9	&  248.4 $\pm$ 12.5 & 70.0 $\pm$ 0.0 & $-15.9$	& 8.4 $\pm$ 0.08	& $-1.86 \pm 0.01$	& $-1.55 \pm 0.01$  \\
\hline
\end{tabular}}
\begin{minipage}{205mm}
\scriptsize{Note. 
{\bf (1)(2):} Kinematic center position derived from the tilted-ring analysis in Section~\ref{TR_FITS};
{\bf (3):} Distance as given in \cite{Hunter_2012};
{\bf (4):} Systemic velocity derived from the tilted-ring analysis in Section~\ref{TR_FITS}; 
{\bf (5):} Average value of the kinematic position angle (PA) derived from the tilted-ring analysis in Section~\ref{TR_FITS}. PA is the angle measured counter-clockwise from the north direction in the sky to the major axis of the receding half of the galaxy;
{\bf (6):} Average value of the kinematic inclination derived from the tilted-ring analysis in Section~\ref{TR_FITS};
{\bf (7):} Absolute $V$ magnitude as given in \cite{Hunter_2012};
{\bf (8):} Oxygen abundance taken from the literature as compiled in \cite{Hunter_2012}. `$...$' indicates no uncertainty available on the measurements;
{\bf (9)(10):} The $H\alpha$ and {\it GALEX FUV} star formation rates normalized
to the area ($\pi R_{d}^{2}$ where $R_{d}$ is the disk scale length) as given in
\cite{Hunter_2004} and \cite{Hunter_2010}, respectively. `$...$' indicates that no measurements are available.}
\end{minipage}
\end{sidewaystable*}

\begin{sidewaystable*}[!htbp]
\caption{Mass modelling results of the LITTLE THINGS sample galaxies}
\label{MD_results_LT}
\begin{center}
\scalebox{0.7}{
\begin{tabular}{@{}lrrrrrrrrrrrrrrrrr}
\hline
\hline
Name        & $R_{\rm max}$ & $R_{0.3}$ & $V_{\rm max}$ & $V_{\rm ISO}(R_{\rm max})$ & $R_{\rm max}$H{\sc i}$_{\rm beam}^{-1}$ & $z_{0}$ & $c$ & $V_{\rm 200}$
& $R_{\rm C}$ & $\rho_{0}$ & $\alpha_{\rm min}$ &$\alpha_{\ML.pdf}$ & $\rm M_{\rm gas}$ & $\rm M_{\rm star}^{\rm KIN}$ & $\rm
M_{\rm star}^{SED}$ & $\rm log(M_{\rm dyn})$ & $\rm log(M_{\rm 200})$ \\
            & (kpc) & (kpc) & (\kms) & (\kms) & & (kpc) &  & (\kms) & (kpc) & ($10^{-3}$ \cubedens)
& & & ($10^{7}\,M_{\odot}$) & ($10^{7}\,M_{\odot}$) & ($10^{7}\,M_{\odot}$) & ($M_{\odot}$) & ($M_{\odot}$) \\
            & (1) & (2) & (3) & (4) & (5) & (6) & (7) & (8) & (9) & (10) & (11)
& (12) & (13) & (14) & (15) & (16) & (17)\\
 
\hline
CVnIdwA			&   2.59	&   2.27 & 23.5 & 23.7	&  13.5 & 0.24  & $-0.4$ (11.0) & 478.5 ($16.1 \pm 1.6$) & $2.01 \pm 0.52$ & $8.19 \pm 1.62$							&$-1.25 \pm 0.21$		&$+0.03 \pm 0.27$	& 2.91 & 0.41 & 0.49 & 8.529 & 9.138 \\
DDO 43			&   4.19	&   4.19 & 33.2 & 38.5	&  13.7 & $...$ & $5.6 \pm 2.6$ (10.2) & $39.8 \pm 14.9$ ($31.4 \pm 1.8$) & $0.94 \pm 0.18$ & $33.22 \pm 8.78$			&$-0.15 \pm 0.13$		&$-0.25^{*} \pm 0.16$	& 23.26  & $...$ & $...$ & 9.159 & 10.317  \\ 
DDO 46			&   2.92	&   2.04 & 66.1 & 77.4	&  15.7 & $...$ & $18.7 \pm 3.5$ (9.0) & $61.2 \pm 9.7$ ($136.7 \pm 6.1$) & $0.51 \pm 0.04$ & $517.82 \pm 65.46$		&$-0.41 \pm 0.01$		&$-0.42^{*} \pm 0.02$	& 22.08  & $...$ & $...$ & 9.609 & 11.925  \\
DDO 47			&   7.71	&   1.55 & 64.7 & 68.9	&  29.3 & $...$ & $-0.4$ (9.2) & 1359.0 ($51.3 \pm 2.5$) & $3.72 \pm 0.48$ & $12.15 \pm 1.49$							&$-1.24 \pm 0.20$		&$-1.25^{*} \pm 0.22$	& 46.80 & $...$ & $...$ & 9.930 & 10.648  \\
DDO 50			&   9.81	&   1.67 & 35.7 & 35.7	&  84.4 & 0.28  & $22.2$ (10.3) & 15.4 ($28.8 \pm 0.2$) & $0.15 \pm 0.08$ & $379.64 \pm 381.14$							&$-0.41 \pm 0.41$		&$+0.10 \pm 0.41$	& 132.52 & 9.79 & 10.60 & 9.463 & 9.895 \\
DDO 52			&   5.43	&   5.43 & 60.2 & 60.5	&  15.9 & 0.26  & $6.1 \pm 1.1$ (9.4) & $71.9 \pm 11.8$ ($57.8 \pm 1.3$) & $1.33 \pm 0.07$ & $48.81 \pm 3.63$			&$-0.55 \pm 0.04$		&$-0.49 \pm 0.02$	& 33.43 & 7.20 & 5.31 & 9.664 & 10.803 \\
DDO 53			&   1.45	&   0.62 & 28.6 & 33.1	&  13.1 & 0.14  & $0.0 \pm 4.4$ (10.4) & $425.8 \pm 2531.9$ ($27.8 \pm 3.2$) & $2.22 \pm 1.95$ & $25.10 \pm 5.63$		&$-0.27 \pm 0.48$		&$+0.14 \pm 0.80$	& 7.00 & 0.96 & 0.97 & 8.567 & 9.849 \\
DDO 70			&	2.00	&	0.55 & 43.9 & 35.5	&  22.9 & 0.11  & $0.0 \pm 10.2$ (10.3) & $866.1 \pm 11717.4$ ($38.6 \pm 1.8$) & $0.51 \pm 0.11$ & $119.95 \pm 34.69$							&$-0.43 \pm 0.02$		&$-0.48 \pm 0.03$	& 3.80  & 1.24  & 1.96 & 8.768 & 10.277  \\
DDO 87			&   7.39	&   4.13 & 56.6 & 56.2	&  26.1 & 0.48  & $0.0 \pm 0.2$ (9.5) & $511.4 \pm 143.8$ ($40.7 \pm 1.6$) & $2.46 \pm 0.11$ & $13.91 \pm 0.77$			&$-0.01 \pm 0.44$		&$-0.01 \pm 0.47$	& 29.12 & 6.18 & 3.27 & 9.734 & 10.346 \\
DDO 101			&   1.95	&   1.05  & 64.5 & 63.6	&  7.5  & 0.13  & $25.0 \pm 1.3$ (9.4) & $43.9 \pm 1$ ($115.7 \pm 6.4$) & $0.32 \pm 0.01$ & $849.14 \pm 77.23$ &$-1.00 \pm 0.15$		&$-1.02 \pm 0.12$	& 3.48 & 5.79 & 6.54 & 9.279 &  510.18 \\
DDO 126			&   3.99	&   2.88 & 37.2 & 40.5	&  24.3 & 0.23  & $-0.2 \pm 44.4$ (10.1) & $626.6 \pm 48442.1$ ($31.4 \pm 1.2$) & $1.33 \pm 0.10$ & $21.59 \pm 2.00$	&$-0.41 \pm 0.15$		&$-0.39 \pm 0.15$	& 16.36 & 2.27 & 1.62 & 9.182 & 10.008 \\
DDO 133			&   3.48	&   3.48 & 46.1 & 47.5	&  16.5 & 0.22  & $4.9 \pm 2.3$ (9.8) & $74.5 \pm 36.7$ ($46.6 \pm 1.6$) & $0.83 \pm 0.06$ & $73.69 \pm 7.61$			&$-0.11 \pm 0.14$		&$-0.11 \pm 0.15$	& 12.85 & 2.62 & 3.04 & 9.261 & 10.522 \\
DDO 154			&   7.32	&   2.59 & 47.8 & 49.2	&  51.3 & 0.14  & $6.4 \pm 0.5$ (9.8) & $48.7 \pm 2.9$ ($41.1 \pm 0.6$) & $0.95 \pm 0.03$ & $53.21 \pm 3.19$			&$-0.39 \pm 0.11$		&$-0.41 \pm 0.13$	& 35.27 & 1.31 & 0.83 & 9.614 & 10.359 \\
DDO 168			&   3.14	&   3.14 & 60.3 & 67.4	&  19.2 & 0.22  & $-0.3 \pm 1171.6$ (9.2) & $2290.8 \pm ...$ ($63.9 \pm 4.0$) & $2.81 \pm 0.83$ & $39.81 \pm 6.37$		&$-0.28 \pm 0.28$		&$+0.97 \pm 0.18$	& 25.94 & 5.13 & 5.85 & 9.520 & 10.934 \\
DDO 210			&   0.31	&   0.31 & 12.0 & 11.3	&  6.0  & 0.10  & $0.0 \pm 202.3$ (12.2) & $339.4 \pm ...$ ($12.4 \pm 1.0$) & $0.20 \pm 0.05$ & $116.43 \pm 23.05$		&$-0.30 \pm 0.07$		&$-0.70 \pm 0.02$	& 0.14 & 0.04 & 0.06 & 6.964 & 8.797 \\
DDO 216			&   1.12	&   0.48 & 18.9 & 15.7	&  13.0 & 0.27  & $13.2 \pm 2.7$ (11.7) & $8.8 \pm 1.0$ ($14.7 \pm 0.4$) & $0.15 \pm 0.03$ & $127.02 \pm 43.35$			&$-0.17 \pm 0.47$		&$-0.30 \pm 0.62$	& 0.49 & 1.60 & 1.51 & 7.807  & 9.019 \\
F564-V3			&   3.71	&   1.83 & 28.8 & 34.2	&  7.0  & $...$ & $10.8 \pm 4.5$ (10.4) & $26.2 \pm 7.1$ ($29.6 \pm 2.1$) & $0.55 \pm 0.20$ & $74.24 \pm 40.17$			&$-0.66^{*} \pm 0.14$	&$-0.68^{*} \pm 0.18$	& 4.37 & $...$ & $...$ & 9.003 & 9.931 \\
IC 10			&   0.54	&   0.20 & 36.5 & 36.1	&  27.0 & 0.07  & $0.2 \pm 951.3$ (18.6) & $2077.1 \pm ...$ ($36.8 \pm 2.2$) & $0.27 \pm 0.12$ & $190.4 \pm 76.4$		&$-1.19 \pm 0.01$		&$-0.25 \pm 0.32$ 		& 1.65  & 11.81  & $...$ & 8.213  & 10.215 \\
IC 1613			&   2.71	&   0.36 & 16.9 & 21.1	&  103.8& 0.15  & $5.0 \pm ...$ (11.2) & $5.4 \pm ...$ ($12.5 \pm 0.4$) & $0.20 \pm 0.04$ & $19.25 \pm 3.45$			&$+1.16 \pm 1.22$		&$-0.10 \pm 0.92$		& 5.93 & 1.94 & 2.88 & 8.448 & 8.808 \\
NGC 1569		&   3.05	&   2.81 & 29.1 & 46.9	&  31.5 & 0.19  & $-0.3 \pm 179.2$ (9.8) & $736.6 \pm ...$ ($40.9 \pm 2.2$) & $2.71 \pm 0.81$ & $15.23 \pm 2.45$		&$-0.77 \pm 0.17$		&$-0.23 \pm 0.67$		& 20.24  & 20.69  & 36.29  & 9.193  & 10.352 \\
NGC 2366		&   8.08	&   2.75 & 58.2 & 60.8	&  70.7 & 0.29  & $5.2 \pm 0.6$ (9.4) & $66.0 \pm 6.1$ ($52.6 \pm 0.8$) & $1.21 \pm 0.04$ & $43.89 \pm 2.51$			&$-0.52 \pm 0.13$		&$-0.53 \pm 0.12$		& 108.24 & 10.81 & 6.94 & 9.841 & 10.680 \\
NGC 3738		&   1.75	&   1.10 & 125.5& 130.2	&  11.7 & 0.12  & $12.4 \pm 15.4$ (8.1) & $310.4 \pm 536.2$ ($598.2 \pm 30.4$) & $0.45 \pm 0.04$ & $2132.36 \pm 277.75$	&$-0.42 \pm 0.03$		&$-0.44 \pm 0.02$		& 12.58 & 12.48 & 46.62 & 9.838 & 13.848 \\
UGC 8508		&   1.86	&   0.46 & 46.0 & 43.5	&  25.0 & 0.09  & $7.8 \pm 1.4$ (10.0) & $54.7 \pm 7.4$ ($44.2 \pm 2.2$) & $1.95 \pm 0.21$ & $45.29 \pm 2.38$ 			&$-0.31 \pm 0.17$		&$-0.38 \pm 0.16$		& 1.19  & 0.30  & 0.77 & 8.913 & 10.453  \\
WLM				&   3.04	&   1.60 & 37.3 & 37.7	&  82.9 & 0.17  & $3.0 \pm 0.8$ (10.2) & $90.8 \pm 26.1$ ($33.3 \pm 0.5$) & $0.74 \pm 0.01$ & $57.46 \pm 1.57$ 			&$+0.03 \pm 0.01$		&$-0.02 \pm 0.01$		& 7.96 & 1.23 & 1.62 & 9.002 & 10.085 \\
Haro 29			&   5.03	&   0.64 & 30.5 & 35.0	&  25.7 & $...$ & $19.5 \pm 1.9$ (10.4) & $23.4 \pm 0.9$ ($32.7 \pm 0.9$) & $0.23 \pm 0.04$ & $414.68 \pm 137.20$ 		&$-0.50 \pm 0.12$		&$-0.50^{*} \pm 0.12$	& 9.35  & $...$ & 1.43 & 9.156 & 10.061 \\
Haro 36			&   3.16	&   3.16 & 58.2 & 52.6	&  10.0 & 0.13  & $-0.4 \pm 1503.4$ (9.7) & $1649.2 \pm ...$ ($47.9 \pm 3.5$) & $8.40 \pm 12.17$ & $16.99 \pm 2.78$ 	&$-0.48 \pm 0.06$		&$-0.50 \pm 0.01$		& 11.16 & 5.81 & $...$ & 9.308 & 10.558 \\
\hline                                                                           
\end{tabular}}
\begin{minipage}{225mm}
\scriptsize{
{\bf (1):} The radius where the outermost part of the rotation curve is measured;
{\bf (2):} The radius where the logarithmic slope of the total rotation curve (DM + baryons) $d{\rm log}V/d{\rm log}R=0.3$. See Section~\ref{RC_SHAPE};
{\bf (3):} The rotation velocity (asymmetric drift corrected) at $\rm R_{\rm max}$;
{\bf (4):} The rotation velocity of the best fitted pseudo-isothermal halo model at $R_{\rm max}$;
{\bf (5):} The ratio of $R_{\rm max}$ to H{\sc i} beam size;
{\bf (6):} The vertical scale height of the stellar disk;
{\bf (7)(8):} Concentration parameter $c$ and rotation velocity $V_{200}$ for a  NFW
halo model (\citeauthor{NFW_1995} \citeyear{NFW_1995}; \citeyear{NFW_1996});
$c$ values in brackets are derived using an empirical relationship between $c -
V_{200}$ as given by \cite{McGaugh_2007}. The corresponding $V_{200}$ values are fitted after
fixing $c$ to the ones in brackets assuming a `minimum disk'. See
Section~\ref{DISK_HALO_DECOMPOSITION} for a detailed description. `$...$' indicates that the uncertainty is unphysically large;
{\bf (9)(10):} Core-radius and core-density of a pseudo-isothermal halo model
(\citeauthor{Begeman_1991} \citeyear{Begeman_1991});
{\bf (11):} The logarithmic inner slope $\alpha_{\rm min}$ of the total matter (DM halo + baryons) density profiles measured in Section~\ref{DM_PROFILE}.
{\bf (12):} The logarithmic inner slope $\alpha_{\ML.pdf}$ of the DM density profiles measured in Section~\ref{DM_PROFILE}.
{\bf (13):} The gas mass derived in Section~\ref{GAS_DISTRIBUTION};
{\bf (14):} The stellar mass derived from the kinematic analysis in Section~\ref{STELLAR_DISTRIBUTION};
{\bf (15):} The stellar mass derived using a spectral energy distribution (SED) fitting technique in \cite{Zhang_2012};
{\bf (16):} The dynamical mass measured using $V_{\rm ISO}(R_{\rm max})$ and $R_{\rm max}$;
{\bf (17):} CDM halo mass $\rm M_{\rm 200}$ estimated from the $V_{\rm 200}$ in
brackets of column (5) and using Eq. 3 given in \cite{Oh_2011b}. See
Section~\ref{DISK_HALO_DECOMPOSITION} for more details. \\
$^{*}$The dynamical contribution by the stellar component is included in the DM density profile as no {\it Spitzer} 3.6$\mu$m image is available for this galaxy.}

\end{minipage}
\end{center}
\end{sidewaystable*}

\section{Mass models} \label{MASS_MODELS}
In this Section, we perform mass modeling of the 26 LITTLE THINGS dwarf galaxies using the
high-resolution VLA H{\sc i} data and {\it Spitzer IRAC} 3.6$\mu$m images as described in
Section~\ref{DATA}. As mentioned earlier, the simple dynamical structure and dominant
circular rotation in the disk of the sample galaxies help to reduce
uncertainties stemming from the decomposition of galaxy rotation
curves into contributions due to baryonic and non--baryonic matter.

The first step in the mass modelling includes deriving rotation curves of the sample
galaxies using the 2D H{\sc i} velocity fields which reflect the total kinematics
including both baryons and a DM halo. 
H{\sc i} is mostly distributed in the disk of a galaxy where the circular rotation is
dominant and is a useful kinematic tracer for deriving the galaxy rotation curve.
This is mainly due to the larger radial extent of H{\sc i} in the disk
compared to stellar components (e.g., 3$\--$4 times; \citeauthor{Sofue_2001}
\citeyear{Sofue_2001}). We proceed to derive  mass models of the
baryons (gas and stars) using H{\sc i} integrated intensity maps and {\it Spitzer IRAC}
3.6$\mu$m images, and subtract their contribution from the total kinematics. Lastly, we
quantify the kinematic residuals in order to examine the DM 
distribution near the centers of the galaxies. In the following sections, we
describe these mass modelling procedures in more detail.

\subsection{Rotation curves: {\it total kinematics}} \label{ROTATION_CURVE}

\subsubsection{Tilted-ring fits} \label{TR_FITS}

For the derivation of the rotation curves of our sample galaxies, we fit
a 2D tilted-ring model which consists of a series of concentric ellipses to the
2D velocity fields extracted from the H{\sc i} data cubes of the sample
galaxies (\citeauthor{Rogstad_1974} \citeyear{Rogstad_1974}). Each ellipse has
its own geometric and kinematic parameters, such as center position (XPOS,
YPOS), position angle (PA), inclination (INCL), systemic velocity (VSYS), and
rotation velocity (VROT). This so-called `tilted-ring analysis' of 2D velocity
fields obtained from H{\sc i}, CO or H$\alpha$ spectroscopic observations has been widely
used for deriving rotation curves of disk-dominated galaxies (e.g., rotation
curves of the THINGS galaxies sample; \citeauthor{deBlok_2008}
\citeyear{deBlok_2008}).

Tilted-ring models, however, only hold for those cases where the velocity
field is a reliable representation of the overall kinematics of a galaxy.
Non-circular motions caused by star-forming activity like
stellar winds, SNe, etc., as well as spiral arms, a bar-like or triaxial galaxy
potential, galaxy mergers or tidal interactions disturb gas motions in galaxies
on small and large scales, resulting in distorted velocity fields.

Low resolution H{\sc i} data usually smooth any features related to small-scale
non-circular motions. This is known as beam smearing and tends to yield a velocity 
gradient along the major axis of a galaxy that is less steep, particularly in the
central regions (e.g., \citeauthor{deBlok_1997} \citeyear{deBlok_1997};
\citeauthor{Swaters_2000} \citeyear{Swaters_2000};
\citeauthor{van_den_Bosch_2001} \citeyear{van_den_Bosch_2001};  
\citeauthor{McGaugh_2001} \citeyear{McGaugh_2001}).  
Beam smearing is significantly reduced in high-resolution
data such as THINGS and LITTLE THINGS (see \citeauthor{deBlok_2001}
\citeyear{deBlok_2001}; \citeauthor{McGaugh_2001} \citeyear{McGaugh_2001};
\citeauthor{Kuzio_de_Naray_2006} \citeyear{Kuzio_de_Naray_2006}).

The LITTLE THINGS sample dwarf galaxies selected in this study, with few exceptions, appear 
to show no significant large-scale kinematic features, such as bars,
spiral arms or warps in their H{\sc i} or {\it Spitzer IRAC} 3.6$\mu$m images.
However, our sample galaxies do suffer from the effect of small-scale turbulent 
gas motions due to stellar winds and SNe. In general, the derived velocity
field in dwarf galaxies is more vulnerable to the impact of stellar activity due to
their low gravitational potential (\citeauthor{Walter_1998} \citeyear{Walter_1998};
\citeauthor{Walter_2001} \citeyear{Walter_2001}; see also
\citeauthor{Bagetakos_2011} \citeyear{Bagetakos_2011}).
Therefore, the extraction of a robust velocity field is essential if one wants to
derive the undisturbed underlying kinematics of a galaxy.

In order to correct for small-scale random motions and extract only the
component due to circular rotating velocity  in a galaxy, we derive the 
`bulk velocity field' as proposed by \citeauthor{Oh_2008} (\citeyear{Oh_2008}; \citeyear{Oh_2011a}).
Compared with other typical types of velocity fields, such as intensity-weighted
mean (IWM), single Gaussian fit and hermite h3, the bulk velocity field has been
found to be ideally suited to extracting the underlying bulk rotation of a galaxy in 
the presence of random non-circular motions. We point out, as an example, 
how the `kinks' or `wiggles' of the iso-velocity contours of the extracted bulk velocity
fields displayed in panel (e) of Fig. A.1
are weaker than those of the IWM velocity fields (panel b). The bulk
velocity field appears to much better represent the overall kinematics of the galaxy
and hence the underlying gravitational potential.

Following the standard procedure described in \cite{Begeman_1989}, we fit tilted-ring
models to the bulk velocity fields of the sample galaxies. For this, we use the
`rotcur' task in GIPSY\footnote{The Groningen Image Processing System} \citep{1992ASPC...25..131V}. The
derived rotation curves of the galaxies are
presented in the figures of the Appendix (e.g., A.2). As seen from the scatter
in the fits made with all ring parameters free (open circles) in the figures,
the extracted bulk velocity fields are not completely free from the effect of small-scale random motions in the
galaxies. However, they are relatively insignificant and are averaged out after
several iterations as shown in the final rotation curves (solid lines) in the
figures.

This is also confirmed in the harmonic analysis of the velocity fields.
As described in \cite{Schoenmakers_1997}, we perform harmonic decompositions of
the bulk and IWM velocity fields of the sample galaxies.
For this, we use the task `{\sc reswri}'  in GIPSY. We expand the velocity fields
into {\it sine} and {\it cosine} terms up to  3$^{\rm{rd}}$ order (i.e.,
$c_{m}$ and $s_{m}$ where $m=1, 2$ and $3$) after fixing the center position, PA
and INCL with those derived from the tilted-ring analysis.  
If we allow {\sc reswri} to fit a velocity field with center position, PA
and INCL as free parameters, non-circular motions tend to be absorbed into
variations in these geometrical parameters, underestimating the amount of
non-circular motions. 
As an example, streaming non-circular motions in a barred galaxy are mainly
responsible for the radial motions that are typically reflected in the $s_{1}$ and
$s_{3}$ terms (\citeauthor{Schoenmakers_1997} \citeyear{Schoenmakers_1997};
\citeauthor{Wong_2004} \citeyear{Wong_2004}; \citeauthor{Spekkens_2007}
\citeyear{Spekkens_2007}). However, these radial motions can  also be modelled by a radial
variation of PA, without the need for $s_{1}$ and $s_{3}$ terms. Similarly, other ring parameters
can affect the harmonic analysis in the same way if they are kept as free
parameters in the fit. To quantify and describe non-circular motions, we
calculate the absolute amplitudes $\langle A \rangle$ and the phases of each component
decomposed. For the amplitudes, we take the median of
$A_m(R)$ as described in \cite{Schoenmakers_1997} (see also
\citeauthor{Trachternach_2008} \citeyear{Trachternach_2008}),

for $m=1$, 
\begin{equation}
\label{eq:1}
A_1(R)=\sqrt{s_1(R)^2},
\end{equation}

for $m>1$,
\begin{equation}
\label{eq:2}
A_m(R)=\sqrt{c_m(R)^2+s_m(R)^2},
\end{equation}
where $R$ is the galaxy radius.

As shown in the section labelled `{\it Harmonic Analysis}' in the Appendix, the amplitudes of
harmonic terms (e.g., $c_{2}$, $s_{1}$ and $s_{2}$ which are corrected for
inclination) derived from the bulk velocity fields (black dots in the Harmonic
Analysis panels) are lower than those derived using the IWM velocity fields over
all radii. This shows that the effect of random non-circular motions are largely
reduced in the bulk velocity fields. The tilted-ring parameters of the
sample galaxies derived using the bulk velocity fields are given in
Table~\ref{LT_TR_properties}. We note that the kinematic center positions given
in Table~\ref{LT_TR_properties} that are derived from the tilted-ring analysis
are offset from the optical ($V$-band, mostly) central isophot used by
\cite{Hunter_2012}.

\subsubsection{Asymmetric drift correction} \label{ASD}

Pressure support caused by random gas motions in the gaseous disk of a galaxy tends to
lower the rotation velocity, which results in an underestimate of the dynamical
mass of the galaxy. In general the dynamical effect of the pressure support is higher in
the outer region of a galaxy where the gas density is low. In particular, this, so-called
`asymmetric drift' is significant in dwarf galaxies whose maximum rotation
velocities are comparable to the velocity dispersions in the gas disk. The
asymmetric drift correction hence should be made to derive more reliable
rotation curves of galaxies where the dynamical support by random motions to its
gas disk is significant. This is the case of some of our sample galaxies whose
velocity dispersions are comparable to their maximum rotation velocities in the
outer region. For the asymmetric drift correction, we follow the method described
in \citeauthor{Bureau_2002} (\citeyear{Bureau_2002}; see also \citeauthor{Oh_2011a} \citeyear{Oh_2011a}) as
follows:

The asymmetric drift correction $\sigma_{\rm D}$ is given as,
\begin{eqnarray}
\label{eq:3}
\sigma^{2}_{\rm D} &=& -R\sigma^{2}\frac{\partial\rm ln(\rho
\sigma^{2})}{\partial R} \nonumber \\
&=& -R\sigma^{2}\frac{\partial\rm ln(\Sigma_{\rm HI} \sigma^{2})}{\partial R},
\end{eqnarray}
\noindent where $\sigma_{\rm D}$ is the asymmetric drift correction, $R$ is the
galaxy radius, $\sigma$ is the velocity dispersion, and $\rho$ is the volume
density of gas disk. In general, $\rho$ can be approximated as the gas surface
density $\Sigma_{\rm HI}$ for a gas disk with an exponential distribution in the vertical
distribution and a constant scale height $z_{0}$ (i.e., $\rm d(ln({\it
z_{0}}))/\rm dr$ = 0). Large fluctuations in the derivative in Eq.~\ref{eq:3}
can be smoothed by fitting an analytical function which has three free parameters,
$I_{0}$ [$\rm M_{\odot}\,{pc}^{-2}\,km^{2}\,s^{-2}$], $R_{0}$ [arcsec], and
$\alpha$ [arcsec$^{-1}$] to the numerator as follows,
\begin{eqnarray}
\label{eq:4}
\Sigma_{\rm HI}\sigma^{2}(R) = \frac{I_{0}(R_{0}+1)}{R_{0}+e^{\alpha R}}.
\end{eqnarray}

Lastly, the corrected rotation velocity $V_{\rm cor}$ is
derived by adding the asymmetric drift correction $\sigma_{\rm D}$ to the
rotation velocity $V_{\rm rot}$ derived from tilted-ring fits,
quadratically,
\begin{eqnarray}
\label{eq:5}
V^{2}_{\rm cor} = V^{2}_{\rm rot} + \sigma^{2}_{\rm D}.
\end{eqnarray}

The analytical function given in Eq~\ref{eq:4} provides a good fit to most sample
galaxies except for DDO 52, IC 10, NGC 3738 and UGC 8508 where a significant
degree of radial fluctuation is present in $\Sigma_{\rm HI}$$\sigma^{2}$ at small radii.
However, the correction at small radii is insignificant and thus will not affect
significantly the final results. 

\cite{2007ApJ...657..773V} have shown that the standard asymmetric correction
can be underestimated without considering the gas pressure gradients triggered
by star formation and feedback. However, the effect of additional thermal pressure gradients in the gas is most likely
insignificant for the sample galaxies in this study. The thermal pressure is
already included in the observed velocity dispersion used for the asymmetric
drift correction as part of its broadening. In addition, as shown in the
velocity dispersion map (i.e., moment 2) in the Appendix, significant anisotropy
in the dispersion is not found in the galaxies. This implies that any separation
of turbulent and thermal pressures, with explicit calculation of asymmetries in
the turbulent component, would have insignificant effect on the asymmetric drift
corrected rotation curves. The asymmetric drift corrected rotation curves
of the sample galaxies which are used for the mass modelling are presented in
the Appendix.

\begin{figure*}
\epsscale{1.0}
\includegraphics[angle=0,width=1.0\textwidth,bb=40 180 540 660,clip=]{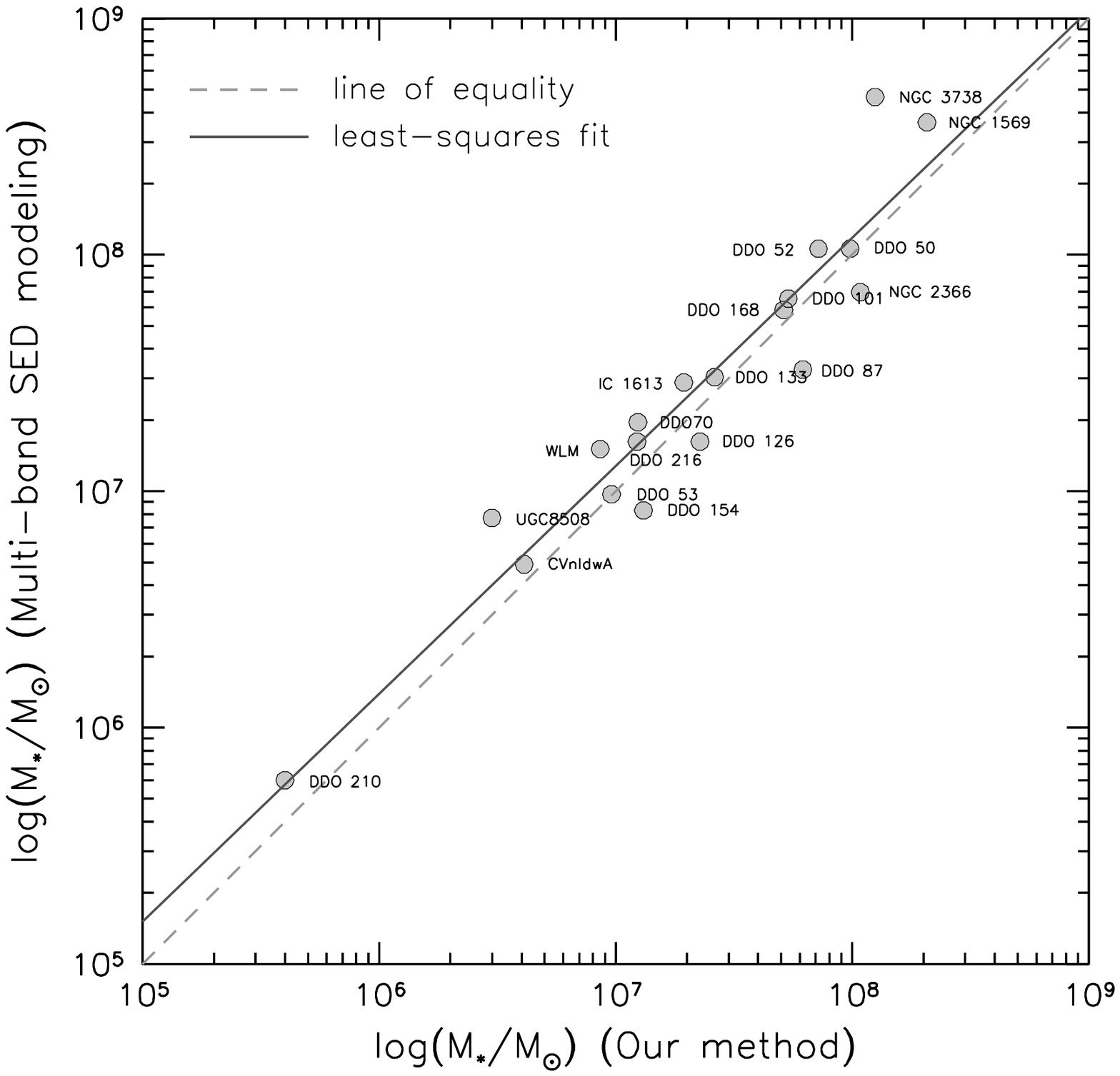}
\caption{
Comparison of the stellar masses derived using our method
with those derived using the multi-band spectral energy distribution (SED) fitting
technique described in \cite{Zhang_2012}. The solid and dashed lines indicate a
least-squares fit with a slope of 0.96, and line of
equality, respectively. The 1$\sigma$ scatter of the fit to the data is 0.20 dex. 
\label{Figure1}}
\end{figure*}

\subsection{Gas distribution} \label{GAS_DISTRIBUTION}
The rotation curves derived in Section~\ref{ROTATION_CURVE} already provide a good approximation
of the DM halos' kinematics of the sample galaxies given the dominant
contribution of the DM halo to the total kinematics of dwarf galaxies.
However, to derive more accurate DM distributions of the galaxies, we
construct mass models of their gaseous and stellar components which account
for most of the baryons in dwarf galaxies. 

We use total integrated H{\sc i} intensity maps (moment 0) of the galaxies to
derive the mass model of the gaseous component. For consistency with the
rotation curves in Section~\ref{ROTATION_CURVE}, we apply the derived tilted-ring models to the
H{\sc i} intensity maps, and obtain gas surface density profiles of the galaxies which
are scaled up by a factor of 1.4 to take Helium and metals into account. We then
convert the gas surface density profiles to the corresponding gas rotation
velocities assuming that gas components are mainly distributed in a thin disk. As an
example, the derived gas surface density profile and the corresponding rotation
velocity of CVnIdwA are shown in the panels (g) and (h) of Fig. A.3,
respectively.
Here, we do not correct for the effect of molecular hydrogen ($\rm H_{2}$) since low
metallicities in dwarf galaxies can induce only a small fraction of the gaseous
component in the form of $\rm H_{2}$ (e.g., \citeauthor{Leroy_2007}
\citeyear{Leroy_2007}; \citeauthor{Leroy_2011} \citeyear{Leroy_2011};
\citeauthor{2012AJ....143..138S} \citeyear{2012AJ....143..138S}).

\subsection{Stellar distribution} \label{STELLAR_DISTRIBUTION}

We use {\it Spitzer IRAC} 3.6$\mu$m images to derive mass models of the stellar
components of the galaxies. Compared with optical images, the {\it Spitzer IRAC}
3.6$\mu$m image is less affected by dust and less sensitive to young stellar populations
which usually emit most energy in the optical regime but occupy only a small
fraction of the total stellar mass. Instead, the {\it Spitzer IRAC} 3.6$\mu$m image is
useful for tracing old stellar populations that are dominant in late-type dwarf
galaxies. This enables us to derive a robust estimate of the stellar mass of
our sample galaxies as used for deriving the
mass models of the stellar components of THINGS galaxies
(\citeauthor{deBlok_2008} \citeyear{deBlok_2008}; \citeauthor{Trachternach_2008}
\citeyear{Trachternach_2008}; \citeauthor{Oh_2008} \citeyear{Oh_2008},
\citeyear{Oh_2011a}).

Like we did for the gas component, we derive 3.6$\mu$m surface brightness profiles
of the stellar components of the galaxies by applying the derived tilted-ring
parameters to the {\it Spitzer IRAC} 3.6$\mu$m images as shown in the figures in the Appendix (e.g.,
Fig. A.3). In general, as discussed in \cite{Walter_2008}, the {\it Spitzer
IRAC} 3.6$\mu$m image provides a pseudo dust free picture of old stellar populations in
galaxies. However, unlike the case of the gas component whose mass can be
directly estimated from H{\sc i} observations, estimating the stellar mass in
galaxies is critically dependent on the assumed stellar mass-to-light ratio (\ML)
value which usually gives rise to the largest uncertainty when converting the
luminosity profile to the mass density profile. In order to derive more reliable
\ML\ values in the 3.6$\mu$m band, we use an empirical relation between galaxy optical
colors and \MLsps\ values based on stellar population synthesis models
(\citeauthor{Bruzual_Charlot_2003} \citeyear{Bruzual_Charlot_2003};
\citeauthor{Bell_2001} \citeyear{Bell_2001}) as given in \cite{Oh_2008}. Hereafter, we call these as model \MLsps\ values.
Using the derived \MLsps\ values, we convert the 3.6$\mu$m surface brightness profiles of our
sample galaxies to stellar surface density profiles (see the figures in the
Appendix, e.g., Fig. A.3).

As discussed in \cite{Oh_2011b}, the kinematic method combined with the model
\MLsps\ values has been found to be reliable for estimating stellar masses of
late-type dwarf galaxies based on a comparison of the derived stellar masses of the
simulated dwarf galaxies to the input ones. This robustly supports the veracity
of the methodology used for measuring the stellar masses of our sample galaxies.
In addition, we also refer to the stellar masses of the sample galaxies derived
using a spectral energy distribution (SED) fitting technique (\citeauthor{Zhang_2012}
\citeyear{Zhang_2012}). As shown in Fig.~\ref{Figure1}, the stellar masses
derived using these two independent methods show good agreement within the scatter. 

Following \cite{Oh_2011a}, we calculate the corresponding rotation velocities
of the stellar components of the sample galaxies from the derived surface
density profiles. For this, we assume a vertical $\rm {sech}^{2}({\it z})$ scale height
distribution of stars with a ratio of $h/z_{0}$=$5$ where $h$ and $z_{0}$ are the radial
scale length and the vertical scale height of stellar disk in the 3.6$\mu$m surface
brightness profiles, respectively (\citeauthor{van_der_Kruit_1981}
\citeyear{van_der_Kruit_1981}; \citeauthor{Kregel_2002} \citeyear{Kregel_2002}).
Although it may overestimate the rotation velocities of the stellar
components of galaxies with a fatter stellar disk, this is a valid assumption
for most disk-dominated dwarf galaxies like our sample galaxies. The
derived rotation velocities of the stellar components of the sample galaxies are
shown in the figures in the Appendix (e.g., panel (d) of Fig. A.3), and the
stellar masses estimated in this paper are given in Table~\ref{MD_results_LT}.

The mean ratio of the masses between the gas and stellar components of the
sample galaxies, $\rm <M_{gas}/M_{star}^{KIN}>$ is $\sim 5.6$ which is consistent
with $\rm <M_{gas}/M_{star}^{SED}> \sim7.2$ (where $\rm M_{star}^{SED}$ is derived
using a SED fitting technique) of 34 LITTLE THINGS sample galaxies in \cite{Zhang_2012}. This indicates that the
majority of the baryons in our sample galaxies is in the form of gaseous
components. As discussed earlier, given that the mass of gaseous components in
galaxies can be reliably measured from H{\sc i} observations without any critical
assumption, the mass models of the baryons derived in this study are likely to
provide a good description of the distribution of baryons in the
galaxies. Therefore, any remaining uncertainties in the mass models of baryons
are most likely to be insignificant and thus will not affect significantly the final mass
models of the DM halos of our sample galaxies.

\subsection{Disk\--halo decomposition} \label{DISK_HALO_DECOMPOSITION}
In this Section, we decompose the total kinematics of the sample galaxies into
the dynamical contributions of the baryonic disks and DM halos by
disentangling the mass models of baryons from the total rotation curves. For
this, we subtract in quadrature the rotation velocities of the gas and stellar
components from the total rotation velocities, and obtain implied rotation
curves for the DM halos as shown in the figures in the Appendix
(e.g., see the left-lower panel of Fig. A.3). For a quantitative analysis of the
DM distribution in the galaxies, we fit CDM (Navarro, Frenk \&
White \citeyear{NFW_1996}, \citeyear{NFW_1997}; hereafter NFW) and
spherical pseudo-isothermal halo models (e.g., \citeauthor{Begeman_1991}
\citeyear{Begeman_1991}), the two representative cusp- and core-like halo
models, to the DM halo rotation curve, respectively. 
An Einasto profile which is a Sersic function in the context of CDM halos has
been found to provide an equal or better fit to the halos in pure DM simulations
compared to an NFW profile \cite{2004MNRAS.349.1039N}. In this
work, we use NFW profiles to quantify the DM halos of the sample galaxies for
consistency with the previous DM mass modelling of THINGS and simulated dwarf
galaxies in \citeauthor{Oh_2011a} (\citeyear{Oh_2011b}, \citeyear{Oh_2011a}).
A relative comparison of the two halo models in terms of fit quality enables us
to examine which model best describes the DM component of the sample galaxies,
especially towards the centers of the galaxies.

\subsubsection{Cusp-like halo model} \label{CUSP_MODEL}
The cosmologically motivated NFW halo model, the so-called `universal density
profile' which describes the cusp-like radial DM distribution found in
DM-only \LCDM\ simulations is given as, 
\begin{equation}
\label{eq:6}
\rho_{\rm{NFW}}(R) = \frac{\rho_{i}}{(R/R_{s})(1+R/R_{s})^{2}},
\end{equation}
where $\rho_{i}$ is correlated with the mean density of the Universe at the time
of the collapse of the halo and $R_{s}$ is the characteristic radius of the DM
halo \citep{NFW_1996}. This profile has been widely adopted to account for the DM distribution
which steeply increases towards the centers of the halos in the simulations.
This, a so-called cusp feature, can be well approximated by a power law,
$\rho$$\sim$$R^{\alpha}$ with $\alpha$$\sim$$-1.0$ near the central region of
the halos, giving a DM halo
rotation velocity as follows,
\begin{equation}
\label{eq:7}
V_{\rm{NFW}}(R) =
V_{200}\sqrt{\frac{{\rm{ln}}(1+cx)-cx/(1+cx)}{x[{\rm{ln}}(1+c)-c/(1+c)]}},
\end{equation}
where $c$ is the concentration parameter defined as $R_{200}/R_{s}$. $V_{200}$ is the
rotation velocity at a radius $R_{200}$ where the mass density contrast with the critical density of the Universe exceeds 200,
and $x$ is defined as $R/R_{200}$ (\citeauthor{NFW_1996} \citeyear{NFW_1996}). In particular, the concentration
parameter c is useful for quantifying the degree of DM concentration in
galaxies. The measurement of the value of c in nearby galaxies provides an
observational constraint on the central cusps predicted from \LCDM\ simulations
(\citeauthor{McGaugh_2007} \citeyear{McGaugh_2007}; see also
\citeauthor{deBlok_2003} \citeyear{deBlok_2003}).

\subsubsection{Core-like halo model} \label{CORE_MODEL}
As an alternative way to describe the DM distribution in a galaxy observationally motivated,
spherical pseudo-isothermal halo models with a central constant-density core have been
used in studies of galaxy rotation curves. The
form of this core-like halo model is given as follows:
\begin{equation}
\label{eq:8}
\rho_{\rm{ISO}}(R) = \frac{\rho_{0}}{1+(R/R_C)^{2}},
\end{equation}
where $\rho_{0}$ and $R_C$ are the core-density and core-radius of a halo, respectively.
This halo model is employed to describe the mass distribution of a DM 
halo with a sizeable constant density-core ($\rho$$\propto$$\rho_{0}$). 
Similarly, the corresponding rotation velocity to the pseudo-isothermal halo
potential is given by,
\begin{equation}
\label{eq:9}
V_{\rm{ISO}}(R) = \sqrt{4\pi
G\rho_{0}R_C^{2}\Biggl[1-\frac{R_C}{R}{\rm{atan}}\Biggl(\frac{R}{R_C}\Biggr)\Biggr]}.
\end{equation}

In order to quantify the DM distribution in the sample galaxies, we fit
the model rotation velocities of the two halo models as given in Eqs.~\ref{eq:7}
and \ref{eq:9} to the kinematic DM signature of the galaxies. As shown in the figures in
the Appendix (e.g., the left-lower panel of Fig.~A.3), the core-like
pseudo-isothermal halo models are mostly preferred over the cusp-like NFW models
in terms of the quality of the fit (i.e., based on reduced $\chi^{2}$ values).
The fitted parameters of the halo models are presented in
Table~\ref{MD_results_LT}.

As given in Table~\ref{MD_results_LT},
we derive the dynamical masses ($M_{\rm dyn}$) of the sample galaxies using
$V_{\rm ISO}(R_{\rm max})$ and $R_{\rm max}$ at which the outermost part of the
rotation curve is measured. Given that rotation curves at the adopted
$R_{\rm max}$ are mostly rising in our sample galaxies, the derived dynamical
mass, $M_{\rm dyn}$ with $V_{\rm ISO}(R_{\rm max})$ and $R_{\rm max}$ gives a lower limit of
the halo mass. For reference, we also estimate the halo masses, $M_{200}$ of our
sample galaxies assuming an NFW halo model. However, the fitted values of $c$
are unphysical (i.e., negative or close to zero) for a large fraction (11/26
galaxies) of the sample galaxies. We therefore derive a $c$ value using an empirical relationship
between $c - V_{200}$ from the WMAP\footnote{The Wilkinson Microwave Anisotropy Probe
(\citeauthor{Spergel_2003} \citeyear{Spergel_2003};
\citeauthor{Spergel_2007} \citeyear{Spergel_2007})} observations in \cite{McGaugh_2007} by
substituting $V_{200}$ with the $V_{\rm ISO}(R_{\rm max})$ adopted. We then fit the NFW halo
model after fixing $c$ and leaving only $V_{200}$ as a free parameter, assuming a minimum
disk (where the rotation curve is attributed to the DM halo only and
the dynamical contribution of baryons is ignored). These $c$ and
$V_{200}$ values are given in brackets in Table~\ref{MD_results_LT}. Lastly, we derive the
resulting halo mass, $M_{\rm 200}$ with the newly estimated $V_{200}$ using 
Eq. 3 in \cite{Oh_2011b}. As presented in Table~\ref{MD_results_LT}, $M_{\rm 200}$
values of the sample galaxies are larger than $M_{\rm dyn}$ values, which
implies that our observations most likely do not reach the flat part of the
rotation curves. This is consistent with the fact that the rotation curves
of most sample galaxies are still rising at the last measured points.

\begin{figure*}
\epsscale{1.0}
\includegraphics[angle=0,width=1.0\textwidth,bb=30 189 575 715,clip=]{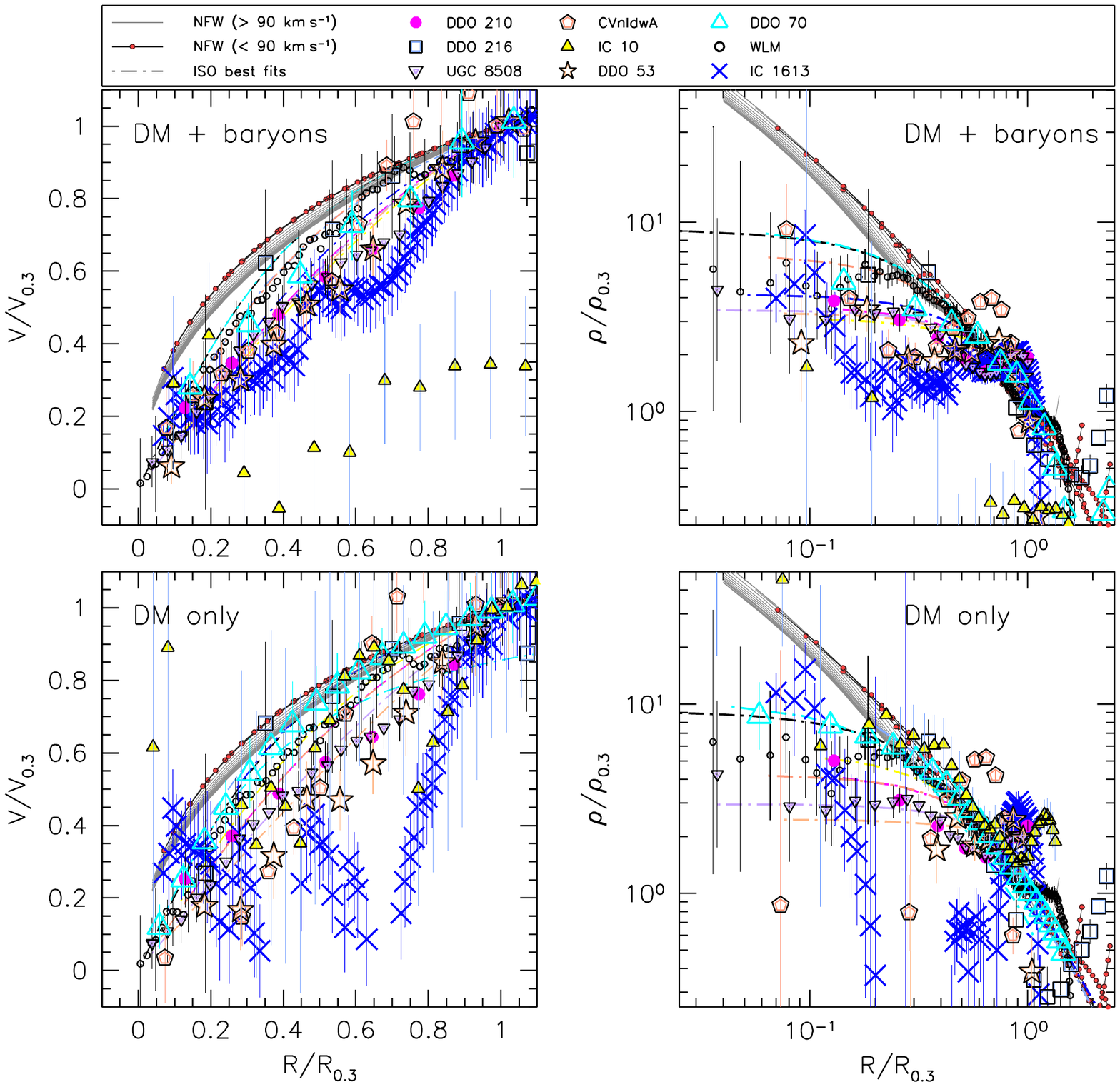}
\caption{{\bf Left panels:} The rotation curves of the first nine galaxies of the 26
LITTLE THINGS (in dynamical mass order) which are all scaled with respect to the
rotation velocity $V_{0.3}$ at $R_{0.3}$ where the logarithmic slope of the
rotation curve is $d{\rm
log}V/d{\rm log}R=0.3$ as described in \cite{Hayashi_2006}. The upper (DM +
baryons) and lower (DM only) panels show the ones including and excluding the
dynamical contribution by baryons, respectively. The grey solid and black solid
lines with small dots indicate the CDM NFW dark matter rotation curves with
$V_{200}$ which is $>$ 90 \kms\, and $<$ 90 \kms,
respectively. The dashed lines (denoted as ISO) show the best fitted
pseudo-isothermal halo models to the galaxies.
{\bf Right panels:} The corresponding dark matter density profiles derived using
the scaled rotation curves in the left panels. The grey ($V_{200}>90$ \kms) and black solid lines
with small dots ($V_{200}<90$ \kms) represent the CDM NFW models with the inner density slope
$\alpha$$\sim$$-1.0$. The dashed lines indicate the best fitted
pseudo-isothermal halo models with $\alpha$$\sim$$0.0$. See 
Section~\ref{DARK_MATTER_DISTRIBUTION} for more details.
\label{Figure2}}
\end{figure*}

Except in those few cases, such as DDO 70, DDO 101, DDO 154, DDO 210, DDO 216, and 
Haro 36 where CDM NFW halo models provide comparable fits to the DM rotation curves, the fitted
values of the NFW halo parameters are unphysical (i.e., negative concentration
parameter $c$, unphysically large values of \VNFW). 
This is consistent with results previously found in other nearby
dwarf and LSB galaxies (e.g., \citeauthor{deBlok_2002}
\citeyear{deBlok_2002}; \citeauthor{Kuzio_de_Naray_2008}
\citeyear{Kuzio_de_Naray_2008}; \citeauthor{2009A&A...505....1V}
\citeyear{2009A&A...505....1V}; \citeauthor{Oh_2011a} \citeyear{Oh_2011a}).
The slowly increasing DM rotation curves in the inner region of the sample galaxies reflect a halo whose
gravitational potential is not deep enough to sustain the power-law DM 
density cusps that are as steep as $\rho$$\propto$$R^{-1.0}$.
Hence, the cosmologically motivated cusp-like halo models
are not able to adequately describe the observed solid-body rotation curves
of our sample dwarf galaxies.

It may be argued that those galaxies that are equally well fitted by the two halo models
indicate the possibility of a kinematic signature of central cusps in dwarf
galaxies being consistent with \LCDM\ simulations. Despite the high-resolution of
LITTLE THINGS VLA H{\sc i} observations, it is, however, most likely that the central
regions of the galaxies are not fully sampled with a sufficient number of
independent synthesized beams needed for distinguishing the inner steepness of
the two halo models, accurately. As quantified in the parameter, $R_{\rm
max}$H{\sc i}$_{\rm beam}^{-1}$ in Table~\ref{MD_results_LT}, this is mainly due
to the small size of the rotating disk (e.g., DDO 210) or the relatively large
distance (e.g., Haro 36). Given that the difference between the cusp- and
core-like halo models is the most prominent in the central regions of galaxies,
higher sampling of the inner regions of the galaxies is required before making a
firmer conclusion on the signature of the potential central cusps. We will
discuss this matter in a more quantitative way in the following Section. 

\begin{figure*}
\epsscale{1.0}
\includegraphics[angle=0,width=1.0\textwidth,bb=30 189 575 715,clip=]{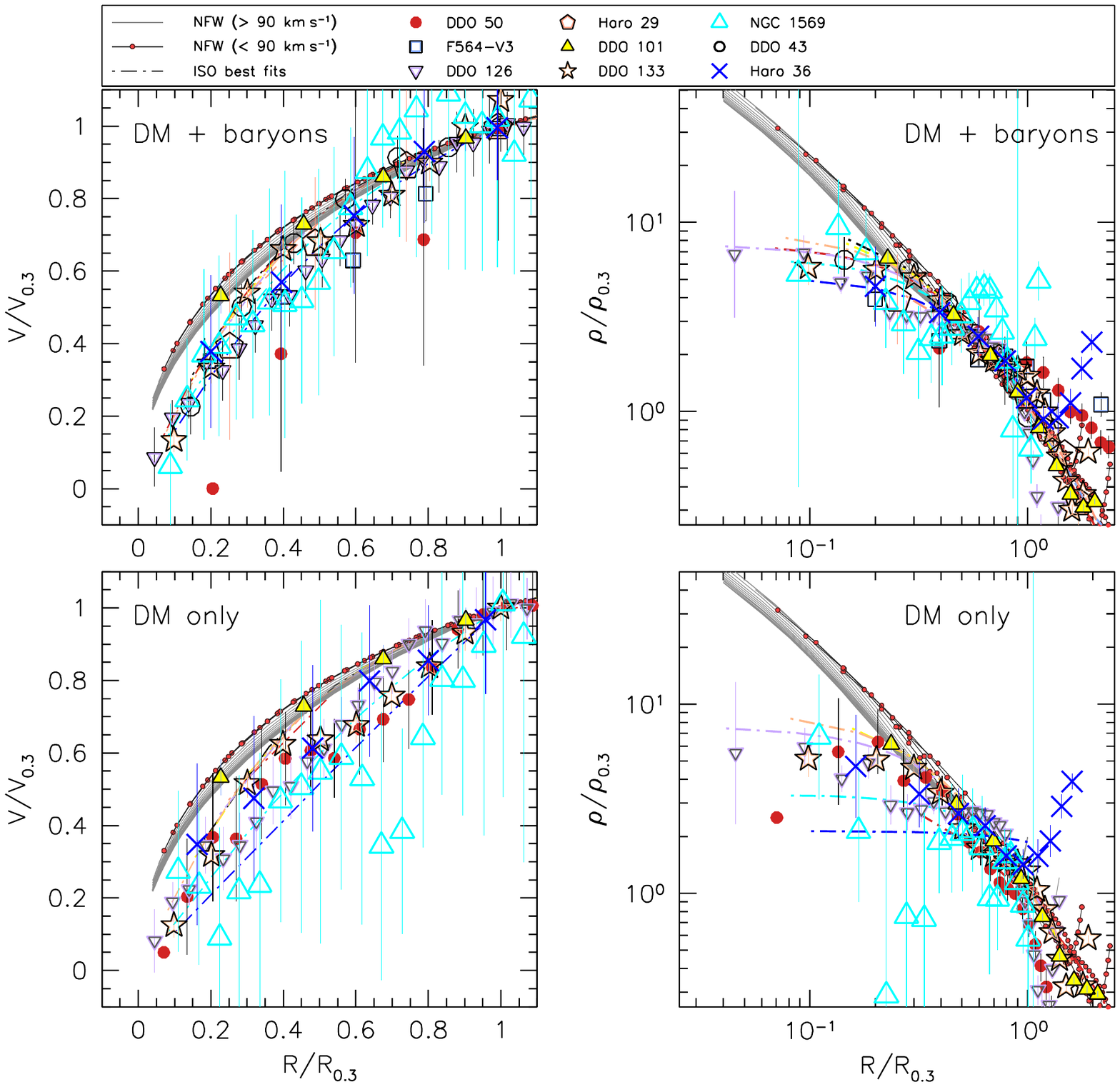}
\caption{{\bf Left panels:} The rotation curves of the other nine galaxies of the 26
LITTLE THINGS (in dynamical mass order) which are all scaled with respect to the
rotation velocity $V_{0.3}$ at $R_{0.3}$ where the logarithmic slope of the
rotation curve is $d{\rm 
log}V/d{\rm log}R=0.3$ as described in \cite{Hayashi_2006}. The upper (DM +
baryons) and lower (DM only) panels show the ones including and excluding the
dynamical contribution by baryons, respectively. The grey solid and black solid
lines with small dots indicate the CDM NFW dark matter rotation curves with
$V_{200}$ which is $>$ 90 \kms\, and $<$ 90 \kms,
respectively. The dashed lines (denoted as ISO) show the best fitted
pseudo-isothermal halo models to the galaxies.
{\bf Right panels:} The corresponding dark matter density profiles derived using
the scaled rotation curves in the left panels. The grey ($V_{200}>90$ \kms) and black solid lines
with small dots ($V_{200}<90$ \kms) represent the CDM NFW models with the inner density slope
$\alpha$$\sim$$-1.0$. The dashed lines indicate the best fitted
pseudo-isothermal halo models with $\alpha$$\sim$$0.0$. See 
Section~\ref{DARK_MATTER_DISTRIBUTION} for more details (continued).
\label{Figure3}}
\end{figure*}

\section{Dark matter distribution} \label{DARK_MATTER_DISTRIBUTION}
In this Section, we compare the inner shape of rotation curves and DM
density profiles of the sample galaxies with those of simulated dwarf galaxies
from N\--body+SPH \LCDM\ simulations in order to examine their DM 
distributions near the centers. These comparisons between observations
and simulations allow us to estimate the degree of cuspiness of the central
DM distribution in a qualitative way. In addition, we also measure the inner
DM density slopes of the galaxies to make a more quantitative
comparison to the simulations. 

\begin{figure*}
\epsscale{1.0}
\includegraphics[angle=0,width=1.0\textwidth,bb=30 189 575 715,clip=]{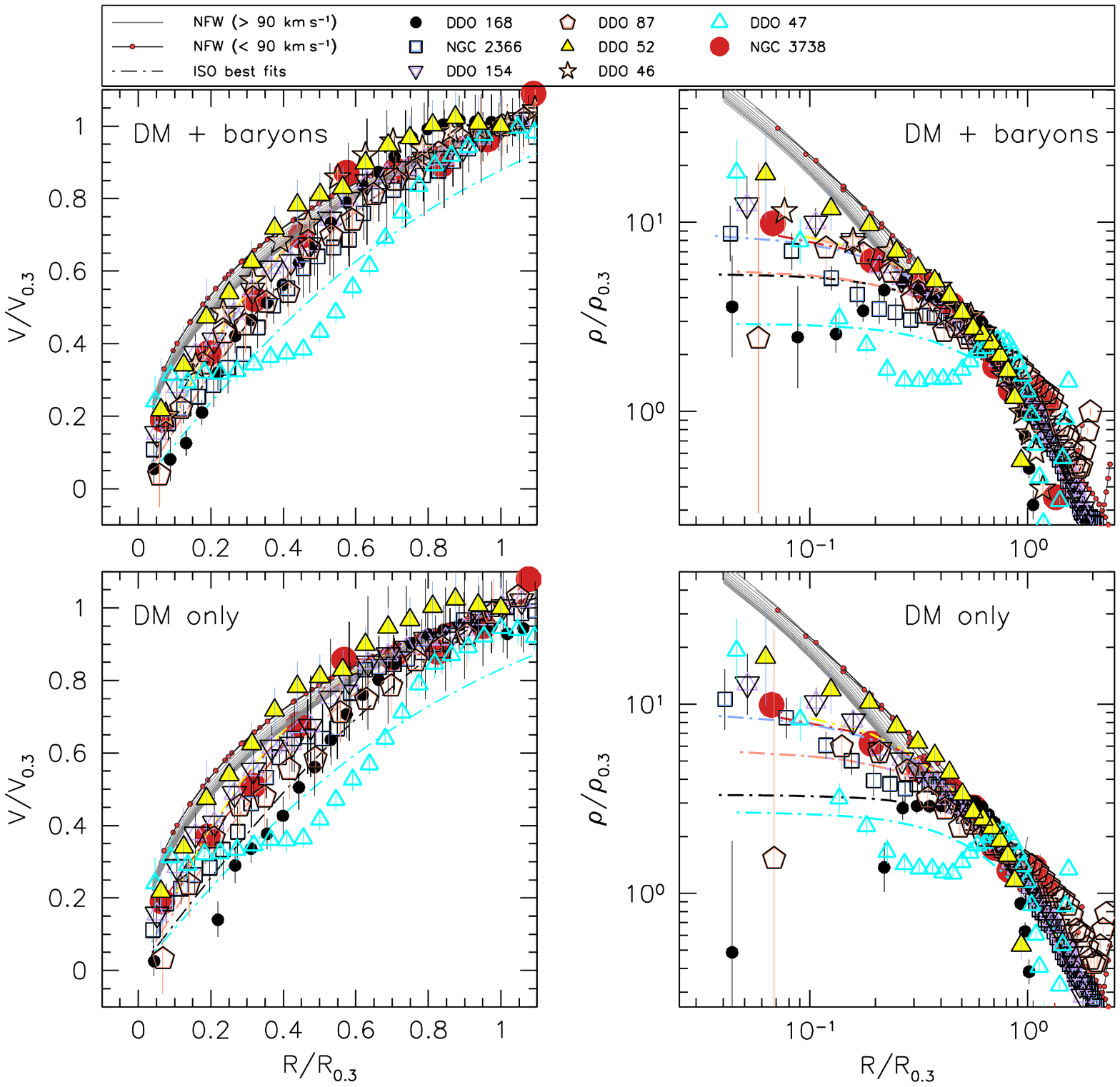}
\caption{{\bf Left panels:} The rotation curves of the remaining eight galaxies of the 26
LITTLE THINGS (in dynamical mass order) which are all scaled with respect to the
rotation velocity $V_{0.3}$ at $R_{0.3}$ where the logarithmic slope of the
rotation curve is $d{\rm
log}V/d{\rm log}R=0.3$ as described in \cite{Hayashi_2006}. The upper (DM +
baryons) and lower (DM only) panels show the ones including and excluding the
dynamical contribution by baryons, respectively. The grey solid and black solid
lines with small dots indicate the CDM NFW dark matter rotation curves with
$V_{200}$ which is $>$ 90 \kms\, and $<$ 90 \kms,
respectively. The dashed lines (denoted as ISO) show the best fitted
pseudo-isothermal halo models to the galaxies.
{\bf Right panels:} The corresponding dark matter density profiles derived using
the scaled rotation curves in the left panels. The grey ($V_{200}>90$ \kms) and black solid lines
with small dots ($V_{200}<90$ \kms) represent the CDM NFW models with the inner density slope
$\alpha$$\sim$$-1.0$. The dashed lines indicate the best fitted
pseudo-isothermal halo models with $\alpha$$\sim$$0.0$. See 
Section~\ref{DARK_MATTER_DISTRIBUTION} for more details (continued).
\label{Figure4}}
\end{figure*}

\subsection{Rotation curve shape} \label{RC_SHAPE}
The characteristic shape of the steeply rising rotation curve inherent in the
cusp-like DM distribution near the centers of simulated dwarf galaxies
based on the \LCDM\ paradigm can be used for a qualitative test of the
simulations (\citeauthor{Hayashi_2006} \citeyear{Hayashi_2006}; see also \citeauthor{Oh_2011a}
\citeyear{Oh_2011a}, \citeyear{Oh_2011b}). This
qualitative test is particularly useful in that a direct comparison between the
observed and predicted DM rotation curves can be made without any
additional assumption on the shape of the DM halo (e.g., a spherical or triaxial halo
potential) which is needed for converting rotation curves to the corresponding density
profiles, and the associated additional uncertainties this might introduce.

This is done in the left panel of Figs.~\ref{Figure2}, \ref{Figure3} and \ref{Figure4} where we scale the rotation curves
of both our sample galaxies and \LCDM\ NFW halos with respect to the rotation
velocity $V_{0.3}$ at a radius $R_{0.3}$ which is where the logarithmic slope of the
curve is $d{\rm log}V/d{\rm log}R=0.3$ (see \citeauthor{Hayashi_2006} \citeyear{Hayashi_2006}).
This enables us not only to make a relative comparison of the rotation
curves between observations and simulations but also to accentuate the inner
rotation curve shape which is sensitive to the degree of central DM 
concentration. 

\begin{figure*}
\epsscale{1.0}
\includegraphics[angle=0,width=1.0\textwidth,bb=30 189 575 715,clip=]{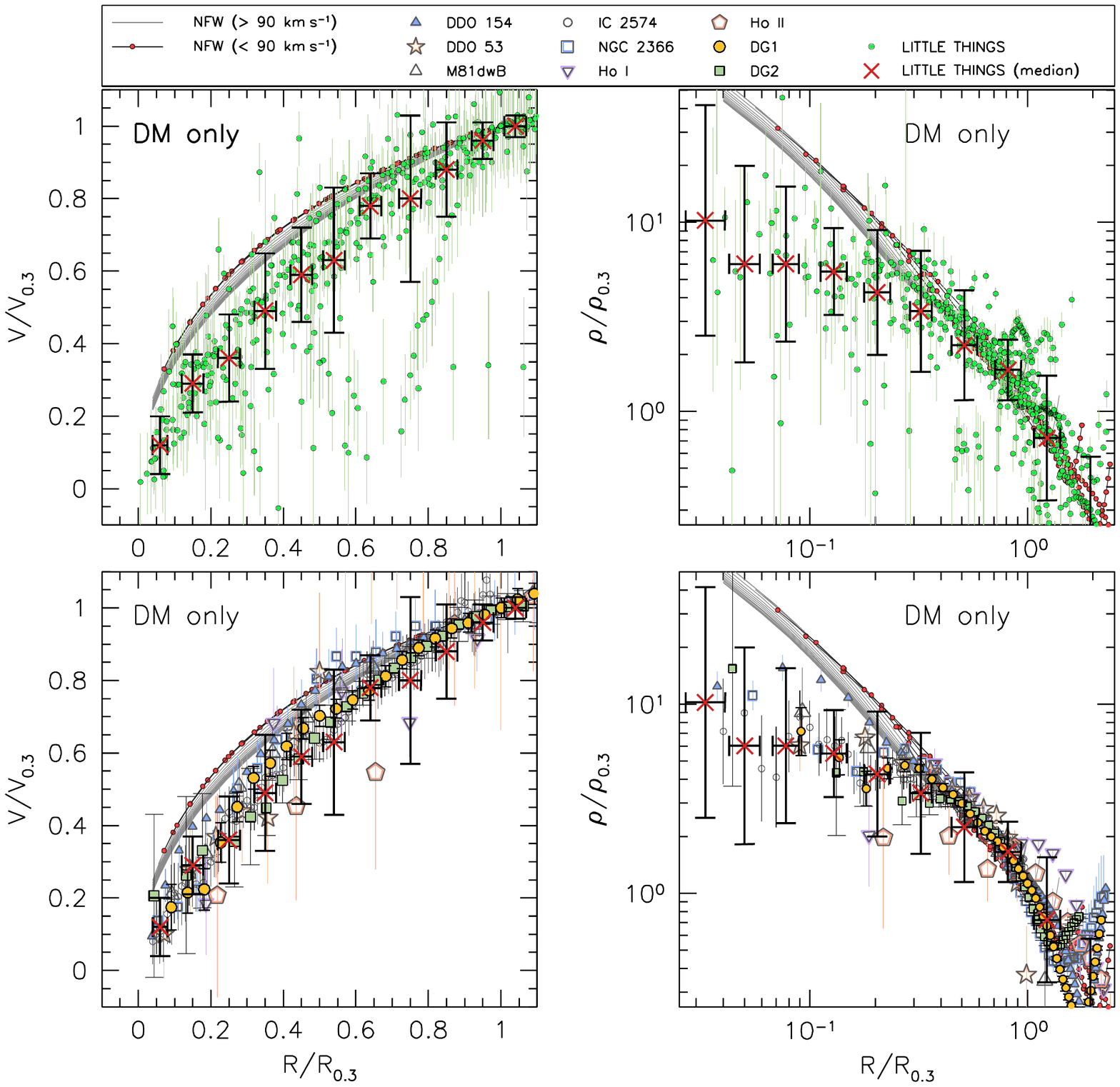}
\caption{{\bf Upper-left panel:} The (DM only) rotation curves (small dots) of the 21 LITTLE
THINGS (including 3 THINGS galaxies) for which {\it Spitzer} 3.6$\mu$m image is
available. These are all scaled with respect to the rotation velocity
$V_{0.3}$ at $R_{0.3}$ where the logarithmic slope of the rotation curve is
$d{\rm log}V/d{\rm log}R=0.3$ as described in \cite{Hayashi_2006}.
The `$\times$' symbol represents the median values of the rotation curves
in each 0.1$\rm R/R_{0.3}$ bin. The error bars show the 1$\sigma$ scatter.
{\bf Lower-left panel:} The scaled rotation curves of the seven THINGS, and the two
simulated dwarf galaxies (DG1 and DG2 in \citeauthor{Governato_2010}
\citeyear{Governato_2010}) which are overplotted to the median values of the
LITTLE THINGS rotation curves.
The grey solid and black solid lines with small dots indicate the CDM NFW dark matter rotation curves with
$V_{200}$ which is $>$ 90 \kms\, and $<$ 90 \kms,
respectively. 
{\bf Right panels:} The corresponding dark matter density profiles derived using
the scaled rotation curves in the left panels. The grey ($V_{200}>90$ \kms) and black solid lines
with small dots ($V_{200}<90$ \kms) represent the CDM NFW models with the inner density slope
$\alpha$$\sim$$-1.0$. 
See Section~\ref{DARK_MATTER_DISTRIBUTION} for more details.
\label{Figure5}}
\end{figure*}

All the scaled (DM only) rotation curves of the 21 LITTLE THINGS sample galaxies
for which {\it Spitzer} 3.6$\mu$m image is available are overplotted in the
upper-left panel of Fig.~\ref{Figure5}. We also overplot the median values of
the rotation curves in each 0.1$\rm R/R_{0.3}$ bin. In the lower-left panel
of Fig.~\ref{Figure5}, we also overplot the scaled rotation 
curves of seven dwarf galaxies from THINGS (three of them are also in LITTLE
THINGS) as well as the two simulated dwarf galaxies presented in
\cite{Governato_2010} which all show a linear increase in their inner regions to
the median values of the LITTLE THINGS rotation curves.
In particular, the two simulated dwarf galaxies
were affected by baryonic feedback processes (mainly repeated gas outflows
driven by SN explosions) in such a way that the central
cusps predicted from DM-only simulations are flattened
(\citeauthor{Governato_2010} \citeyear{Governato_2010}). The flattened DM
distribution results in slowly increasing rotation curves in the inner region of
the simulated galaxies (see \cite{Oh_2011b} for further discussion).

In line with the results in \cite{Oh_2011b}, the
inner shape of the scaled rotation curves of the LITTLE THINGS sample galaxies
falls mostly below that of the DM-only NFW halo models, indicating a
shallower DM distribution near the centers. As shown in the lower-left
panel of Figs.~\ref{Figure2}, \ref{Figure3} and \ref{Figure4}, the discrepancy with NFW models becomes more 
pronounced when comparing these models to DM rotation curves derived after 
subtracting the contribution from baryons from the total
kinematics for each of the sample galaxies. Instead, they are more in line with those
of both the THINGS dwarf galaxies and simulations (DG1 and DG2) where the effect of
baryonic feedback processes is included. The linearly (or less steeply) rising
rotation curves in the inner region of the galaxies indicate a nearly constant
or shallower mass distribution towards the centers as found in the majority of
nearby dwarf galaxies. The LITTLE THINGS sample galaxies give no clear
indication of the central cusps in their DM halos unlike the predictions from
\LCDM\ DM-only simulations. 

\subsection{Dark matter density profiles} \label{DM_PROFILE}
As a more direct way to examine the central DM distribution of the
sample galaxies and compare them with \LCDM\ simulations, we derive their DM
density profiles from the DM rotation curves decomposed in
Section~\ref{DISK_HALO_DECOMPOSITION}. A direct conversion of the rotation curve
to the corresponding DM density profile can be made by the following formula (see
\citeauthor{deBlok_2001} \citeyear{deBlok_2001} for more details),
\begin{equation}
\label{eq:10}
\rho(R) = \frac{1}{4\pi G}\Biggl[2\frac{V}{R}\frac{\partial V}{\partial R} +
\Biggl(\frac{V}{R}\Biggr)^{2}\Biggr],
\end{equation}
where $V$ is a rotation velocity observed at a radius $R$, and $G$ is the
gravitational constant. For this conversion, we assume a spherical halo
potential which is valid for most nearby galaxies
(\citeauthor{Trachternach_2008} \citeyear{Trachternach_2008}). This method has
been used for deriving DM density profiles of
dwarf and LSB disk galaxies, and proved to be reliable
as found in the comparison of the derived DM density profiles of
simulated dwarf galaxies with their input ones (\citeauthor{Oh_2011b}
\citeyear{Oh_2011b}).

In the right panel of Figs.~\ref{Figure2}, \ref{Figure3} and \ref{Figure4}, we present the derived (dark)
matter density profiles of the LITTLE THINGS sample galaxies. In addition, we
calculate the median values of the DM density profiles as shown in the
upper-right panel of Fig.~\ref{Figure5}. We also overplot the DM density profiles of
the THINGS and simulated dwarf galaxies in the lower-right panel of
Fig.~\ref{Figure5}. The radial fluctuation shown in some of the DM density
profiles (e.g., IC 1613 etc.) is largely due to fluctuations in the derived
rotation curves at the relevant radius. These are mainly because of either the
effect of non-circular motions, the noise in velocity profiles with low S/N
values, or both in the region. 

As already implied by the solid-body
like inner rotation curve shape of the LITTLE THINGS sample galaxies in
Section~\ref{RC_SHAPE}, their central DM density profiles are
systematically shallower than those of the cusp-like DM
density profiles predicted from DM-only \LCDM\ simulations. As
shown in the lower-right panel of Figs.~\ref{Figure2}, \ref{Figure3} and \ref{Figure4}, the difference is even more pronounced in the
comparison of the DM density profiles corrected for the baryons
although the dynamical contribution by baryons is rather insignificant. The sample dIrr 
galaxies are more consistent with the core-like DM density profiles (dot-dashed
lines in the right-hand frames of Figs.~\ref{Figure2}, \ref{Figure3} and \ref{Figure4}) than
NFW--type profiles. This is much like the THINGS dwarf galaxies, and the simulated dwarfs (DG1 and DG2) with baryonic
feedback processes as shown in the lower-right panel of Fig.~\ref{Figure5}.

We also measure the inner density slopes $\alpha$ of the DM density
profiles to quantify the cuspiness of the central DM distribution. This
yields a more quantitative comparison between the observations and simulations.
As shown in the figures in the Appendix (e.g., panel (f) of Fig. A.3), we
perform a least squares fit (dotted lines) to the inner
data points (grey dots) within a `break radius'. As described in
\citeauthor{deBlok_2002} (\citeyear{deBlok_2002}; see also \citeauthor{Oh_2011a} \citeyear{Oh_2011a}), we
determine a break radius of a DM density profile where the slope changes most
rapidly in the inner region of the profile. Following
\cite{deBlok_2002}, we adopt the mean difference between the slopes which are
measured including the first data point outside the break-radius and excluding
the data point at the break radius, respectively, as an errorbar $\Delta\alpha$
of the inner density slope. We measure the inner density slopes $\alpha$, of the galaxies from their
total matter (including both DM halo and baryons) as well as DM-only density
profiles. The former, a so-called `minimum disk assumption' that attributes the total
rotation curve to the DM component only, gives a steeper inner density slope.
Meanwhile, the latter where the dynamical
contribution by baryons is subtracted from the total rotation curve allows us to
examine the effect of the model \MLsps\ on the measured inner density slope. The
mean values of the slopes of the 26 LITTLE THINGS dwarf galaxies are
$\alpha_{\rm min}$$=$$-0.42\pm 0.21$ and $\alpha_{\ML.pdf}$$=$$-0.32\pm
0.24$\footnote{The dynamical contribution by the stellar component is included
in the DM density profiles of five galaxies where no {\it Spitzer} 3.6$\mu$m
image is available.} for the minimum disk and the model \MLsps\ disk assumptions, respectively.

As expected, the slopes measured assuming the minimum
disk are slightly steeper than those measured using the model \MLsps\ disk for most
sample galaxies. However, the difference between the two slopes is largely
insignificant since most of the sample galaxies are DM dominated as
indicated by their low baryonic fraction. The measured logarithmic inner density
slopes $\alpha$ of our sample galaxies are listed in Table~\ref{MD_results_LT}.

The LITTLE THINGS sample dwarf galaxies do not in general agree with the steep
logarithmic inner slope ($\sim$$1.0$) of the DM density profiles predicted
from \LCDM\ DM-only simulations. Instead, they show a range of shallower
slopes being consistent with a core-like DM distribution at the centers,
which supports the previous results found in nearby dwarf and
LSB disk galaxies.
Our results ($\alpha_{\rm min}$$=$$-0.42\pm 0.21$; $\alpha_{\rm
\ML.pdf}$$=$$-0.32\pm 0.24$) are consistent with the mean logarithmic slope,
$\alpha_{\rm min}$$=$$-0.29\pm 0.07$, of the seven THINGS dwarf galaxies derived
assuming a minimum disk in \cite{Oh_2011a}. Moreover, if we combine the sample
dwarf galaxies from LITTLE THINGS and THINGS which have a similar data quality
and whose inner density slopes are derived in exactly the same way, the mean
value of the slopes of the 29 dwarf galaxies is $\alpha_{\rm min}$$=$$-0.40\pm
0.24$. This shows good agreement within the error bars with $\alpha_{\rm
min}$$=$$-0.2\pm 0.2$ derived from LSB galaxies in \cite{deBlok_2001} but a
clear deviation from the $\alpha$$\sim$$-1.0$ predicted from \LCDM\ DM-only
simulations.

A few galaxies in our sample, such as DDO 101 and DDO 210 whose
rotation curves are equally fitted by both NFW and pseudo-isothermal halo models
appear to have relatively steeper slopes compared to the other ones. 
However, as discussed in Section~\ref{DISK_HALO_DECOMPOSITION}, this could be
due to insufficient sampling of the dark matter density profiles in the
inner region. The gradient of the logarithmic density slope, $d{\rm log}\rho/d{\rm
log}R$ gradually decreases towards the outer region of a galaxy, giving a steeper
slope $\alpha$. Therefore, the steeper a slope $\alpha$ is, the more data points in the outer
regions are included when measuring the logarithmic slope of a DM 
density profile. This could, conceivably, be the case for galaxies with
insufficient spatial sampling. As discussed above, the insufficient sampling
mainly arises from either the smaller size of the H{\sc i} disk (e.g., DDO 210) or the larger
distances of the galaxies (e.g., Haro 36).

\begin{figure*}
\epsscale{1.0}
\includegraphics[angle=0,width=1.0\textwidth,bb=10 150 580 690,clip=]{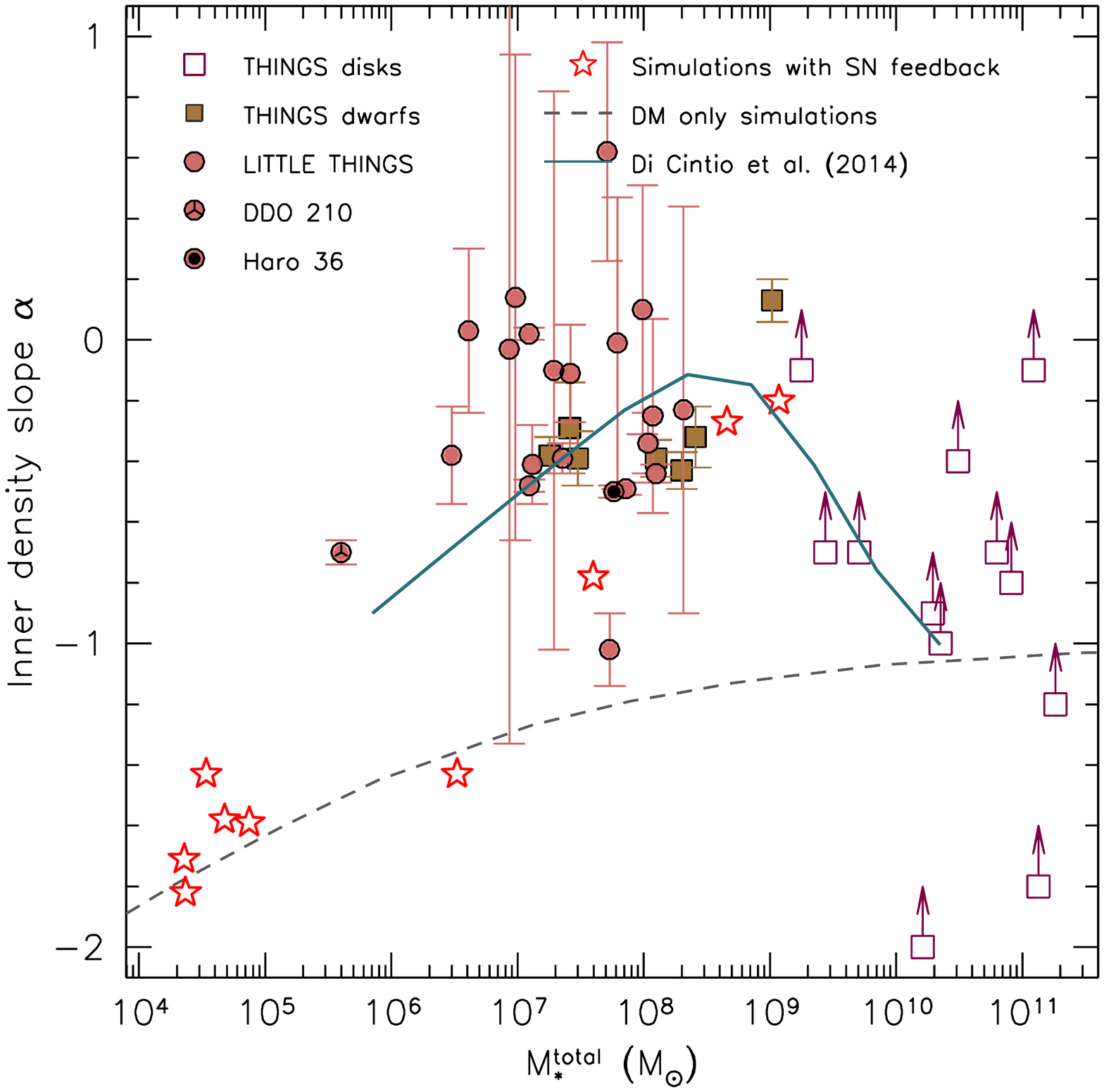}
\caption{The inner dark matter density slope $\alpha$ of the sample galaxies
from LITTLE THINGS (filled circles) and THINGS (filled squares, THINGS dwarfs;
open squares, THINGS disk galaxies) against their total stellar masses. The dashed
line indicates the $\alpha$\--$\rm M_{*}^{total}$ prediction from \LCDM\ dark
matter only simulations. The open stars and solid line represent the $\alpha$ vs. $\rm
M_{*}^{total}$ of the resolved halos from the \LCDM\ SPH simulations with
baryonic feedback processes which are measured at 500 pc and z=0
(\citeauthor{Governato_2012} \citeyear{Governato_2012};
\citeauthor{2014MNRAS.437..415D} \citeyear{2014MNRAS.437..415D}). 
See Section~\ref{SN_FEEDBACK} for more discussions.
\label{Figure6}}
\end{figure*}

\section{Effect of SN feedback on the central cu.pdf} \label{SN_FEEDBACK}

As discussed in \cite{Governato_2010}, the constant-density cores
observed near the centers of dwarf galaxies can be reconciled with simulations 
by taking the effect of baryonic feedback processes into account without a need 
for any explicit modification of the current \LCDM\ paradigm. In
particular, repeated gas outflows driven by SNe have been found to be efficient
enough to redistribute the matter in dwarf galaxies, resulting in shallower DM
density profiles as observed in nearby dwarf galaxies
(\citeauthor{Oh_2011b} \citeyear{Oh_2011b}).

The investigation of the effect of SN-driven gas outflows on the central DM
distribution has been extended to low mass field dwarf galaxies
using  GASOLINE (\citeauthor{Wadsley_2004} \citeyear{Wadsley_2004}), a
parallel SPH tree-code with multistepping (\citeauthor{Governato_2012}
\citeyear{Governato_2012}; \citeauthor{Pontzen_2012}
\citeyear{Pontzen_2012}; \citeauthor{2014MNRAS.437..415D}
\citeyear{2014MNRAS.437..415D}; see also \citeauthor{Pontzen_2014}
\citeyear{Pontzen_2014}). In the simulations,
the present-day stellar mass of galaxies ranges from $10^{9.8}$ down to
$10^{4.5}$ \Msun\ where the energy transfer from repeated gas outflows to the DM
component becomes inefficient. More specifically, the SN feedback in small halos
where less than 0.03\% of the total amount of baryons is converted into stars is
less effective at removing the central cusps and turning the cusp-like DM
density profiles into core-like ones. According to this scenario, it is expected that the central DM
distribution in these systems remains cuspy, and the inner slopes of their DM
density profiles are steeper than those of higher mass counterparts
(\citeauthor{Governato_2012} \citeyear{Governato_2012};
\citeauthor{Zolotov_2012} \citeyear{Zolotov_2012}). It implies that the central
cusps predicted from \LCDM\ DM-only simulations should survive in low mass halos.
This underlines the cosmological importance of low mass dwarf galaxies in the
local Universe for testing the `cusp/core' problem.

As discussed earlier, the systematic uncertainties caused by low resolution
radio observations are significantly reduced in the high-resolution H{\sc i} data
from LITTLE THINGS, which allows us to derive more accurate rotation curves of
the sample galaxies and thus their central DM distributions. In
addition, the {\it Spitzer IRAC} 3.6$\mu$m data combined with model \MLsps\ values based on
stellar population synthesis models and galaxy colors provide more reliable
stellar masses of the sample galaxies. 

In Fig.~\ref{Figure6}, we plot the inner density slopes $\alpha$ of the sample dwarf galaxies
from both LITTLE THINGS and THINGS against their stellar masses $\rm M_{*}$ on a
logarithmic scale. In addition, we also add those of a sample of THINGS disk
galaxies whose mass models were derived in \cite{deBlok_2008} in order to
examine the $\alpha \-- \rm M_{*}$ relationship in the higher mass regime. Compared to dwarf
galaxies, the central kinematics of disk galaxies are usually dominated by
a bulge component. It is therefore not trivial to perform a reliable disk-halo
decomposition of the disk galaxies despite using the multi-wavelength data from
THINGS whose data quality is comparable to that of LITTLE THINGS. For this
reason, in Fig.~\ref{Figure6}, we use the inner density slopes of
the disk galaxies measured assuming a minimum disk. As discussed in
Section~\ref{DM_PROFILE}, it should be noted that the minimum disk assumption attributes the total
kinematics of a galaxy to the DM only, ignoring the contribution of baryonic
components, giving a lower limit on the inner density slope (i.e., a steeper slope). 

As shown in Fig.~\ref{Figure6}, the \LCDM\ SPH simulations including the effect
of baryonic feedback processes predict cusp-like DM distributions with steeper inner density
slopes ($\alpha < -1.0$) in DM halos whose stellar mass is less
than about $10^6$ \Msun. As discussed in \cite{Governato_2012}, the lower the
mass of the stellar component in a galaxy, the less the dynamical effect of SN feedback on
the DM potential. Consequently, this results in that the central DM
distribution in the low mass halo regime remains cuspy. According to this,
the initial cusps formed in the early Universe would still exist today in low mass dwarf
galaxies which have stellar masses less than $\sim10^6$ \Msun\ where the repeated thermal
energy injection from SN explosions becomes substantially inefficient, mainly
due to the rapidly decreasing star formation efficiency in these systems
(\citeauthor{Governato_2012} \citeyear{Governato_2012};
\citeauthor{Pontzen_2012} \citeyear{Pontzen_2012}).

However, in Fig.~\ref{Figure6}, as the stellar mass of a galaxy increases, the inner DM
density slope $\alpha$ becomes shallower in the \LCDM\ SPH simulations
including baryonic feedback processes (open stars) with respect to the counterpart in the
DM-only simulations (dashed line). As discussed in \cite{Governato_2010} and
\cite{Pontzen_2012} (see also \citeauthor{2014MNRAS.437..415D}
\citeyear{2014MNRAS.437..415D}), the central DM cusps can be disrupted by the rapid gas
injection caused by SN-driven gas outflows into the
central region of galaxies, resulting in a shallower DM density distribution.
This shows that DM-baryon interactions in dwarf galaxies through gas outflows
play a critical role not only in forming bulgeless dwarf galaxies but also in
flattening central cusps predicted from DM-only \LCDM\ simulations. This
demonstrates that proper modeling of DM-baryon interactions in hydrodynamical \LCDM\ 
galaxy simulations is able to alleviate the long-standing tension associated
with the central DM distribution in dwarf galaxies between simulations
and observations. We refer to \cite{Governato_2012} (see also \citeauthor{Pontzen_2012}
\citeyear{Pontzen_2012}; \citeyear{Pontzen_2014} and
\citeauthor{2014MNRAS.437..415D} \citeyear{2014MNRAS.437..415D}) for a detailed
discussion of the effect of SN feedback on the central DM distribution in dwarf
galaxies.

Meanwhile, in Fig.~\ref{Figure6}, the trend of slope change $d{\rm
log}\alpha/d{\rm log}M_{*}$ predicted from dwarf galaxy simulations with
baryonic feedback processes is reversed in massive disk galaxies (open boxes)
where a bulge component becomes dominant in the central kinematics. As discussed
in \cite{2014MNRAS.437..415D}, the effect of SN feedback can be surpassed by the
deep gravitational potential which is caused by the bulge component in the central
region. However, as discussed earlier, the steep slopes of the THINGS disk
galaxies could be partially affected by the minimum disk assumption used for
deriving their DM density profiles. The central kinematics of a disk galaxy with
a substantial bulge component is sensitive to even small uncertainties in \ML\
when converting the luminosity profile to the mass density profile. This makes
it difficult to perform an accurate disk-halo decomposition of bulge-dominated
disk galaxies. This stresses the usefulness of bulge--less dwarf galaxies in
testing the effect of SN feedback on the central DM distribution of galaxies.

As shown in Fig.~\ref{Figure6}, the simulated dwarf galaxies which have
comparable stellar masses ranging from $10^6$ to $10^9$ \Msun\ show good
agreement with the majority of the sample dwarf galaxies from THINGS and LITTLE
THINGS. However, as already discussed in Section~\ref{DM_PROFILE}, the rotation curves of some LITTLE
THINGS sample galaxies (e.g., DDO 70, DDO 101, DDO 154, DDO 210 and Haro 36) are equally
well fitted by CDM NFW and pseudo-isothermal halo models in terms of
$\chi^{2}_{red}$ values. Moreover some of them, such as DDO 101, DDO 210 and Haro 36,
appear to have relatively steep inner density slopes (although Haro 36 is
defined as a blue compact dwarf galaxy which usually shows a steep increase in
rotation velocity in the inner region mainly due to young starburst components
formed during galaxy interaction or mergers; e.g., \citeauthor{Bekki_2008}
\citeyear{Bekki_2008}). In particular, DDO 210 is a good
candidate for testing the SN feedback efficiency scenario by Governato et al.
(2012) given that its stellar mass falls within the regime where according to
the simulations primordial CDM cusps are expected to survive.
In Fig.~\ref{Figure6}, DDO 210 shows
no distinct signature of the central cusp ($\alpha$$\sim$$-1.0$) given its corresponding
stellar mass, deviating from the prediction of \LCDM\ DM-only simulations
(dashed line). Nevertheless, in a qualitative sense, its relatively steep inner
density slope $\alpha$$\sim$$-0.70$ compared to the others still leaves a room for a
potential signature of the central cusp, which supports the lower SN energy
injection scenario in low mass dwarf galaxies. However, as noted earlier, the
resolution of LITTLE THINGS H{\sc i} observations is not high enough to 
resolve the small H{\sc i} disk ($\sim$60\arcsec\ diameter) and distinguish 
between cusp- and core-like DM behaviour near the center of DDO 210. 
We note that DDO 210 has the most compact H{\sc i} disk (in terms of the beam size,
e.g., $R_{\rm max}$H{\sc i}$_{\rm beam}^{-1}$$\sim$$6.0$ as given in
Table~\ref{MD_results_LT}) of the sample galaxies. Likewise, some of our sample
galaxies may still suffer from beam smearing which is discussed in the following section.

\begin{figure*}
\epsscale{1.0}
\includegraphics[angle=0,width=1.0\textwidth,bb=10 160 580 690,clip=]{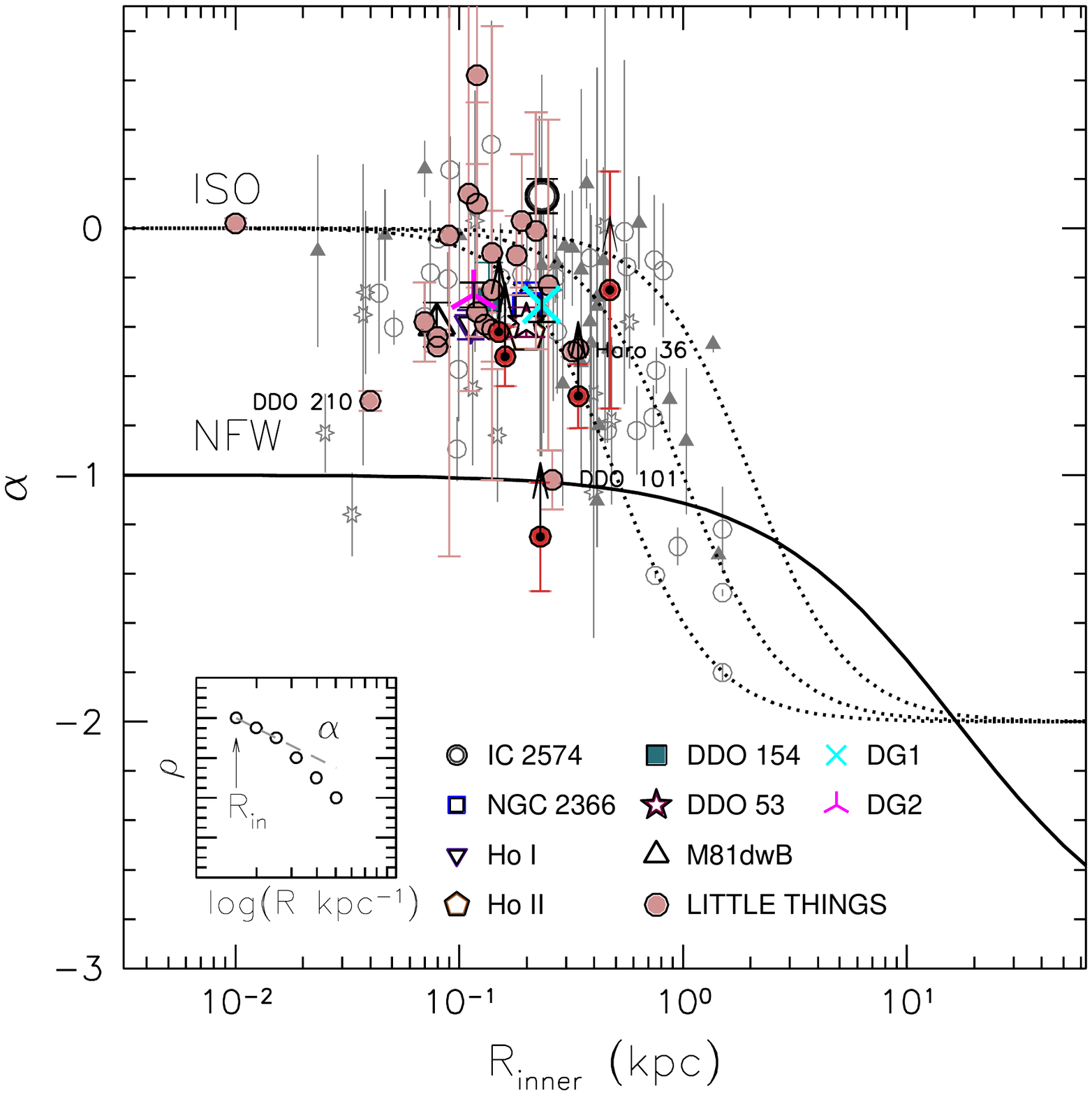}
\caption{The inner slope of the dark matter density profiles $\alpha$ vs. the
radius $\rm R_{in}$ of the innermost point within which $\alpha$ is measured as described in
the small figure (\citeauthor{deBlok_2001} \citeyear{deBlok_2001}). The
$\alpha$\--$\rm R_{in}$ of the sample galaxies from LITTLE THINGS, THINGS and
the two simulated dwarf galaxies (DG1 and DG2: \citeauthor{Governato_2010}
\citeyear{Governato_2010}) as well as the previous measurements (grey symbols)
of LSB galaxies (open circles: \citeauthor{deBlok_2001} \citeyear{deBlok_2001};
triangles: \citeauthor{deBlok_2002} \citeyear{deBlok_2002}; open stars:
\citeauthor{Swaters_2003} \citeyear{Swaters_2003}). Filled circles with arrows
indicate the galaxies of which inner density slopes are measured assuming a `minimum
disk', giving a steeper slope. The solid and dotted lines
represent the $\alpha$\--$\rm R_{in}$ trends of dark-matter-only \LCDM\ NFW
and pseudo-isothermal halo models, respectively. See
Section~\ref{BEAM_SMEARING_EFFECT} for more details.
\label{Figure7}}
\end{figure*}

\section{Inner density slope vs. resolution} \label{BEAM_SMEARING_EFFECT}
As in \cite{deBlok_2001}, for a quantitative examination of the beam smearing
effect on our sample galaxies, we plot the inner density slopes $\alpha$ of the
galaxies including the THINGS sample as well as the two simulated dwarf galaxies (DG1
and DG2) modelled by \cite{Governato_2010} against the observed radii of
their innermost point $R_{\rm inner}$ in Fig.~\ref{Figure7}. For the sample
dwarf galaxies from LITTLE THINGS, we use the slopes derived
assuming the model \MLsps\ disk. We also show the $\alpha$\--$R_{\rm inner}$ relations of the NFW and
pseudo-isothermal halo models as solid and dotted lines, respectively, derived
using their analytical formulas as given in Eqs.~\ref{eq:6} and \ref{eq:8}. 

As shown in Fig.~\ref{Figure7}, most sample galaxies show significant deviations from the
predicted $\alpha$\--$R_{\rm inner}$ trend (solid line) of \LCDM\ NFW halos
at around a $R_{\rm inner}$ of $\sim$0.2 kpc. Instead, they are
more consistent with those of pseudo-isothermal halo models with different
core-radii (dotted lines) as well as the earlier results found from LSB galaxies
(grey symbols) in \cite{deBlok_2002}. However, the clear difference between the
two halo models (i.e., NFW and pseudo-isothermal) at high
resolutions (e.g., $R_{\rm inner}$$<$0.5 kpc) becomes ambiguous as the innermost radius
$R_{\rm inner}$ of a given DM density profile increases. 
For example, a galaxy with a larger $R_{\rm inner}$ (i.e., low resolution) tends
to show a steeper inner slope of DM density profile.
The larger $R_{\rm inner}$ makes it lie in the region where the slopes of the two halo models are
approximately similar to each other. 
In addition, as discussed in Section~\ref{DM_PROFILE},
the derivative $d{\rm log}\rho/d{\rm log}R$ of a DM density profile on a logarithmic scale decreases towards the
outer region of a galaxy. If the DM density profile is affected by beam
smearing, the break radius of the profile which is determined when
measuring the inner slope tends to migrate into the outer regime where
$d{\rm log}\rho/d{\rm log}R$ has a lower value. Therefore, the inner density slope $\alpha$ within the
break radius is most likely to be steeper than the ones derived from well
sampled profiles. This could be the case of DDO 101 and DDO 210. In particular,
DDO 101 is most likely to be affected by the beam smearing effect as shown in
Fig.~\ref{Figure7}.

Yet higher resolution velocity fields obtained with radio
interferometers or using other tracers such as integral field mapping are 
required to study the effect of SN feedback on the central
cusps of the lowest mass dwarf galaxies. Such high--resolution observations of
low mass dwarf galaxies would provide an ultimate test of the \LCDM\ paradigm. Unlike clusters of galaxies
where the depth of the gravitational potential well is deep enough to retain warm DM (WDM)
as well as CDM, there is no room for WDM in dwarf galaxies inhabiting DM halos
with much shallower potential wells. Therefore, finding a signature of a central
cusp in dwarf galaxies will prove that there is at least some cold DM in the
Universe. This again highlights the cosmological importance of low mass dwarf
galaxies, not only for resolving the `cusp/core' controversy in \LCDM\
simulations but also as an indirect proof for the existence of CDM in the Universe.

\section{Conclusion} \label{CONCLUSIONS}

In this paper we derive the rotation curves of 26 dwarf galaxies culled from
LITTLE THINGS, and examine their DM distributions near the centers of
the galaxies. From this, we address the `cusp/core' problem which has been one
of the long-standing problems in \LCDM\ simulations on galactic
scales. The high-resolution LITTLE THINGS H{\sc i} data ($\sim$6\arcsec angular;
$\sim$2.6 \kms\ spectral) complemented with optical and {\it Spitzer IRAC} 3.6$\mu$m images 
are sufficiently detailed to resolve the central region of the sample galaxies where the cusp- and
core-like halo models are clearly distinguished. 

In particular, we use the bulk velocity fields of the galaxies extracted using
the method described in \cite{Oh_2008} to correct for turbulent random non-circular
gas motions. This enables us to derive more reliable rotation curves and thus
more accurate DM distributions in the galaxies. We corrected for the modest dynamical 
contribution by baryons in dwarf galaxies by using {\it Spitzer IRAC} 3.6$\mu$m images combined with 
model \MLsps\ values based on stellar population synthesis models. This allowed us to
derive robust mass models of the stellar components of the galaxies and thus better
constrain their central DM distributions. 

From this, we found that the decomposed DM rotation curves of most
sample galaxies are well matched in shape to those of core-like halos which
are characterised by a linear increase of rotation velocity in the inner region. We also
derive the DM density profiles of the sample galaxies and quantify the
degree of the central DM concentration by measuring the logarithmic
inner slopes of the profiles. The mean value of the inner slopes $\alpha$ of
the 26 sample galaxies is $-0.32$, which indicates a mass distribution with a
sizable constant density-core towards the centers of the galaxies. This is
consistent with that found in most nearby dwarf galaxies (e.g., LSB galaxies in
\citeauthor{deBlok_2002} \citeyear{deBlok_2002}; THINGS dwarfs in
\citeauthor{Oh_2011a} \citeyear{Oh_2011a}) which all show a linear
increase in the inner shapes of their rotation curves resulting in shallower
inner density slopes ($<\alpha>\sim-0.2$). Considering the fact that observational
uncertainties are significantly reduced in the high-resolution LITTLE THINGS
data, the core-like DM distribution found in our sample galaxies
provides a stringent observational constraint on the central DM 
distribution of halos in \LCDM\ simulations. 

We find that the derived slopes of the DM density profiles do not
agree with the cusp predicted by \LCDM\ DM-only simulations.
However, recent cosmological N--body SPH galaxy simulations by
\cite{Governato_2010} (see also \citeauthor{Governato_2012}
\citeyear{Governato_2012}; \citeauthor{Pontzen_2012} \citeyear{Pontzen_2012})
have shown that the discrepancy between observations and
simulations can be reconciled within the \LCDM\ paradigm by considering the
dynamical effect of baryonic feedback processes on the central cusps. According
to the simulations, DM-baryon interactions in dwarf galaxies through gas outflows
driven by SN explosions play a critical role not only in forming bulgeless dwarf
galaxies but also in turning central cusps into cores. 

As discussed in \cite{Oh_2011b}, the slowly rising rotation curves and the
resulting shallower DM density profiles of the simulated dwarf galaxies
with SN feedback are qualitatively similar to those of the dwarf galaxies from
THINGS. This is also the case for our sample galaxies from LITTLE THINGS whose
DM rotation curves in the inner region rise too slowly to match the steep rotation curves
of CDM halos. This shows that proper modelling of DM-baryon interactions in
\LCDM\ galaxy simulations is able to alleviate the long-standing tension between
observations and simulations regarding the central DM distribution in
dwarf galaxies.

Notwithstanding the dominant trend of core-like DM distribution in the
LITTLE THINGS sample galaxies, some of the sample galaxies, such as DDO 210 and
Haro 36 are equally well fitted by core- and cusp-like halo models in
describing their DM rotation curves. They have relatively steeper inner
density slopes with $\alpha$$\approx$ $-0.70$ and $-0.50$ for DDO 210 and Haro 36, respectively,
compared to the mean value (DM only) of the rest of the sample ($\sim-0.29$). It
is possible, however, despite the high angular resolution of the LITTLE
THINGS H{\sc i} data, the inner density slopes of these two galaxies are
affected by beam smearing (the H{\sc i} disk of DDO 210 being intrinsically
small and Haro\,36 being one of the more distant dwarfs). Insufficient spatial
resolution in the inner region of the galaxies acts to results in steeper
observed inner DM density slopes. 

According to the latest N--body SPH simulations of dwarf galaxies with baryonic
feedback processes (\citeauthor{Governato_2012} \citeyear{Governato_2012}), the
SN feedback in low mass dwarf galaxies with a stellar mass less than
$10^6$\Msun\ is not sufficient to disrupt the central cusps, the repeated energy injection 
from SN explosions into DM halos becoming inefficient,
largely due to low star formation efficiencies in these low mass systems. However, previous
observational studies regarding the `cusp/core' problem have mostly focused on
relatively massive dwarf galaxies for which reliable rotation curves are
available. Low mass dwarf galaxies have been usually excluded for the study of
the central DM distribution due to the low amplitude of their maximum
rotation velocities. The kinematics of such low mass dwarf galaxies is
more vulnerable to kinematic disturbances like non-circular motions in galaxies
compared to more massive ones, and also more sensitive to additional corrections
(e.g., asymmetric drift) made when deriving rotation curves. In this respect,
the possibility of a selection effect in favor of relatively massive dwarf
galaxies where the effect of SN feedback is enough to turn the central cusps
into cores should be considered.

It would therefore be worthwhile to perform high-resolution follow-up
observations, for example using an optical integral field unit,
of low mass dwarf galaxies including some of the LITTLE THINGS dwarf galaxies
(e.g., DDO 210) whose stellar masses lie in the regime where
primordial CDM cusps are predicted from the simulations
(\citeauthor{Governato_2012} \citeyear{Governato_2012}).
These high-resolution observations will enable us to
achieve a finer sampling of the central region of the galaxies, and thus more
accurate inner DM density profiles. From this, more stringent
observational constraints on the central cusp of low mass dwarf galaxies could
be provided. Moreover, an accurate measurement of the DM distribution in
these low mass dwarf galaxies will provide an ultimate test for the CDM paradigm
given that dwarf galaxies inhabiting DM halos with a shallow potential well have
only room for CDM, unlike clusters of galaxies whose gravitational potential is
deep enough to retain WDM as well as CDM. 
Therefore, the presence or absence of a signature of the
central cusp in these low mass halos will provide a critical observational
test, either supporting or falsifying the \LCDM\ paradigm.

\acknowledgements
We thank W.~J.~G. de Blok for providing the inner density slopes and stellar
masses of the THINGS disk galaxies and useful comments and discussion. We thank
Fabio Governato, Andrew Pontzen, Chris Brook and Arianna Di Cintio for useful
discussion on the simulations and providing the data. Parts of this research were
conducted by the Australian Research Council Centre of Excellence for All-sky Astrophysics (CAASTRO),
through project number CE110001020. This work was funded in
part by the National Science Foundation through grants AST-0707563 and
AST-0707426 to DAH and BGE.

\bibliography{ms}

\renewcommand{\thesection}{Appendix}
\label{Appendix}
\section{Data and kinematic analysis}

\newcounter{appendix_section}
\renewcommand{\thesection}{A.\arabic{appendix_section}}
\renewcommand{\theequation}{A.\arabic{equation}}
\renewcommand{\thefigure}{A.\arabic{figure}}

In this Appendix, we present the data and kinematic analysis of the 26 LITTLE
THINGS (\citeauthor{Hunter_2012} \citeyear{Hunter_2012}) dwarf galaxies. For
each galaxy, we show the (1) data, (2) kinematic analysis, and (3) mass
modelling with descriptions: 

\noindent {\bf A. Data$\--$} {\bf (a)} Integrated H{\sc i} intensity map (moment
$0$). The contour levels start at $+3\sigma$ in steps of $+3\sigma$. {\bf (b)}
Intensity-weighted mean velocity field (moment 1). {\bf (c)} Velocity dispersion map
(moment 2). {\bf (d)} {\it Spitzer IRAC 3.6}$\mu$m image obtained
from the archives including `{\it Spitzer} Infrared Nearby Galaxies Survey' (SINGS;
\citeauthor{Kennicutt_2003} \citeyear{Kennicutt_2003}) and {\it Spitzer} `Local Volume Legacy'
(LVL; \citeauthor{2009ApJ...703..517D} \citeyear{2009ApJ...703..517D}). {\bf
(e)} Bulk velocity field extracted using the
method described in \cite{Oh_2008}. {\bf (f)} Velocity field of strong
non-circular motions as in \cite{Oh_2008}. {\bf (g)} Model velocity field
of the tilted-ring model derived using the bulk velocity field in the panel
(e). {\bf (h)} Velocity field of weak non-circular motions as in \cite{Oh_2008}.
{\bf (i), (j)} Position-velocity diagram taken along the average position
angle of the major and minor axes as given in Table~\ref{LT_TR_properties}. The
dashed lines indicate the systemic velocity and 
position of the kinematic center derived in this paper. The bulk (black dots)
and asymmetric drift corrected bulk (yellow) rotation curves are overplotted.
The curves are converted back to radial velocities using the geometrical
parameters determined by the tilted-ring analysis as listed in
Table~\ref{LT_TR_properties}. Moment maps and velocity fields are extracted
using the robust-weighted data cubes. The beam size is indicated by the ellipse
in the bottom-right corner of each panel. See \cite{Hunter_2012} for a detailed
description of the data cube.

\noindent {\bf B. Kinematic analysis} \\
\indent $\bullet$ {\bf Rotation curves$\--$} The tilted-ring model derived using
the bulk velocity field as given in the panel {\bf (e)} of {\bf A. Data}. The open
grey circles shown in all panels are the fit results with all ring parameters
(i.e., XPOS, YPOS, VSYS, PA, INCL and VROT) free. The grey filled dots in the VROT panel
indicate the final rotation velocity derived using the entire velocity field
after fixing other ring parameters to the values as shown in other panels (solid
lines). The upright and upside-down triangles show the rotation velocities
derived using the receding and approaching sides, respectively, while keeping
other ring parameters at the values indicated in the other panels with solid
lines.

\indent $\bullet$ {\bf Asymmetric drift correction$\--$} {\bf (a)} grey dots represent the
radial asymmetric drift correction, $\sigma_{\rm D}$. The open circles indicate
the derived rotation velocity from the tilted-ring analysis, and the black dots
show the one corrected for asymmetric drift. {\bf (b)} Azimuthally averaged H{\sc
i} velocity dispersion. {\bf (c)} Radial H{\sc i} surface density derived
applying the kinematic geometry from the tilted-ring analysis. {\bf (d)} The
dashed line shows the fit of the analytical function given in Eq.~\ref{eq:4}
to $\Sigma_{\rm HI}$$\sigma^{2}$. See Section~\ref{ASD} for more details.

\indent $\bullet$ {\bf Harmonic analysis$\--$} Harmonic decompositions of the
bulk (black dots) and intensity-weighted mean (moment 1) velocity fields. The
decomposition is made using the {\sc reswri} task in GIPSY. The solid and dashed
lines in the middle-right panel indicate average global elongations of the
halo potential as described in \citeauthor{Schoenmakers_1997}
(\citeyear{Schoenmakers_1997}; \citeyear{Schoenmakers_1999}) derived using the
bulk and moment 1 velocity fields, respectively. The median absolute amplitudes
($\langle A \rangle$) and the phases ($\phi_{1}$, $\phi_{2}$, and $\phi_{3}$) of
each component are presented in the bottom panels.
 
\noindent {\bf C. Mass modelling}  \\
\indent $\bullet$ {\bf Mass models of baryons$\--$} {\bf (a)} {\it Spitzer} IRAC
3.6$\mu$m surface brightness profile derived applying the kinematic geometry
from the tilted-ring analysis. {\bf (b)} The stellar mass-to-light value in the
3.6$\mu$m, \MLsps\ derived using the empirical relation described in
\cite{Oh_2008} which is based on stellar population synthesis models in
\cite{Bruzual_Charlot_2003} and \cite{Bell_2001}. We refer to \cite{Oh_2008} for
a full description. {\bf (c)} Stellar mass surface density profile in the
3.6$\mu$m. {\bf (d)} The rotation velocity for the stellar component derived
using the mass density profile in panel (c). {\bf (e)} Azimuthally
averaged optical color, $B-V$. {\bf (f)} Gas column density derived applying the
kinematic geometry from the tilted-ring analysis. {\bf (g)} Mass surface density
profile for the gas component which is scaled up by 1.4 to account for Helium
and metals. {\bf (h)} The resulting rotation velocity for the gas component.

\indent $\bullet$ {\bf Disk\--halo decomposition$\--$} The grey dots indicate
the bulk rotation curve which is corrected for asymmetric drift. The dotted
and dash-dotted lines represent the rotation velocities of the gas and stellar
components, respectively. The open circles show the rotation velocity of
dark matter halo only, derived after subtracting the contribution to the rotational 
velocity of the baryons. The dashed and solid lines are the fits of
\LCDM\ NFW and pseudo-isothermal halo models to the dark matter--only curve (open
circles). The reduced $\chi^{2}$ value for each halo model is denoted. The dots
(DM rotation curve $-$ NFW model) and open circles (DM rotation curve $-$
pseudo-isothermal model) in the lower panel represent the velocity differences between
the rotation curve of the dark matter halo and the best fit halo models.

\indent $\bullet$ {\bf Mass density profile$\--$} The circles indicate the mass
density profile derived from the asymmetric drift corrected bulk rotation curve
which includes the dynamical contributions by dark matter halo and baryons
(i.e., minimum disk assumption). The squares show the one derived from the dark
matter rotation curve where the dynamical contribution by baryons is subtracted
(dark matter only). The logarithmic inner slope $\alpha$ of the dark matter
density profile is measured by a least squares fit (dotted line) to the inner
data points (filled squares). The dashed and solid lines show the density
profiles derived from the best fit NFW and pseudo-isothermal halo models,
respectively.

\clearpage

\begin{figure}
\epsscale{1.0}
\figurenum{A.1}
\includegraphics[angle=0,width=1.0\textwidth,bb=50 140 550
745,clip=]{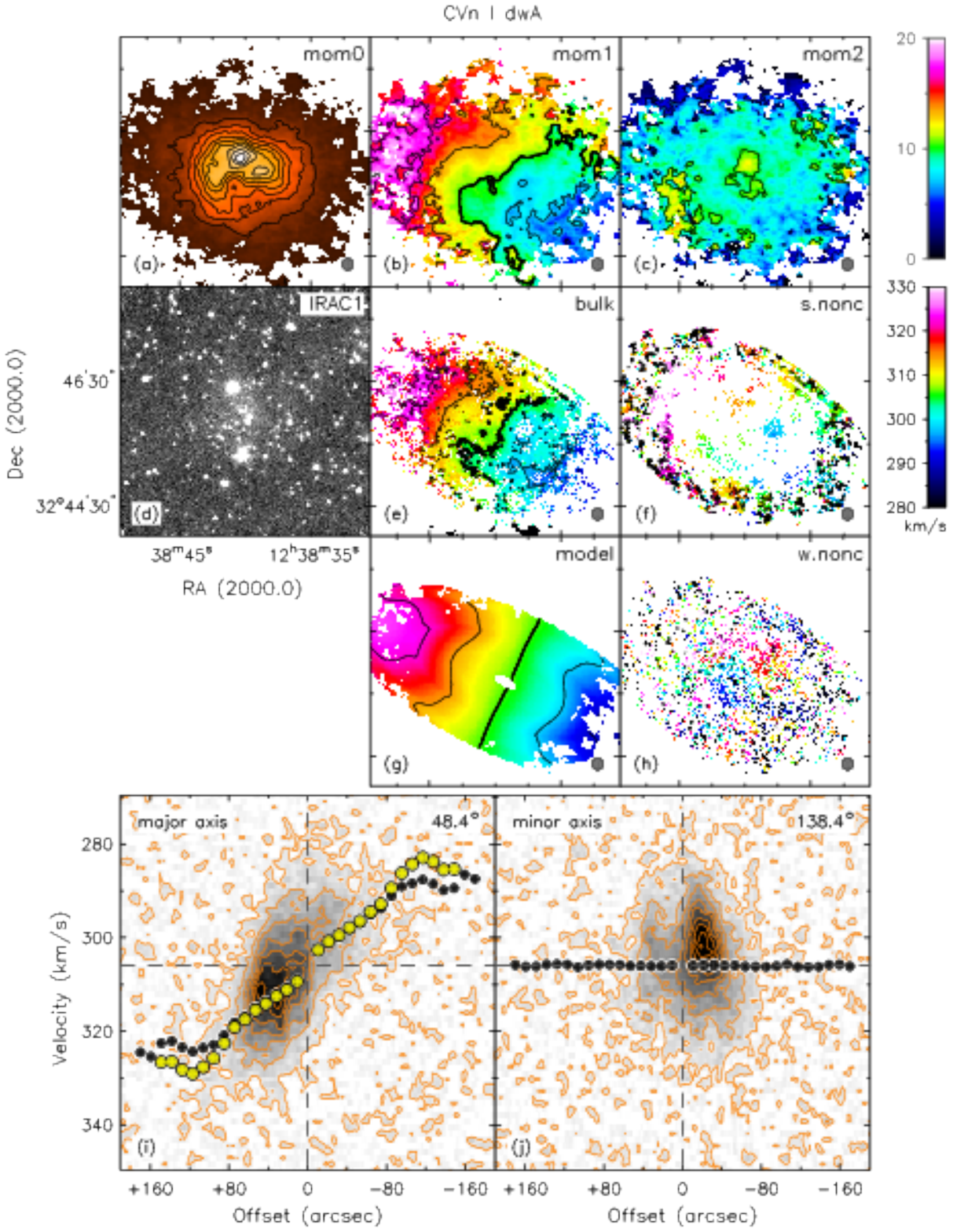}
\caption{H{\sc i} data and {\it Spitzer IRAC} 3.6$\mu$m image of CVnIdwA. The
systemic velocity is indicated by the thick contours in the velocity fields, and
the iso-velocity contours are spaced by 8 \kms. Velocity dispersion contours run
from 0 to 20 \kms\ with a spacing of 5 \kms. See Appendix section A for details.
\label{CVnIdwA_data_PV}}
\end{figure}

\begin{figure}
\epsscale{1.0}
\figurenum{A.2}
\includegraphics[angle=0,width=1.0\textwidth,bb=35 140 570 710,clip=]{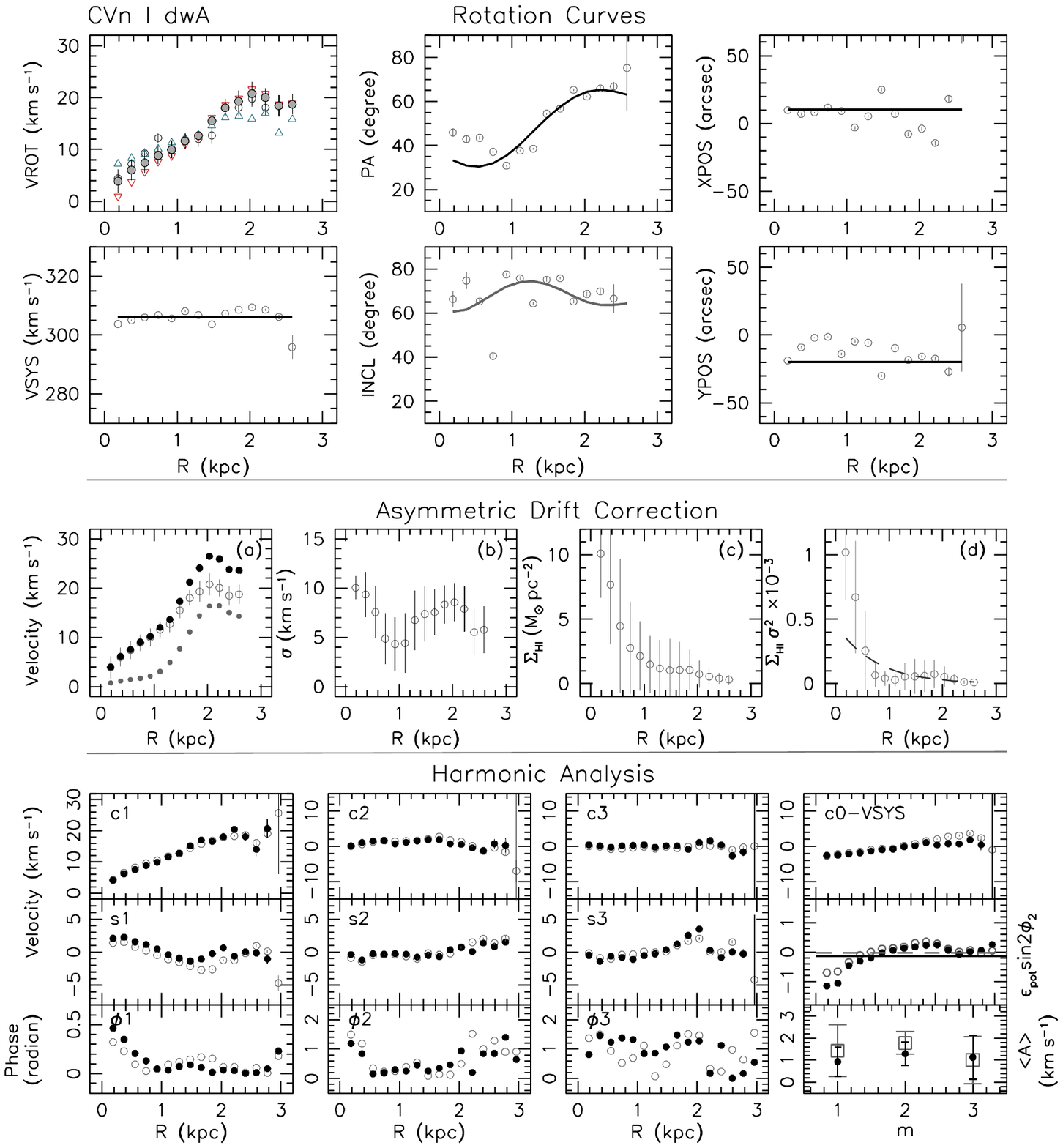}
\caption{Rotation curves, asymmetric drift correction and harmonic analysis
of CVnIdwA. See Appendix section B for details.
\label{CVnIdwA_TR_HD}}
\end{figure}
{\clearpage}

\begin{figure}
\epsscale{1.0}
\figurenum{A.3}
\includegraphics[angle=0,width=1.0\textwidth,bb=40 175 540
690,clip=]{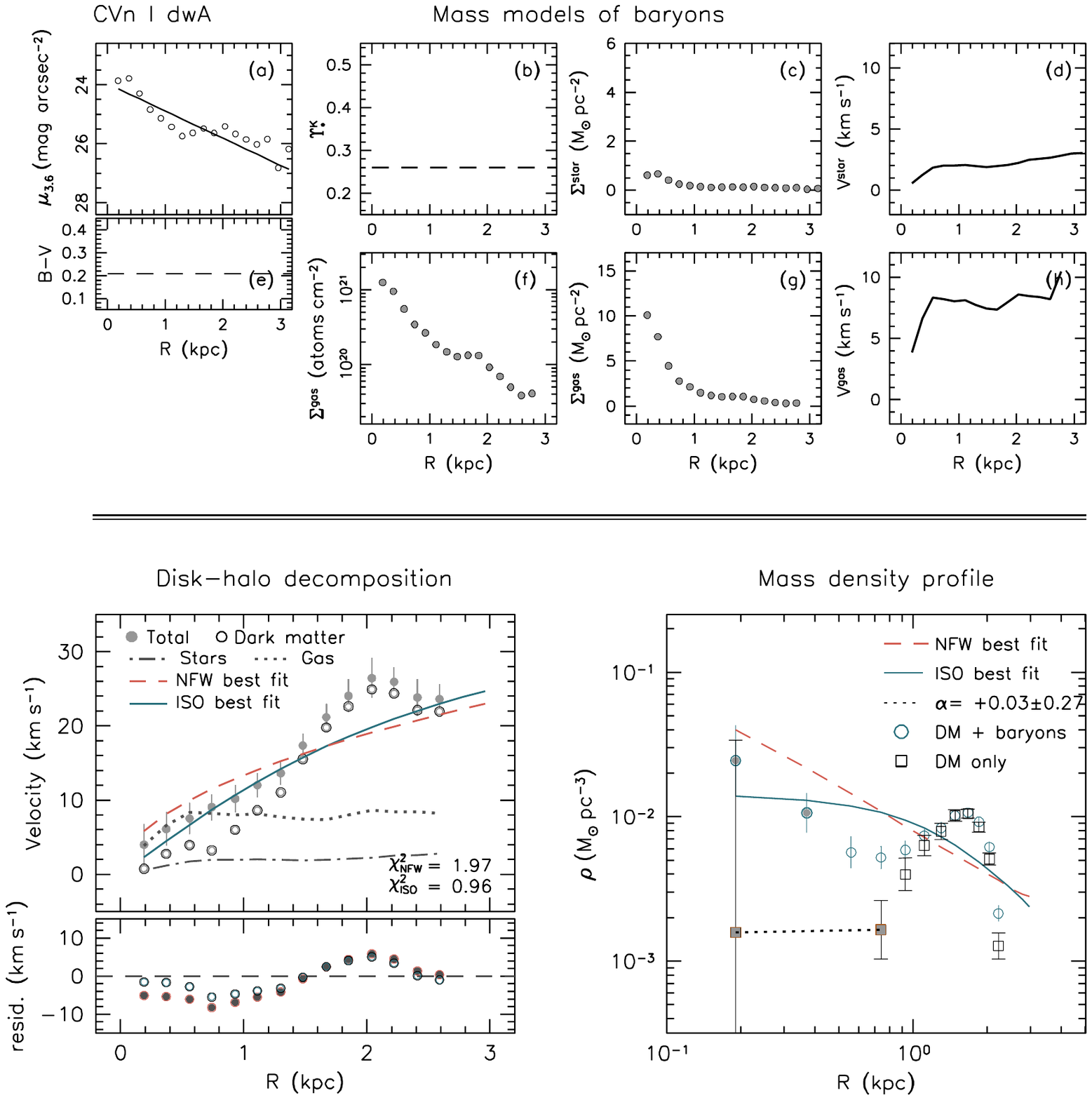}
\caption{The mass models of baryons, disk-halo decomposition and mass density
profile of CVnIdwA. Please refer to the text in Sections~\ref{MASS_MODELS} and
\ref{DARK_MATTER_DISTRIBUTION} for full information.
\label{MD_DH_DM_CVnIdwA}}
\end{figure}
{\clearpage}

\begin{figure}
\epsscale{1.0}
\figurenum{A.4}
\includegraphics[angle=0,width=1.0\textwidth,bb=60 140 540
745,clip=]{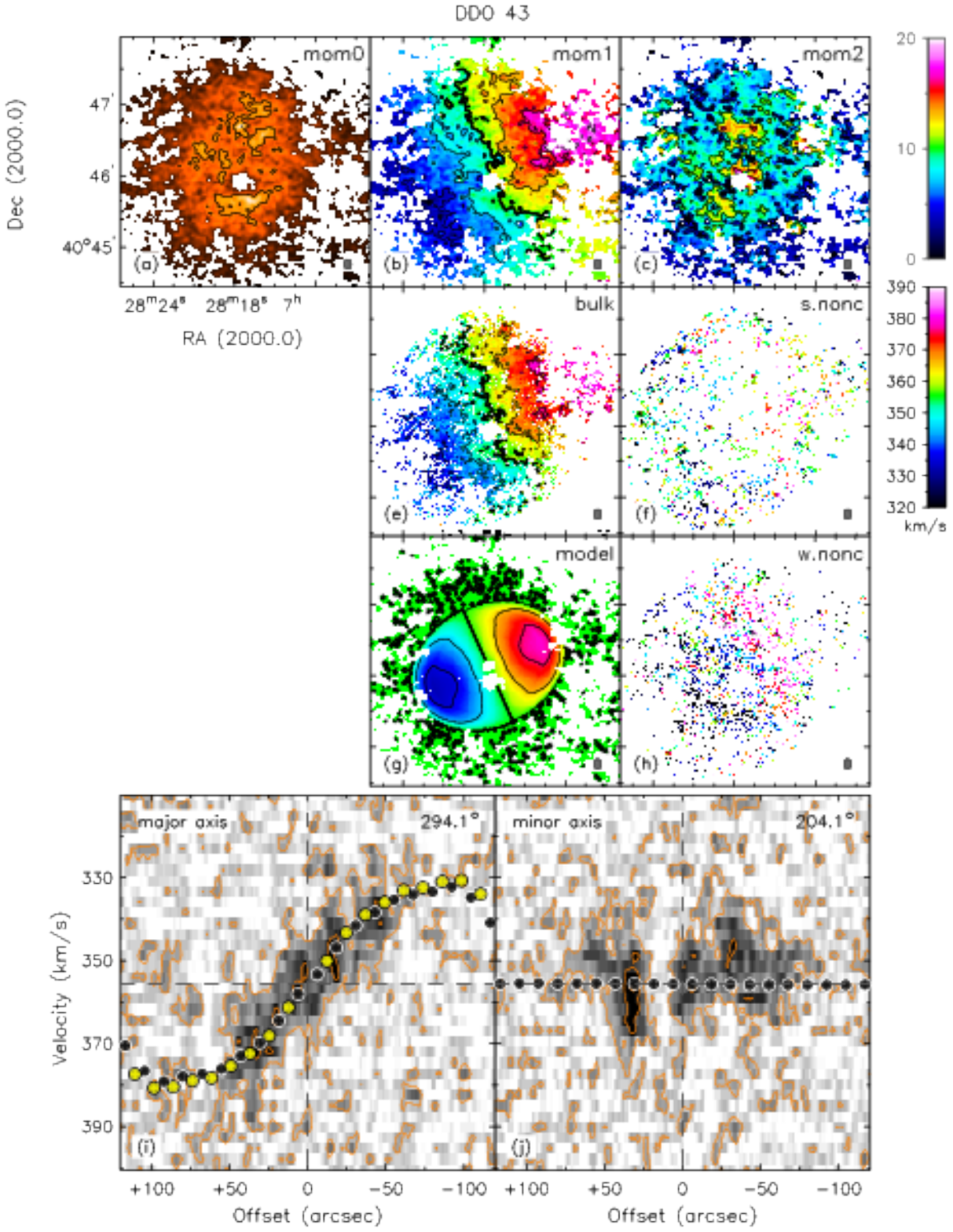}
\caption{H{\sc i} data and {\it Spitzer IRAC} 3.6$\mu$m image of DDO 43. The
systemic velocity is indicated by the thick contours in the velocity fields, and
the iso-velocity contours are spaced by 10 \kms. Velocity dispersion contours run
from 0 to 20 \kms\ with a spacing of 5 \kms. See Appendix section A for details.
\label{ddo43_data_PV}}
\end{figure}
{\clearpage}

\begin{figure}
\epsscale{1.0}
\figurenum{A.5}
\includegraphics[angle=0,width=1.0\textwidth,bb=35 140 570
710,clip=]{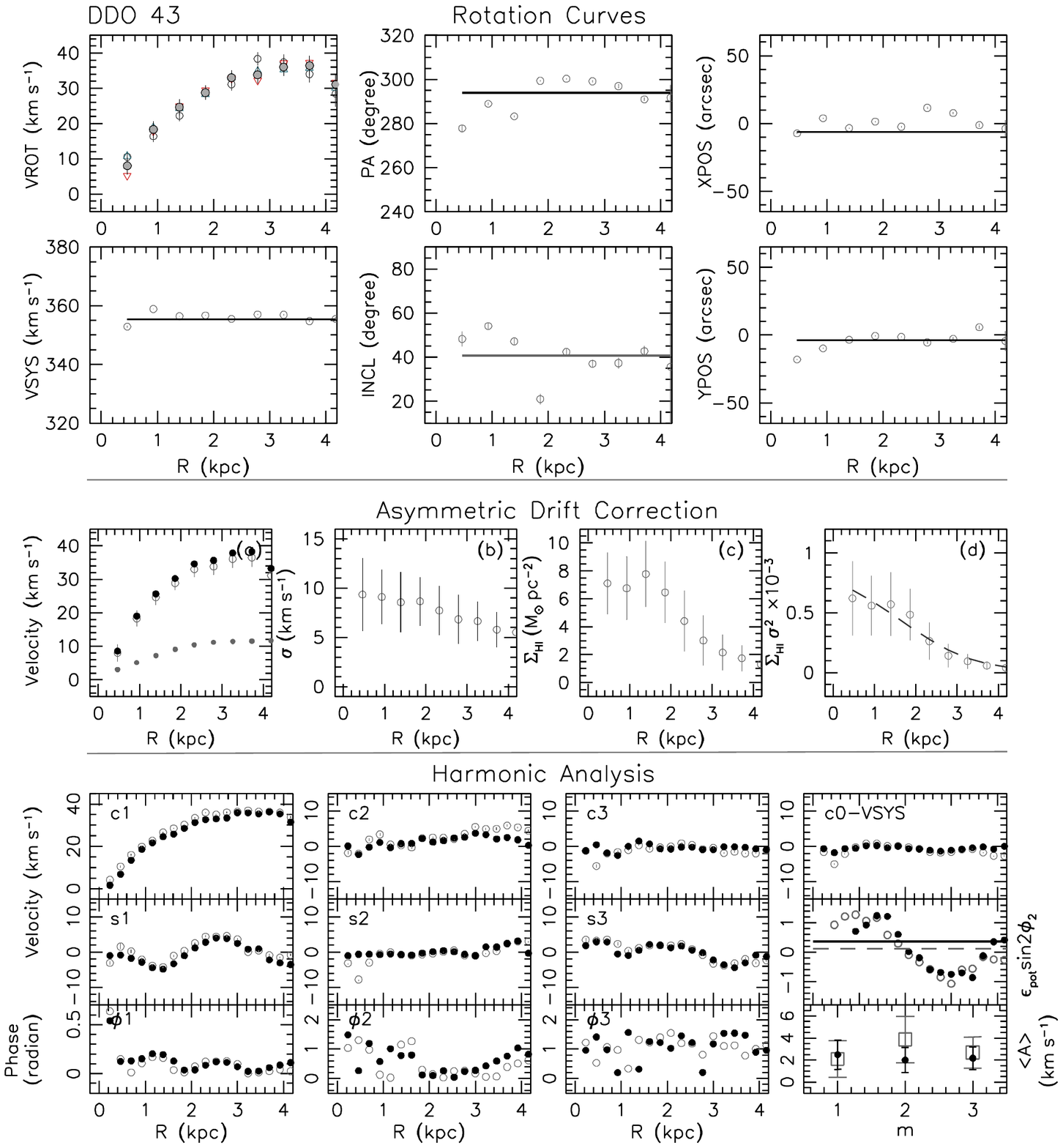}
\caption{Rotation curves, asymmetric drift correction and harmonic analysis
of DDO 43. See Appendix section B for details.
\label{ddo43_TR_HD}}
\end{figure}
{\clearpage}

\begin{figure}
\epsscale{1.0}
\figurenum{A.6}
\includegraphics[angle=0,width=1.0\textwidth,bb=40 175 540
690,clip=]{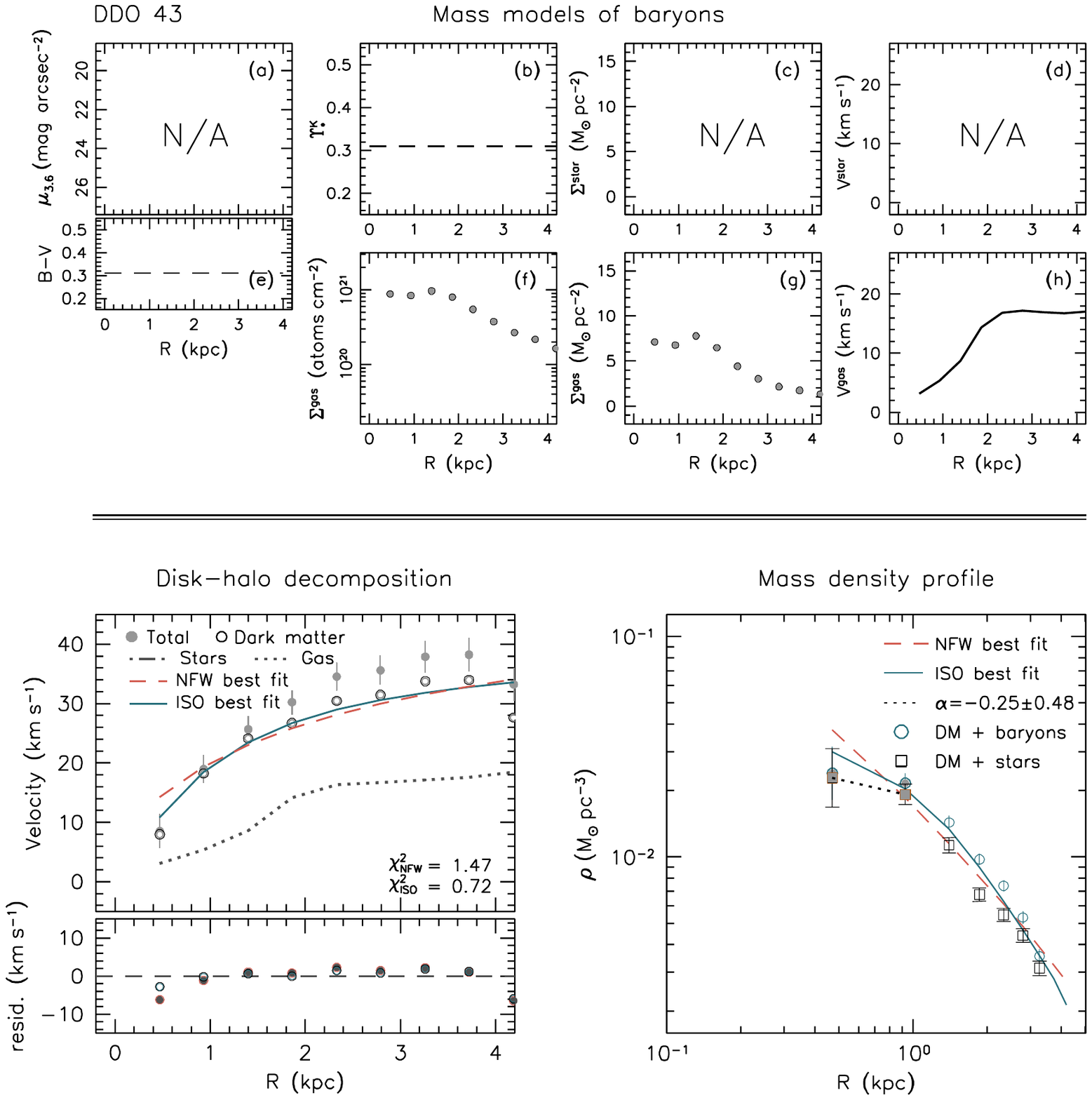}
\caption{The mass models of baryons, disk-halo decomposition and mass density
profile of DDO 43. Please refer to the text in Sections~\ref{MASS_MODELS} and
\ref{DARK_MATTER_DISTRIBUTION} for full information.
\label{MD_DH_DM_ddo43}}
\end{figure}
{\clearpage}

\begin{figure}
\epsscale{1.0}
\figurenum{A.7}
\includegraphics[angle=0,width=1.0\textwidth,bb=60 140 540
745,clip=]{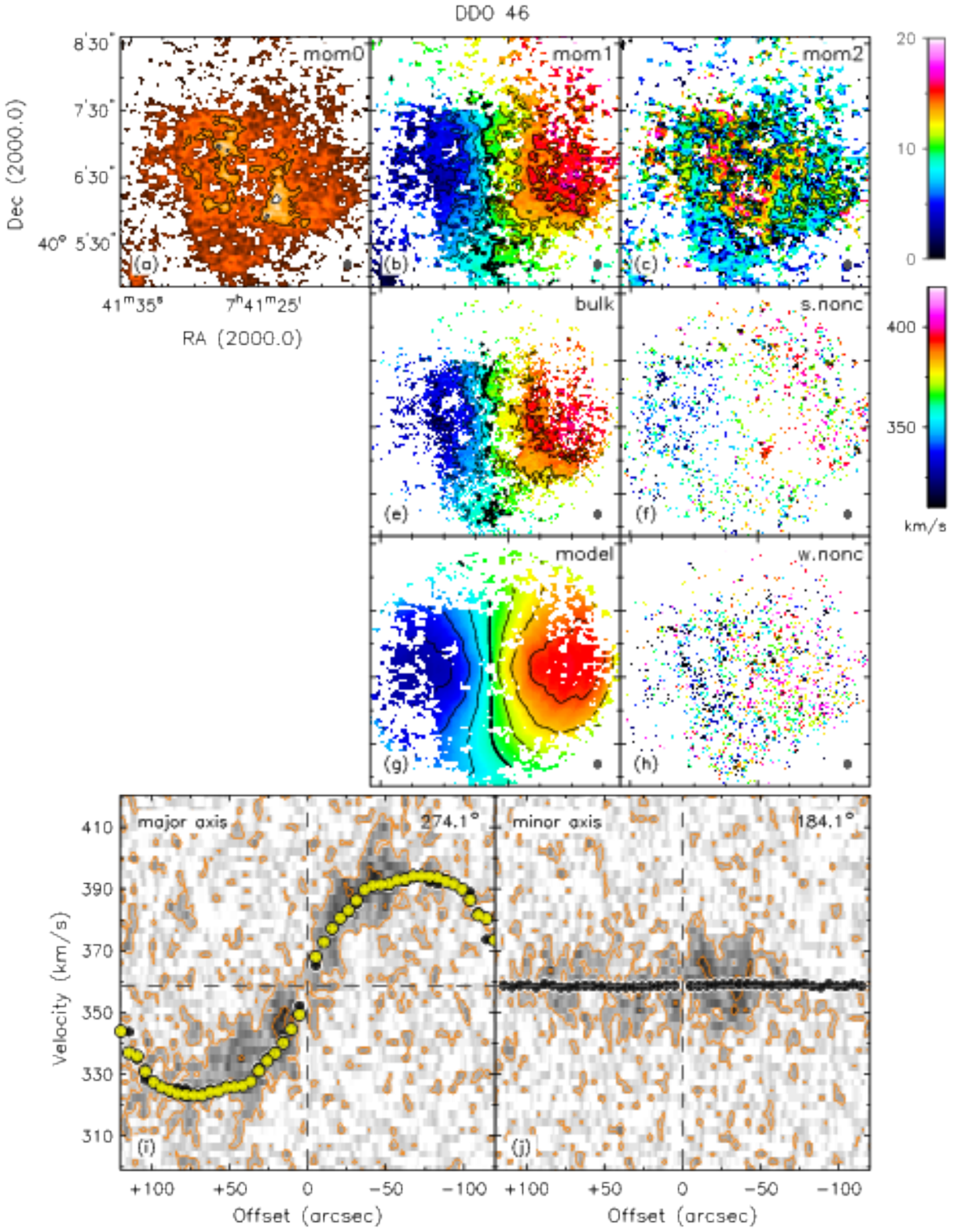}
\caption{H{\sc i} data and {\it Spitzer IRAC} 3.6$\mu$m image of DDO 46. The
systemic velocity is indicated by the thick contours in the velocity fields, and
the iso-velocity contours are spaced by 10 \kms. Velocity dispersion contours run
from 0 to 20 \kms\ with a spacing of 5 \kms. See Appendix section A for details.
\label{ddo46_data_PV}}
\end{figure}
{\clearpage}

\begin{figure}
\epsscale{1.0}
\figurenum{A.8}
\includegraphics[angle=0,width=1.0\textwidth,bb=35 140 570
710,clip=]{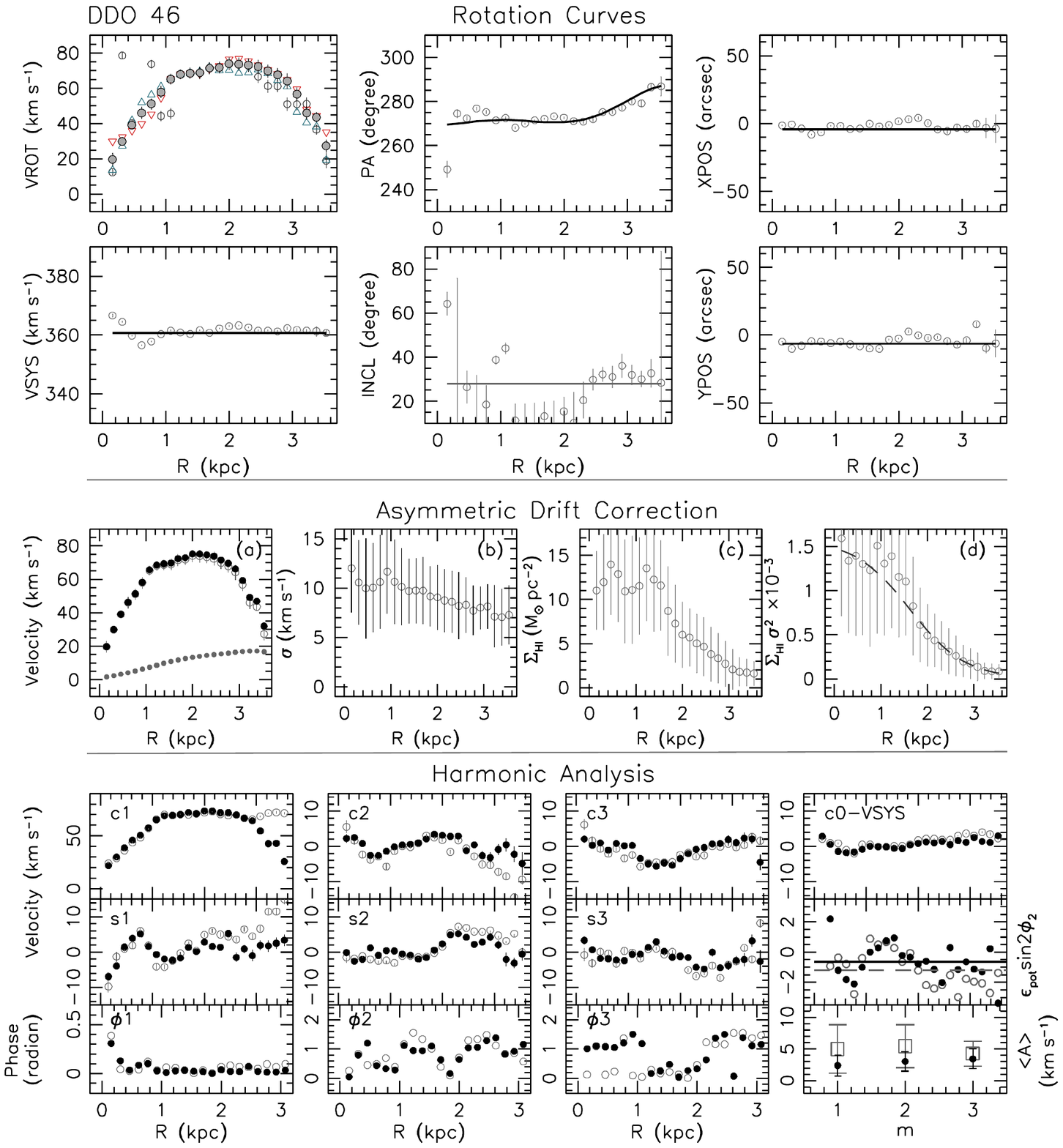}
\caption{Rotation curves, asymmetric drift correction and harmonic analysis
of DDO 46. See Appendix section B for details.
\label{ddo46_TR_HD}}
\end{figure}
{\clearpage}

\begin{figure}
\epsscale{1.0}
\figurenum{A.9}
\includegraphics[angle=0,width=1.0\textwidth,bb=40 175 540
690,clip=]{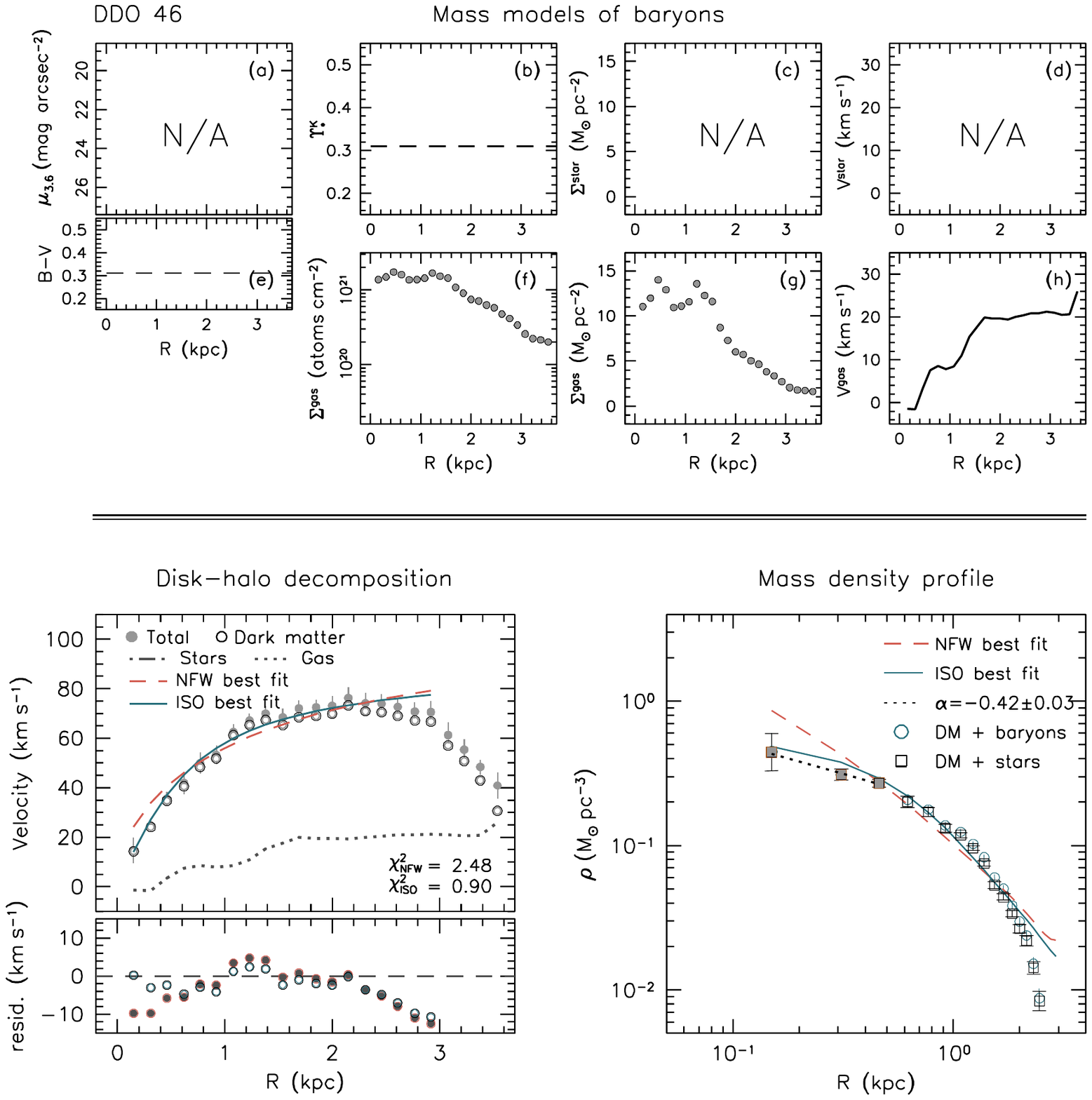}
\caption{The mass models of baryons, disk-halo decomposition and mass density
profile of DDO 46. Please refer to the text in Sections~\ref{MASS_MODELS} and
\ref{DARK_MATTER_DISTRIBUTION} for full information.
\label{MD_DH_DM_ddo46}}
\end{figure}
{\clearpage}

\begin{figure}
\epsscale{1.0}
\figurenum{A.10}
\includegraphics[angle=0,width=1.0\textwidth,bb=60 140 540
745,clip=]{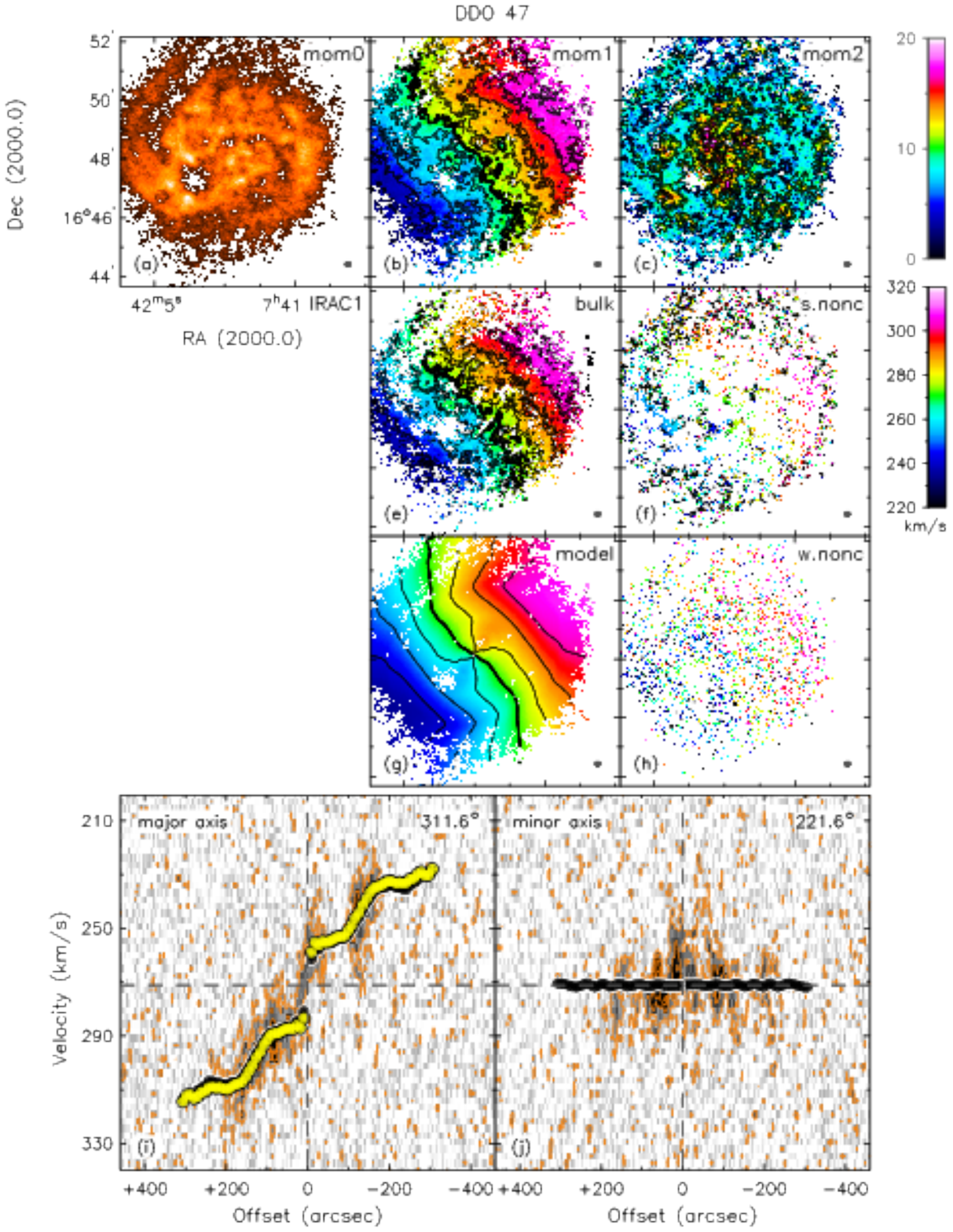}
\caption{H{\sc i} data and {\it Spitzer IRAC} 3.6$\mu$m image of DDO 47. The
systemic velocity is indicated by the thick contours in the velocity fields, and
the iso-velocity contours are spaced by 10 \kms. Velocity dispersion contours run
from 0 to 20 \kms\ with a spacing of 5 \kms. See Appendix section A for details.
\label{ddo47_data_PV}}
\end{figure}
{\clearpage}

\begin{figure}
\epsscale{1.0}
\figurenum{A.11}
\includegraphics[angle=0,width=1.0\textwidth,bb=35 140 570
710,clip=]{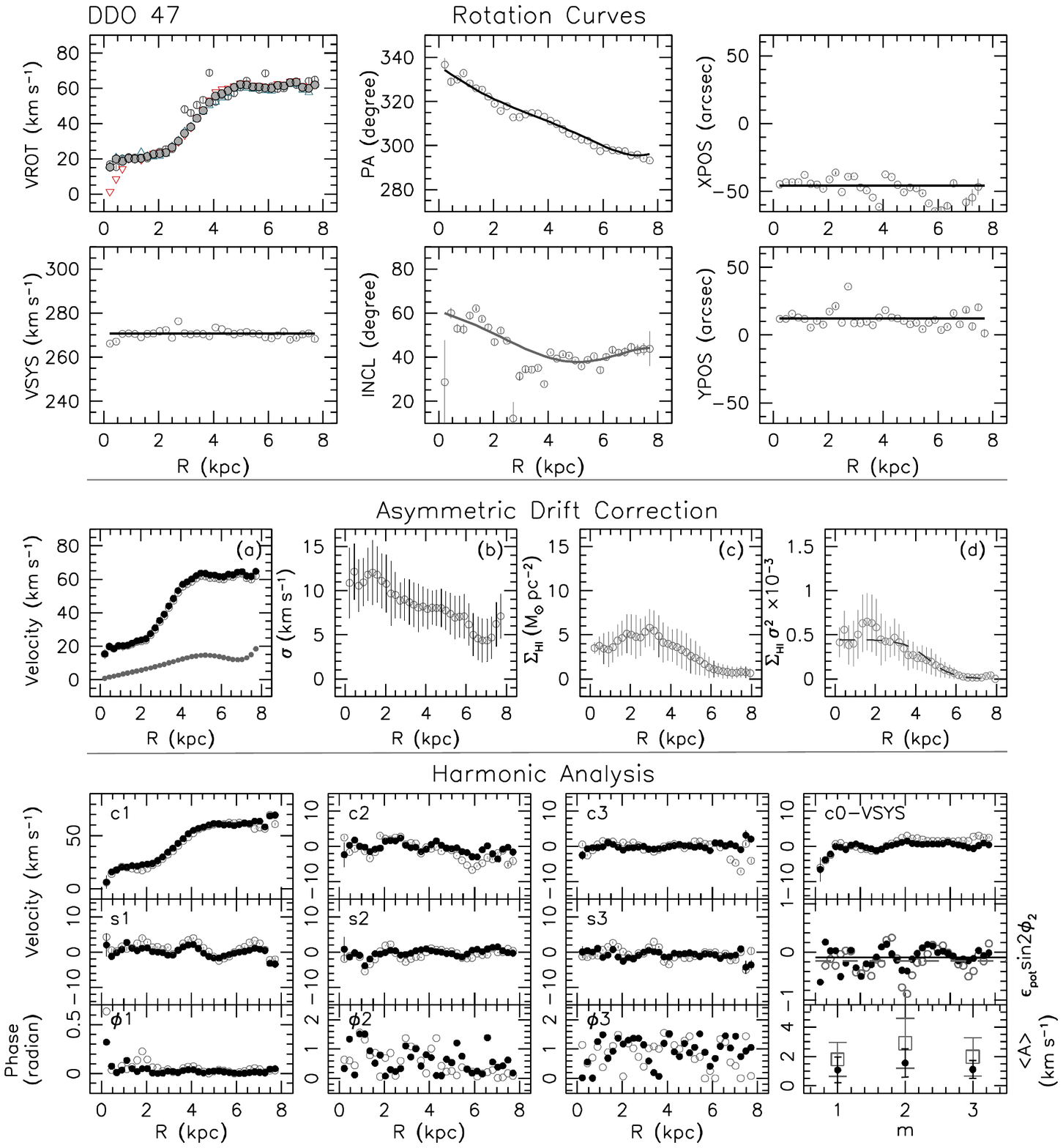}
\caption{Rotation curves, asymmetric drift correction and harmonic analysis
of DDO 47. See Appendix section B for details.
\label{ddo47_TR_HD}}
\end{figure}
{\clearpage}

\begin{figure}
\epsscale{1.0}
\figurenum{A.12}
\includegraphics[angle=0,width=1.0\textwidth,bb=40 175 540
690,clip=]{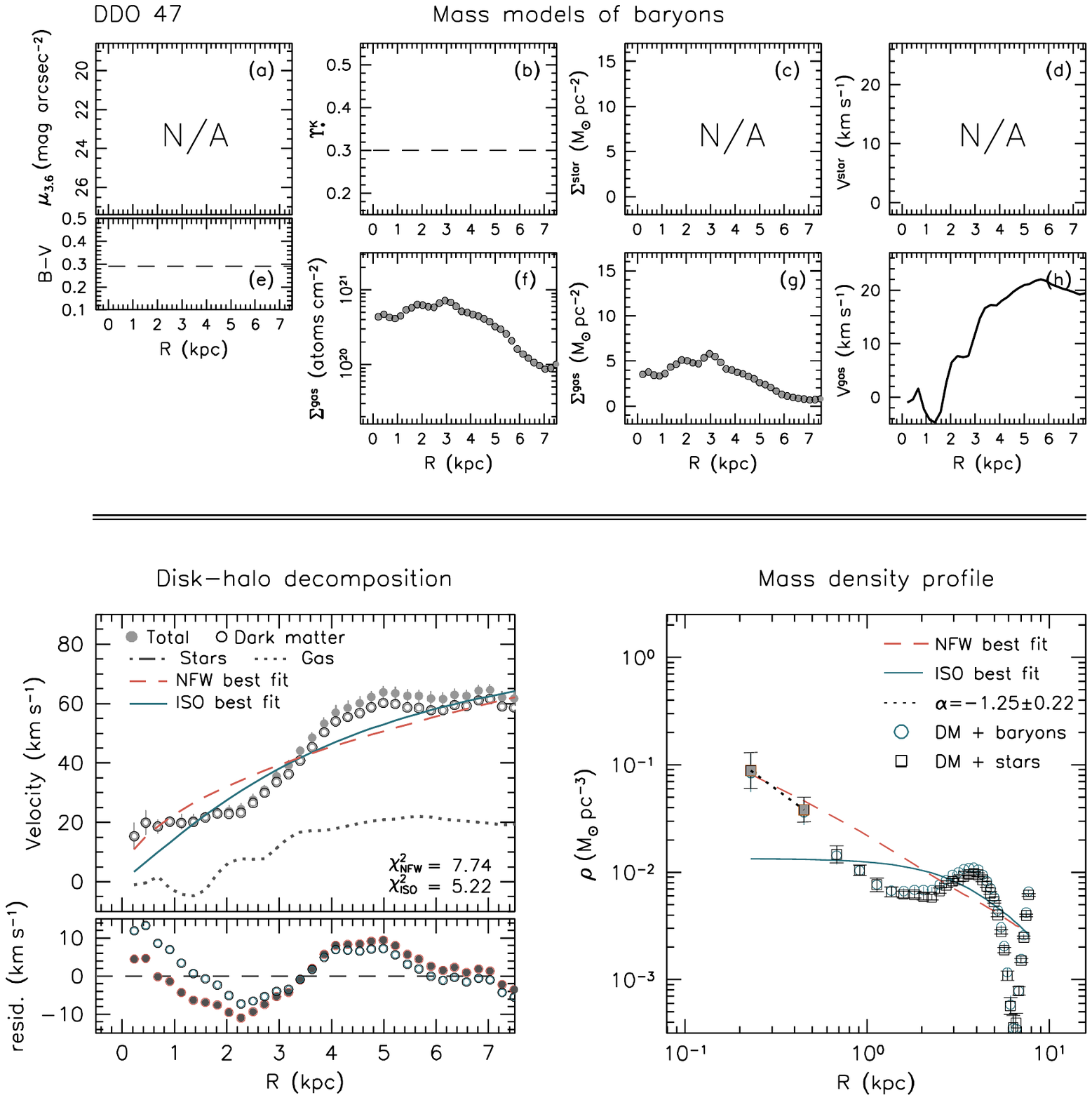}
\caption{The mass models of baryons, disk-halo decomposition and mass density
profile of DDO 47. Please refer to the text in Sections~\ref{MASS_MODELS} and
\ref{DARK_MATTER_DISTRIBUTION} for full information.
\label{MD_DH_DM_ddo47}}
\end{figure}
{\clearpage}

\begin{figure}
\epsscale{1.0}
\figurenum{A.13}
\includegraphics[angle=0,width=1.0\textwidth,bb=60 140 540
745,clip=]{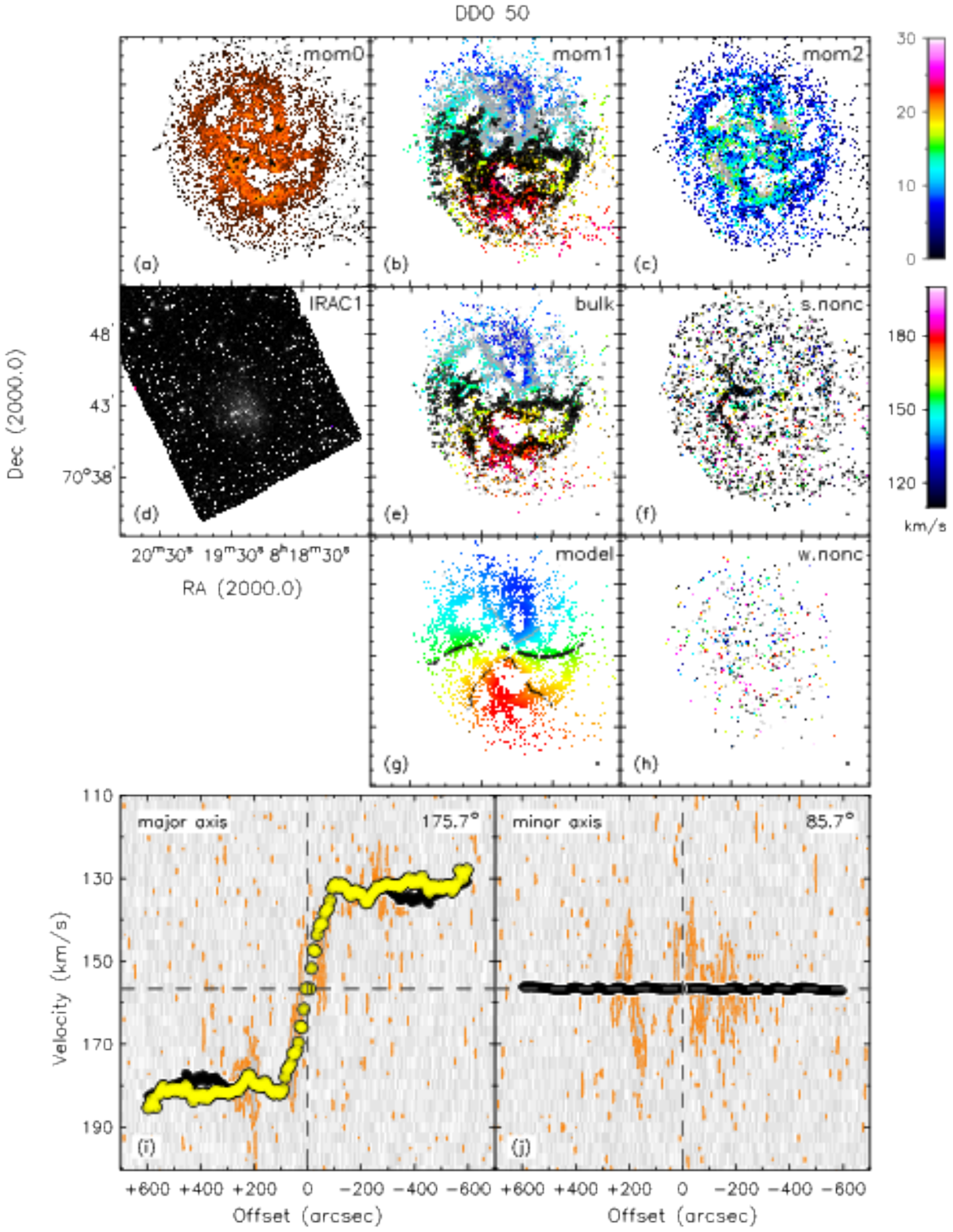}
\caption{H{\sc i} data and {\it Spitzer IRAC} 3.6$\mu$m image of DDO 50. The
systemic velocity is indicated by the thick contours in the velocity fields, and
the iso-velocity contours are spaced by 15 \kms. Velocity dispersion contours run
from 0 to 30 \kms\ with a spacing of 20 \kms. See Appendix section A for details.
\label{ddo50_data_PV}}
\end{figure}
{\clearpage}

\begin{figure}
\epsscale{1.0}
\figurenum{A.14}
\includegraphics[angle=0,width=1.0\textwidth,bb=35 140 570
710,clip=]{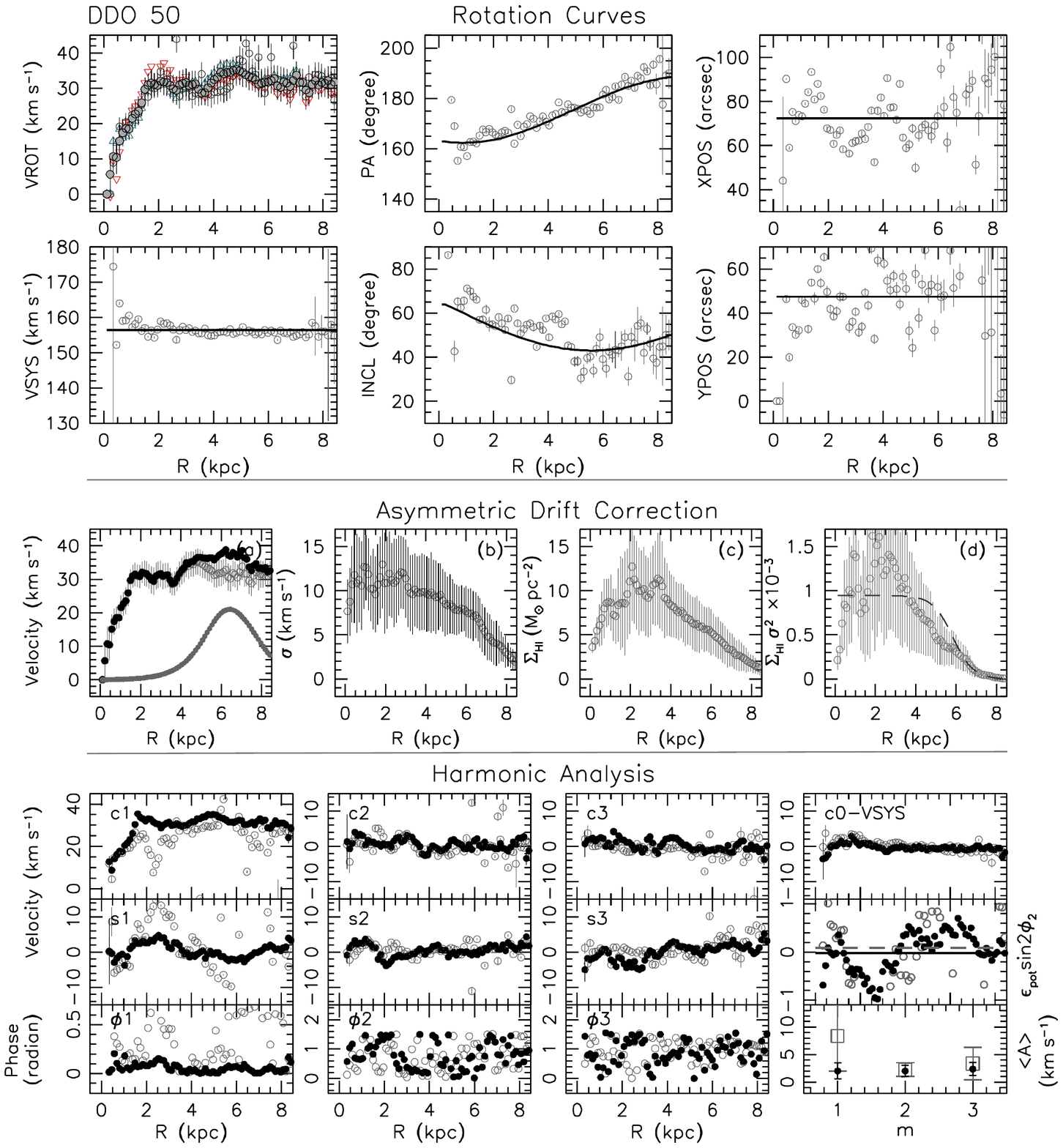}
\caption{Rotation curves, asymmetric drift correction and harmonic analysis
of DDO 50. See Appendix section B for details.
\label{ddo50_TR_HD}}
\end{figure}
{\clearpage}

\begin{figure}
\epsscale{1.0}
\figurenum{A.15}
\includegraphics[angle=0,width=1.0\textwidth,bb=40 175 540
690,clip=]{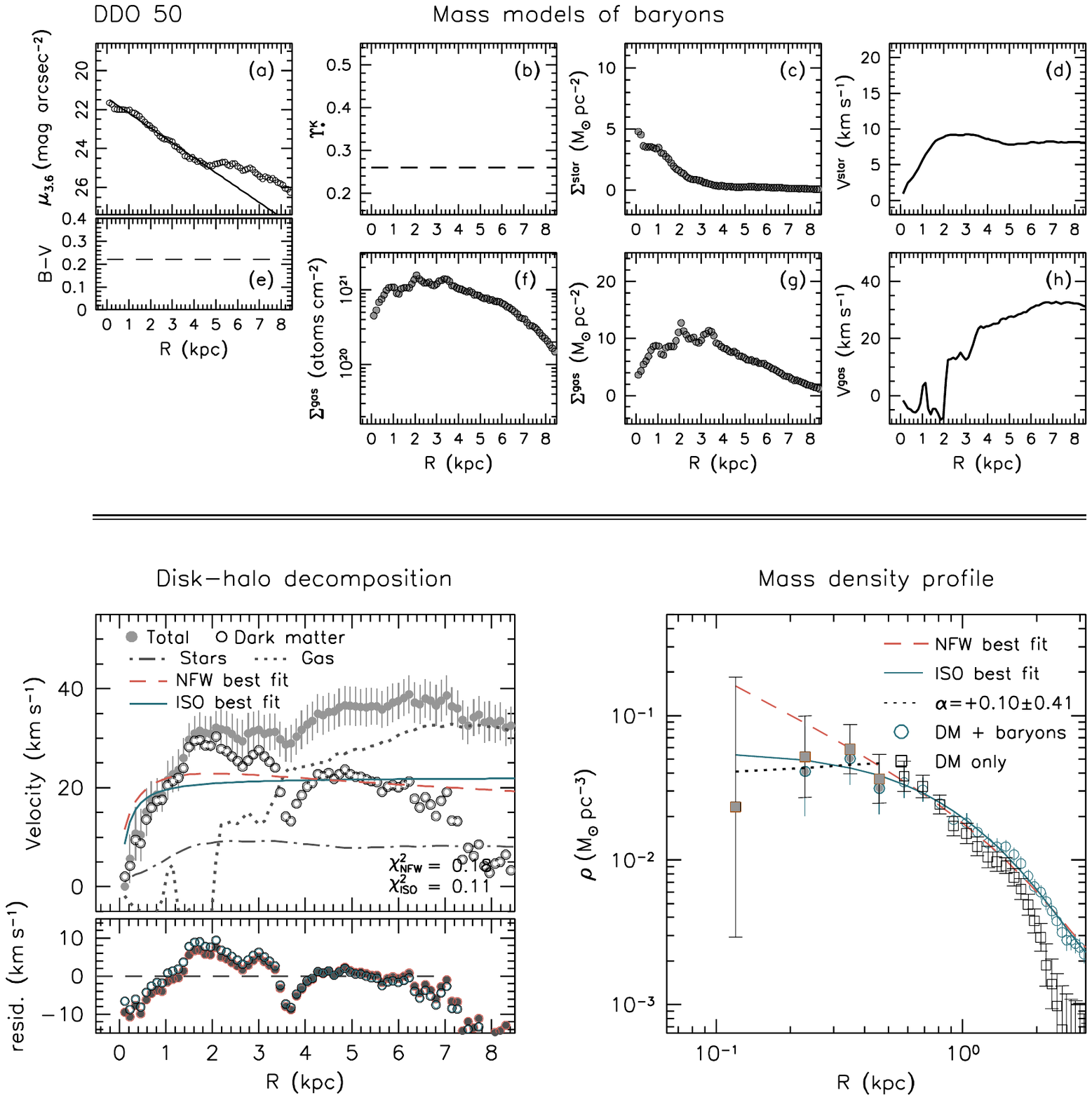}
\caption{The mass models of baryons, disk-halo decomposition and mass density
profile of DDO 50. Please refer to the text in Sections~\ref{MASS_MODELS} and
\ref{DARK_MATTER_DISTRIBUTION} for full information.
\label{MD_DH_DM_ddo50}}
\end{figure}
{\clearpage}

\begin{figure}
\epsscale{1.0}
\figurenum{A.16}
\includegraphics[angle=0,width=1.0\textwidth,bb=60 140 540
745,clip=]{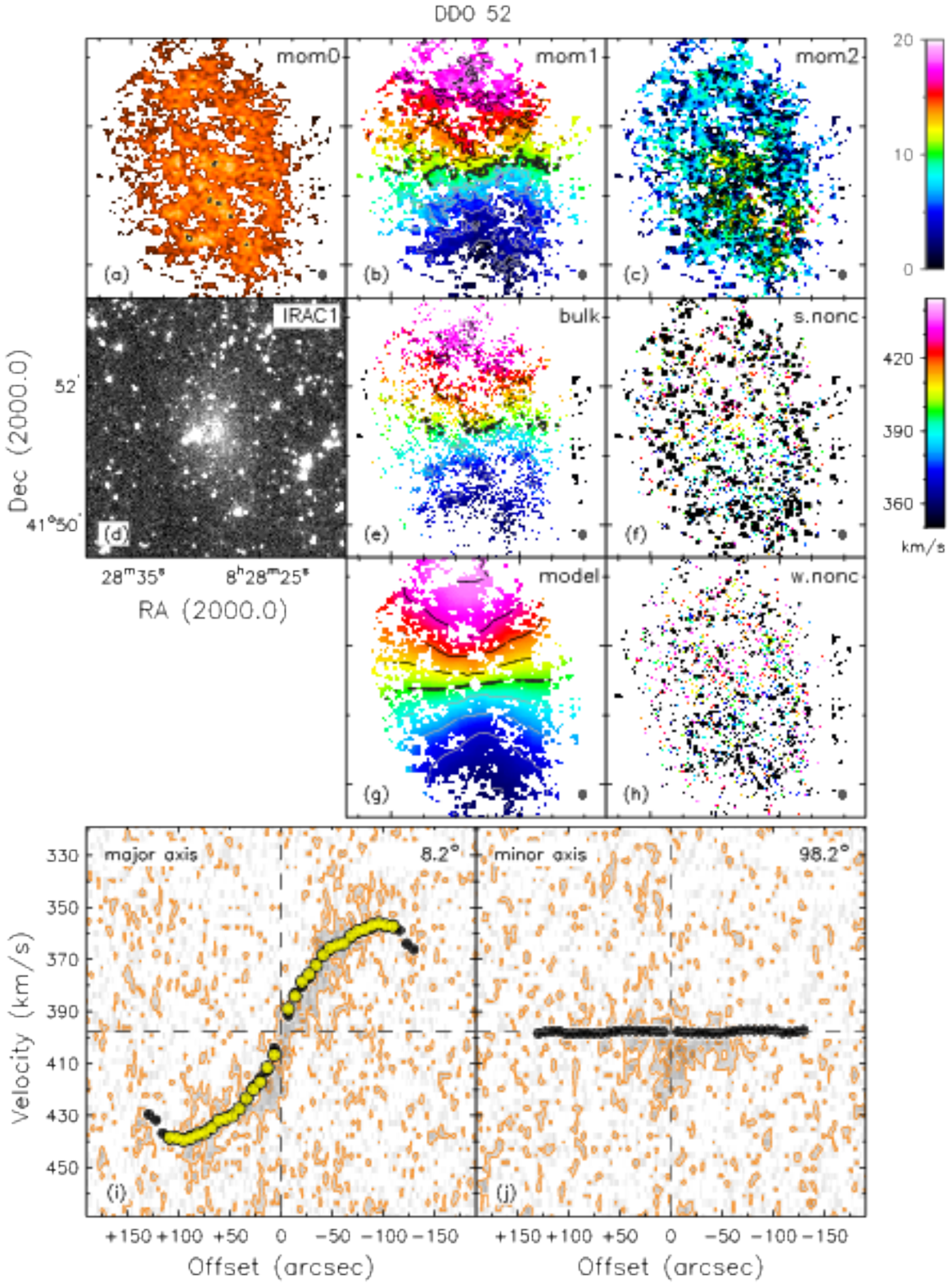}
\caption{H{\sc i} data and {\it Spitzer IRAC} 3.6$\mu$m image of DDO 52. The
systemic velocity is indicated by the thick contours in the velocity fields, and
the iso-velocity contours are spaced by 10 \kms. Velocity dispersion contours run
from 0 to 20 \kms\ with a spacing of 5 \kms. See Appendix section A for details.
\label{ddo52_data_PV}}
\end{figure}
{\clearpage}

\begin{figure}
\epsscale{1.0}
\figurenum{A.17}
\includegraphics[angle=0,width=1.0\textwidth,bb=35 140 570
710,clip=]{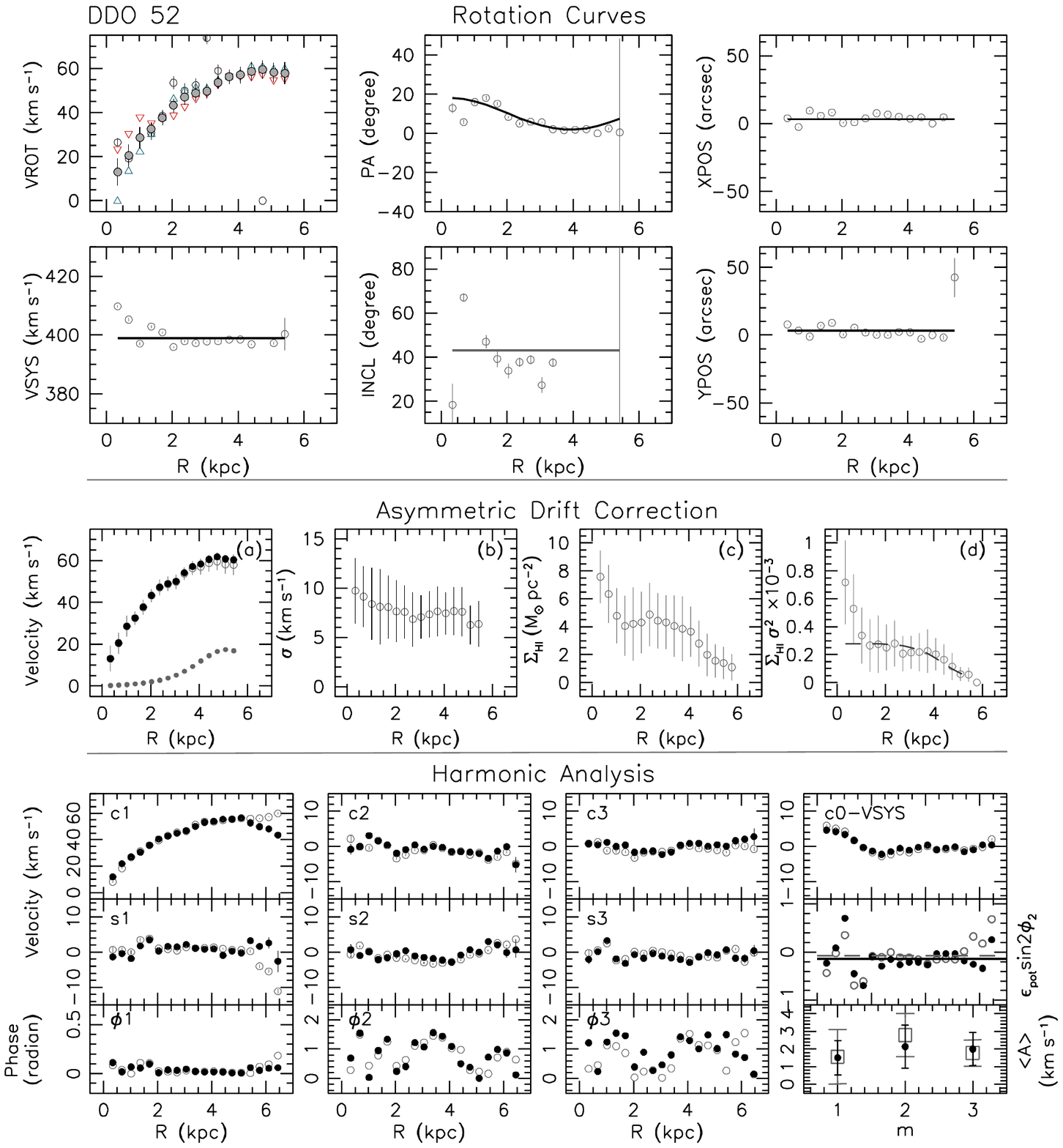}
\caption{Rotation curves, asymmetric drift correction and harmonic analysis
of DDO 52. See Appendix section B for details.
\label{ddo52_TR_HD}}
\end{figure}
{\clearpage}

\begin{figure}
\epsscale{1.0}
\figurenum{A.18}
\includegraphics[angle=0,width=1.0\textwidth,bb=40 175 540
690,clip=]{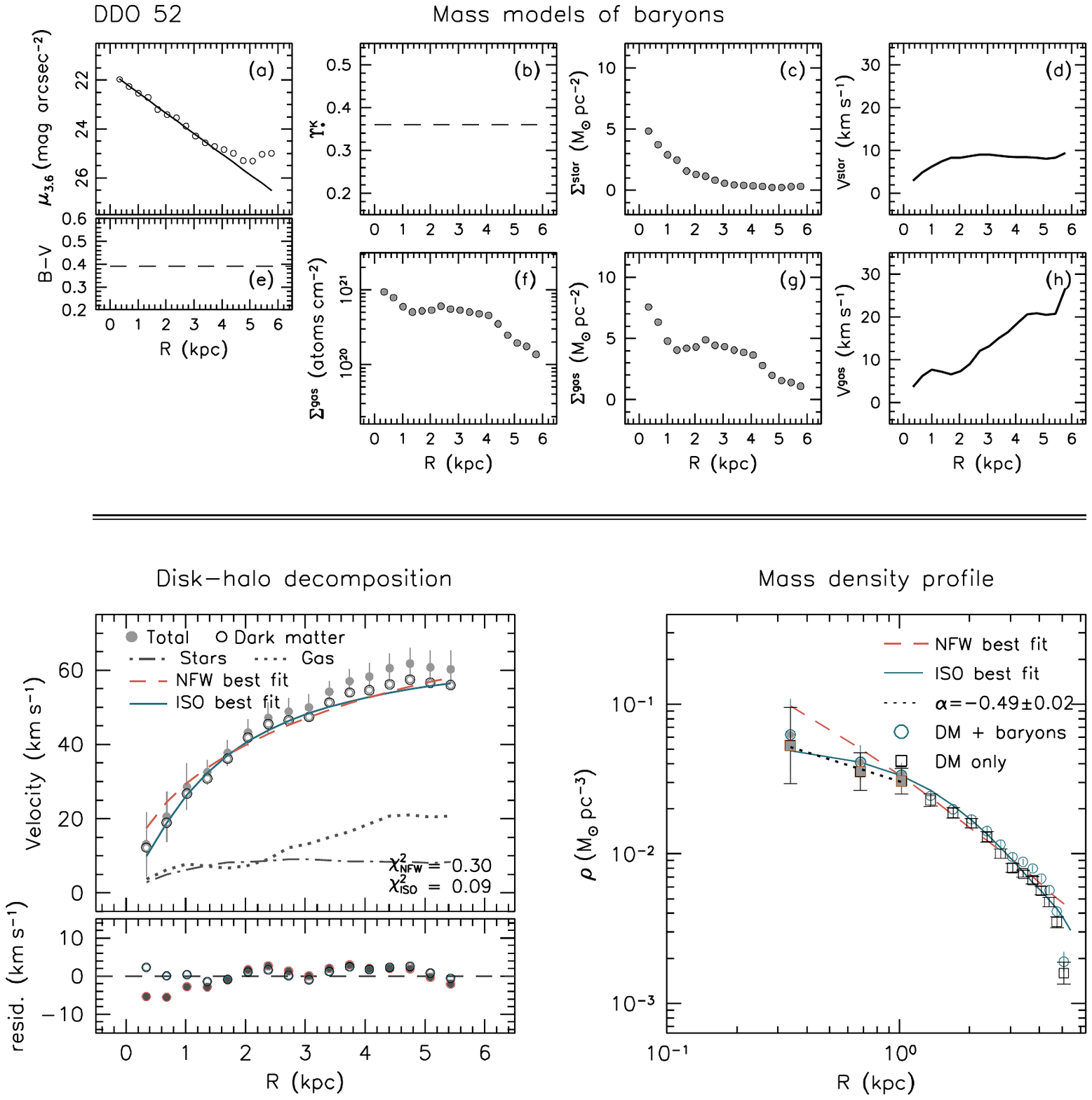}
\caption{The mass models of baryons, disk-halo decomposition and mass density
profile of DDO 52. Please refer to the text in Sections~\ref{MASS_MODELS} and
\ref{DARK_MATTER_DISTRIBUTION} for full information.
\label{MD_DH_DM_ddo52}}
\end{figure}
{\clearpage}

\begin{figure}
\epsscale{1.0}
\figurenum{A.19}
\includegraphics[angle=0,width=1.0\textwidth,bb=60 140 540
745,clip=]{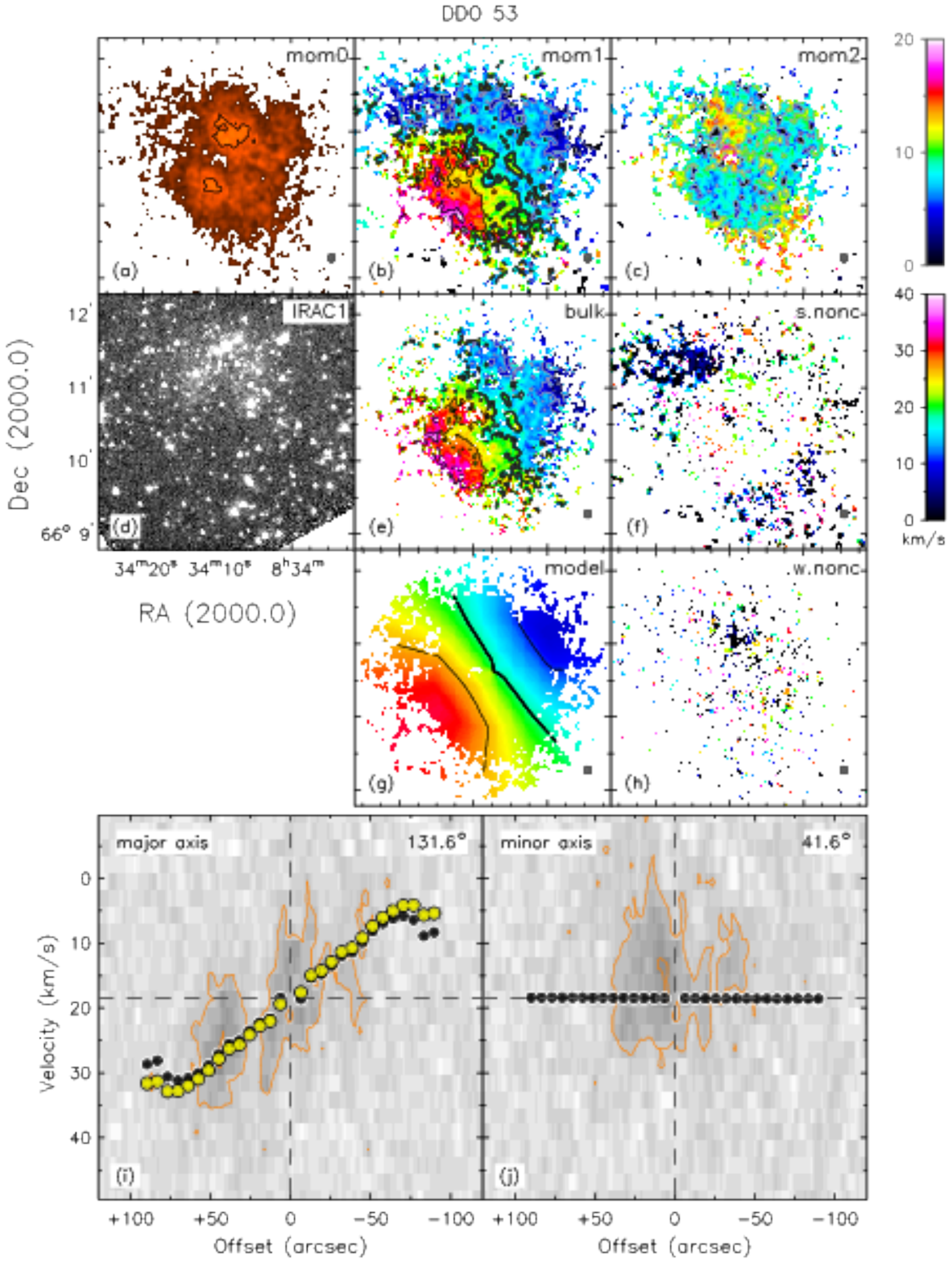}
\caption{H{\sc i} data and {\it Spitzer IRAC} 3.6$\mu$m image of DDO 53. The
systemic velocity is indicated by the thick contours in the velocity fields, and
the iso-velocity contours are spaced by 8 \kms. Velocity dispersion contours run
from 0 to 20 \kms\ with a spacing of 5 \kms. See Appendix section A for details.
\label{ddo53_data_PV}}
\end{figure}
{\clearpage}

\begin{figure}
\epsscale{1.0}
\figurenum{A.20}
\includegraphics[angle=0,width=1.0\textwidth,bb=35 140 570
710,clip=]{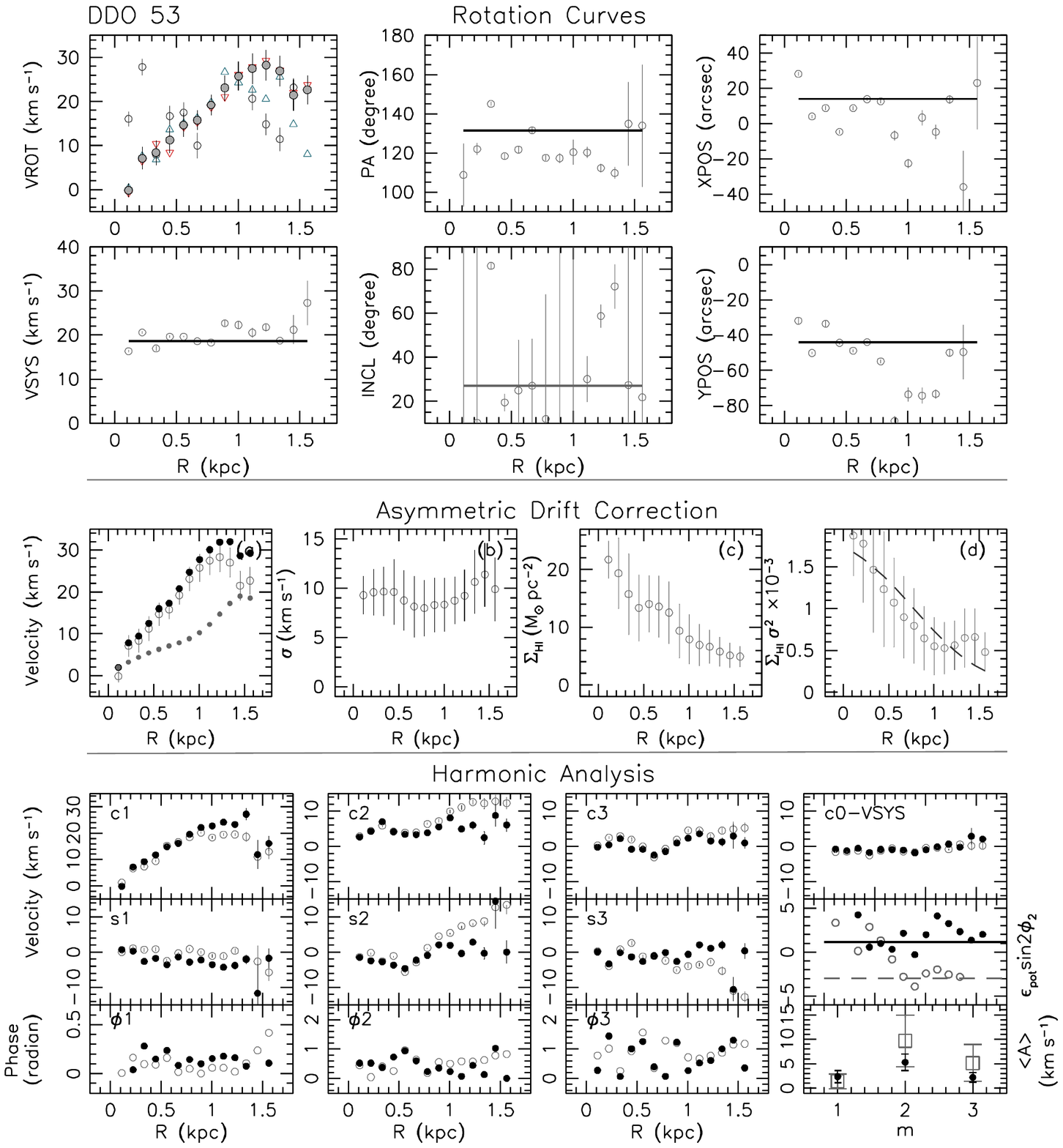}
\caption{Rotation curves, asymmetric drift correction and harmonic analysis
of DDO 53. See Appendix section B for details.
\label{ddo53_TR_HD}}
\end{figure}
{\clearpage}

\begin{figure}
\epsscale{1.0}
\figurenum{A.21}
\includegraphics[angle=0,width=1.0\textwidth,bb=40 175 540
690,clip=]{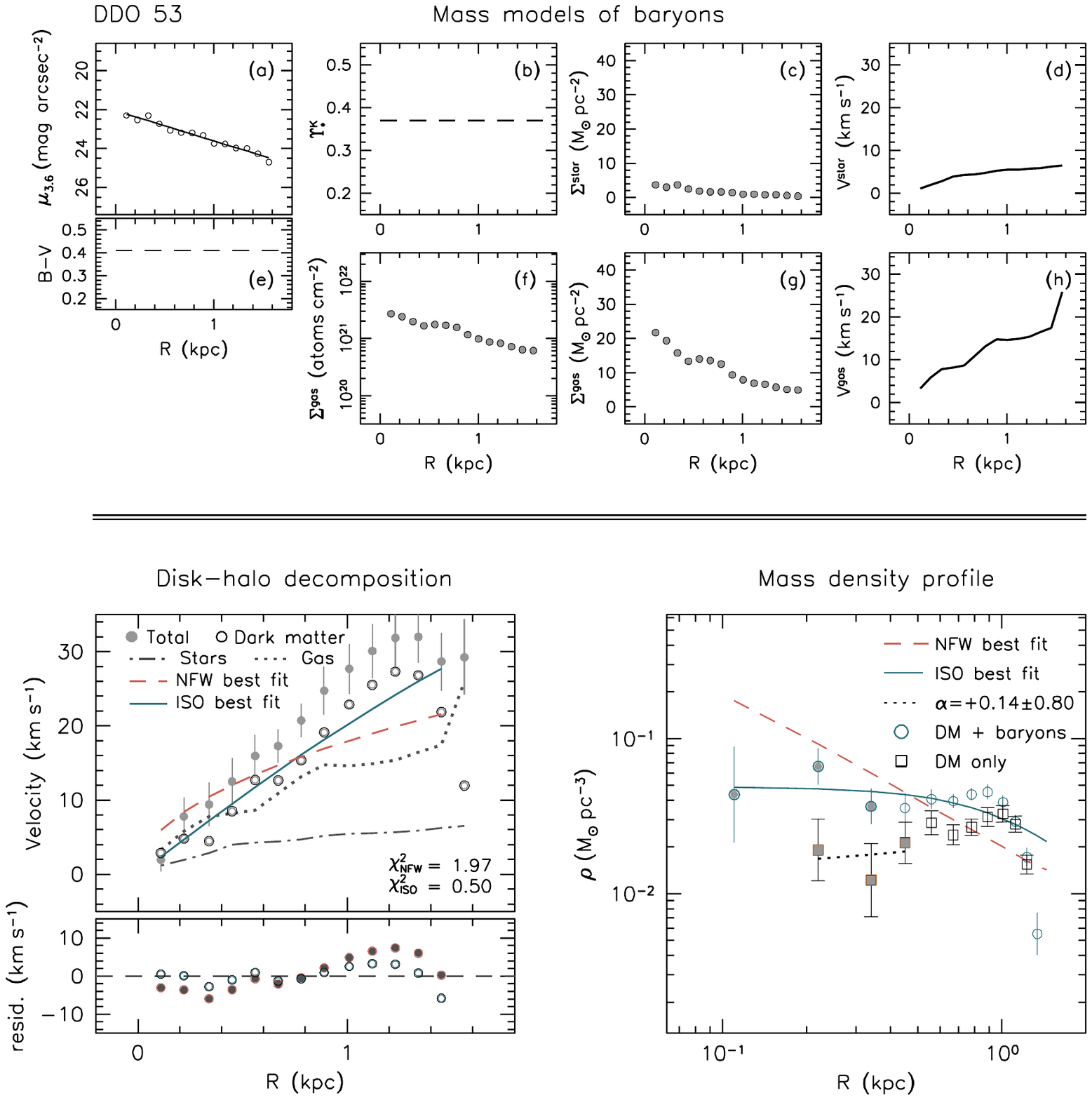}
\caption{The mass models of baryons, disk-halo decomposition and mass density
profile of DDO 53. Please refer to the text in Sections~\ref{MASS_MODELS} and
\ref{DARK_MATTER_DISTRIBUTION} for full information.
\label{MD_DH_DM_ddo53}}
\end{figure}
{\clearpage}

\begin{figure}
\epsscale{1.0}
\figurenum{A.22}
\includegraphics[angle=0,width=1.0\textwidth,bb=60 140 540
745,clip=]{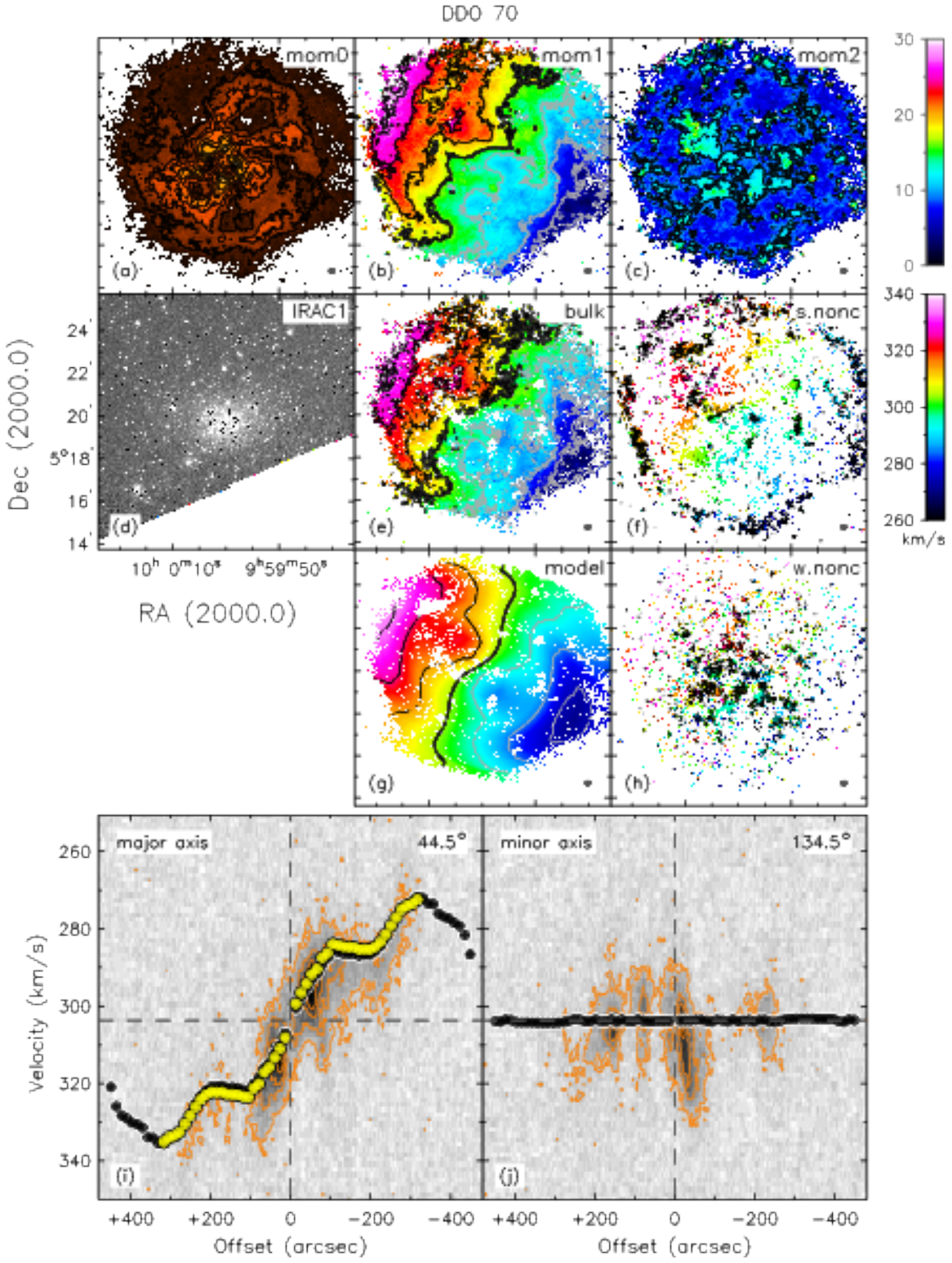}
\caption{H{\sc i} data and {\it Spitzer IRAC} 3.6$\mu$m image of DDO 70. The
systemic velocity is indicated by the thick contours in the velocity fields, and
the iso-velocity contours are spaced by 10 \kms. Velocity dispersion contours run
from 0 to 30 \kms\ with a spacing of 10 \kms. See Appendix section A for details.
\label{ddo70_data_PV}}
\end{figure}
{\clearpage}

\begin{figure}
\epsscale{1.0}
\figurenum{A.23}
\includegraphics[angle=0,width=1.0\textwidth,bb=35 140 570
710,clip=]{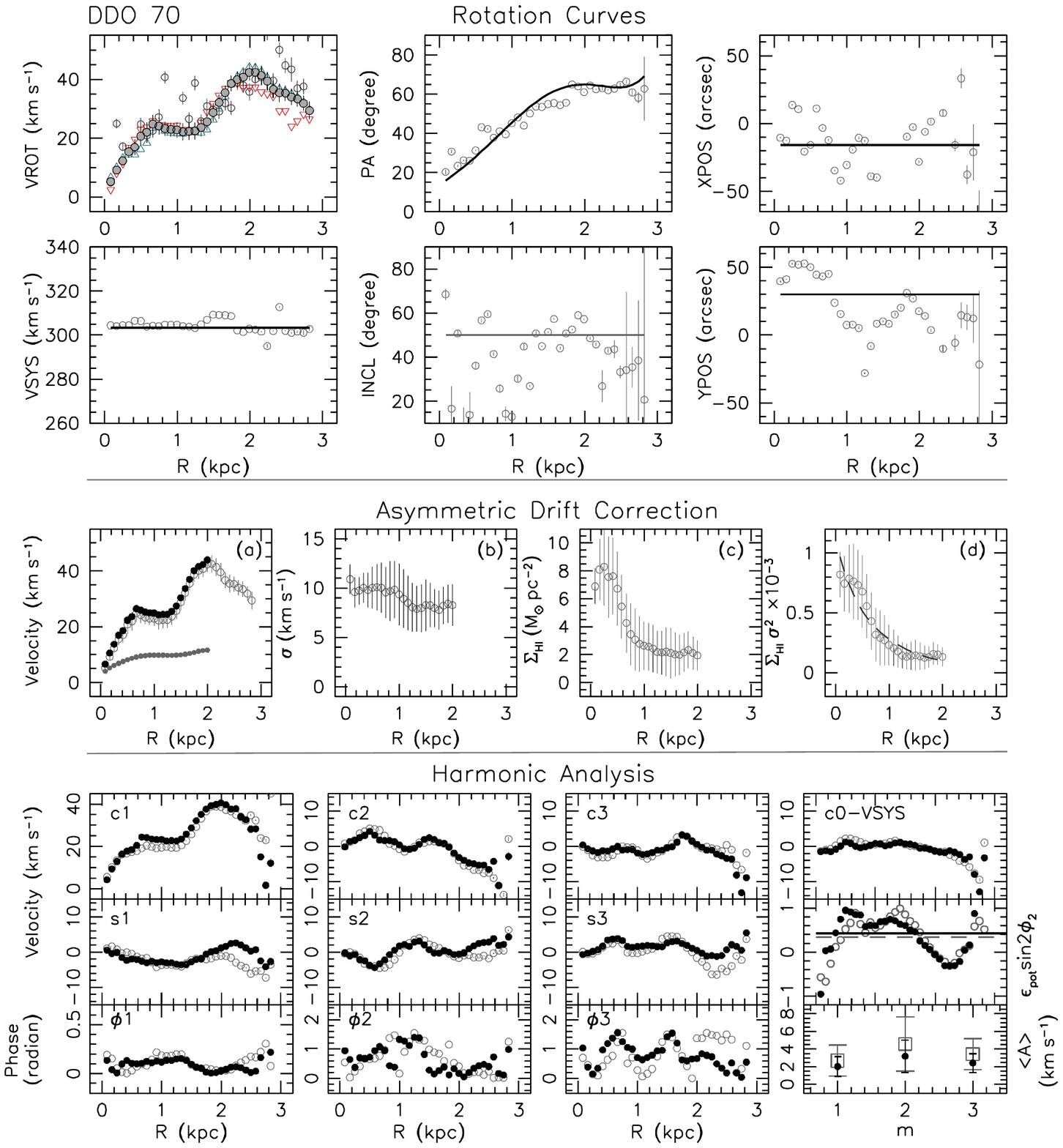}
\caption{Rotation curves, asymmetric drift correction and harmonic analysis
of DDO 70. See Appendix section B for details.
\label{ddo70_TR_HD}}
\end{figure}
{\clearpage}

\begin{figure}
\epsscale{1.0}
\figurenum{A.24}
\includegraphics[angle=0,width=1.0\textwidth,bb=40 175 540
690,clip=]{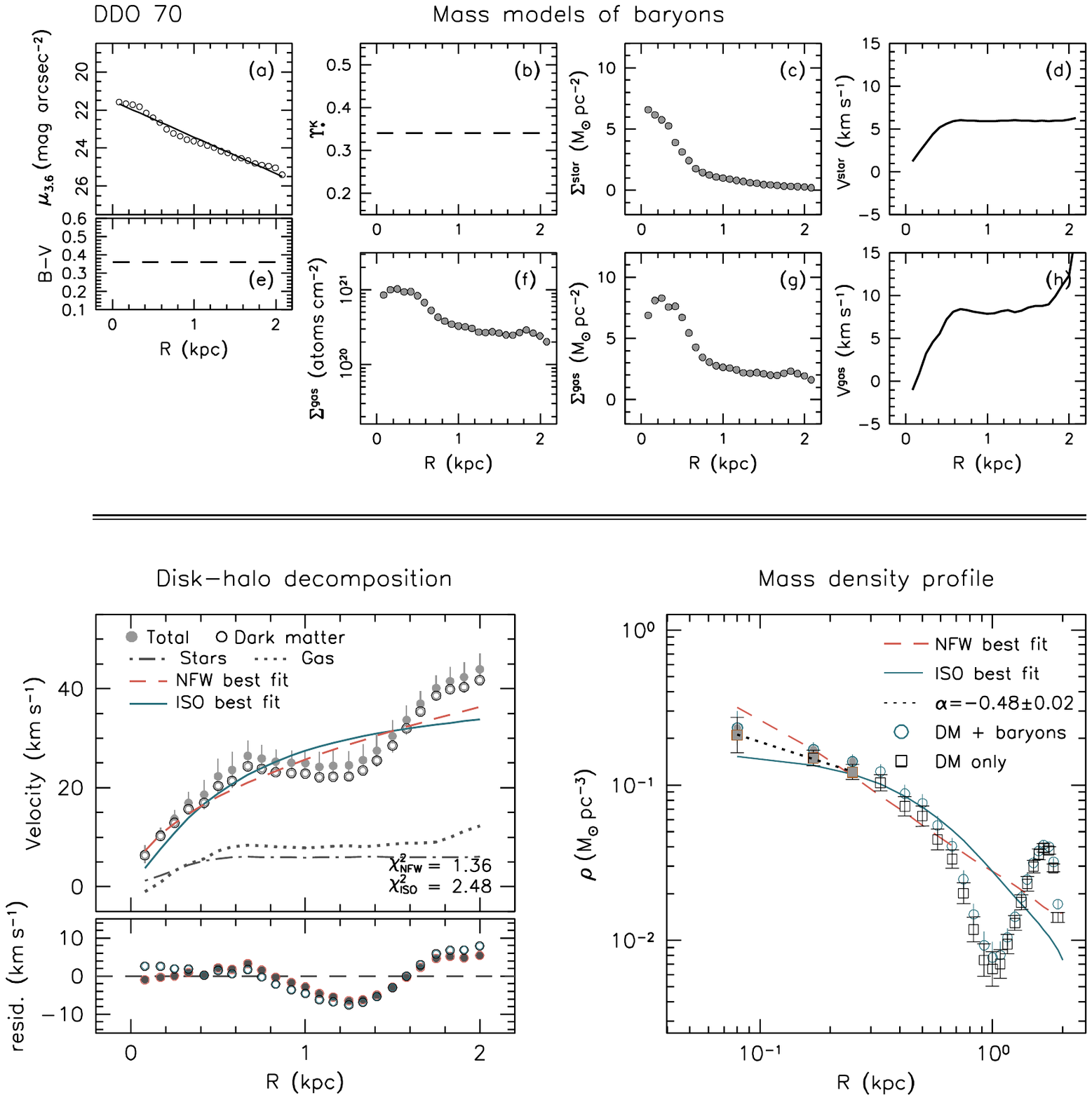}
\caption{The mass models of baryons, disk-halo decomposition and mass density
profile of DDO 70. Please refer to the text in Sections~\ref{MASS_MODELS} and
\ref{DARK_MATTER_DISTRIBUTION} for full information.
\label{MD_DH_DM_ddo70}}
\end{figure}
{\clearpage}

\begin{figure}
\epsscale{1.0}
\figurenum{A.25}
\includegraphics[angle=0,width=1.0\textwidth,bb=60 140 540
745,clip=]{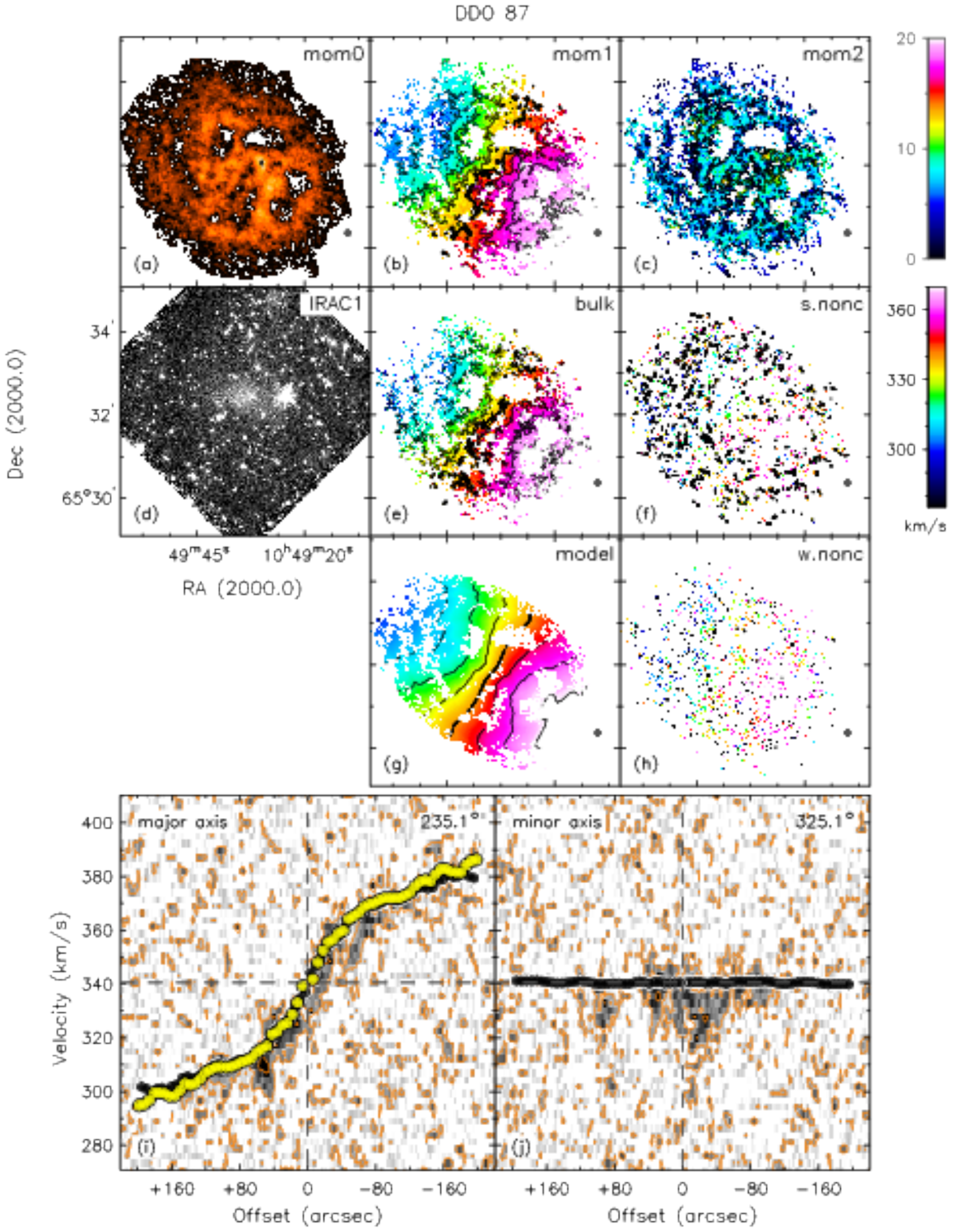}
\caption{H{\sc i} data and {\it Spitzer IRAC} 3.6$\mu$m image of DDO 87. The
systemic velocity is indicated by the thick contours in the velocity fields, and
the iso-velocity contours are spaced by 10 \kms. Velocity dispersion contours run
from 0 to 20 \kms\ with a spacing of 5 \kms. See Appendix section A for details.
\label{ddo87_data_PV}}
\end{figure}
{\clearpage}

\begin{figure}
\epsscale{1.0}
\figurenum{A.26}
\includegraphics[angle=0,width=1.0\textwidth,bb=35 140 570
710,clip=]{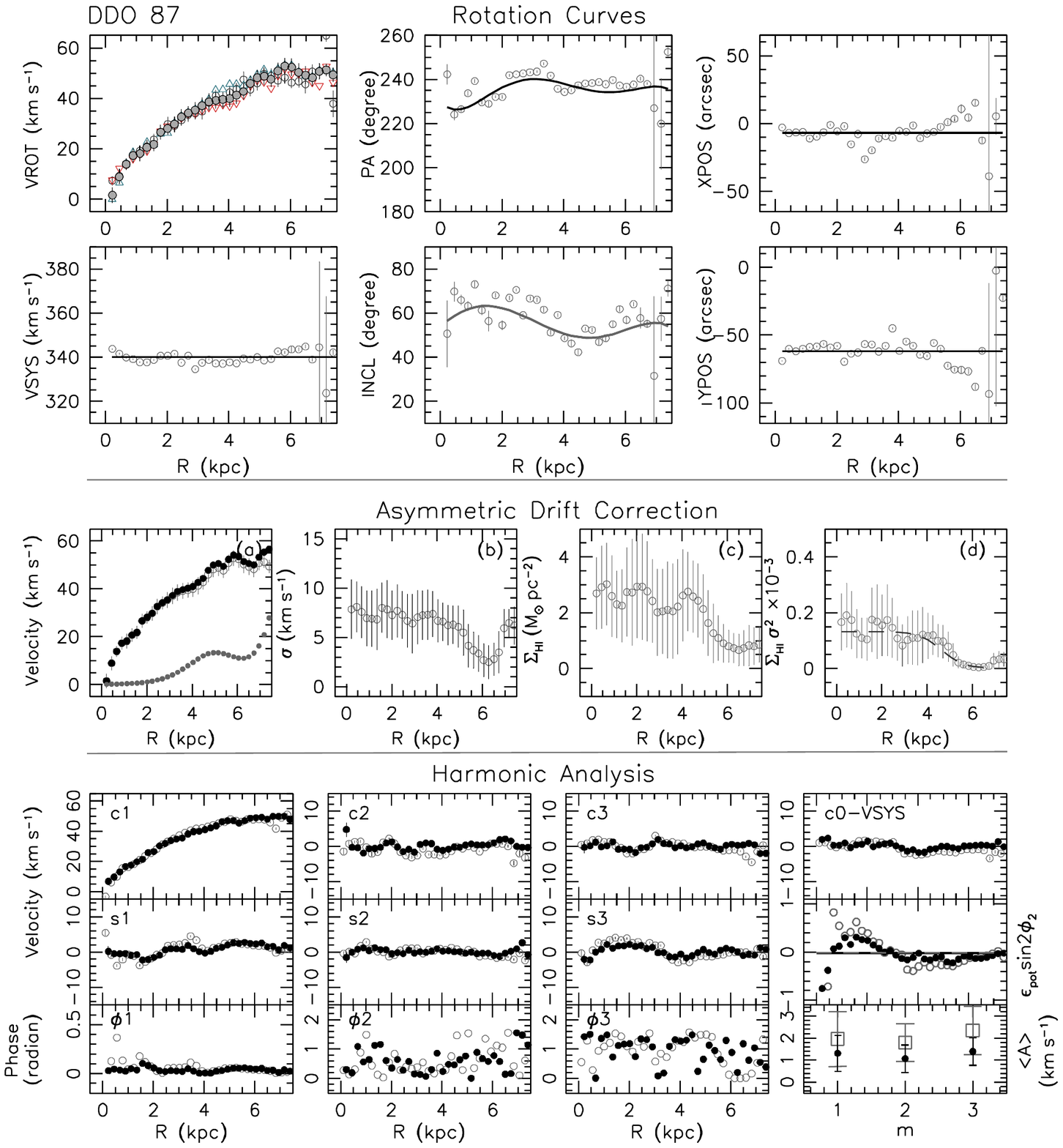}
\caption{Rotation curves, asymmetric drift correction and harmonic analysis
of DDO 87. See Appendix section B for details.
\label{ddo87_TR_HD}}
\end{figure}
{\clearpage}

\begin{figure}
\epsscale{1.0}
\figurenum{A.27}
\includegraphics[angle=0,width=1.0\textwidth,bb=40 175 540
690,clip=]{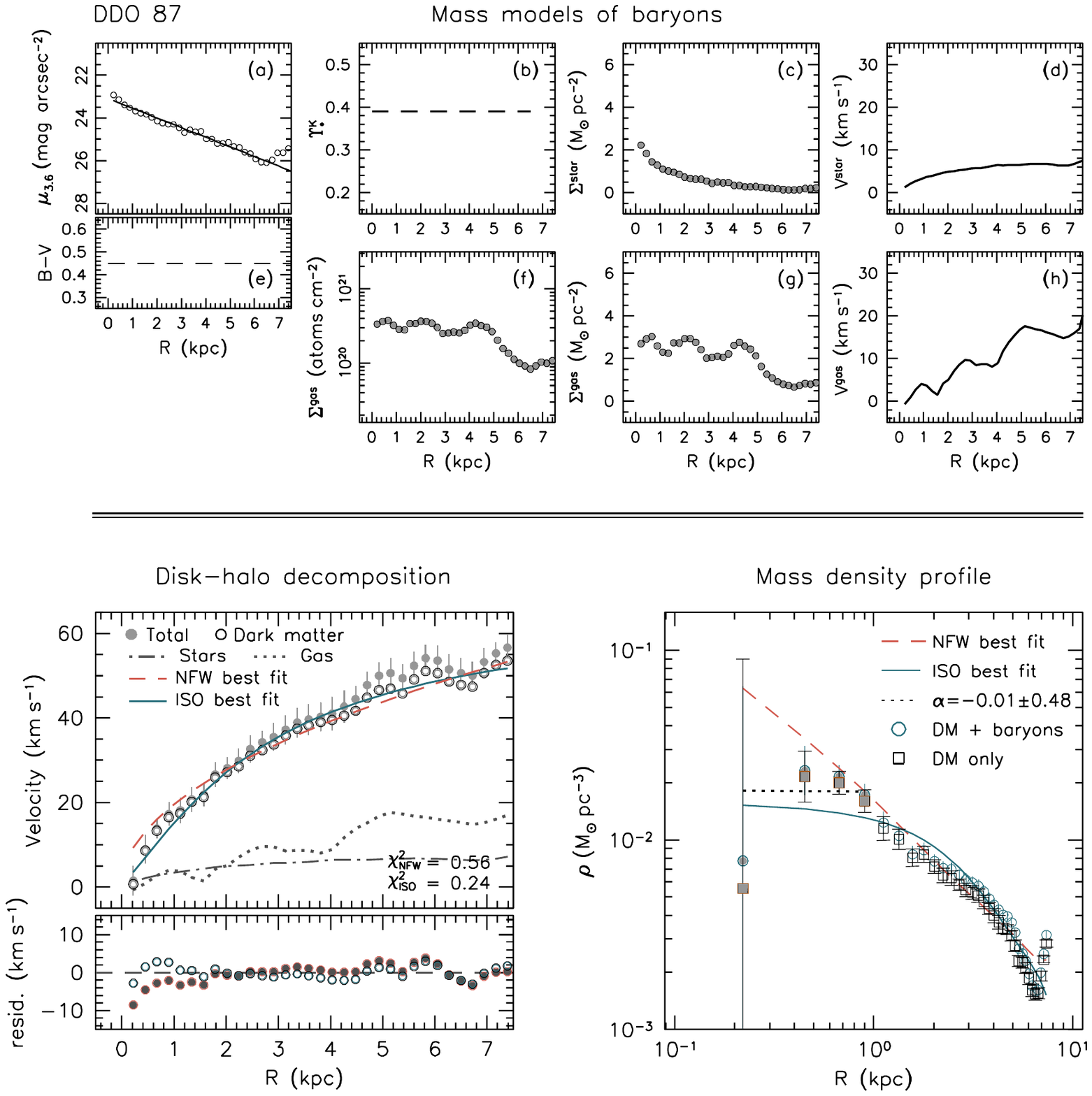}
\caption{The mass models of baryons, disk-halo decomposition and mass density
profile of DDO 87. Please refer to the text in Sections~\ref{MASS_MODELS} and
\ref{DARK_MATTER_DISTRIBUTION} for full information.
\label{MD_DH_DM_ddo87}}
\end{figure}
{\clearpage}

\begin{figure}
\epsscale{1.0}
\figurenum{A.28}
\includegraphics[angle=0,width=1.0\textwidth,bb=60 140 540
745,clip=]{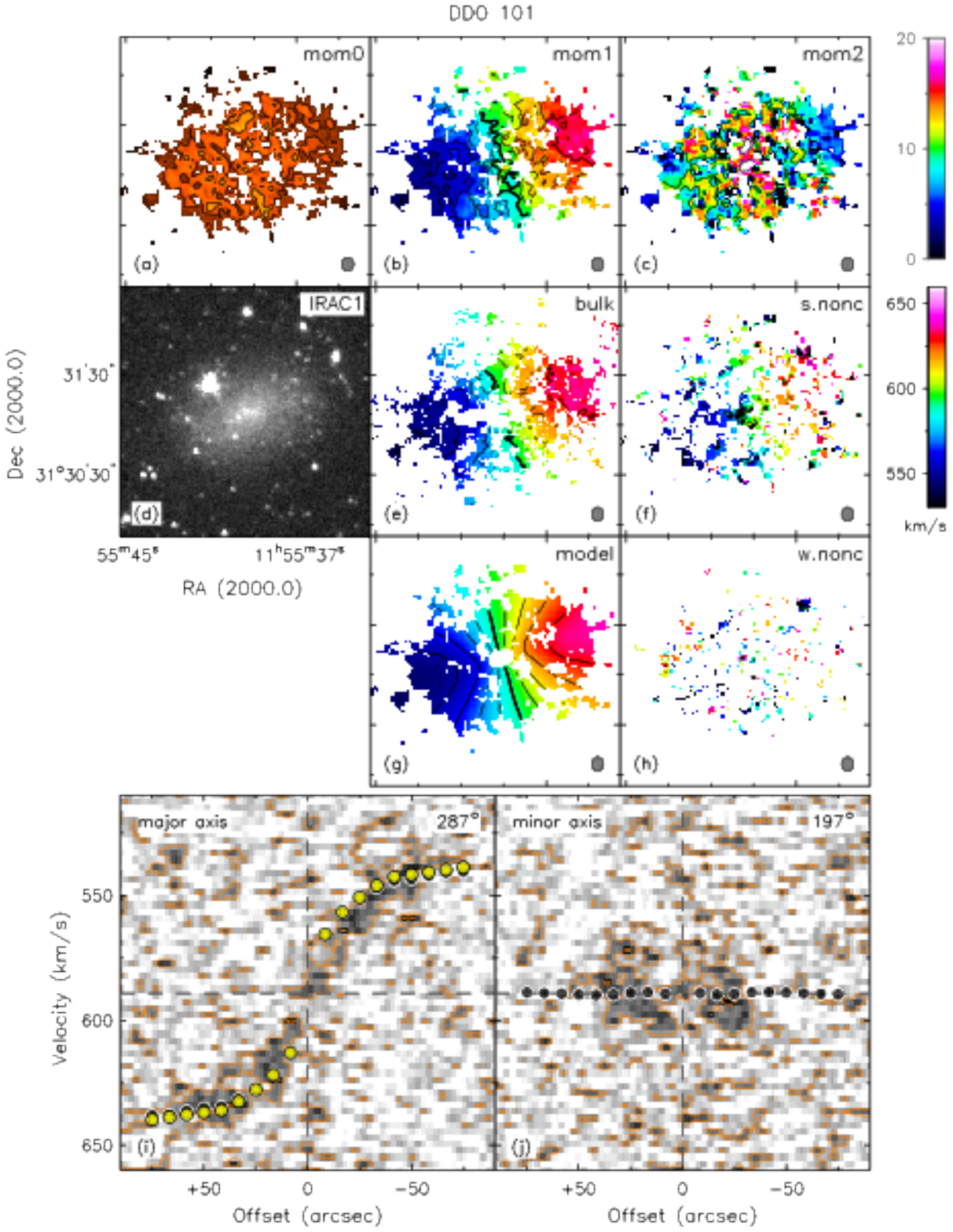}
\caption{H{\sc i} data and {\it Spitzer IRAC} 3.6$\mu$m image of DDO 101. The
systemic velocity is indicated by the thick contours in the velocity fields, and
the iso-velocity contours are spaced by 10 \kms. Velocity dispersion contours run
from 0 to 20 \kms\ with a spacing of 10 \kms. See Appendix section A for details.
\label{ddo101_data_PV}}
\end{figure}
{\clearpage}

\begin{figure}
\epsscale{1.0}
\figurenum{A.29}
\includegraphics[angle=0,width=1.0\textwidth,bb=35 140 570
710,clip=]{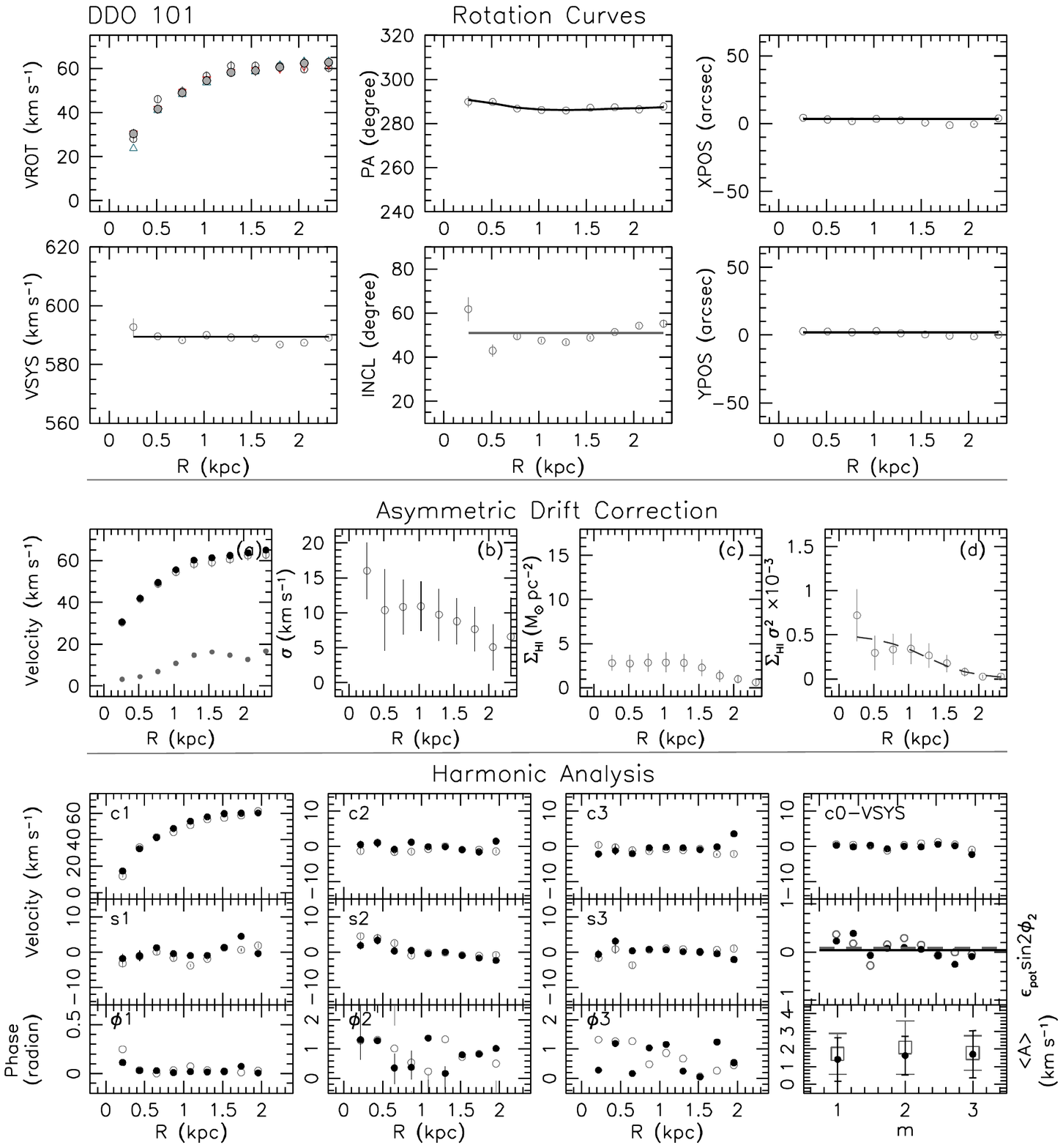}
\caption{Rotation curves, asymmetric drift correction and harmonic analysis
of DDO 101. See Appendix section B for details.
\label{ddo101_TR_HD}}
\end{figure}
{\clearpage}

\begin{figure}
\epsscale{1.0}
\figurenum{A.30}
\includegraphics[angle=0,width=1.0\textwidth,bb=40 175 540
690,clip=]{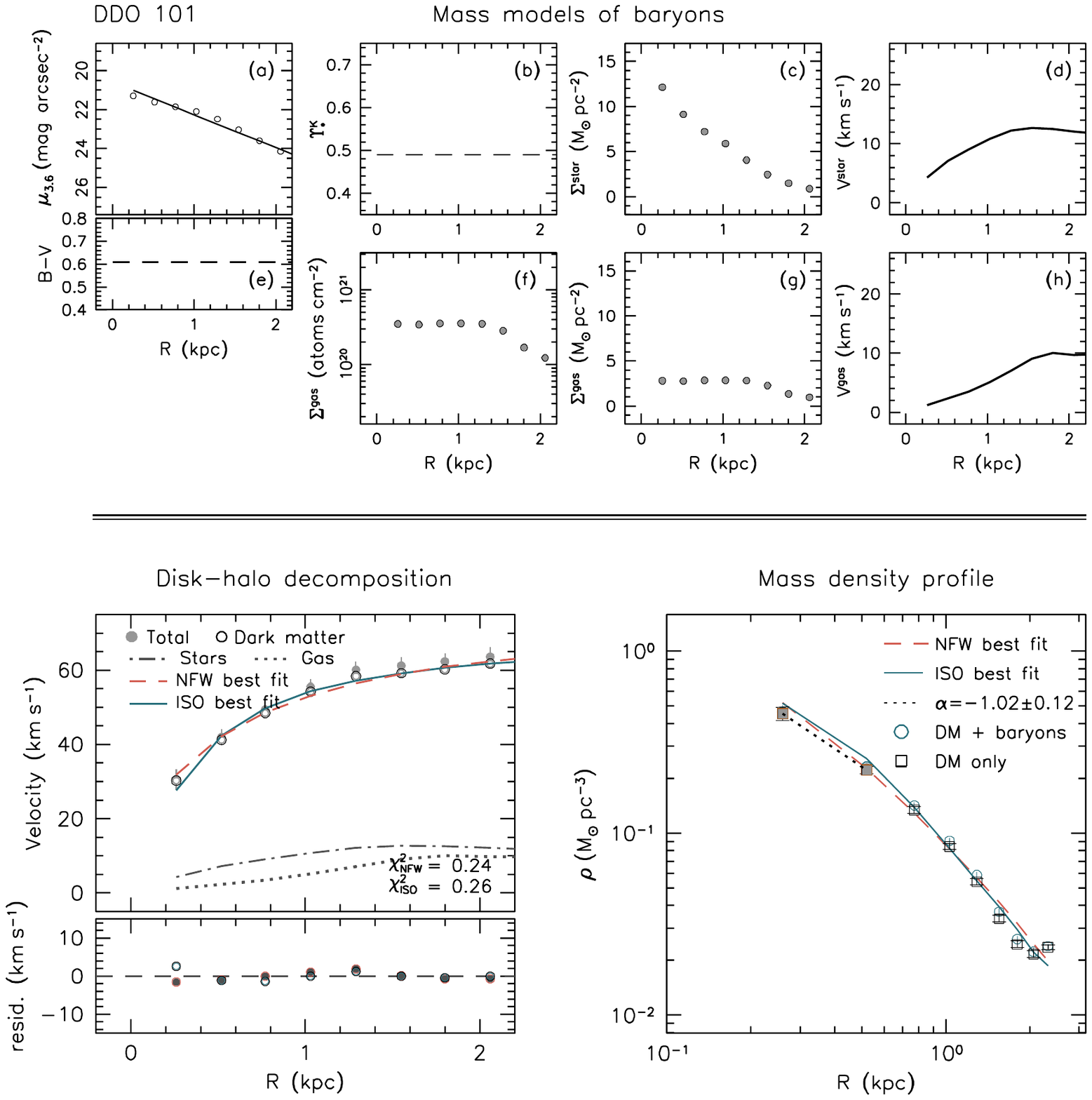}
\caption{The mass models of baryons, disk-halo decomposition and mass density
profile of DDO 101. Please refer to the text in Sections~\ref{MASS_MODELS} and
\ref{DARK_MATTER_DISTRIBUTION} for full information.
\label{MD_DH_DM_ddo101}}
\end{figure}
{\clearpage}

\begin{figure}
\epsscale{1.0}
\figurenum{A.31}
\includegraphics[angle=0,width=1.0\textwidth,bb=60 140 540
745,clip=]{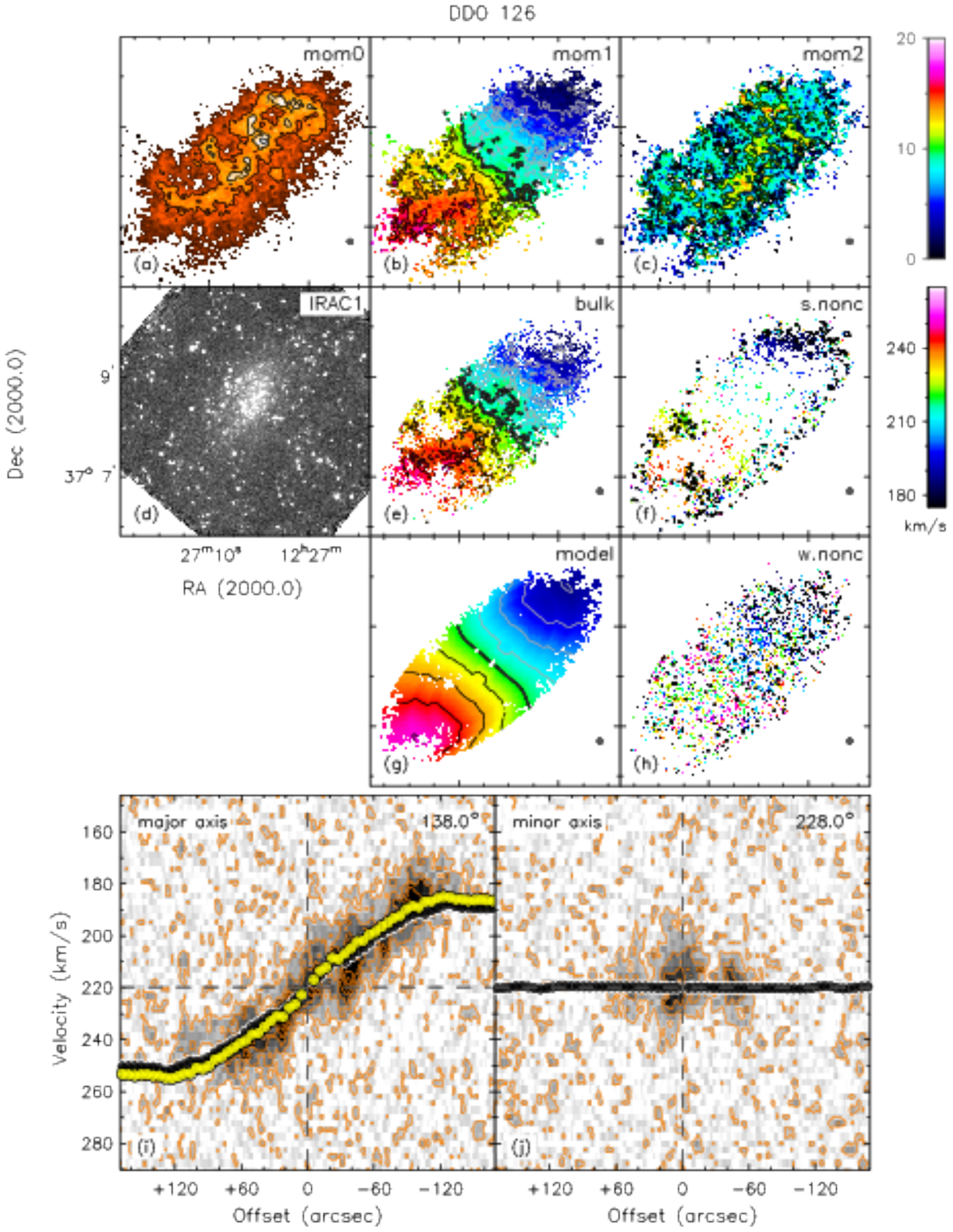}
\caption{H{\sc i} data and {\it Spitzer IRAC} 3.6$\mu$m image of DDO 126. The
systemic velocity is indicated by the thick contours in the velocity fields, and
the iso-velocity contours are spaced by 8 \kms. Velocity dispersion contours run
from 0 to 20 \kms\ with a spacing of 5 \kms. See Appendix section A for details.
\label{ddo126_data_PV}}
\end{figure}
{\clearpage}

\begin{figure}
\epsscale{1.0}
\figurenum{A.32}
\includegraphics[angle=0,width=1.0\textwidth,bb=35 140 570
710,clip=]{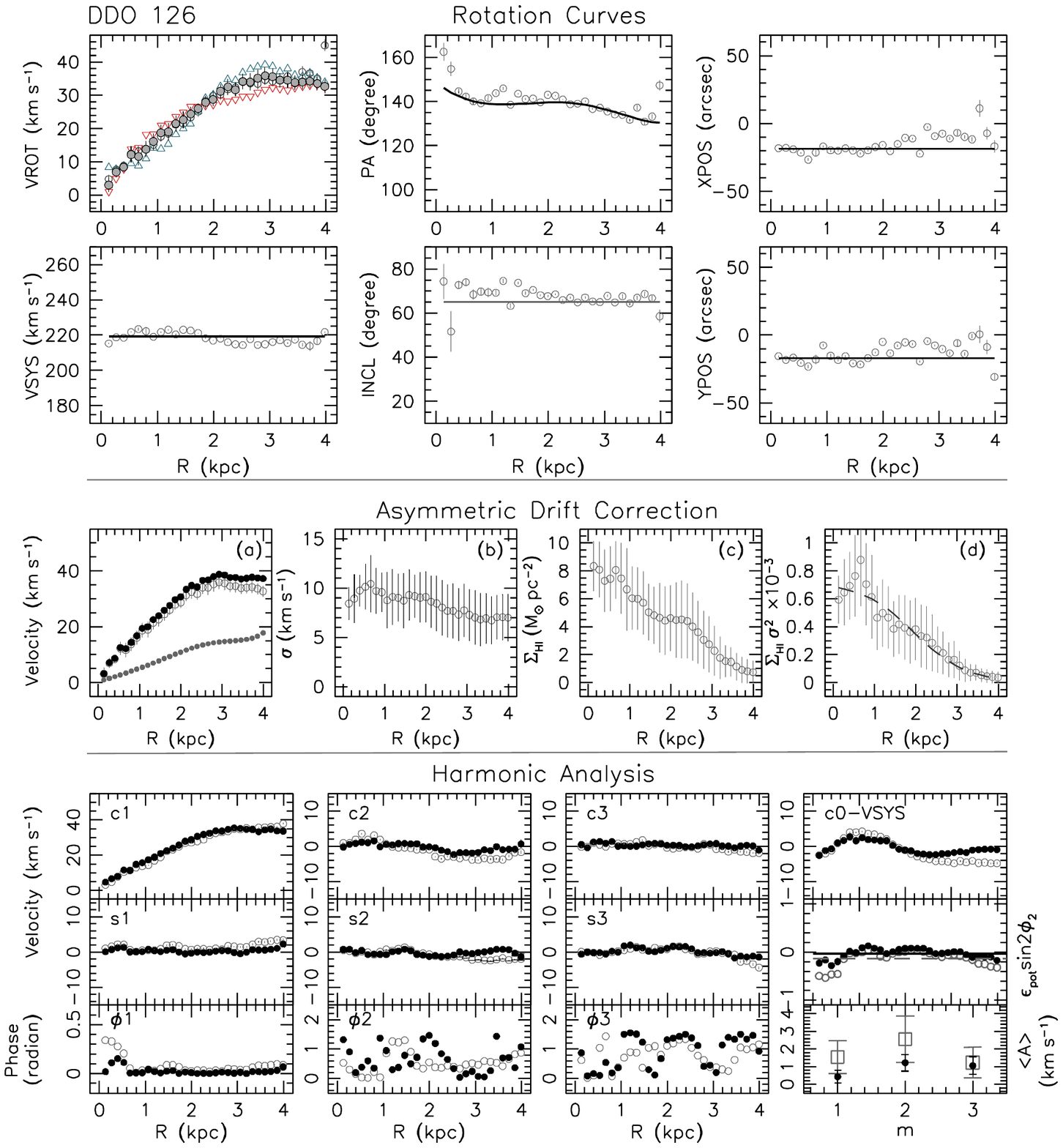}
\caption{Rotation curves, asymmetric drift correction and harmonic analysis
of DDO 126. See Appendix section B for details.
\label{ddo126_TR_HD}}
\end{figure}
{\clearpage}

\begin{figure}
\epsscale{1.0}
\figurenum{A.33}
\includegraphics[angle=0,width=1.0\textwidth,bb=40 175 540
690,clip=]{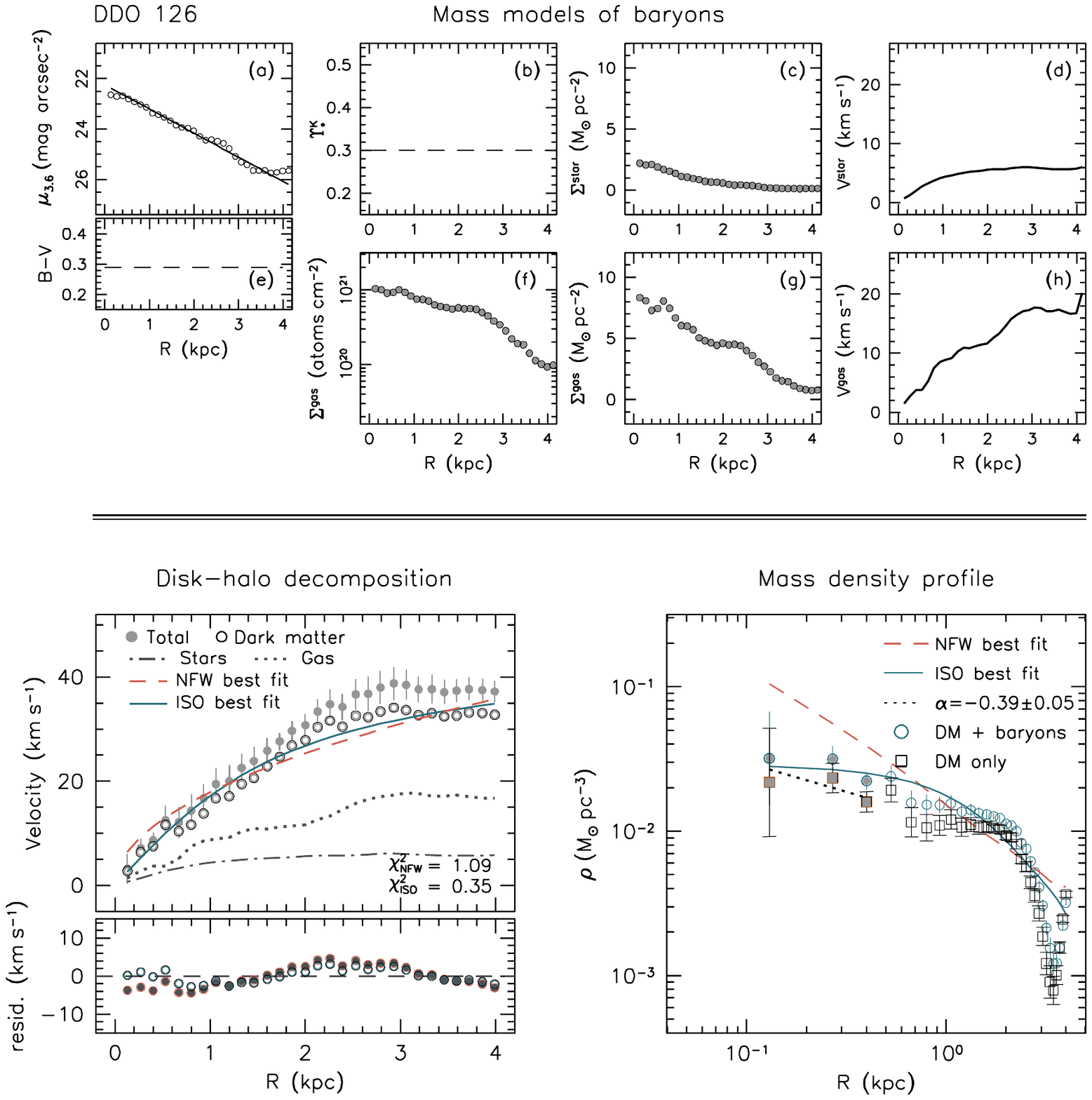}
\caption{The mass models of baryons, disk-halo decomposition and mass density
profile of DDO 126. Please refer to the text in Sections~\ref{MASS_MODELS} and
\ref{DARK_MATTER_DISTRIBUTION} for full information.
\label{MD_DH_DM_ddo126}}
\end{figure}
{\clearpage}

\begin{figure}
\epsscale{1.0}
\figurenum{A.34}
\includegraphics[angle=0,width=1.0\textwidth,bb=60 140 540
745,clip=]{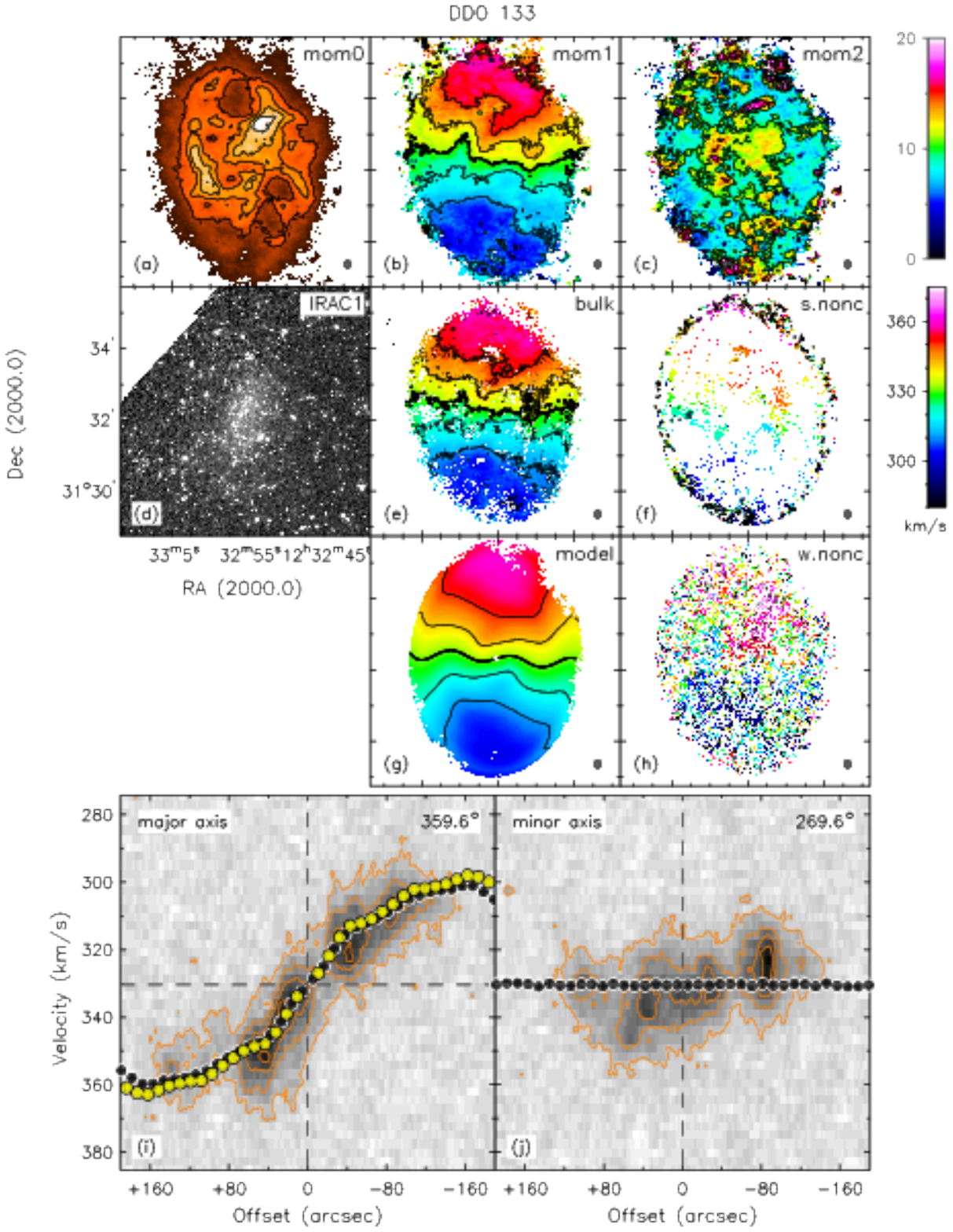}
\caption{H{\sc i} data and {\it Spitzer IRAC} 3.6$\mu$m image of DDO 133. The
systemic velocity is indicated by the thick contours in the velocity fields, and
the iso-velocity contours are spaced by 10 \kms. Velocity dispersion contours run
from 0 to 20 \kms\ with a spacing of 5 \kms. See Appendix section A for details.
\label{ddo133_data_PV}}
\end{figure}
{\clearpage}

\begin{figure}
\epsscale{1.0}
\figurenum{A.35}
\includegraphics[angle=0,width=1.0\textwidth,bb=35 140 570
710,clip=]{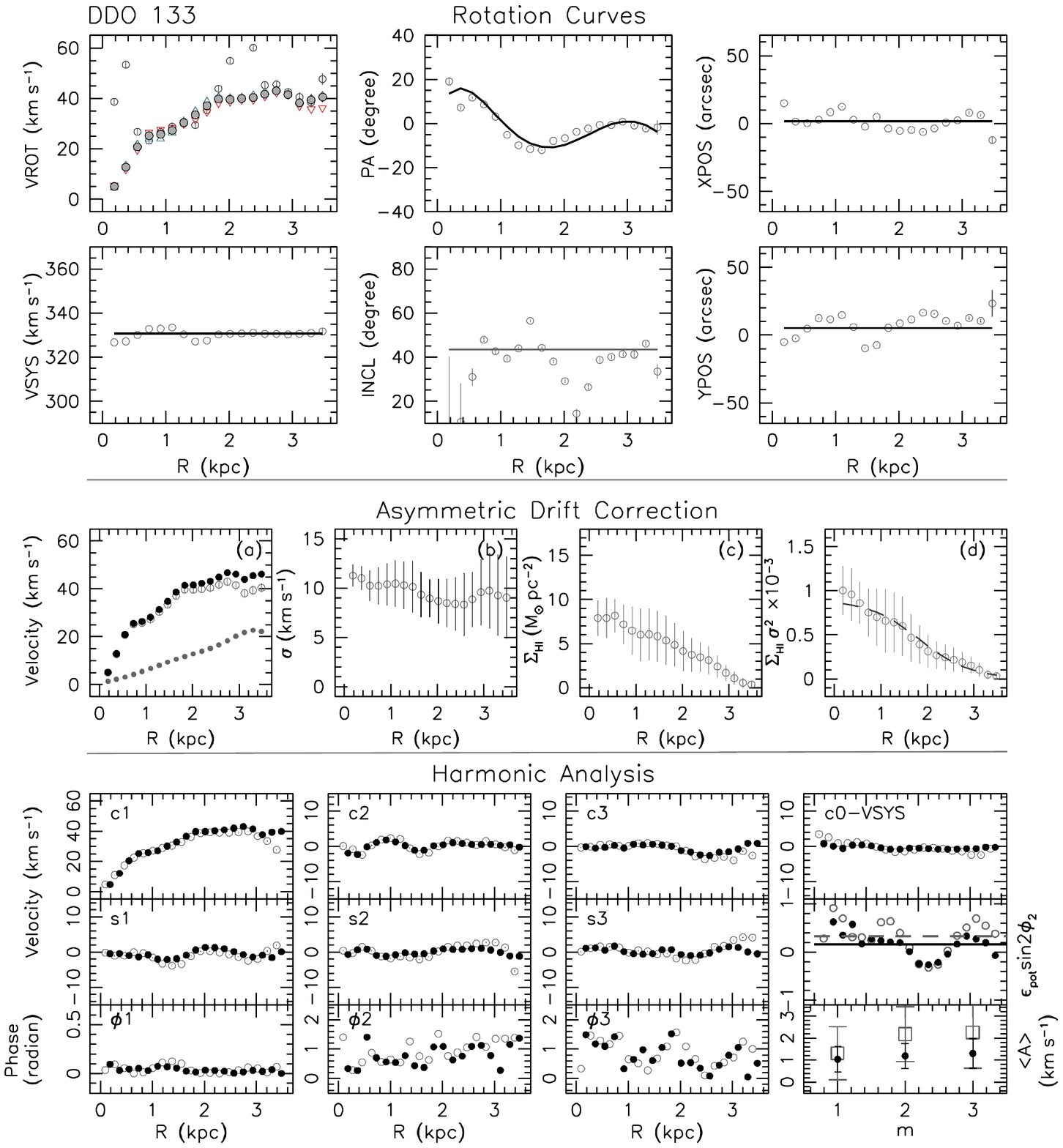}
\caption{Rotation curves, asymmetric drift correction and harmonic analysis
of DDO 133. See Appendix section B for details.
\label{ddo133_TR_HD}}
\end{figure}
{\clearpage}

\begin{figure}
\epsscale{1.0}
\figurenum{A.36}
\includegraphics[angle=0,width=1.0\textwidth,bb=40 175 540
690,clip=]{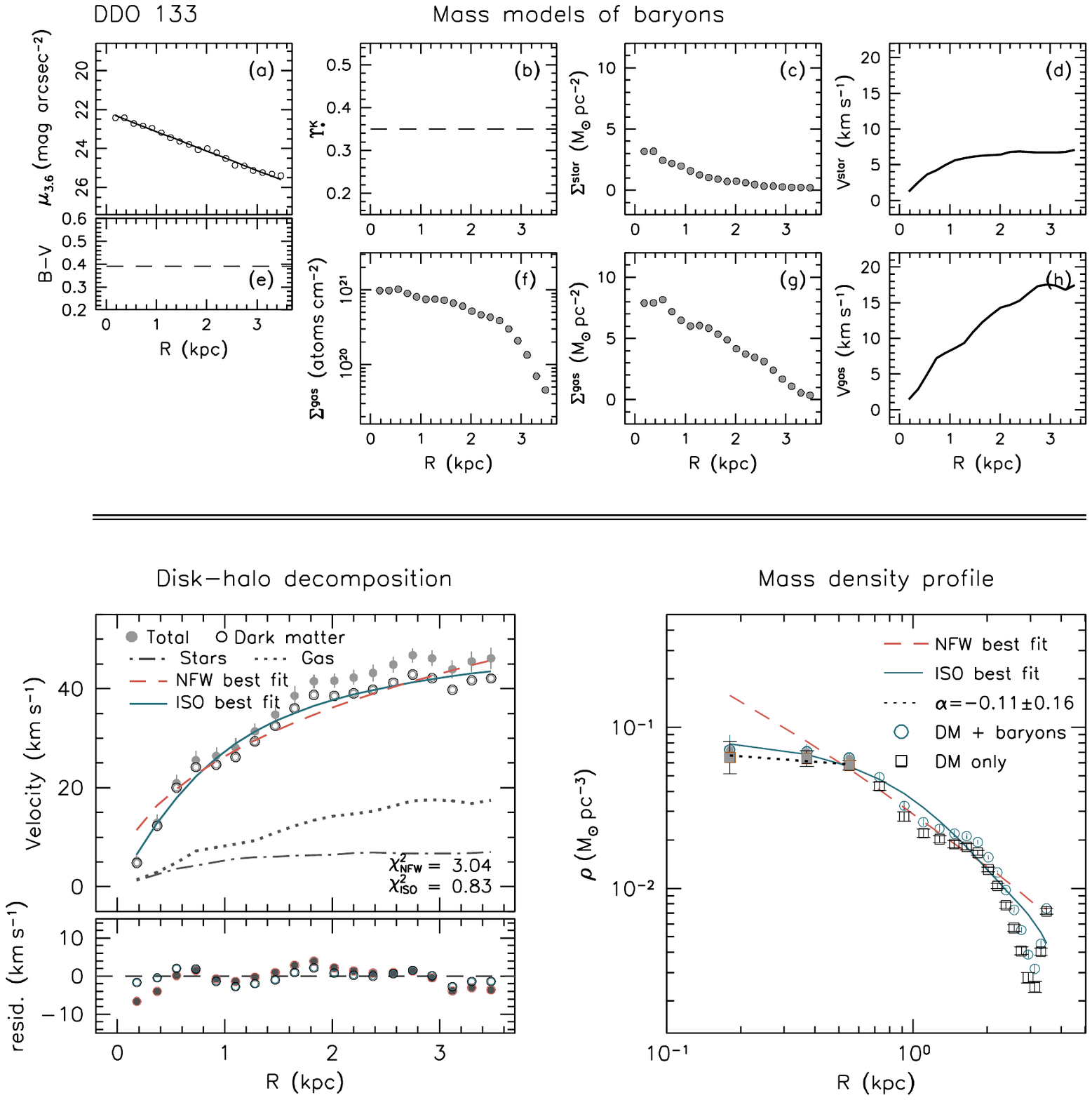}
\caption{The mass models of baryons, disk-halo decomposition and mass density
profile of DDO 133. Please refer to the text in Sections~\ref{MASS_MODELS} and
\ref{DARK_MATTER_DISTRIBUTION} for full information.
\label{MD_DH_DM_ddo133}}
\end{figure}
{\clearpage}

\begin{figure}
\epsscale{1.0}
\figurenum{A.37}
\includegraphics[angle=0,width=1.0\textwidth,bb=60 140 540
745,clip=]{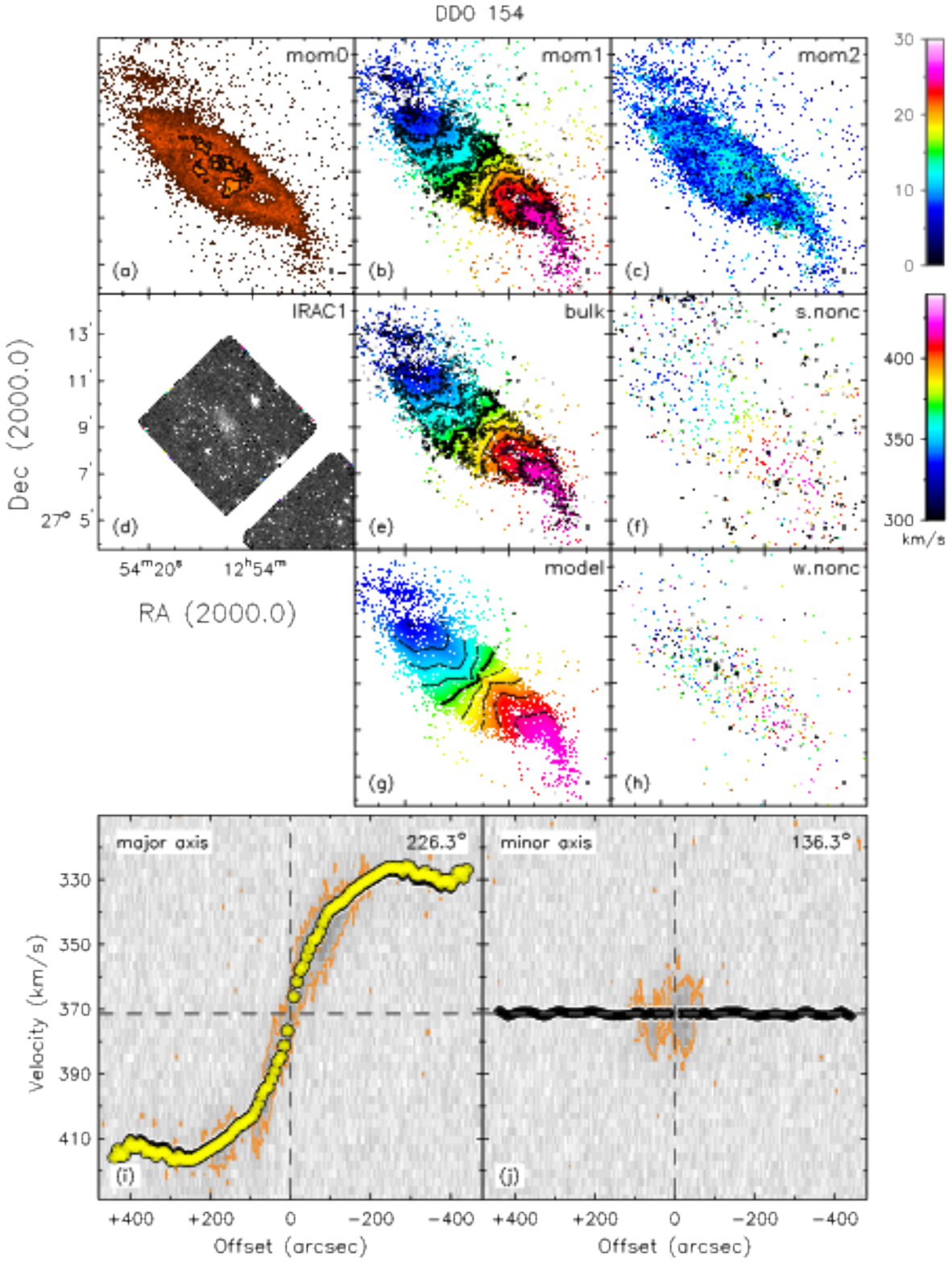}
\caption{H{\sc i} data and {\it Spitzer IRAC} 3.6$\mu$m image of DDO 154. The
systemic velocity is indicated by the thick contours in the velocity fields, and
the iso-velocity contours are spaced by 10 \kms. Velocity dispersion contours run
from 0 to 20 \kms\ with a spacing of 15 \kms. See Appendix section A for details.
\label{ddo154_data_PV}}
\end{figure}
{\clearpage}

\begin{figure}
\epsscale{1.0}
\figurenum{A.38}
\includegraphics[angle=0,width=1.0\textwidth,bb=35 140 570
710,clip=]{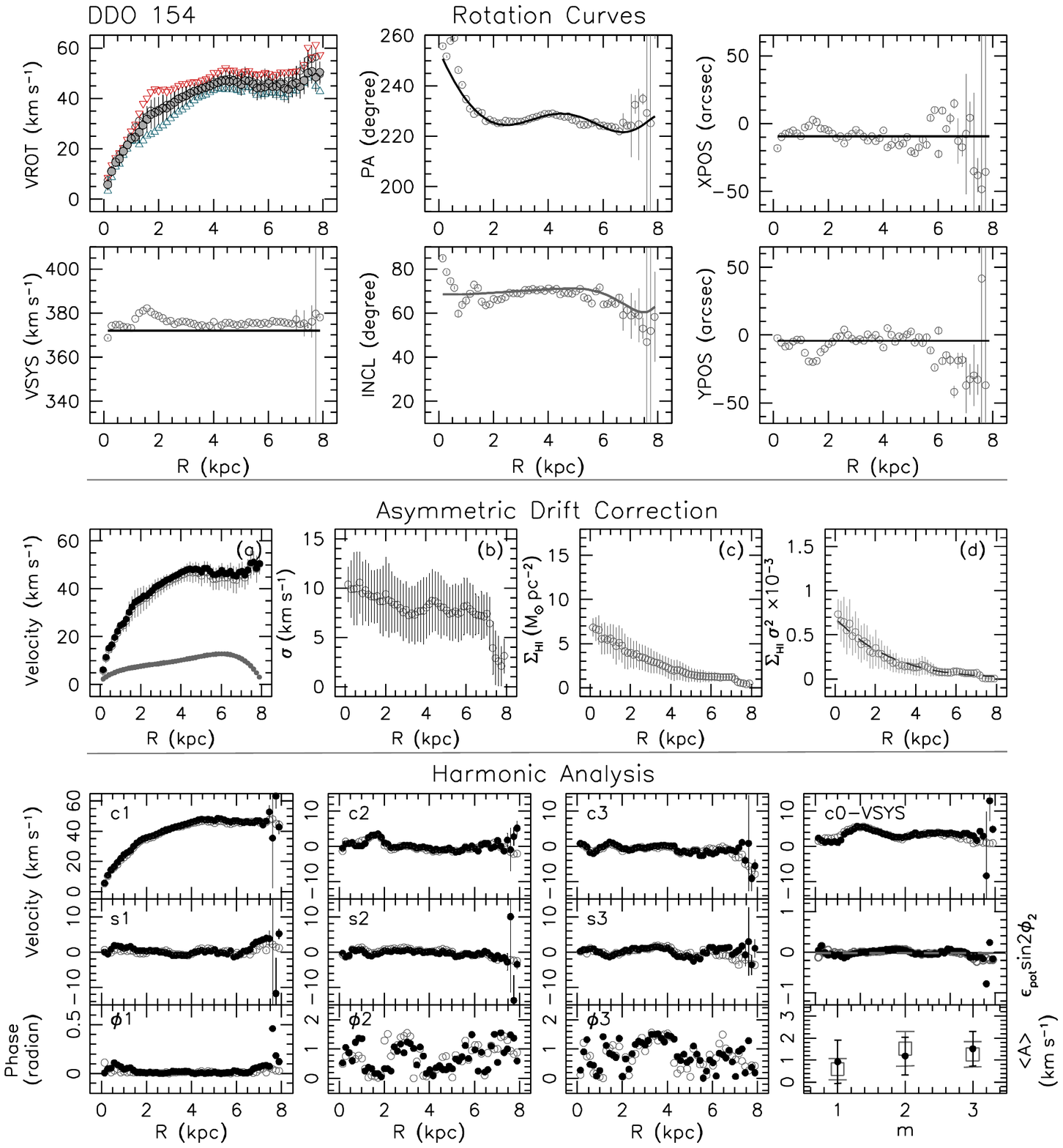}
\caption{Rotation curves, asymmetric drift correction and harmonic analysis
of DDO 154. See Appendix section B for details.
\label{ddo154_TR_HD}}
\end{figure}
{\clearpage}

\begin{figure}
\epsscale{1.0}
\figurenum{A.39}
\includegraphics[angle=0,width=1.0\textwidth,bb=40 175 540
690,clip=]{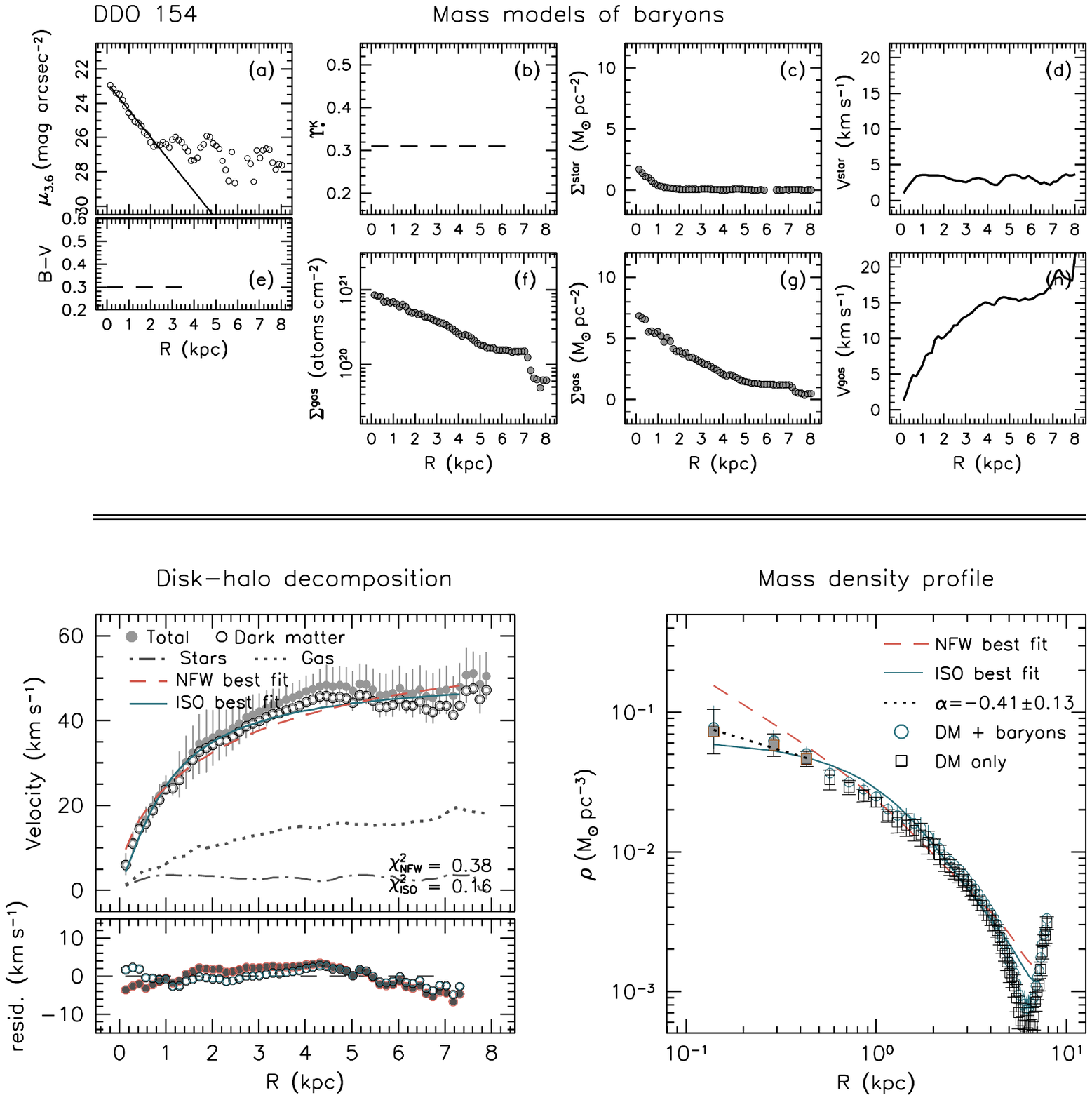}
\caption{The mass models of baryons, disk-halo decomposition and mass density
profile of DDO 154. Please refer to the text in Sections~\ref{MASS_MODELS} and
\ref{DARK_MATTER_DISTRIBUTION} for full information.
\label{MD_DH_DM_ddo154}}
\end{figure}
{\clearpage}

\begin{figure}
\epsscale{1.0}
\figurenum{A.40}
\includegraphics[angle=0,width=1.0\textwidth,bb=60 140 540
745,clip=]{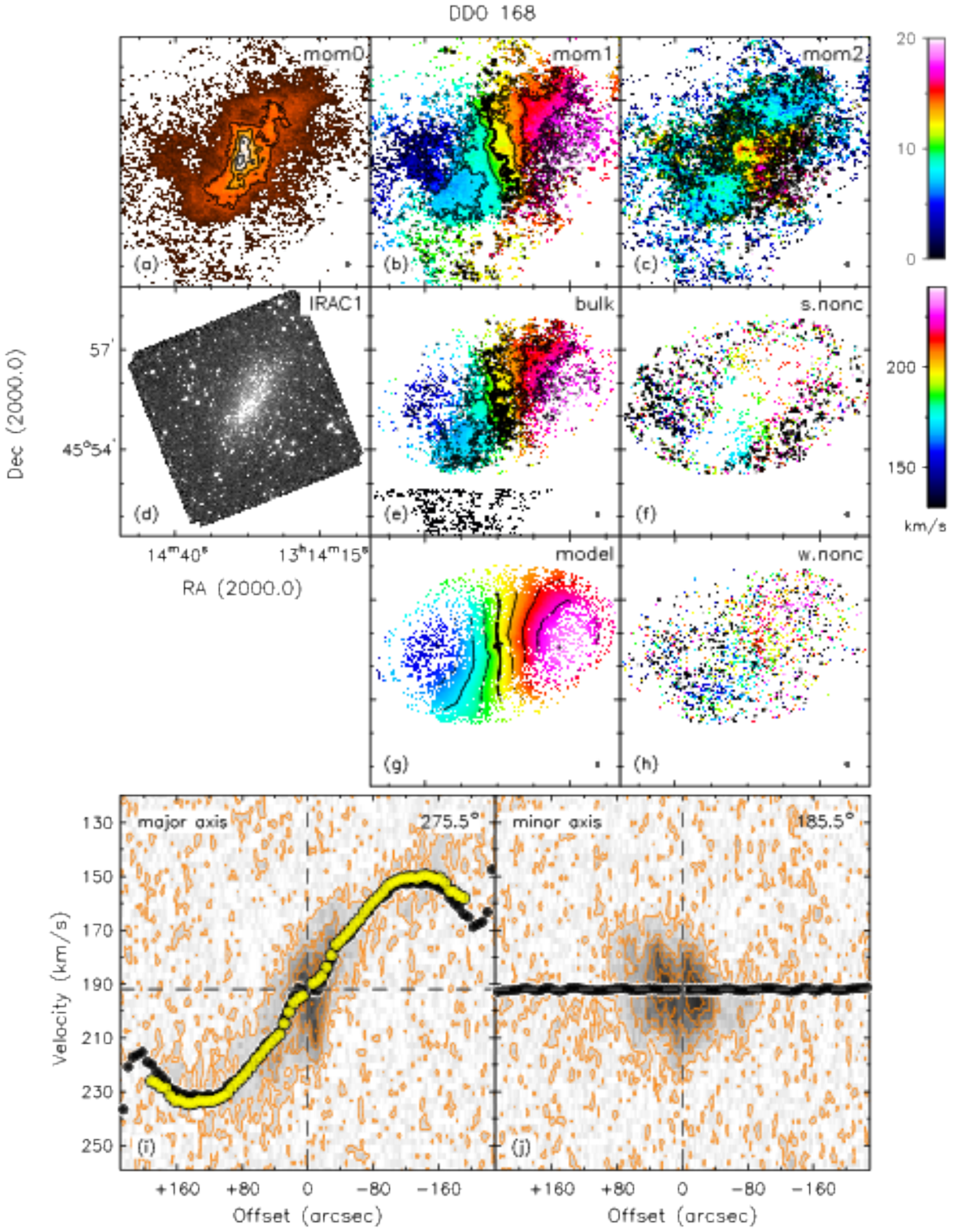}
\caption{H{\sc i} data and {\it Spitzer IRAC} 3.6$\mu$m image of DDO 168. The
systemic velocity is indicated by the thick contours in the velocity fields, and
the iso-velocity contours are spaced by 10 \kms. Velocity dispersion contours run
from 0 to 20 \kms\ with a spacing of 5 \kms. See Appendix section A for details.
\label{ddo168_data_PV}}
\end{figure}
{\clearpage}

\begin{figure}
\epsscale{1.0}
\figurenum{A.41}
\includegraphics[angle=0,width=1.0\textwidth,bb=35 140 570
710,clip=]{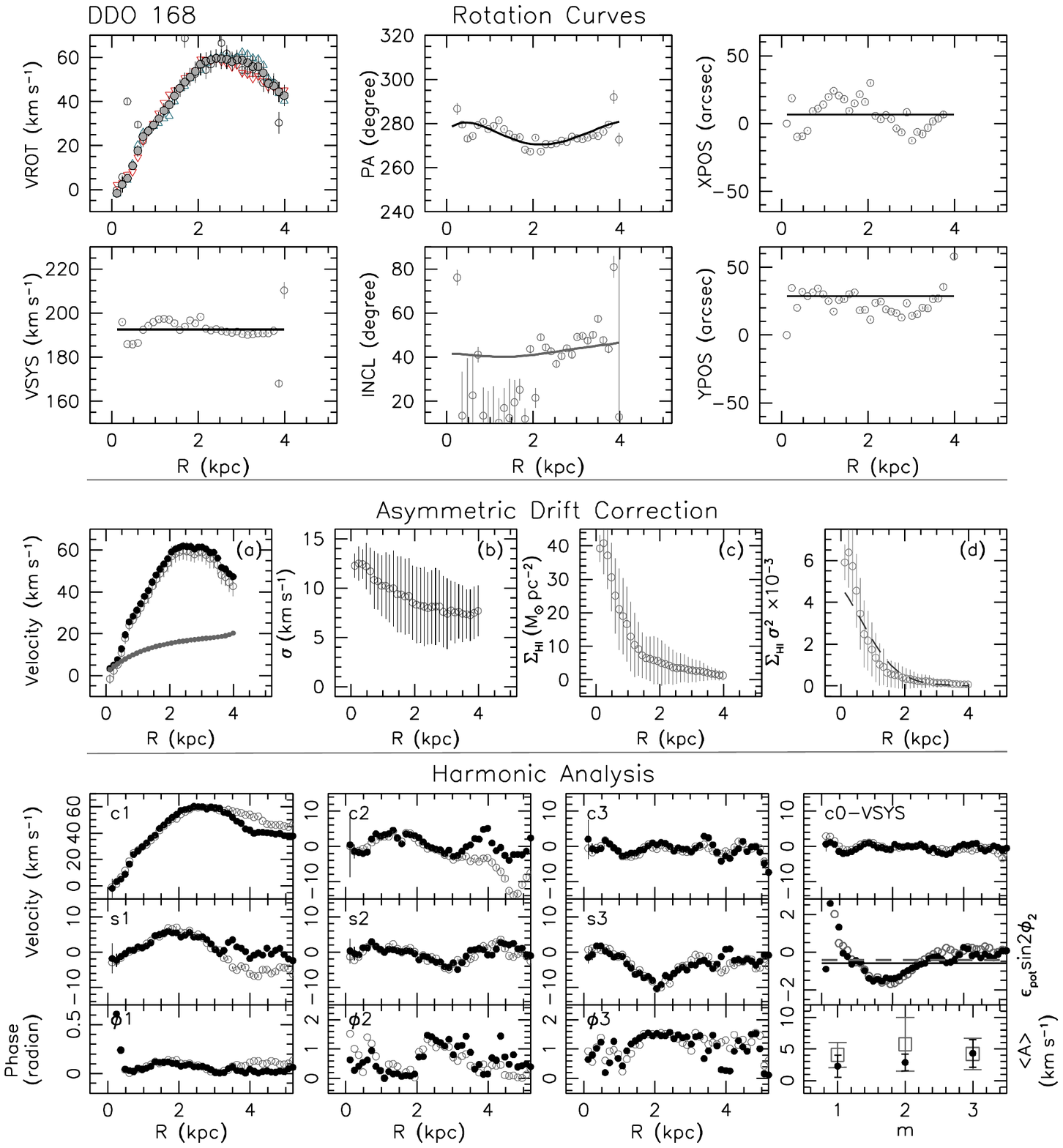}
\caption{Rotation curves, asymmetric drift correction and harmonic analysis
of DDO 168. See Appendix section B for details.
\label{ddo168_TR_HD}}
\end{figure}
{\clearpage}

\begin{figure}
\epsscale{1.0}
\figurenum{A.42}
\includegraphics[angle=0,width=1.0\textwidth,bb=40 175 540
690,clip=]{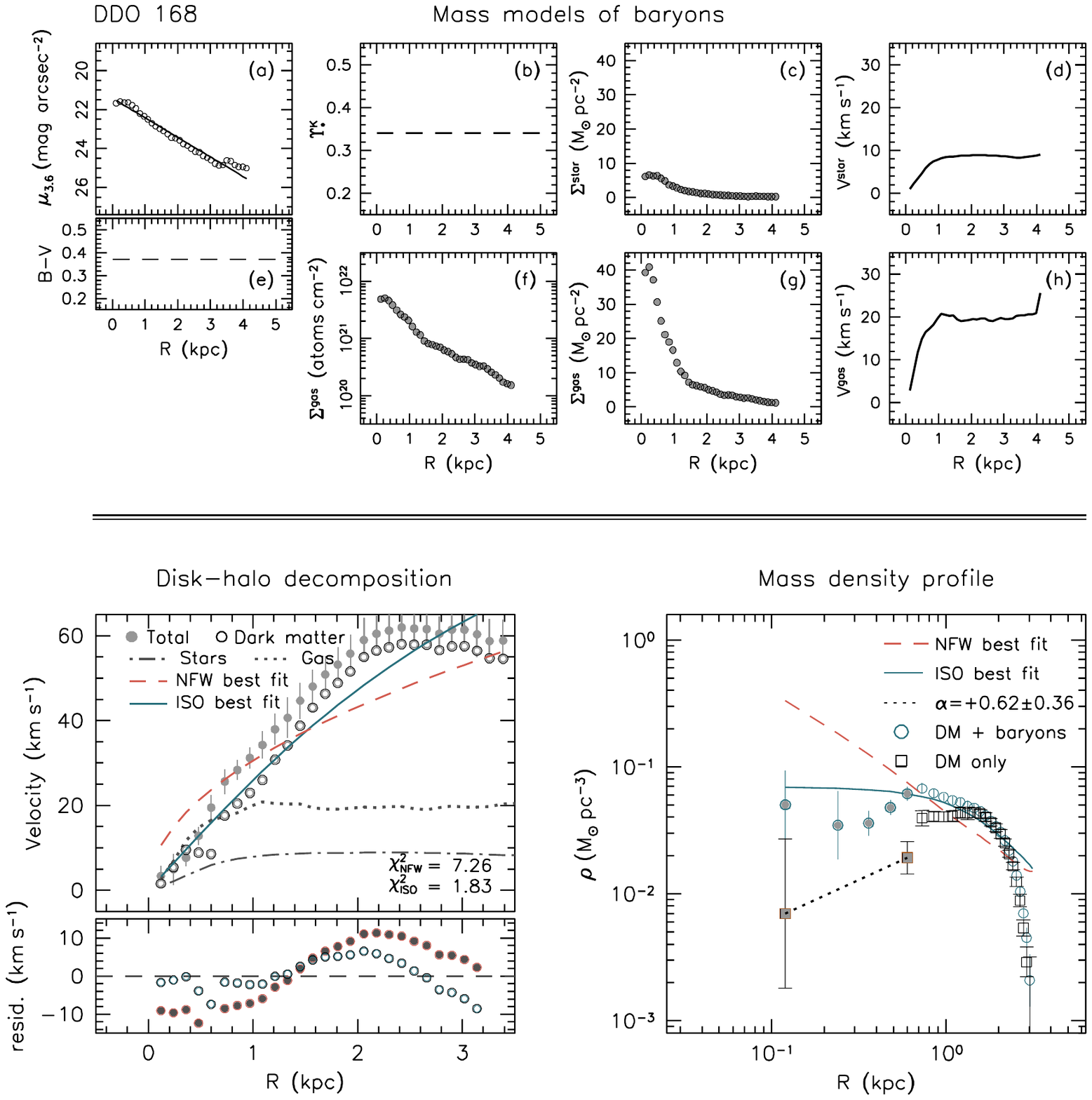}
\caption{The mass models of baryons, disk-halo decomposition and mass density
profile of DDO 168. Please refer to the text in Sections~\ref{MASS_MODELS} and
\ref{DARK_MATTER_DISTRIBUTION} for full information.
\label{MD_DH_DM_ddo168}}
\end{figure}
{\clearpage}

\begin{figure}
\epsscale{1.0}
\figurenum{A.43}
\includegraphics[angle=0,width=1.0\textwidth,bb=60 140 540
745,clip=]{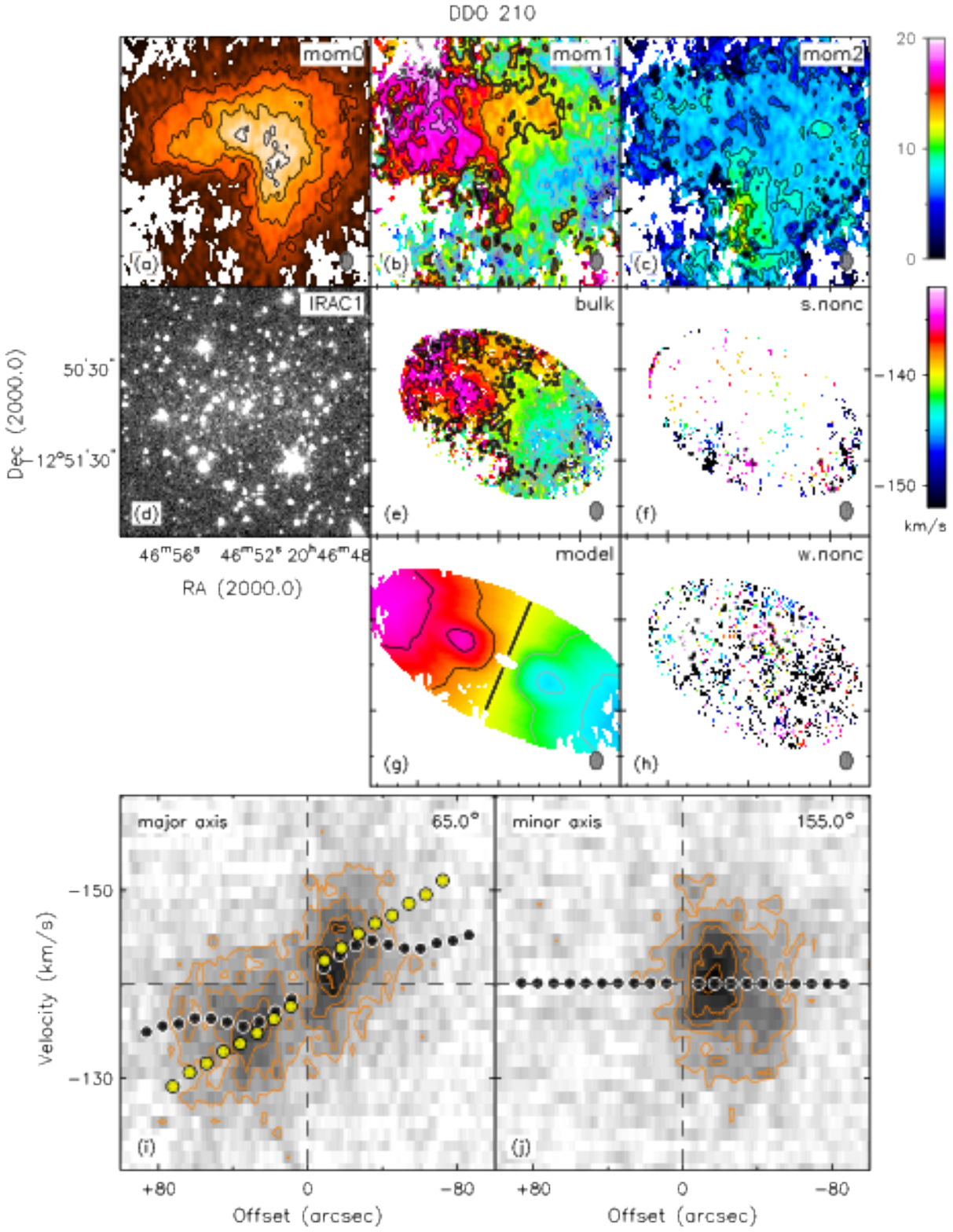}
\caption{H{\sc i} data and {\it Spitzer IRAC} 3.6$\mu$m image of DDO 210. The
systemic velocity is indicated by the thick contours in the velocity fields, and
the iso-velocity contours are spaced by 2 \kms. Velocity dispersion contours run
from 0 to 20 \kms\ with a spacing of 2 \kms. See Appendix section A for details.
\label{ddo210_data_PV}}
\end{figure}
{\clearpage}

\begin{figure}
\epsscale{1.0}
\figurenum{A.44}
\includegraphics[angle=0,width=1.0\textwidth,bb=35 140 570
710,clip=]{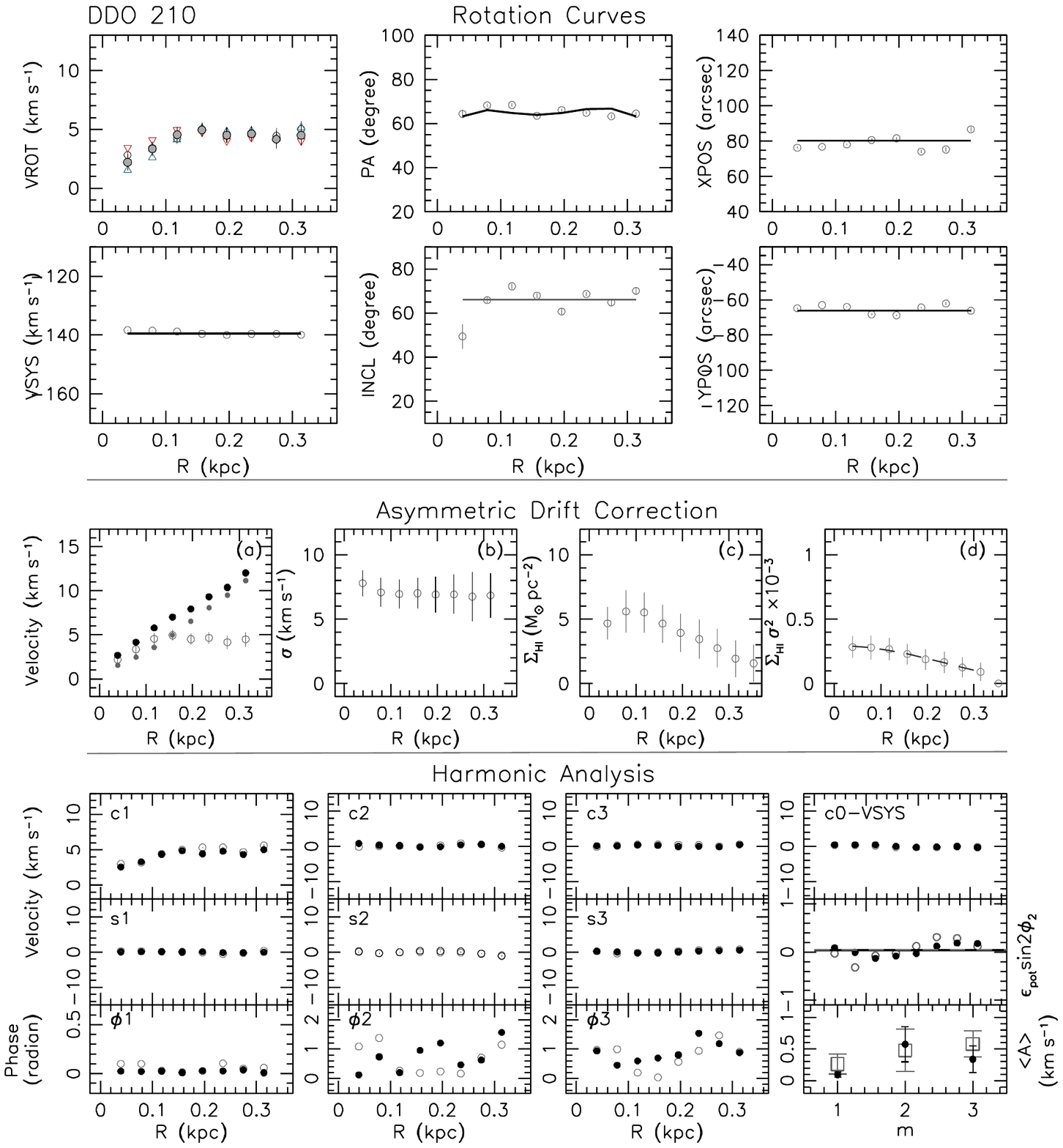}
\caption{Rotation curves, asymmetric drift correction and harmonic analysis
of DDO 210. See Appendix section B for details.
\label{ddo210_TR_HD}}
\end{figure}
{\clearpage}

\begin{figure}
\epsscale{1.0}
\figurenum{A.45}
\includegraphics[angle=0,width=1.0\textwidth,bb=40 175 540
690,clip=]{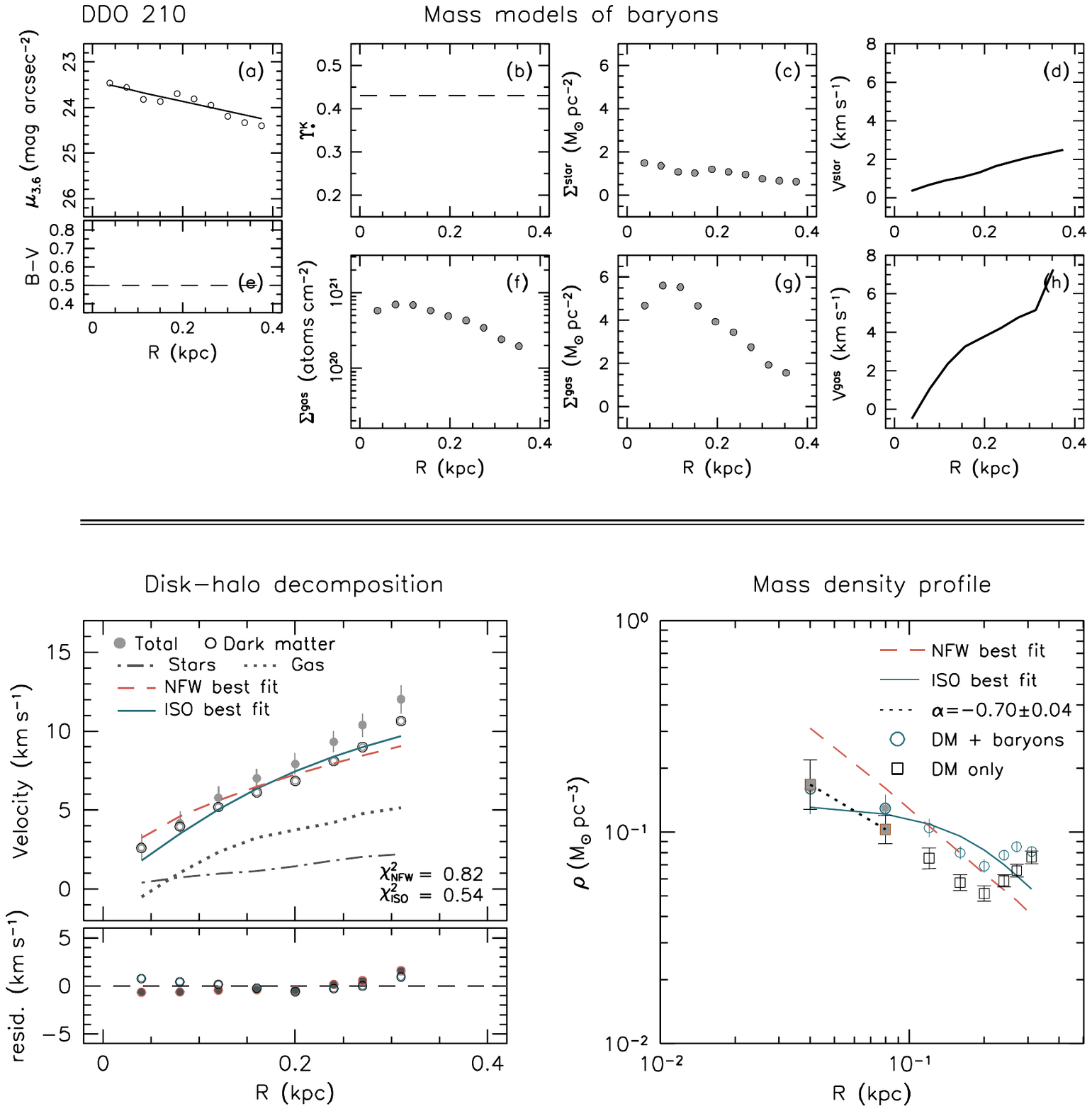}
\caption{The mass models of baryons, disk-halo decomposition and mass density
profile of DDO 210. Please refer to the text in Sections~\ref{MASS_MODELS} and
\ref{DARK_MATTER_DISTRIBUTION} for full information.
\label{MD_DH_DM_ddo210}}
\end{figure}
{\clearpage}

\begin{figure}
\epsscale{1.0}
\figurenum{A.46}
\includegraphics[angle=0,width=1.0\textwidth,bb=60 140 540
745,clip=]{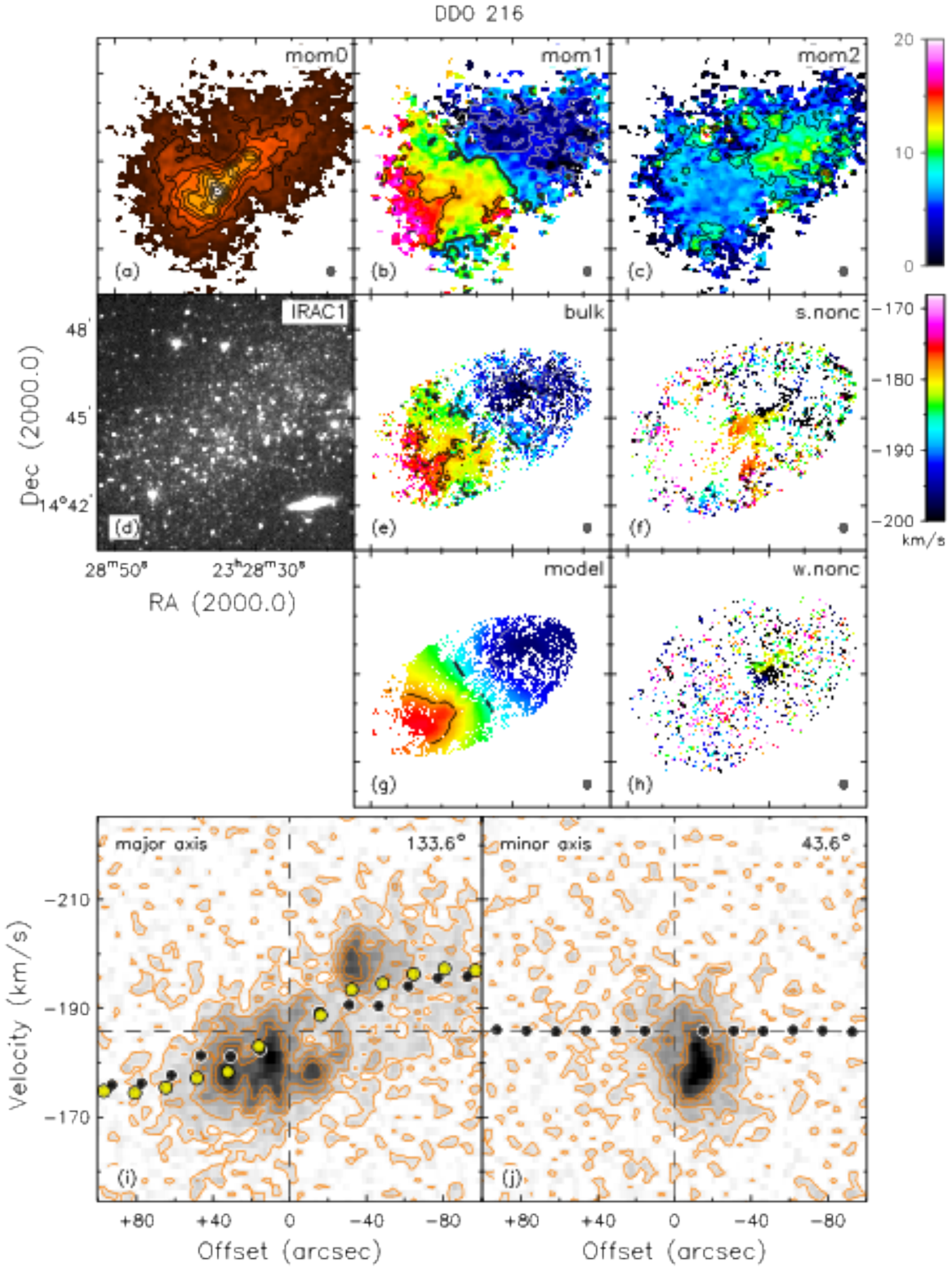}
\caption{H{\sc i} data and {\it Spitzer IRAC} 3.6$\mu$m image of DDO 216. The
systemic velocity is indicated by the thick contours in the velocity fields, and
the iso-velocity contours are spaced by 8 \kms. Velocity dispersion contours run
from 0 to 20 \kms\ with a spacing of 4 \kms. See Appendix section A for details.
\label{ddo216_data_PV}}
\end{figure}
{\clearpage}

\begin{figure}
\epsscale{1.0}
\figurenum{A.47}
\includegraphics[angle=0,width=1.0\textwidth,bb=35 140 570
710,clip=]{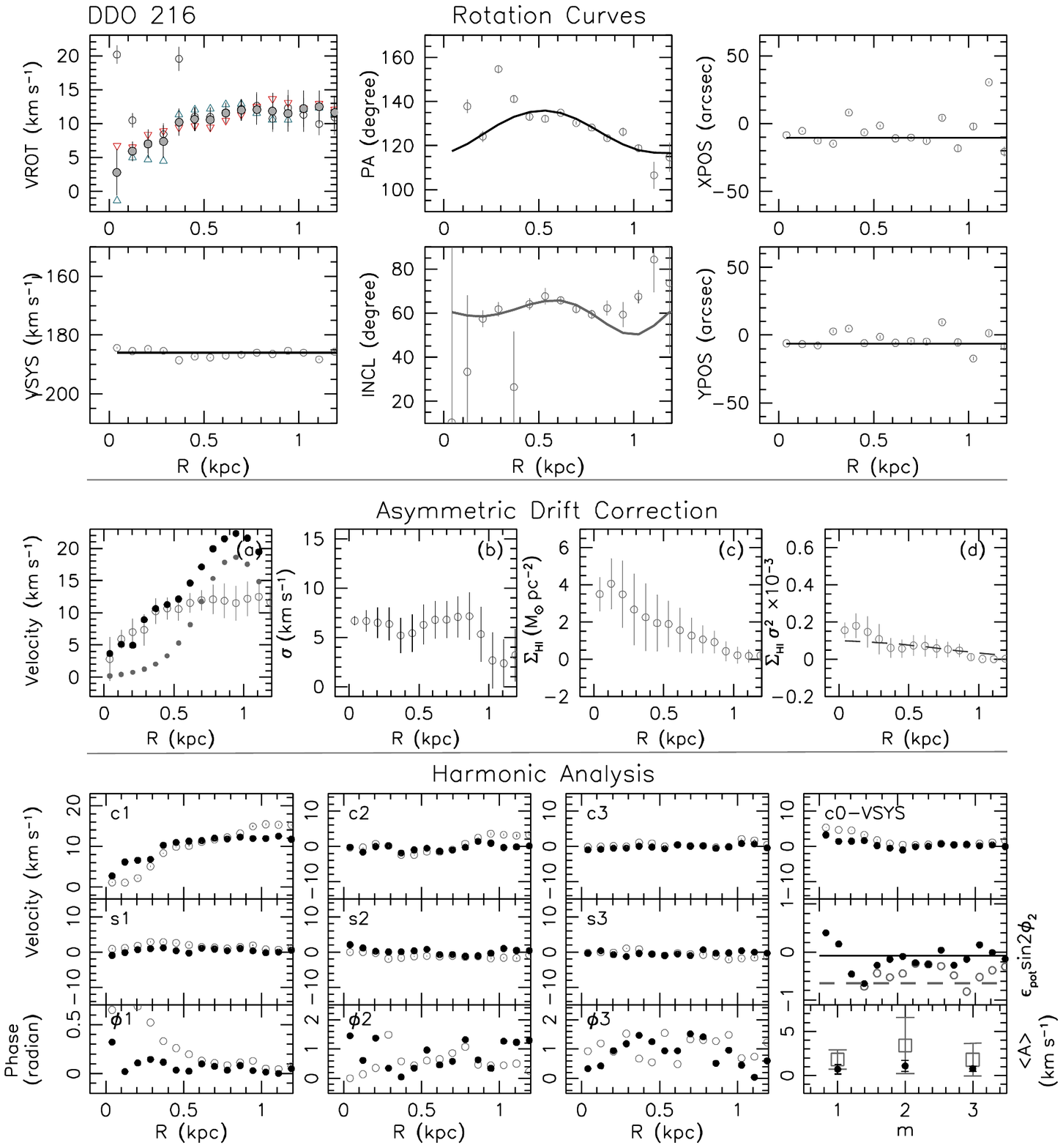}
\caption{Rotation curves, asymmetric drift correction and harmonic analysis
of DDO 216. See Appendix section B for details.
\label{ddo216_TR_HD}}
\end{figure}
{\clearpage}

\begin{figure}
\epsscale{1.0}
\figurenum{A.48}
\includegraphics[angle=0,width=1.0\textwidth,bb=40 175 540
690,clip=]{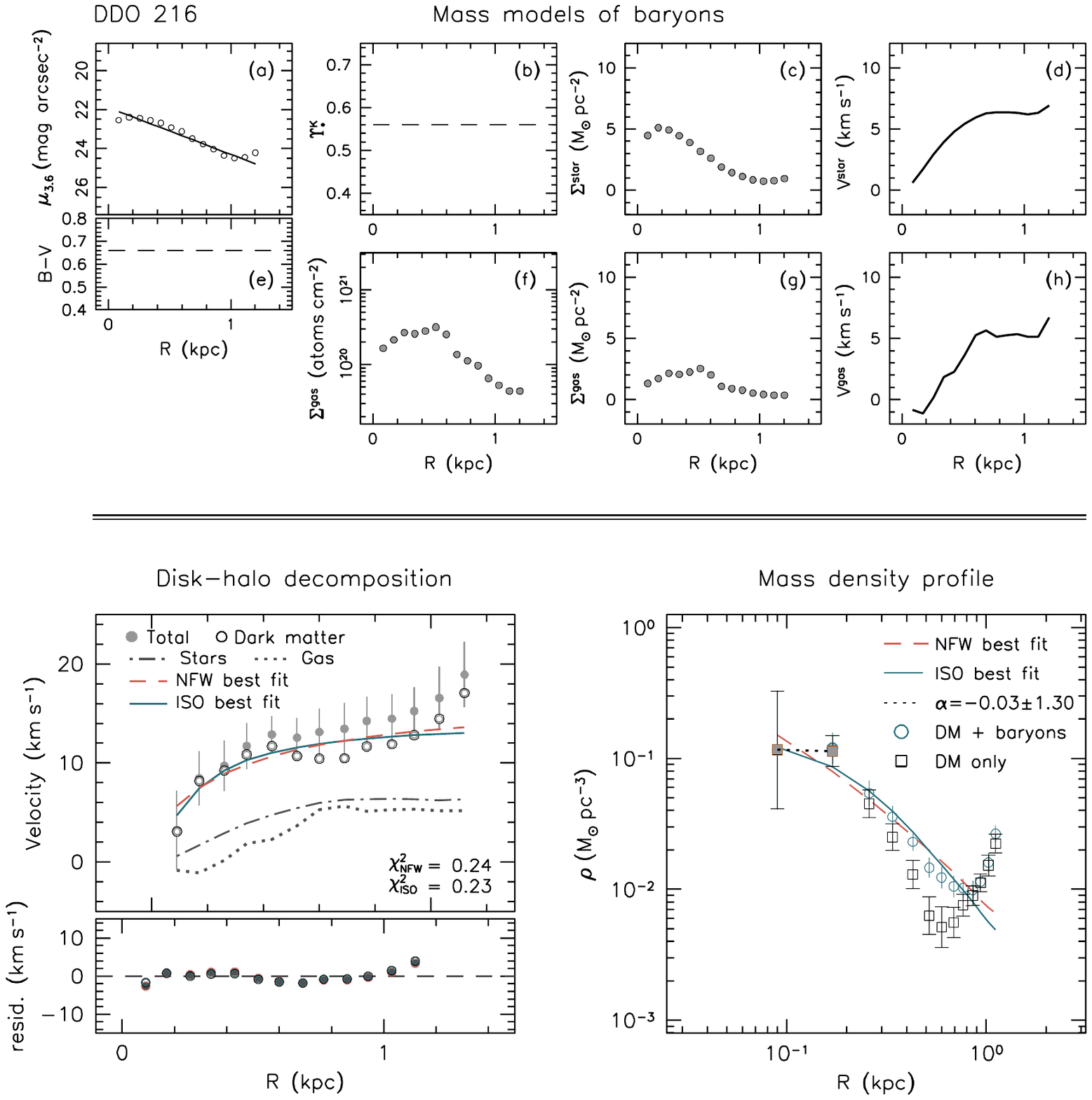}
\caption{The mass models of baryons, disk-halo decomposition and mass density
profile of DDO 216. Please refer to the text in Sections~\ref{MASS_MODELS} and
\ref{DARK_MATTER_DISTRIBUTION} for full information.
\label{MD_DH_DM_ddo216}}
\end{figure}
{\clearpage}

\begin{figure}
\epsscale{1.0}
\figurenum{A.49}
\includegraphics[angle=0,width=1.0\textwidth,bb=60 140 540
745,clip=]{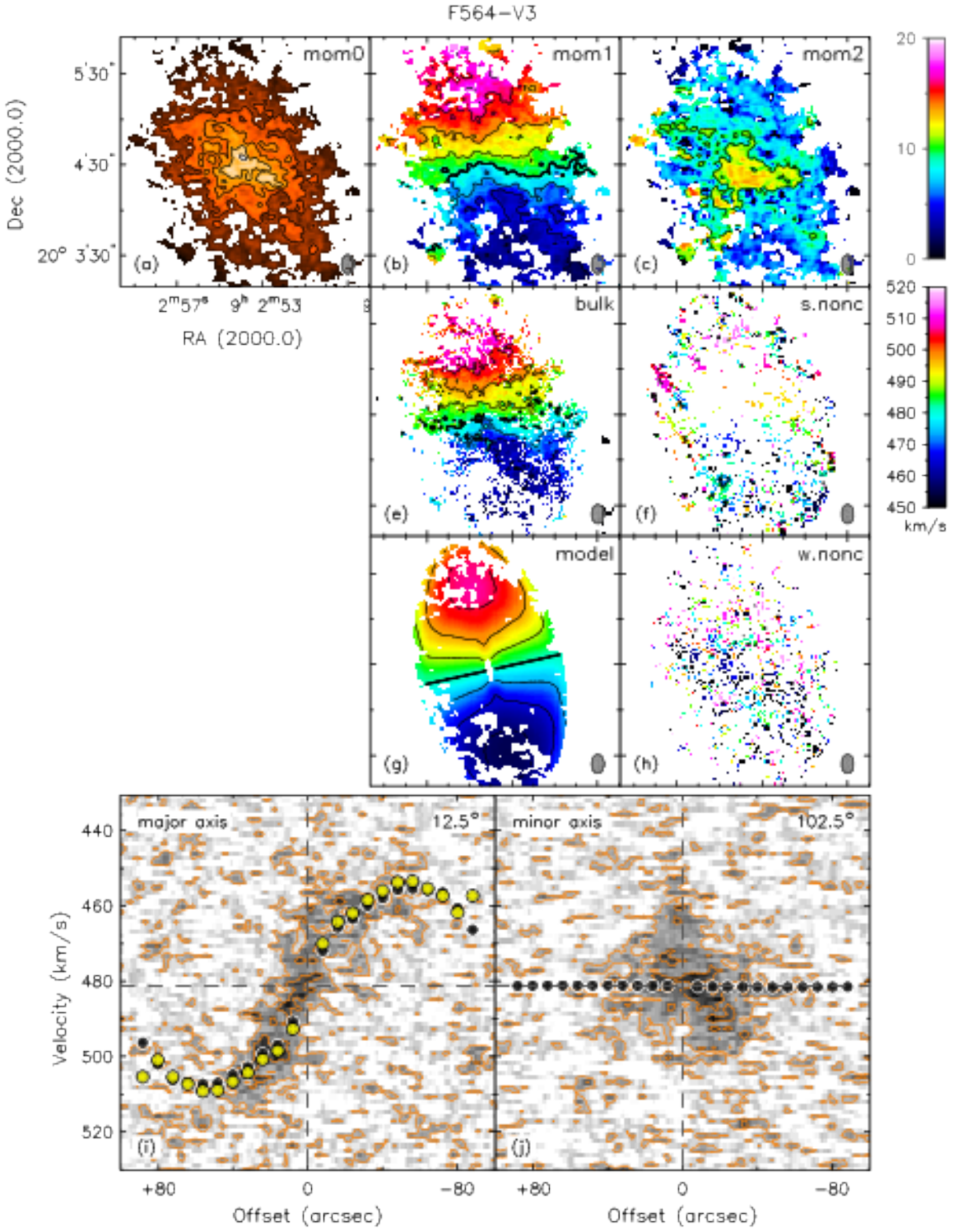}
\caption{H{\sc i} data and {\it Spitzer IRAC} 3.6$\mu$m image of F564-V3. The
systemic velocity is indicated by the thick contours in the velocity fields, and
the iso-velocity contours are spaced by 8 \kms. Velocity dispersion contours run
from 0 to 20 \kms\ with a spacing of 10 \kms. See Appendix section A for details.
\label{f564_v3_data_PV}}
\end{figure}
{\clearpage}

\begin{figure}
\epsscale{1.0}
\figurenum{A.50}
\includegraphics[angle=0,width=1.0\textwidth,bb=35 140 570
710,clip=]{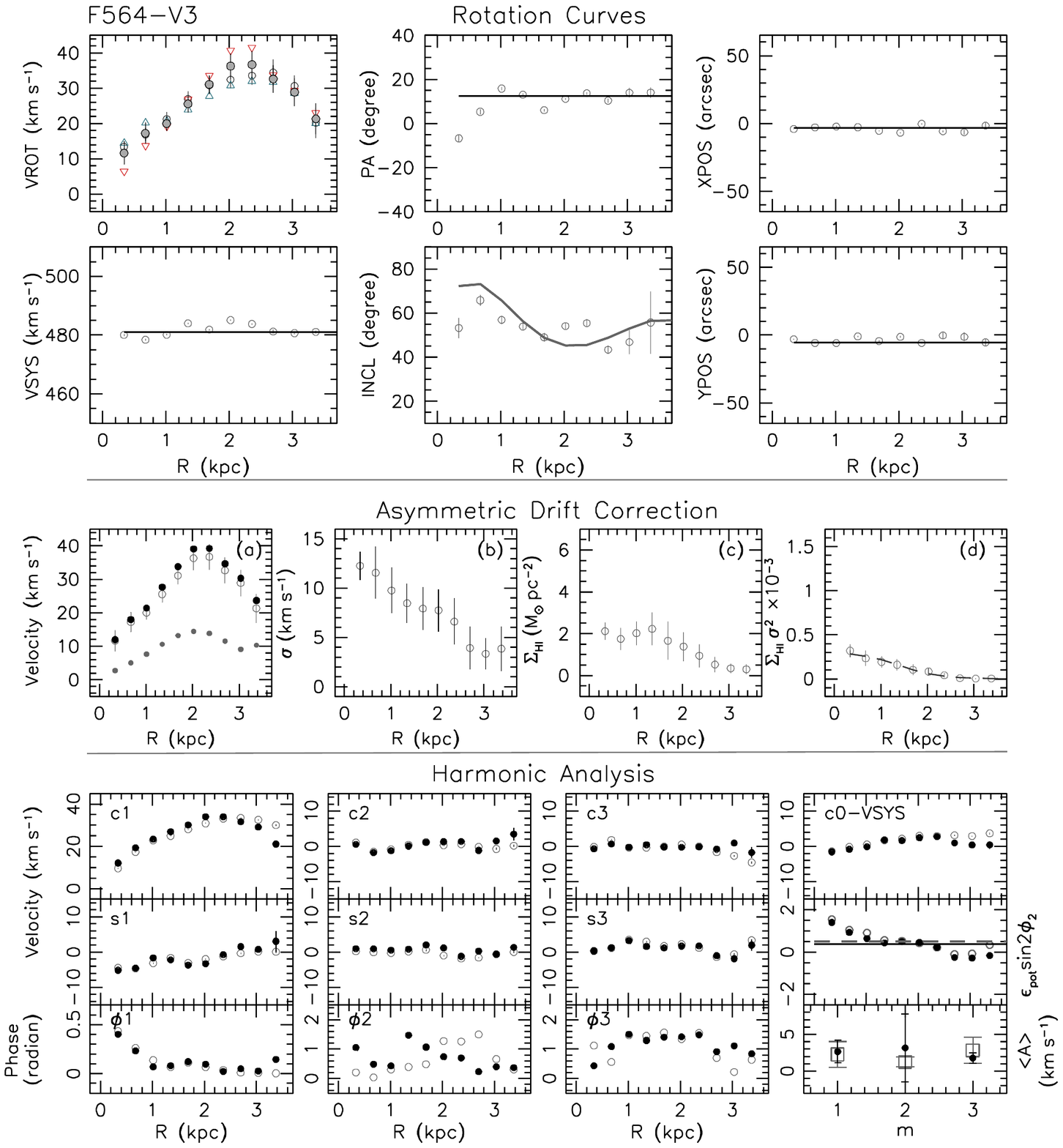}
\caption{Rotation curves, asymmetric drift correction and harmonic analysis
of F564-V3. See Appendix section B for details.
\label{f564_v3_TR_HD}}
\end{figure}
{\clearpage}

\begin{figure}
\epsscale{1.0}
\figurenum{A.51}
\includegraphics[angle=0,width=1.0\textwidth,bb=40 175 540
690,clip=]{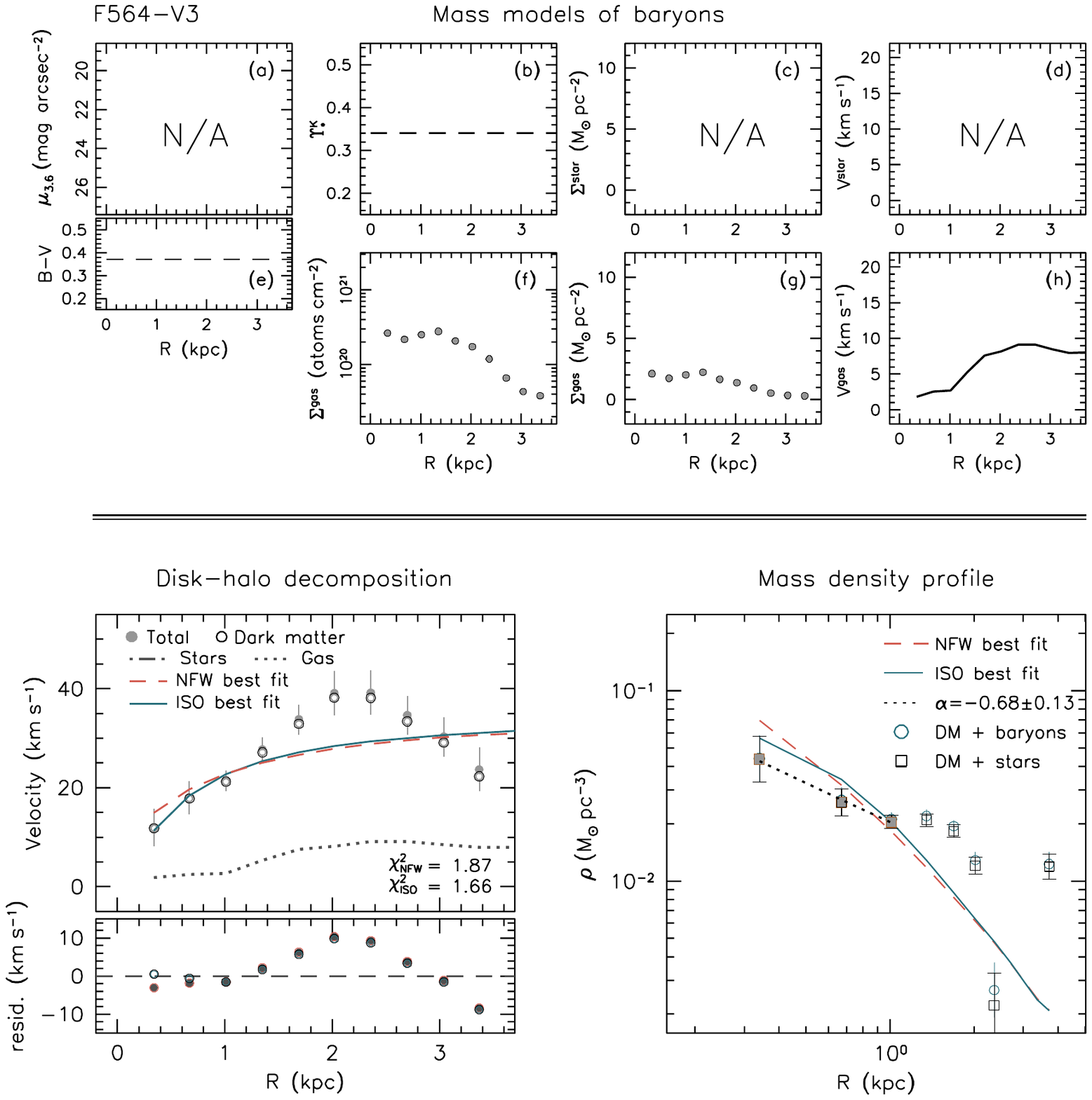}
\caption{The mass models of baryons, disk-halo decomposition and mass density
profile of F564-V3. Please refer to the text in Sections~\ref{MASS_MODELS} and
\ref{DARK_MATTER_DISTRIBUTION} for full information.
\label{MD_DH_DM_f564_v3}}
\end{figure}
{\clearpage}

\begin{figure}
\epsscale{1.0}
\figurenum{A.52}
\includegraphics[angle=0,width=1.0\textwidth,bb=60 140 540
745,clip=]{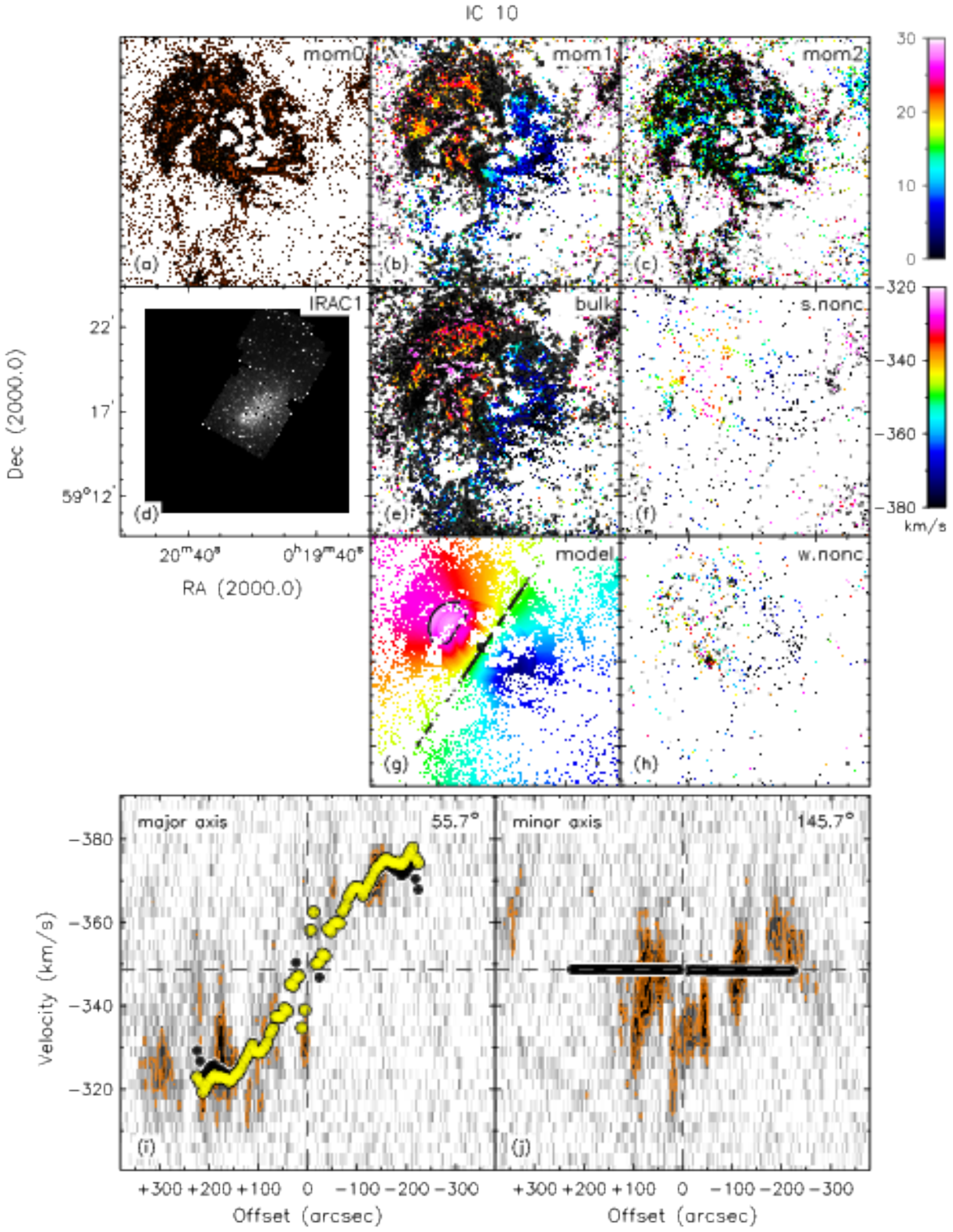}
\caption{H{\sc i} data and {\it Spitzer IRAC} 3.6$\mu$m image of IC 10. The
systemic velocity is indicated by the thick contours in the velocity fields, and
the iso-velocity contours are spaced by 20 \kms. Velocity dispersion contours run
from 0 to 30 \kms\ with a spacing of 20 \kms. See Appendix section A for details.
\label{ic10_data_PV}}
\end{figure}
{\clearpage}

\begin{figure}
\epsscale{1.0}
\figurenum{A.53}
\includegraphics[angle=0,width=1.0\textwidth,bb=35 140 570
710,clip=]{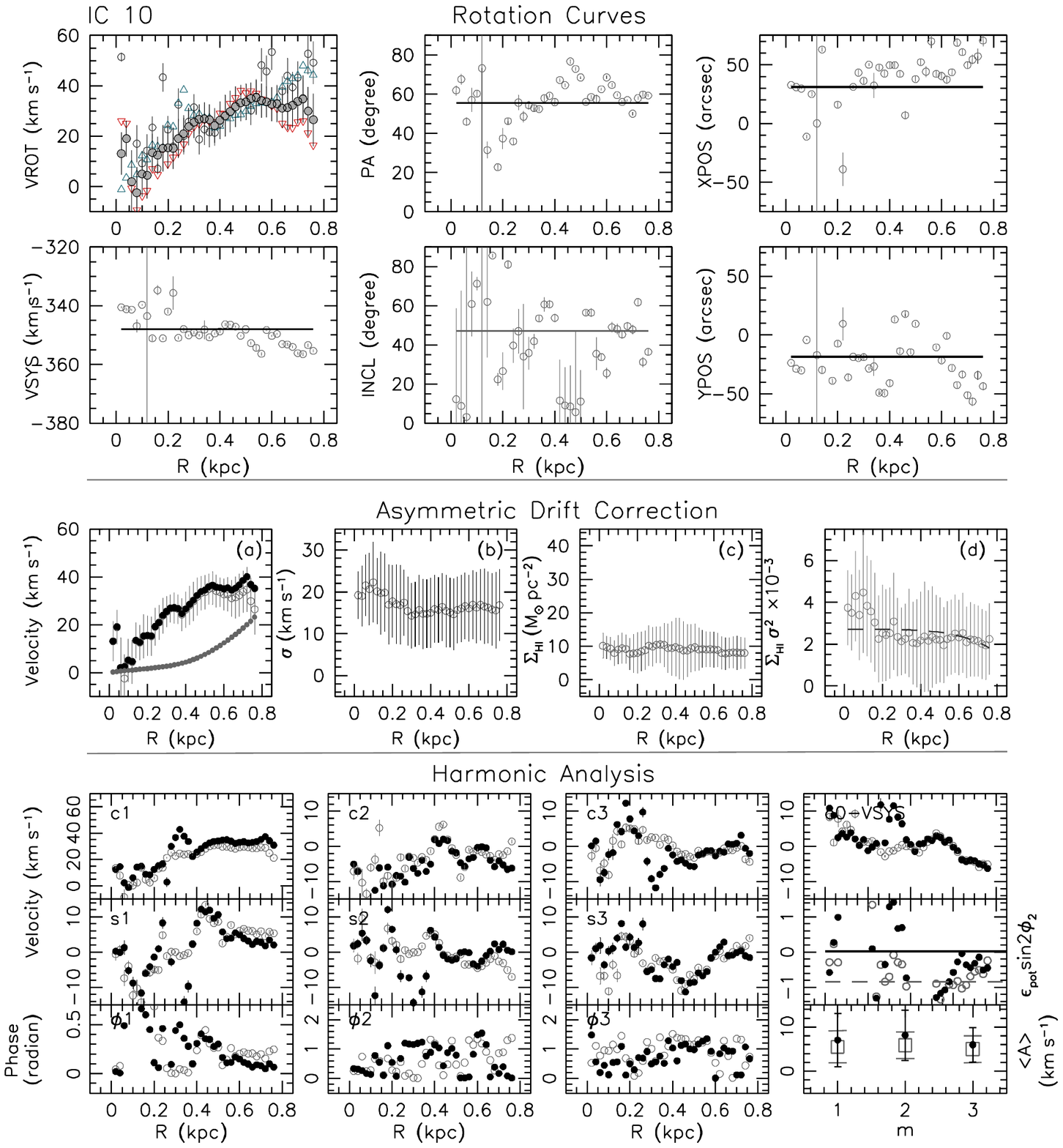}
\caption{Rotation curves, asymmetric drift correction and harmonic analysis
of IC 10. See Appendix section B for details.
\label{ic10_TR_HD}}
\end{figure}
{\clearpage}

\begin{figure}
\epsscale{1.0}
\figurenum{A.54}
\includegraphics[angle=0,width=1.0\textwidth,bb=40 175 540
690,clip=]{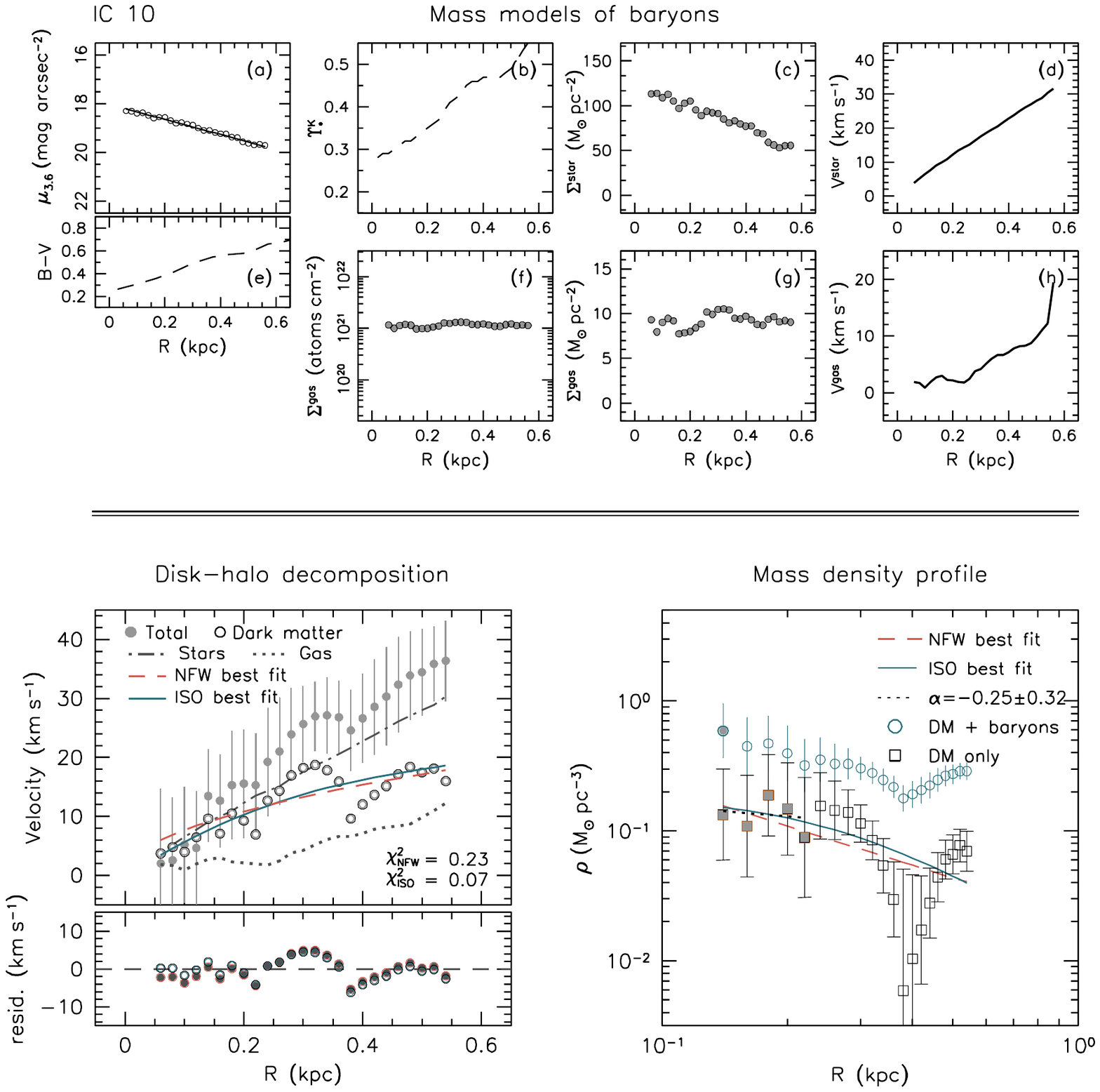}
\caption{The mass models of baryons, disk-halo decomposition and mass density
profile of IC 10. Please refer to the text in Sections~\ref{MASS_MODELS} and
\ref{DARK_MATTER_DISTRIBUTION} for full information.
\label{MD_DH_DM_ic10}}
\end{figure}
{\clearpage}

\begin{figure}
\epsscale{1.0}
\figurenum{A.55}
\includegraphics[angle=0,width=1.0\textwidth,bb=60 140 540
745,clip=]{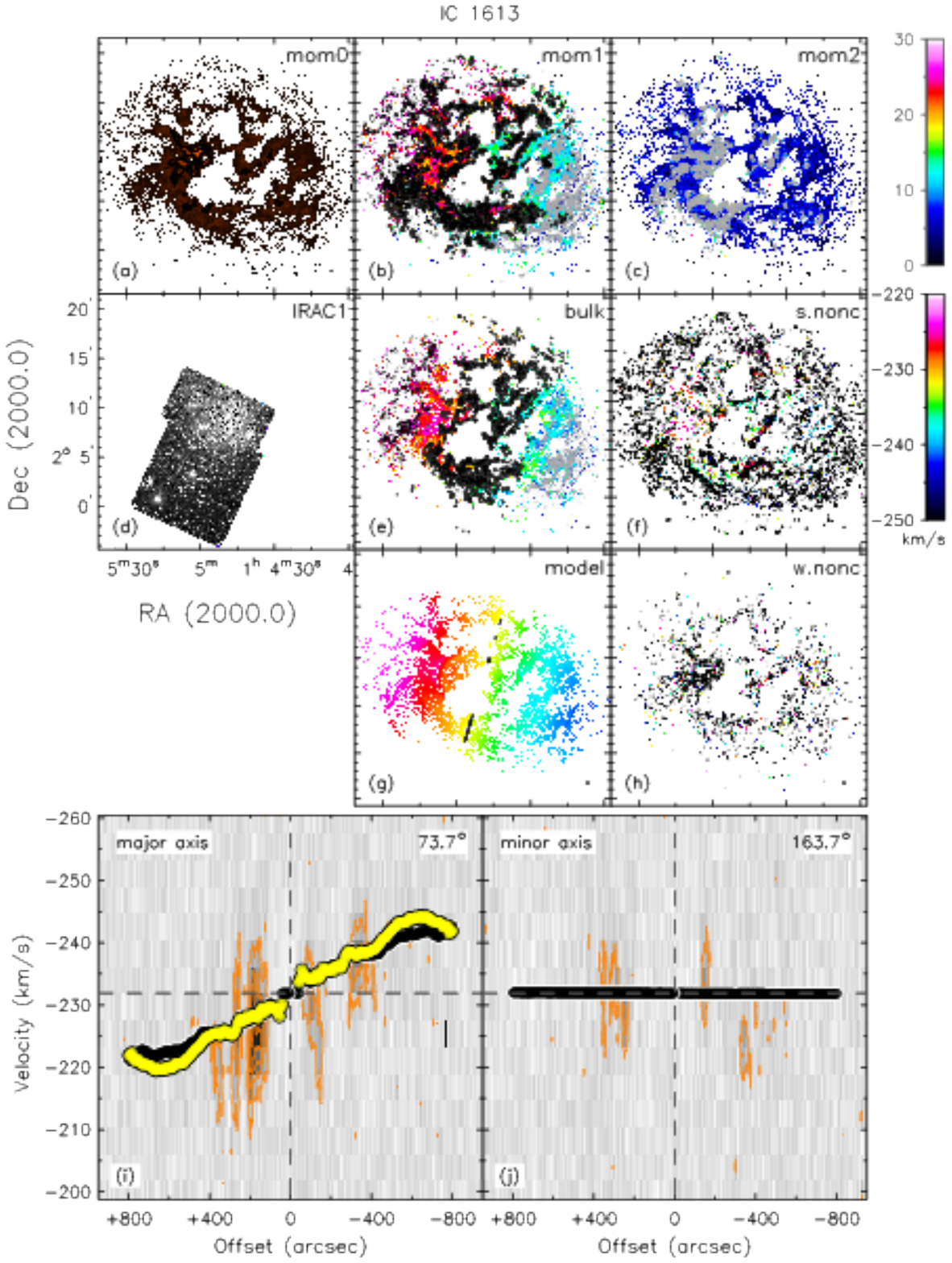}
\caption{H{\sc i} data and {\it Spitzer IRAC} 3.6$\mu$m image of IC 1613. The
systemic velocity is indicated by the thick contours in the velocity fields, and
the iso-velocity contours are spaced by 10 \kms. Velocity dispersion contours run
from 0 to 20 \kms\ with a spacing of 10 \kms. See Appendix section A for details.
\label{ic1613_data_PV}}
\end{figure}
{\clearpage}

\begin{figure}
\epsscale{1.0}
\figurenum{A.56}
\includegraphics[angle=0,width=1.0\textwidth,bb=35 140 570
710,clip=]{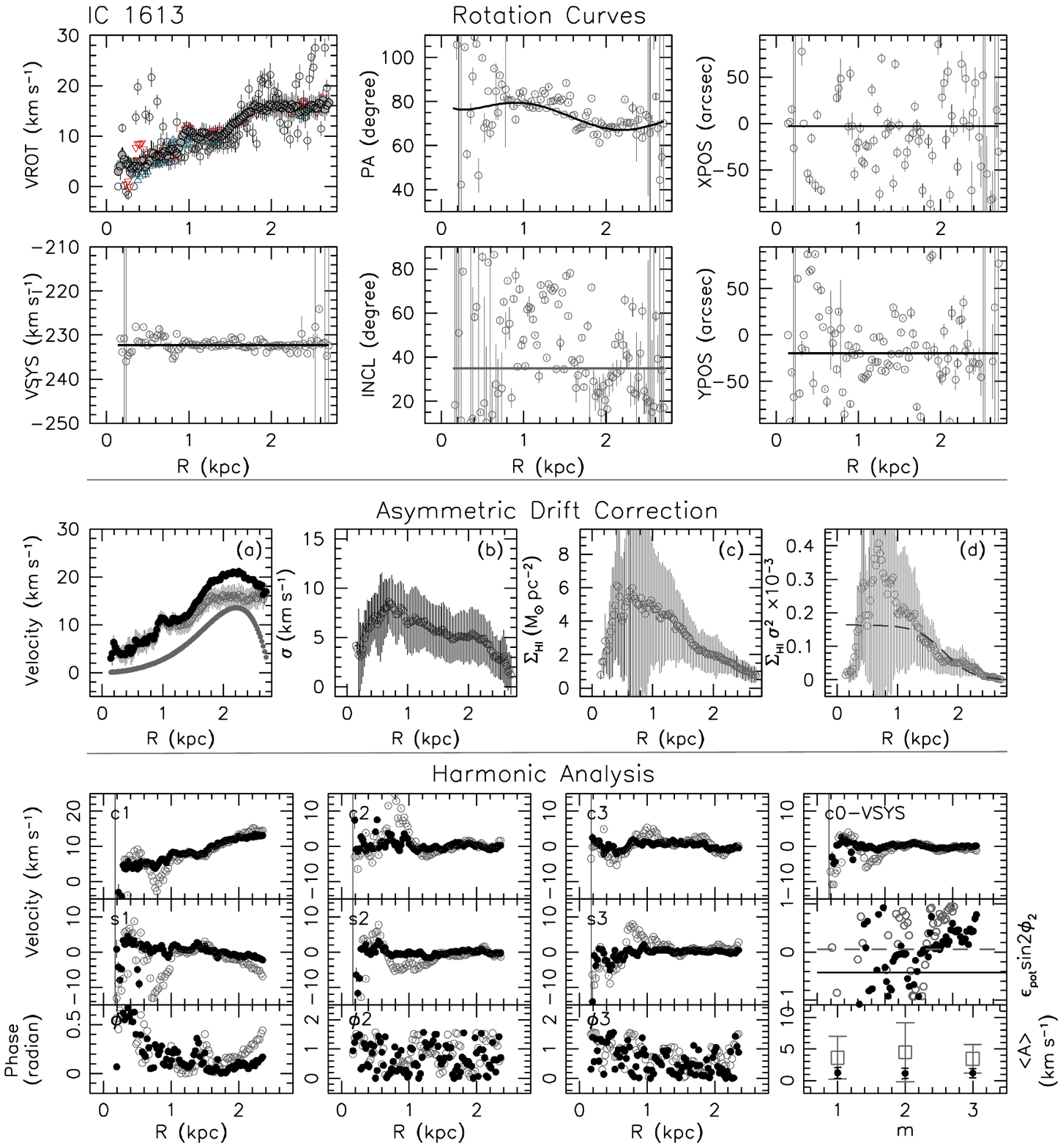}
\caption{Rotation curves, asymmetric drift correction and harmonic analysis
of IC 1613. See Appendix section B for details.
\label{ic1613_TR_HD}}
\end{figure}
{\clearpage}

\begin{figure}
\epsscale{1.0}
\figurenum{A.57}
\includegraphics[angle=0,width=1.0\textwidth,bb=40 175 540
690,clip=]{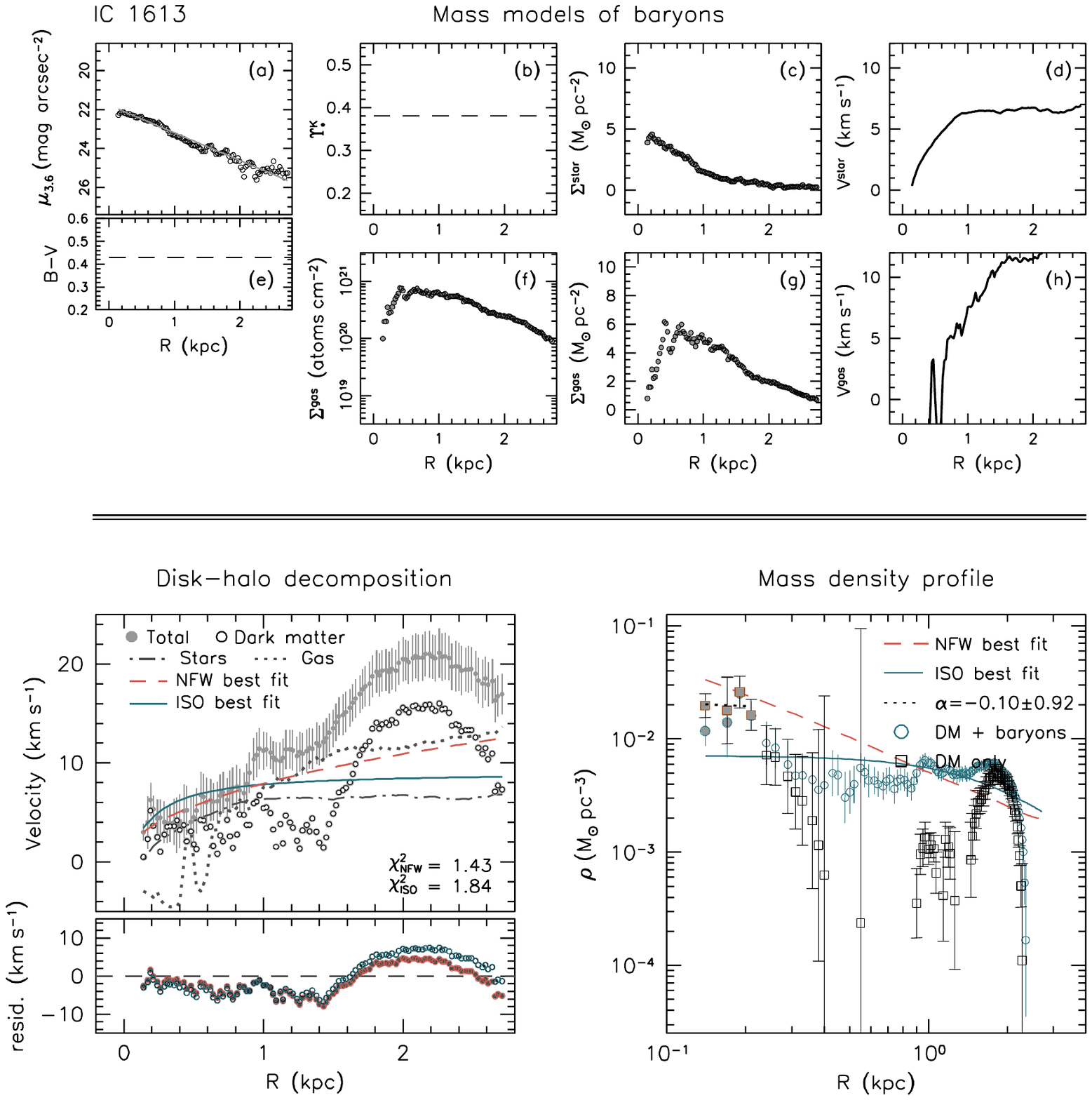}
\caption{The mass models of baryons, disk-halo decomposition and mass density
profile of IC 1613. Please refer to the text in Sections~\ref{MASS_MODELS} and
\ref{DARK_MATTER_DISTRIBUTION} for full information.
\label{MD_DH_DM_ic1613}}
\end{figure}
{\clearpage}

\begin{figure}
\epsscale{1.0}
\figurenum{A.58}
\includegraphics[angle=0,width=1.0\textwidth,bb=60 140 540
745,clip=]{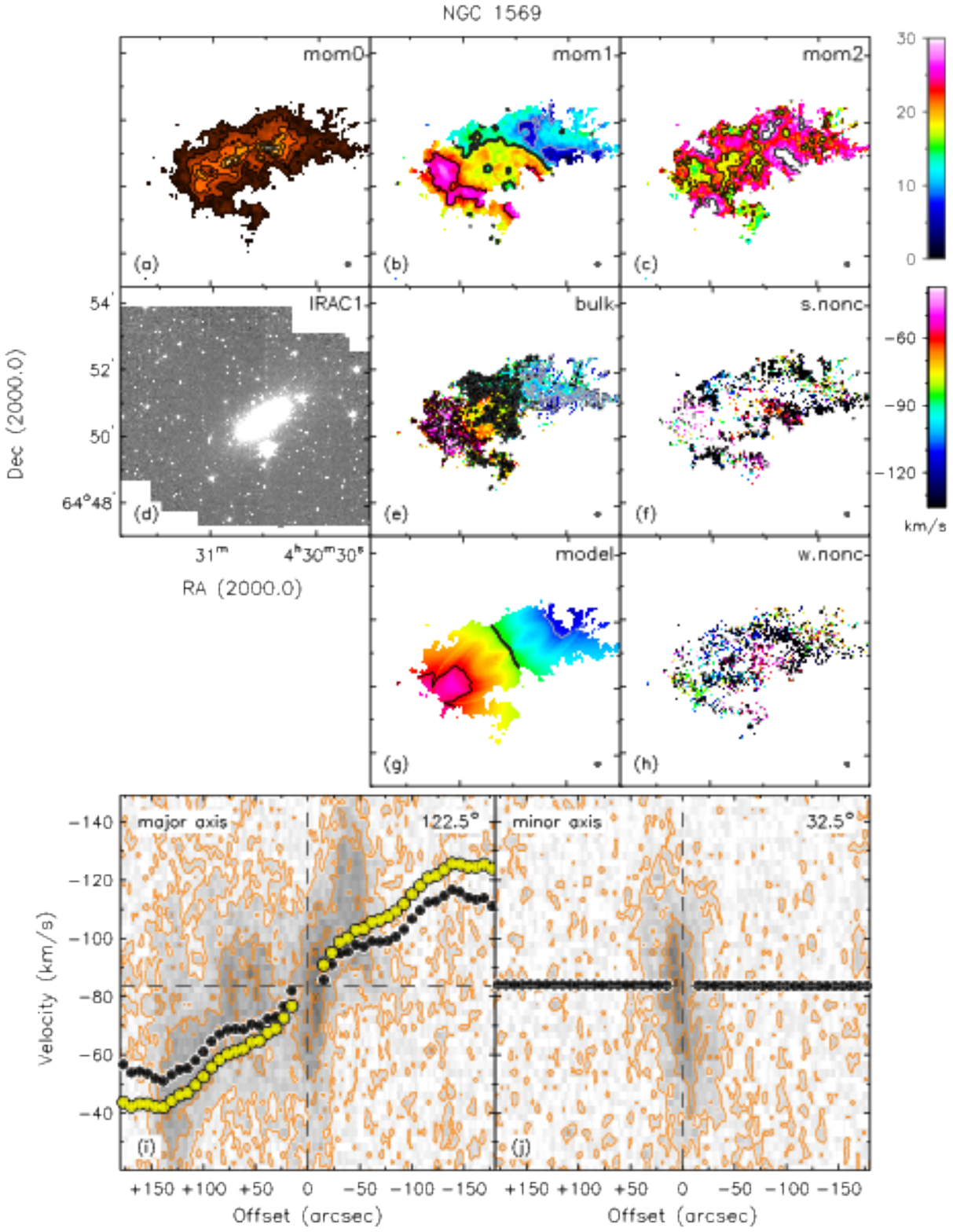}
\caption{H{\sc i} data and {\it Spitzer IRAC} 3.6$\mu$m image of NGC 1569. The
systemic velocity is indicated by the thick contours in the velocity fields, and
the iso-velocity contours are spaced by 25 \kms. Velocity dispersion contours run
from 0 to 20 \kms\ with a spacing of 10 \kms. See Appendix section A for details.
\label{ngc1569_data_PV}}
\end{figure}
{\clearpage}

\begin{figure}
\epsscale{1.0}
\figurenum{A.59}
\includegraphics[angle=0,width=1.0\textwidth,bb=35 140 570
710,clip=]{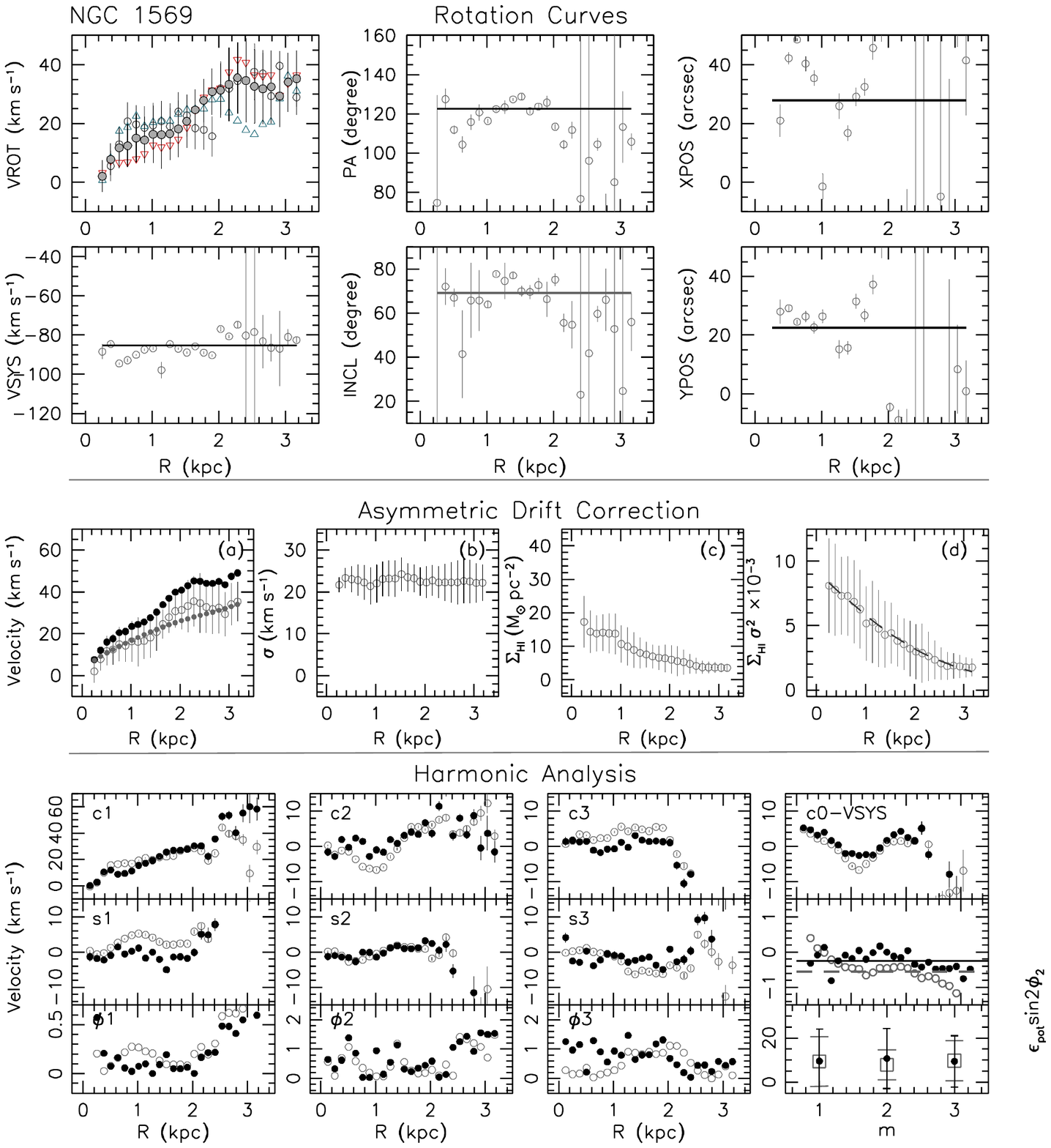}
\caption{Rotation curves, asymmetric drift correction and harmonic analysis
of NGC 1569. See Appendix section B for details.
\label{ngc1569_TR_HD}}
\end{figure}
{\clearpage}

\begin{figure}
\epsscale{1.0}
\figurenum{A.60}
\includegraphics[angle=0,width=1.0\textwidth,bb=40 175 540
690,clip=]{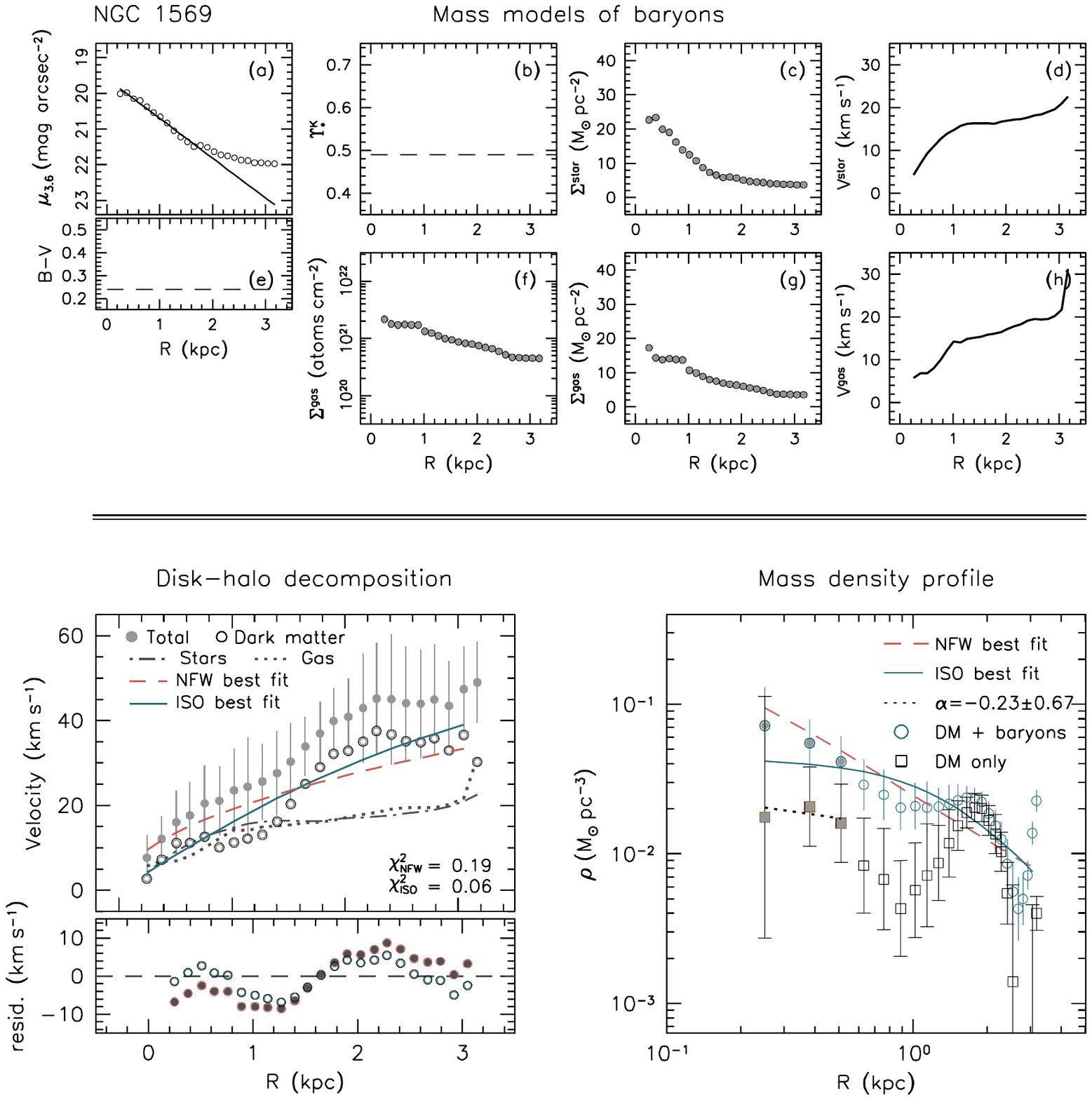}
\caption{The mass models of baryons, disk-halo decomposition and mass density
profile of NGC 1569. Please refer to the text in Sections~\ref{MASS_MODELS} and
\ref{DARK_MATTER_DISTRIBUTION} for full information.
\label{MD_DH_DM_ngc1569}}
\end{figure}
{\clearpage}

\begin{figure}
\epsscale{1.0}
\figurenum{A.61}
\includegraphics[angle=0,width=1.0\textwidth,bb=60 140 540
745,clip=]{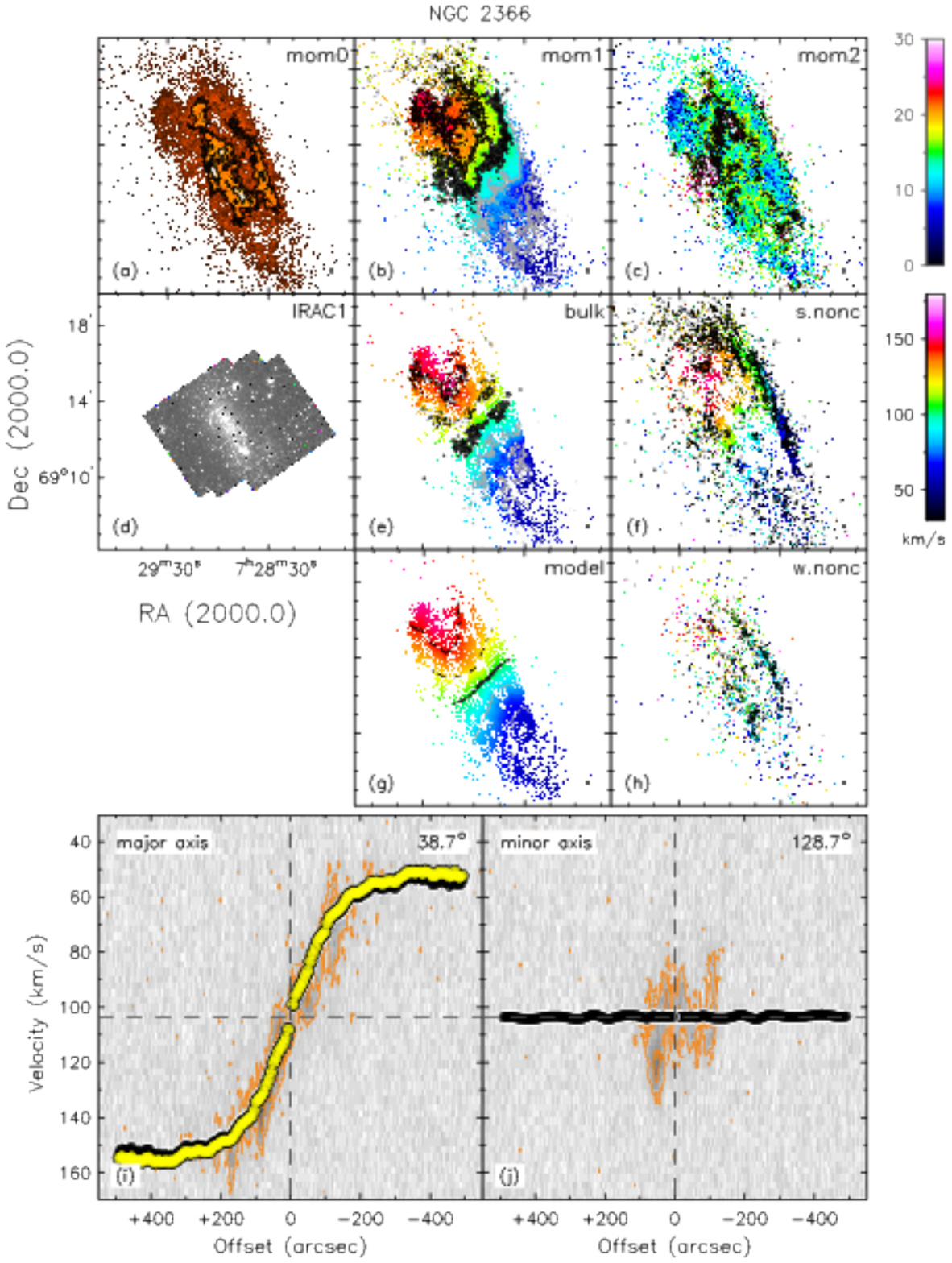}
\caption{H{\sc i} data and {\it Spitzer IRAC} 3.6$\mu$m image of NGC 2366. The
systemic velocity is indicated by the thick contours in the velocity fields, and
the iso-velocity contours are spaced by 20 \kms. Velocity dispersion contours run
from 0 to 20 \kms\ with a spacing of 20 \kms. See Appendix section A for details.
\label{ngc2366_data_PV}}
\end{figure}
{\clearpage}

\begin{figure}
\epsscale{1.0}
\figurenum{A.62}
\includegraphics[angle=0,width=1.0\textwidth,bb=35 140 570
710,clip=]{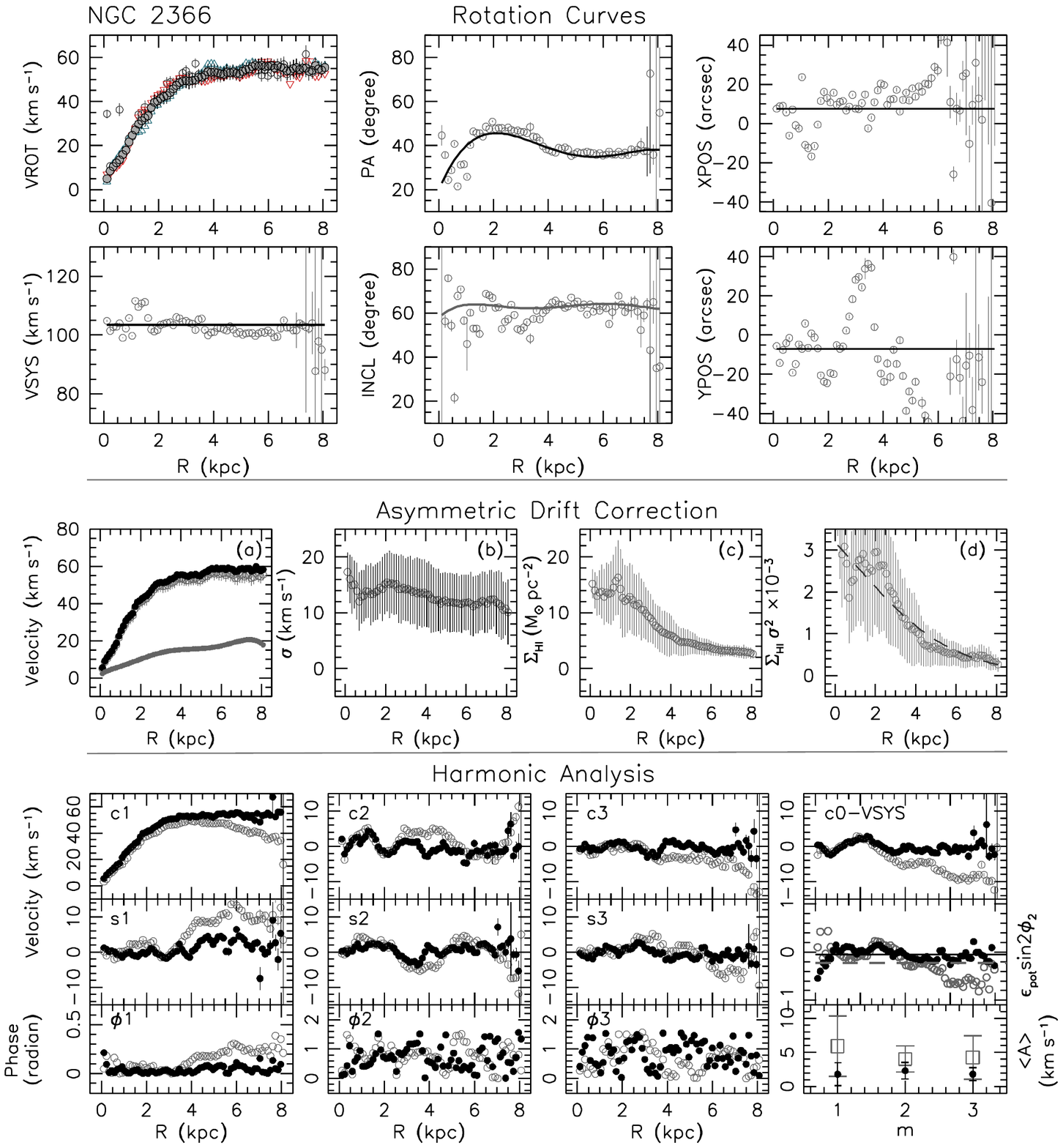}
\caption{Rotation curves, asymmetric drift correction and harmonic analysis
of NGC 2366. See Appendix section B for details.
\label{ngc2366_TR_HD}}
\end{figure}
{\clearpage}

\begin{figure}
\epsscale{1.0}
\figurenum{A.63}
\includegraphics[angle=0,width=1.0\textwidth,bb=40 175 540
690,clip=]{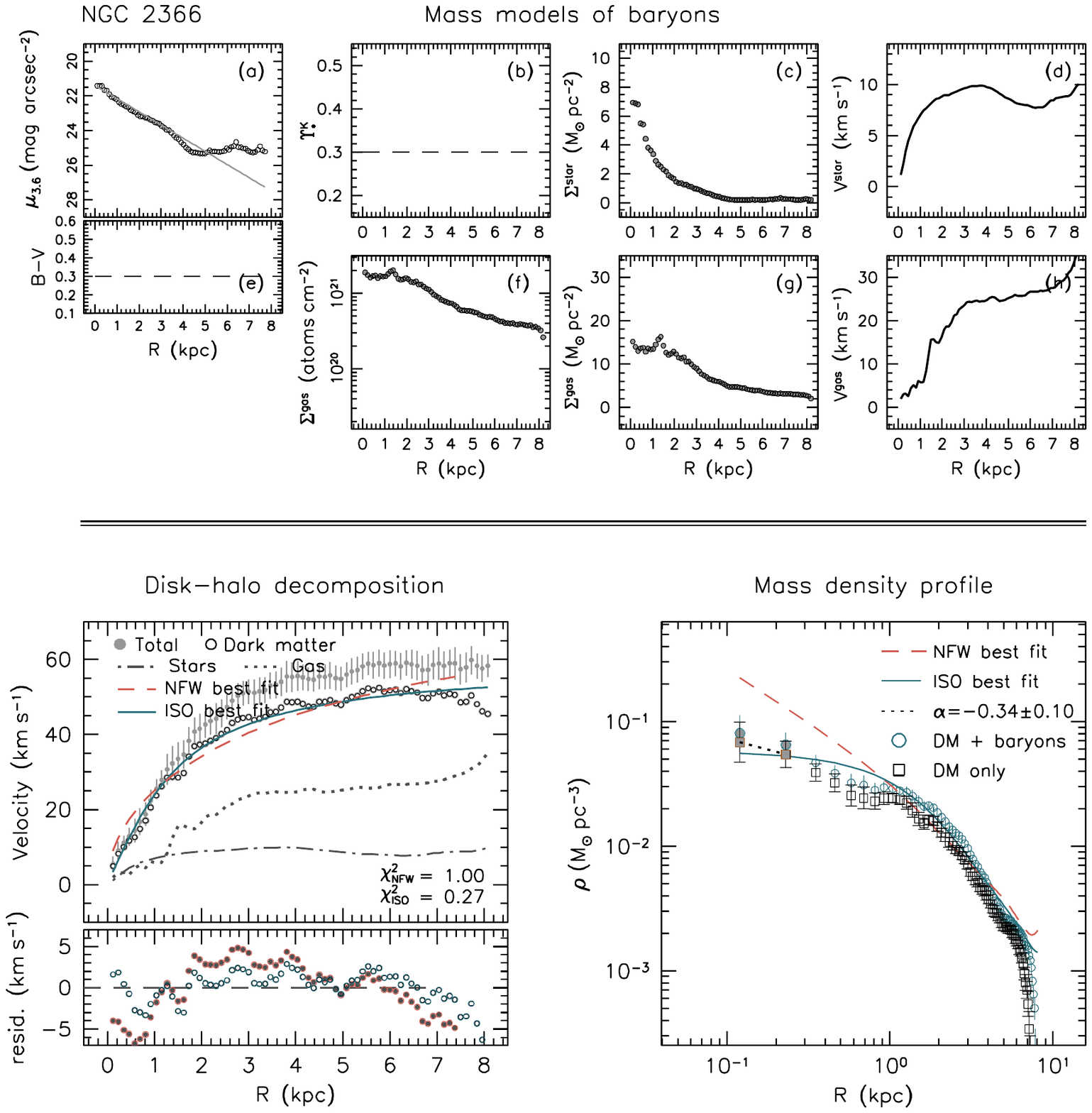}
\caption{The mass models of baryons, disk-halo decomposition and mass density
profile of NGC 2366. Please refer to the text in Sections~\ref{MASS_MODELS} and
\ref{DARK_MATTER_DISTRIBUTION} for full information.
\label{MD_DH_DM_ngc2366}}
\end{figure}
{\clearpage}

\begin{figure}
\epsscale{1.0}
\figurenum{A.64}
\includegraphics[angle=0,width=1.0\textwidth,bb=60 140 540
745,clip=]{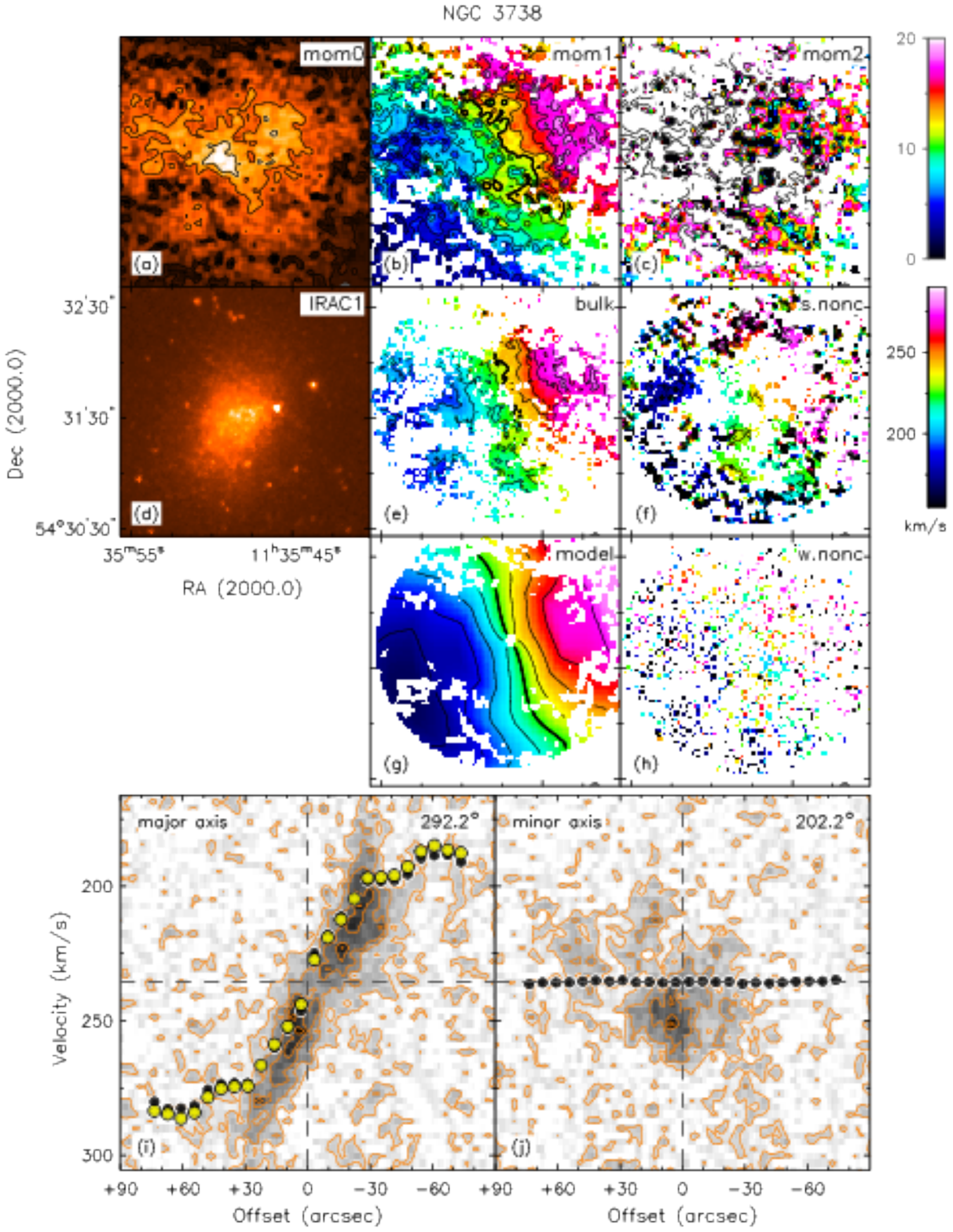}
\caption{H{\sc i} data and {\it Spitzer IRAC} 3.6$\mu$m image of NGC 3738. The
systemic velocity is indicated by the thick contours in the velocity fields, and
the iso-velocity contours are spaced by 10 \kms. Velocity dispersion contours run
from 0 to 20 \kms\ with a spacing of 10 \kms. See Appendix section A for details.
\label{ngc3738_data_PV}}
\end{figure}
{\clearpage}

\begin{figure}
\epsscale{1.0}
\figurenum{A.65}
\includegraphics[angle=0,width=1.0\textwidth,bb=35 140 570
710,clip=]{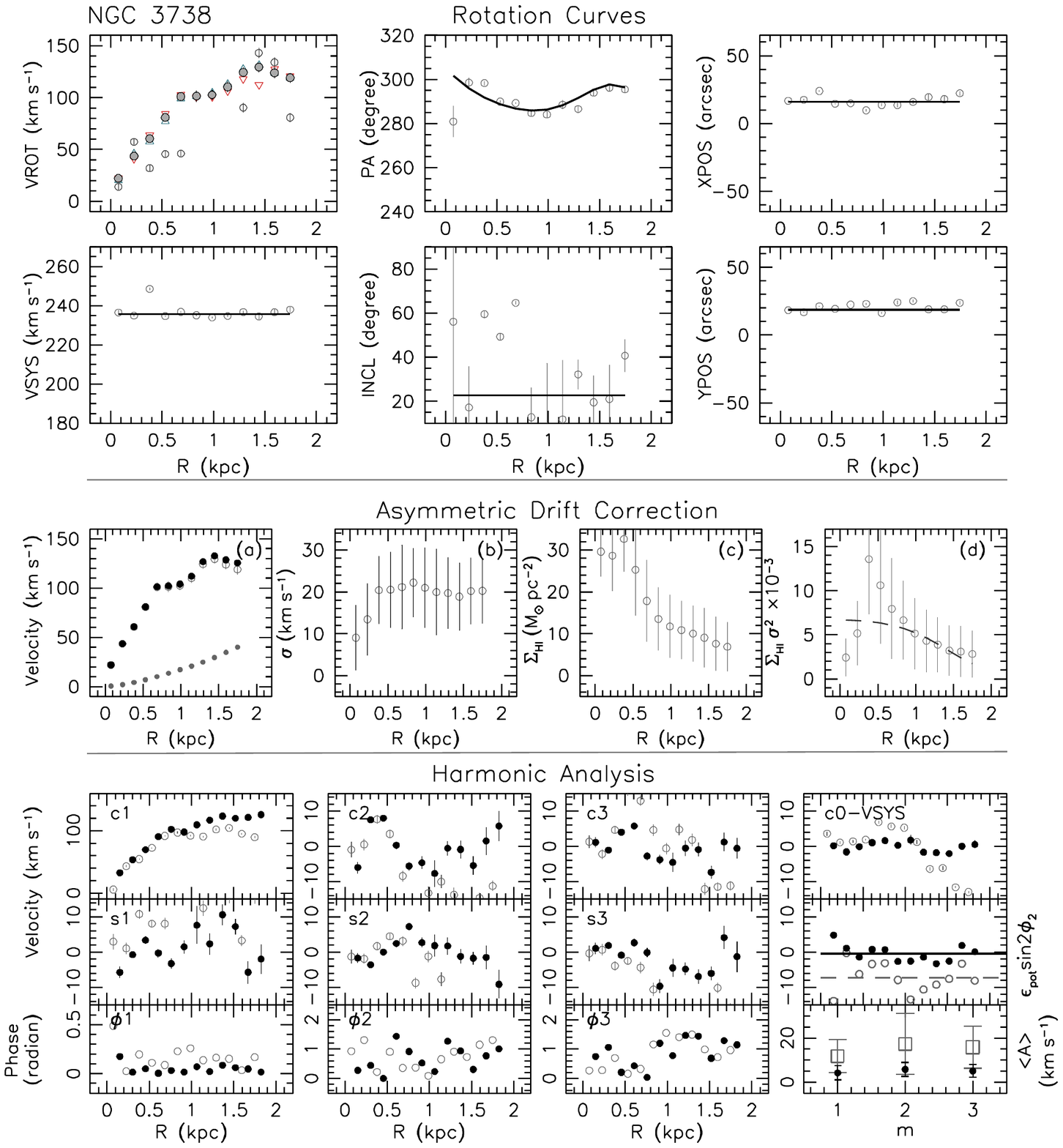}
\caption{Rotation curves, asymmetric drift correction and harmonic analysis
of NGC 3738. See Appendix section B for details.
\label{ngc3738_TR_HD}}
\end{figure}
{\clearpage}

\begin{figure}
\epsscale{1.0}
\figurenum{A.66}
\includegraphics[angle=0,width=1.0\textwidth,bb=40 175 540
690,clip=]{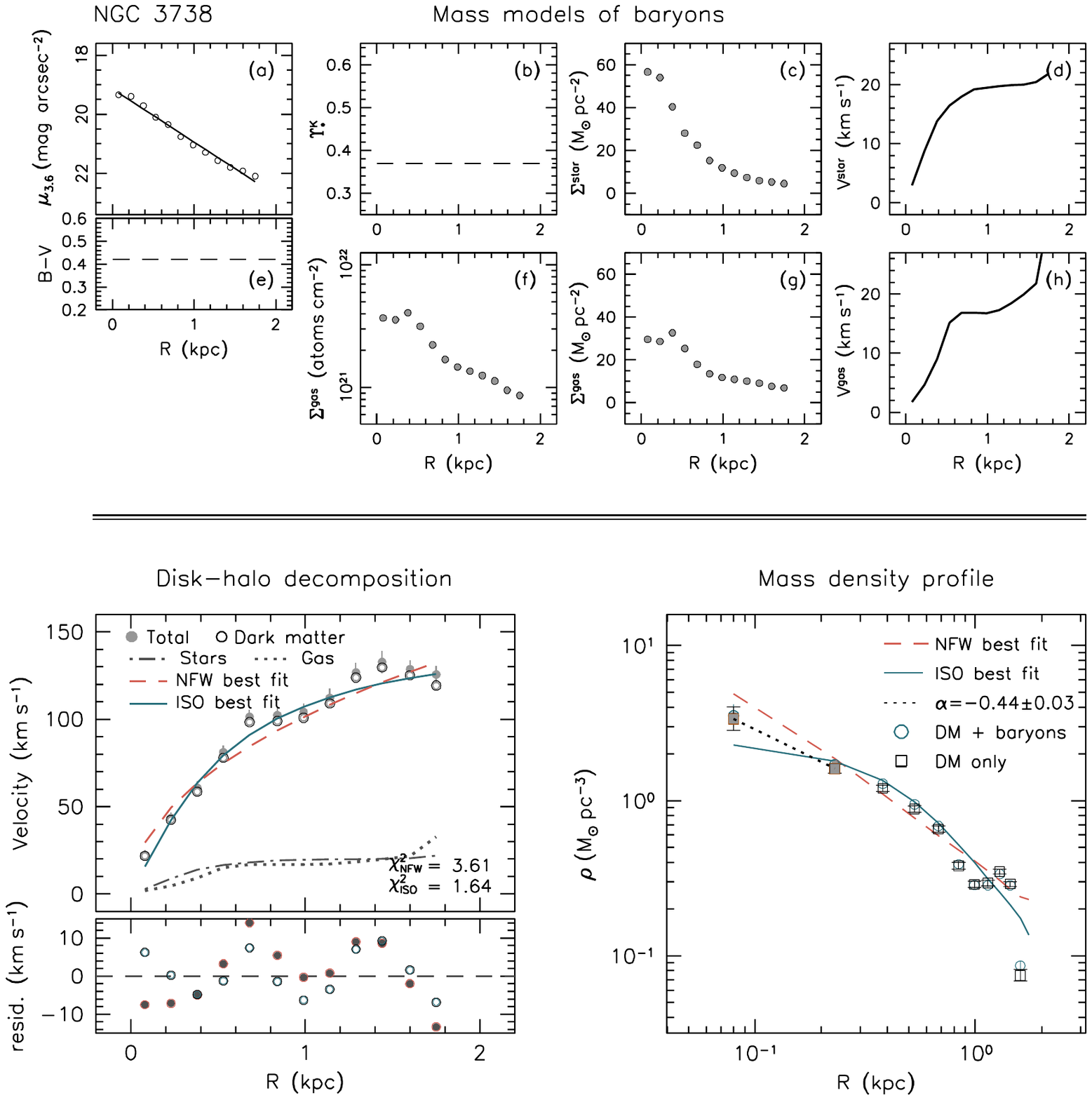}
\caption{The mass models of baryons, disk-halo decomposition and mass density
profile of NGC 3738. Please refer to the text in Sections~\ref{MASS_MODELS} and
\ref{DARK_MATTER_DISTRIBUTION} for full information.
\label{MD_DH_DM_ngc3738}}
\end{figure}
{\clearpage}

\begin{figure}
\epsscale{1.0}
\figurenum{A.67}
\includegraphics[angle=0,width=1.0\textwidth,bb=60 140 540
745,clip=]{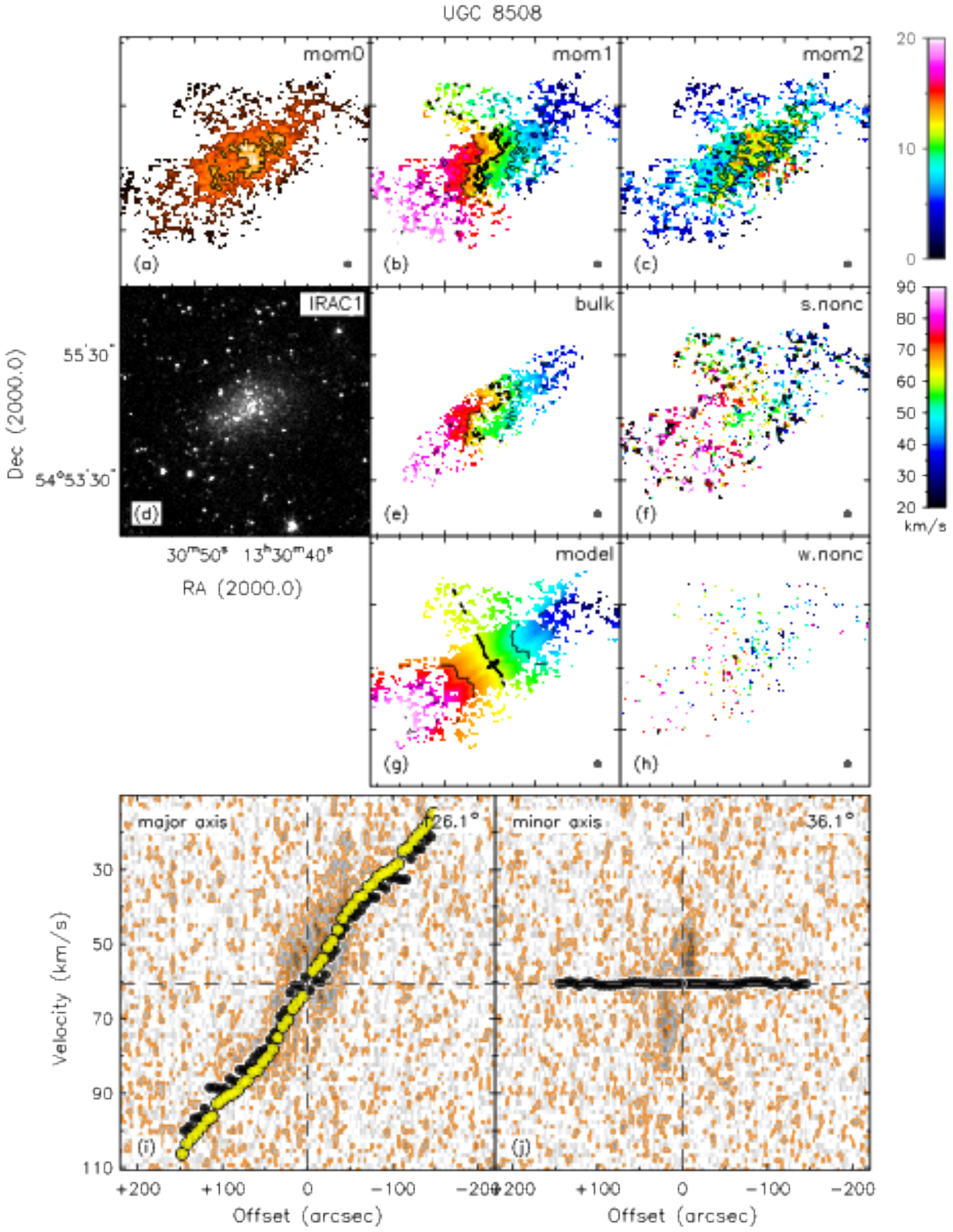}
\caption{H{\sc i} data and {\it Spitzer IRAC} 3.6$\mu$m image of UGC 8508. The
systemic velocity is indicated by the thick contours in the velocity fields, and
the iso-velocity contours are spaced by 10 \kms. Velocity dispersion contours run
from 0 to 20 \kms\ with a spacing of 10 \kms. See Appendix section A for details.
\label{ugc8508_data_PV}}
\end{figure}
{\clearpage}

\begin{figure}
\epsscale{1.0}
\figurenum{A.68}
\includegraphics[angle=0,width=1.0\textwidth,bb=35 140 570
710,clip=]{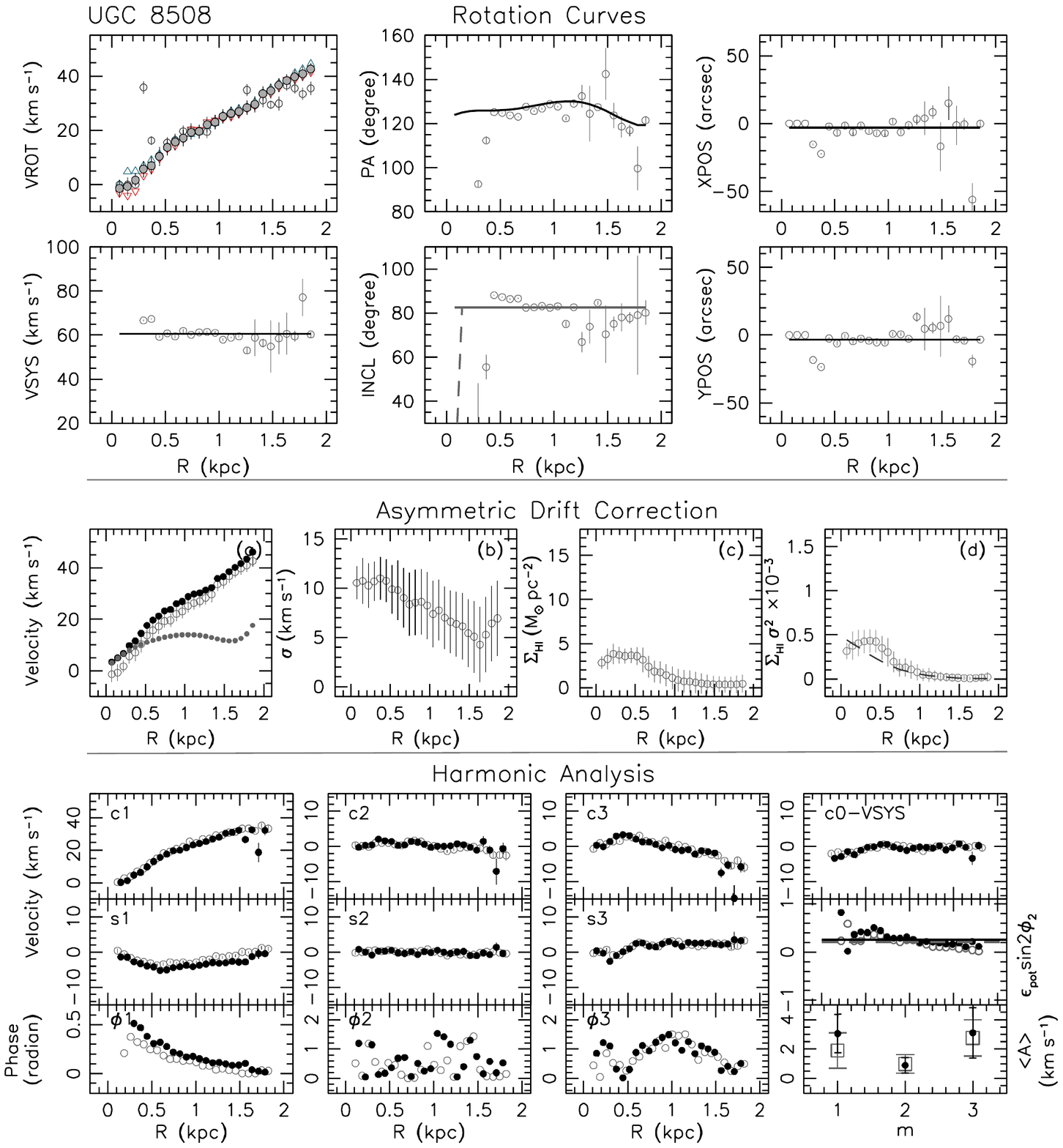}
\caption{Rotation curves, asymmetric drift correction and harmonic analysis
of UGC 8508. See Appendix section B for details.
\label{ugc8508_TR_HD}}
\end{figure}
{\clearpage}

\begin{figure}
\epsscale{1.0}
\figurenum{A.69}
\includegraphics[angle=0,width=1.0\textwidth,bb=40 175 540
690,clip=]{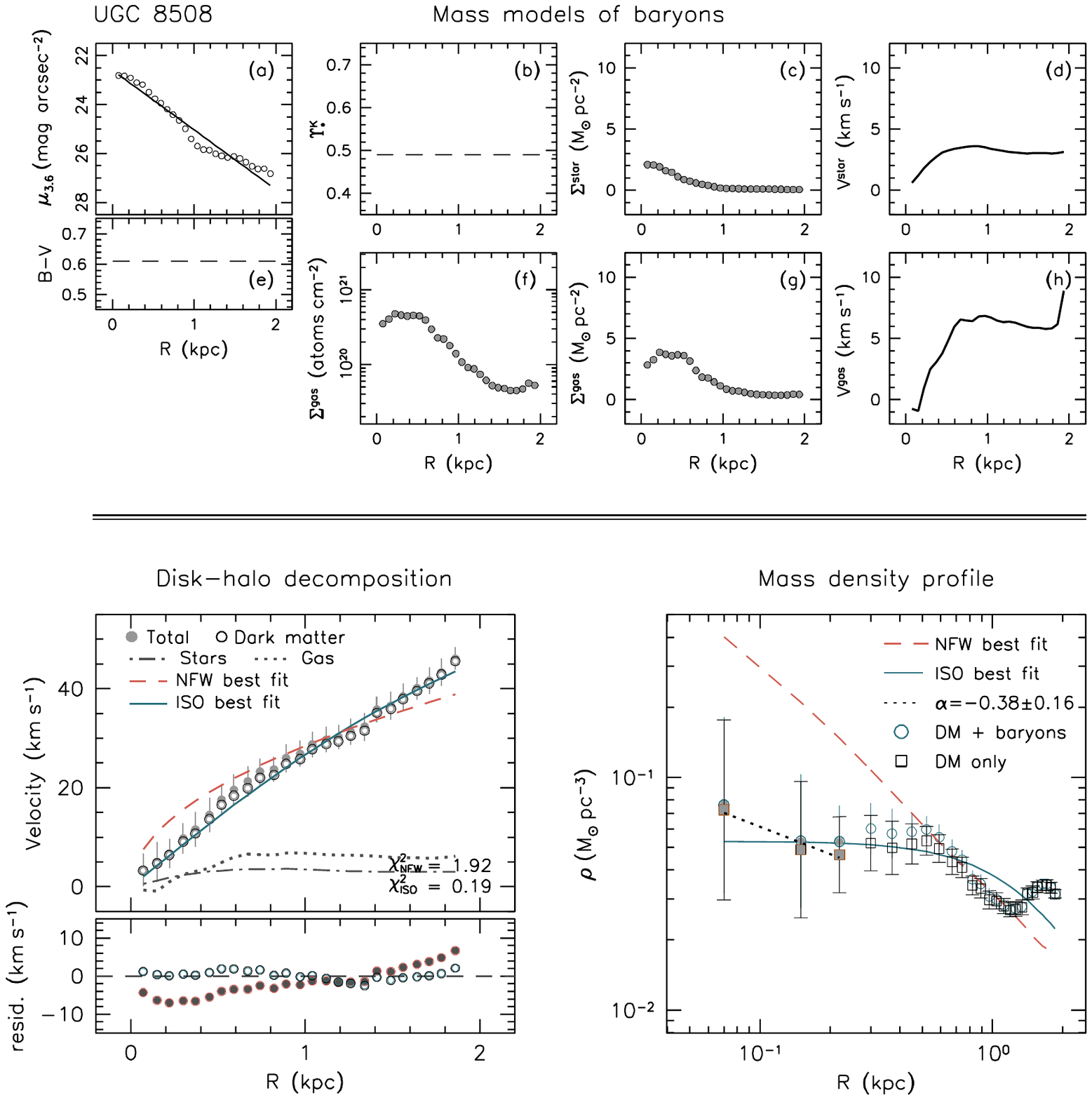}
\caption{The mass models of baryons, disk-halo decomposition and mass density
profile of UGC 8508. Please refer to the text in Sections~\ref{MASS_MODELS} and
\ref{DARK_MATTER_DISTRIBUTION} for full information.
\label{MD_DH_DM_ugc8508}}
\end{figure}
{\clearpage}

\begin{figure}
\epsscale{1.0}
\figurenum{A.70}
\includegraphics[angle=0,width=1.0\textwidth,bb=60 140 540
745,clip=]{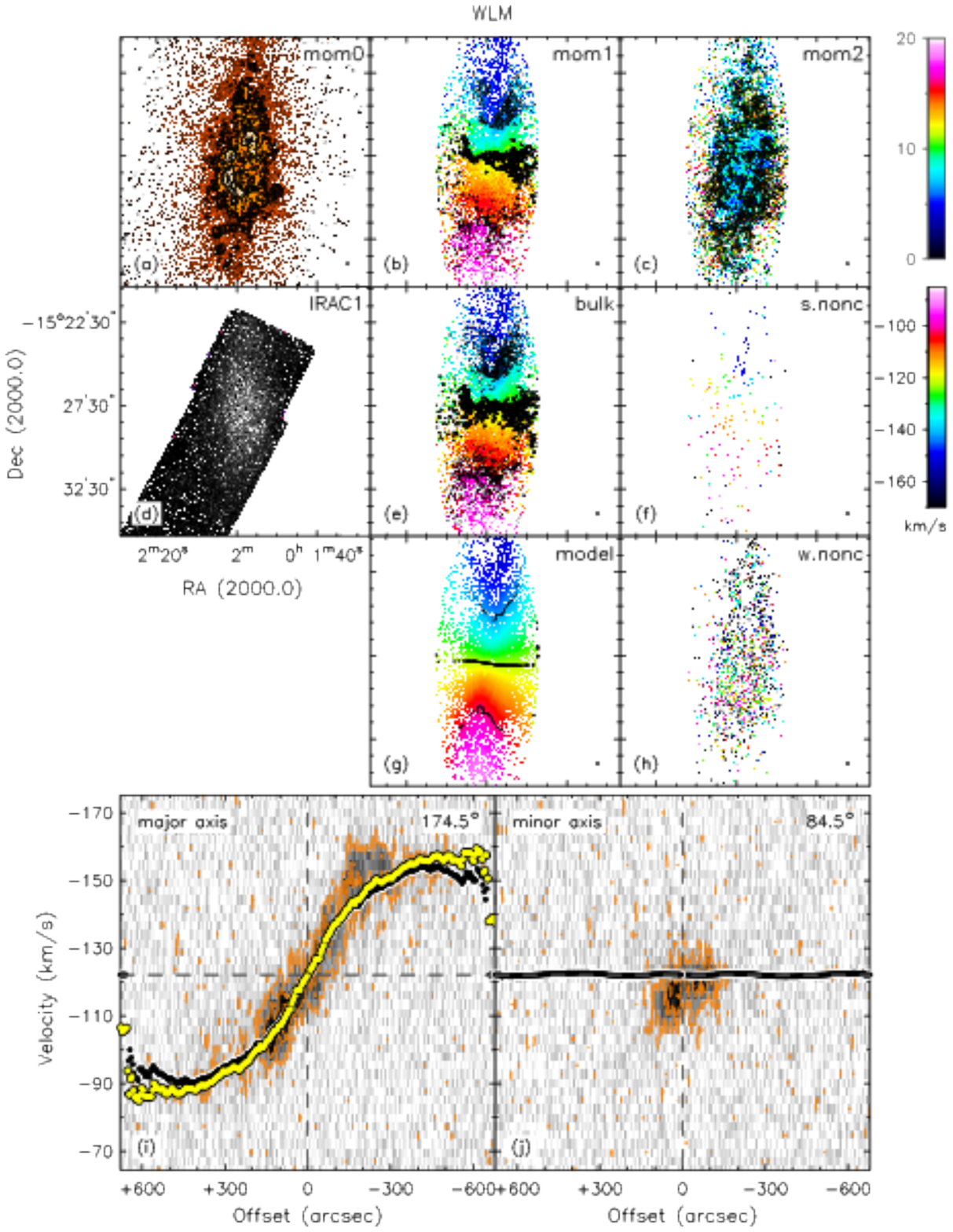}
\caption{H{\sc i} data and {\it Spitzer IRAC} 3.6$\mu$m image of WLM. The
systemic velocity is indicated by the thick contours in the velocity fields, and
the iso-velocity contours are spaced by 20 \kms. Velocity dispersion contours run
from 0 to 20 \kms\ with a spacing of 10 \kms. See Appendix section A for details.
\label{wlm_data_PV}}
\end{figure}
{\clearpage}

\begin{figure}
\epsscale{1.0}
\figurenum{A.71}
\includegraphics[angle=0,width=1.0\textwidth,bb=35 140 570
710,clip=]{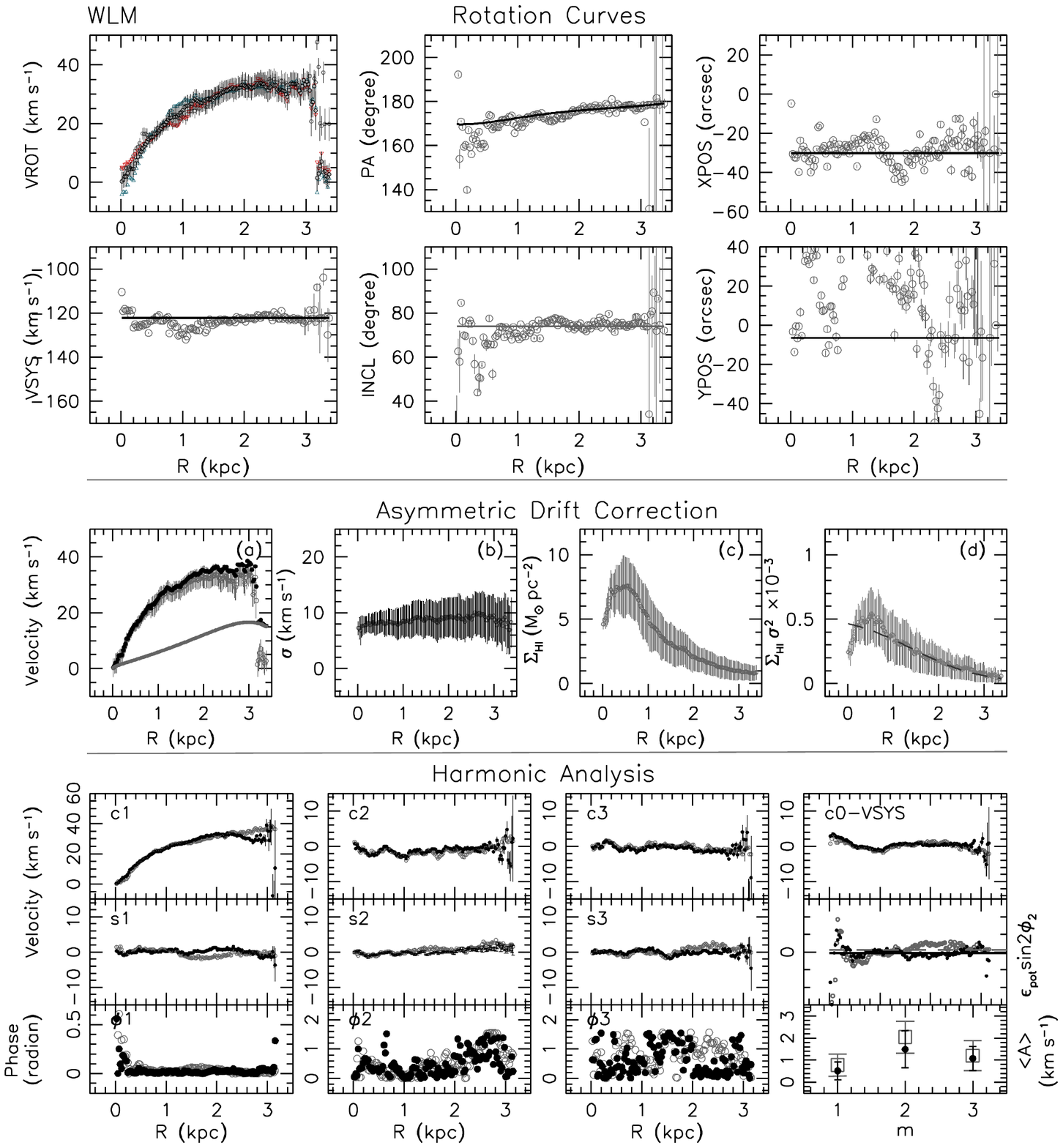}
\caption{Rotation curves, asymmetric drift correction and harmonic analysis
of WLM. See Appendix section B for details.
\label{wlm_TR_HD}}
\end{figure}
{\clearpage}

\begin{figure}
\epsscale{1.0}
\figurenum{A.72}
\includegraphics[angle=0,width=1.0\textwidth,bb=40 175 540
690,clip=]{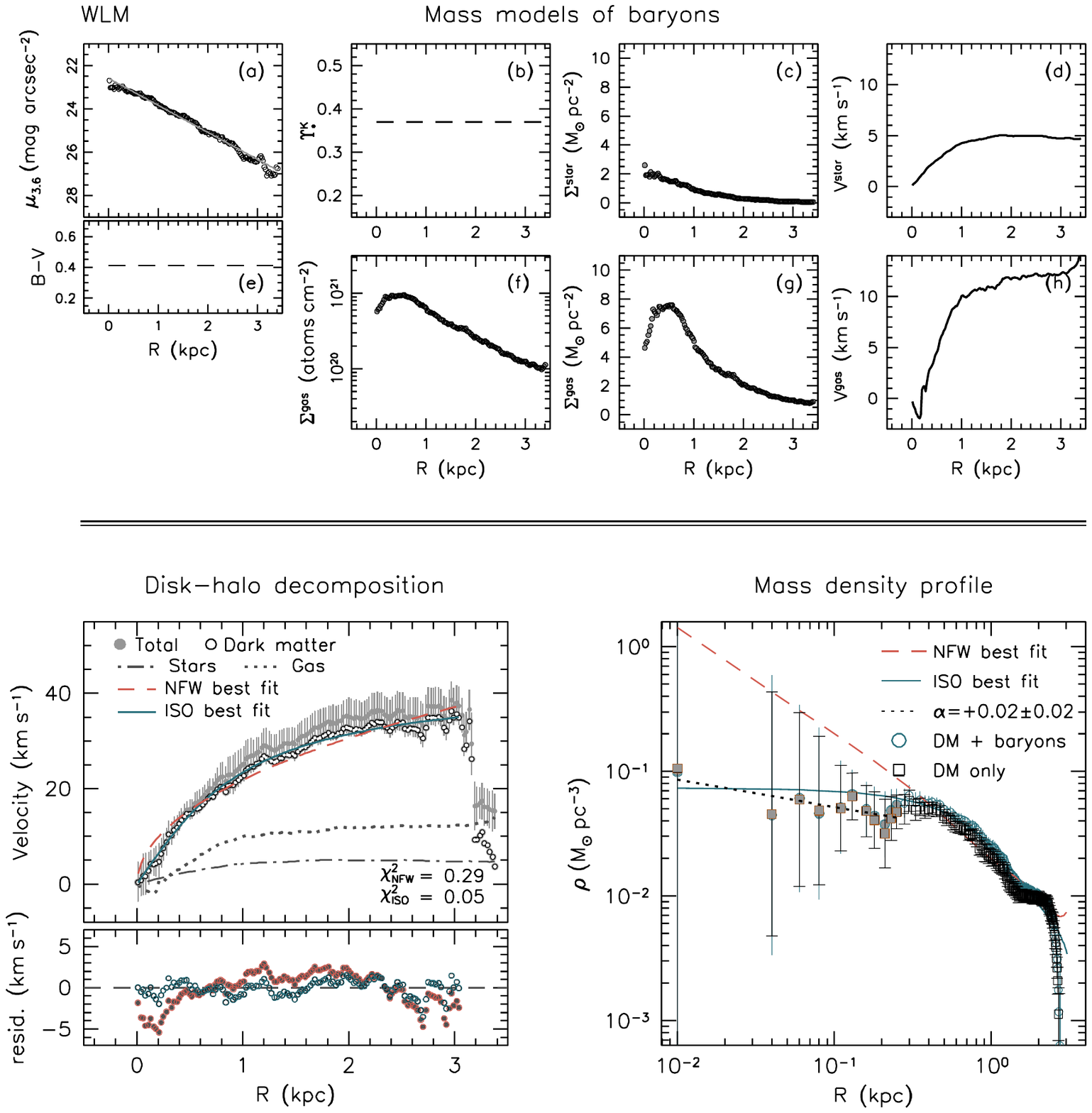}
\caption{The mass models of baryons, disk-halo decomposition and mass density
profile of WLM. Please refer to the text in Sections~\ref{MASS_MODELS} and
\ref{DARK_MATTER_DISTRIBUTION} for full information.
\label{MD_DH_DM_wlm}}
\end{figure}
{\clearpage}

\begin{figure}
\epsscale{1.0}
\figurenum{A.73}
\includegraphics[angle=0,width=1.0\textwidth,bb=60 140 540
745,clip=]{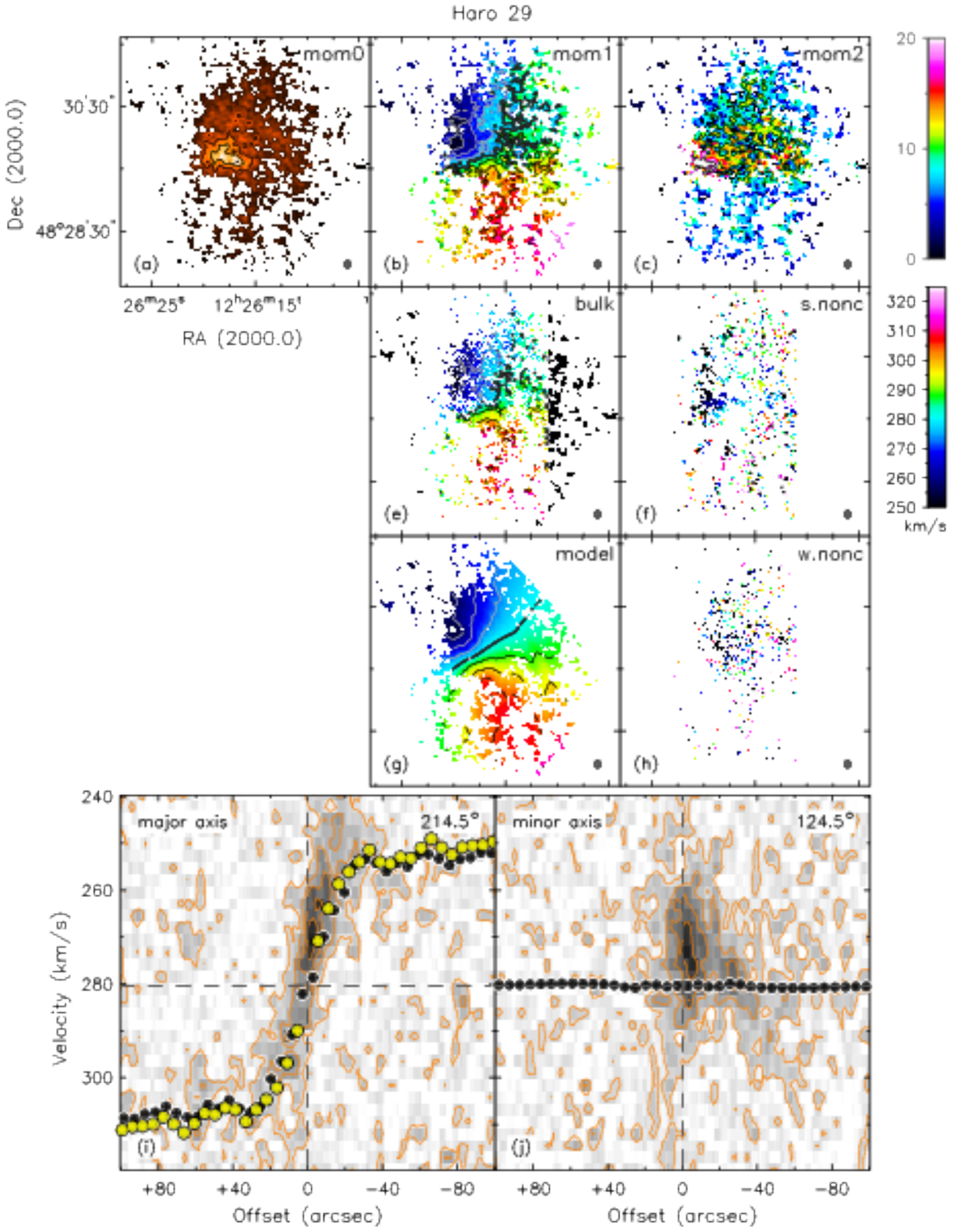}
\caption{H{\sc i} data and {\it Spitzer IRAC} 3.6$\mu$m image of Haro 29. The
systemic velocity is indicated by the thick contours in the velocity fields, and
the iso-velocity contours are spaced by 8 \kms. Velocity dispersion contours run
from 0 to 30 \kms\ with a spacing of 10 \kms. See Appendix section A for details.
\label{haro29_data_PV}}
\end{figure}
{\clearpage}

\begin{figure}
\epsscale{1.0}
\figurenum{A.74}
\includegraphics[angle=0,width=1.0\textwidth,bb=35 140 570
710,clip=]{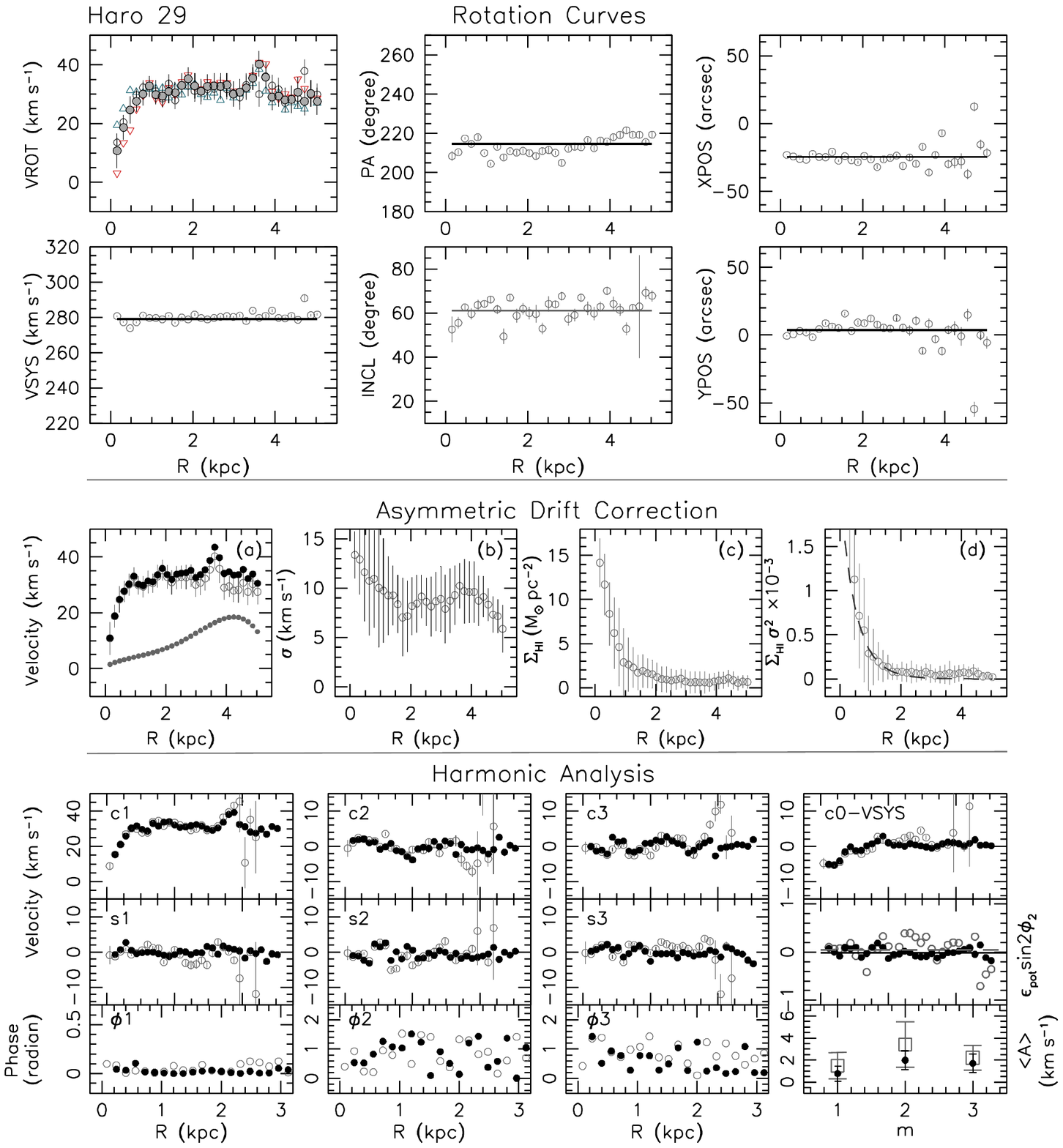}
\caption{Rotation curves, asymmetric drift correction and harmonic analysis
of Haro 29. See Appendix section B for details.
\label{haro29_TR_HD}}
\end{figure}
{\clearpage}

\begin{figure}
\epsscale{1.0}
\figurenum{A.75}
\includegraphics[angle=0,width=1.0\textwidth,bb=40 175 540
690,clip=]{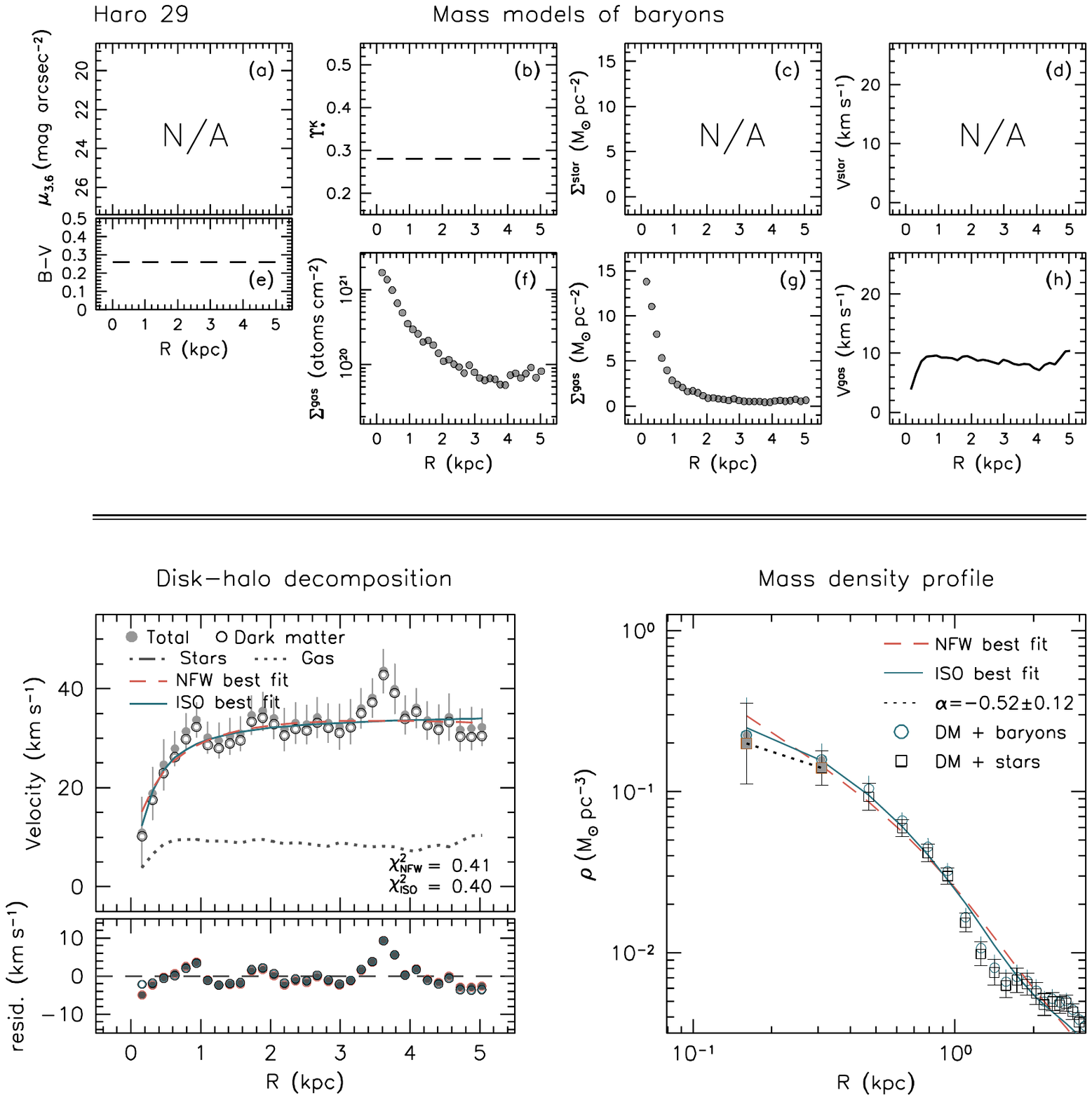}
\caption{The mass models of baryons, disk-halo decomposition and mass density
profile of Haro 29. Please refer to the text in Sections~\ref{MASS_MODELS} and
\ref{DARK_MATTER_DISTRIBUTION} for full information.
\label{MD_DH_DM_haro29}}
\end{figure}
{\clearpage}

\begin{figure}
\epsscale{1.0}
\figurenum{A.76}
\includegraphics[angle=0,width=1.0\textwidth,bb=60 140 540
745,clip=]{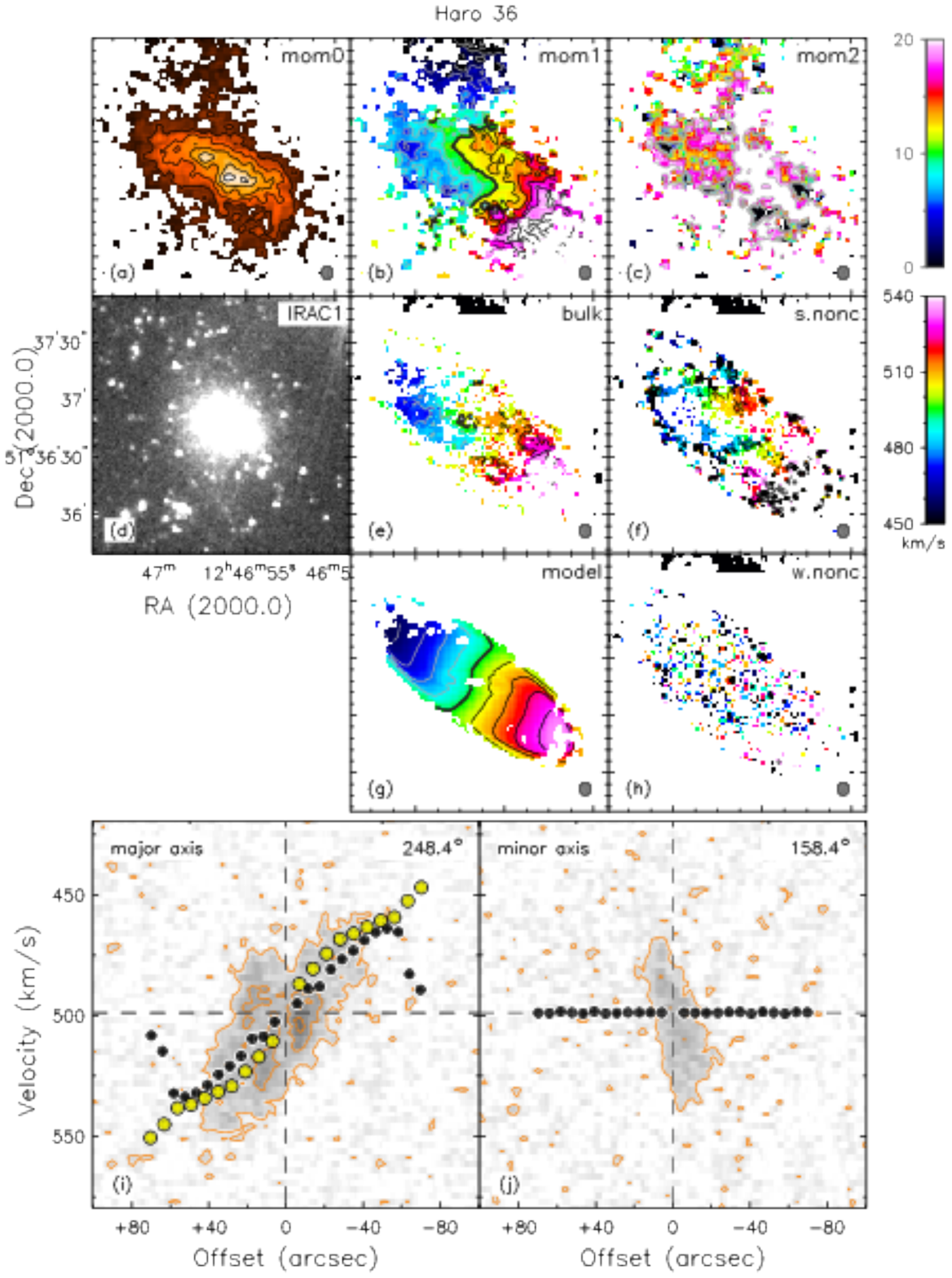}
\caption{H{\sc i} data and {\it Spitzer IRAC} 3.6$\mu$m image of Haro 36. The
systemic velocity is indicated by the thick contours in the velocity fields, and
the iso-velocity contours are spaced by 8 \kms. Velocity dispersion contours run
from 0 to 20 \kms\ with a spacing of 5 \kms. See Appendix section A for details.
\label{haro36_data_PV}}
\end{figure}
{\clearpage}

\begin{figure}
\epsscale{1.0}
\figurenum{A.77}
\includegraphics[angle=0,width=1.0\textwidth,bb=35 140 570
710,clip=]{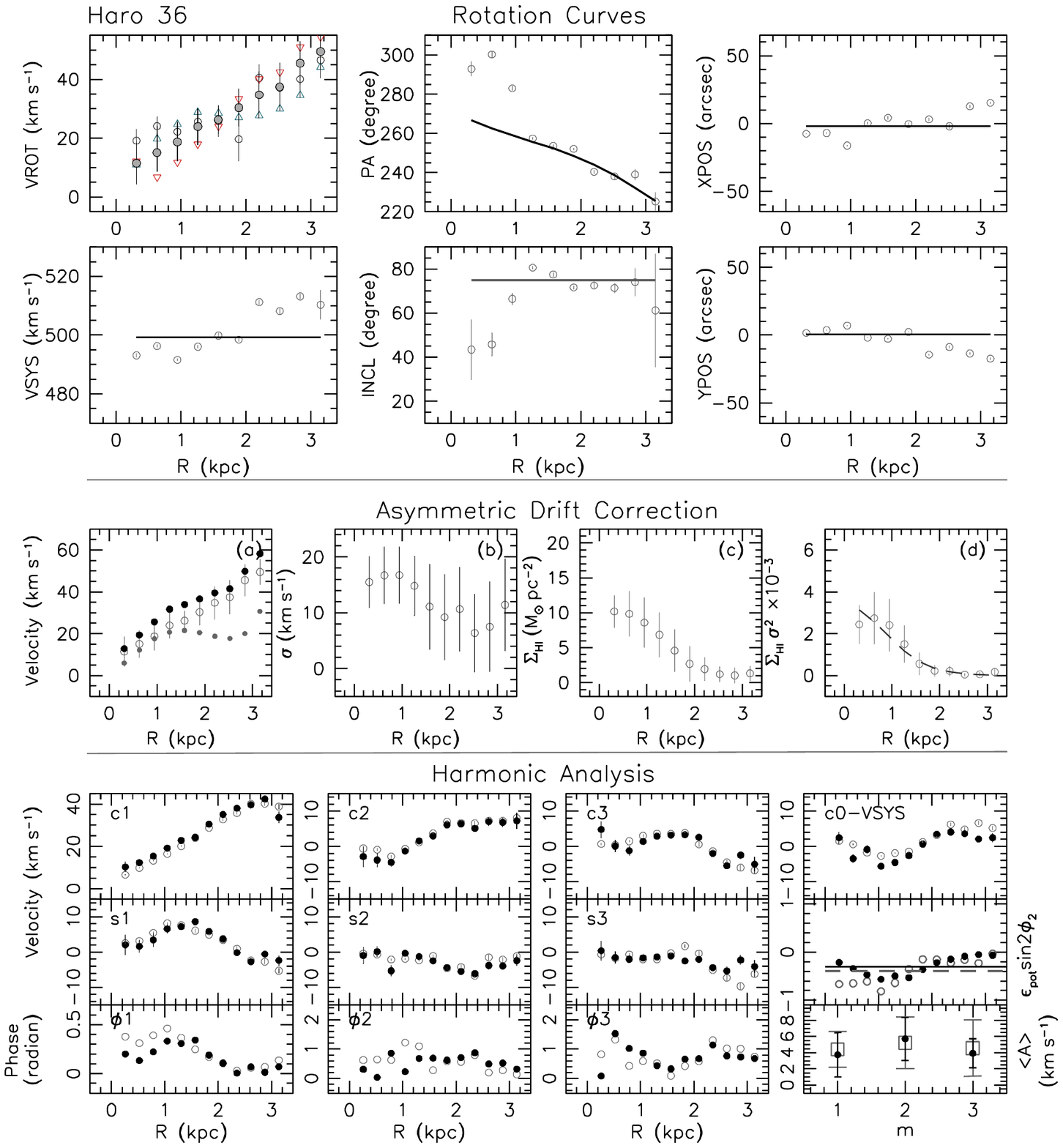}
\caption{Rotation curves, asymmetric drift correction and harmonic analysis
of Haro 36. See Appendix section B for details.
\label{haro36_TR_HD}}
\end{figure}
{\clearpage}

\begin{figure}
\epsscale{1.0}
\figurenum{A.78}
\includegraphics[angle=0,width=1.0\textwidth,bb=40 175 540
690,clip=]{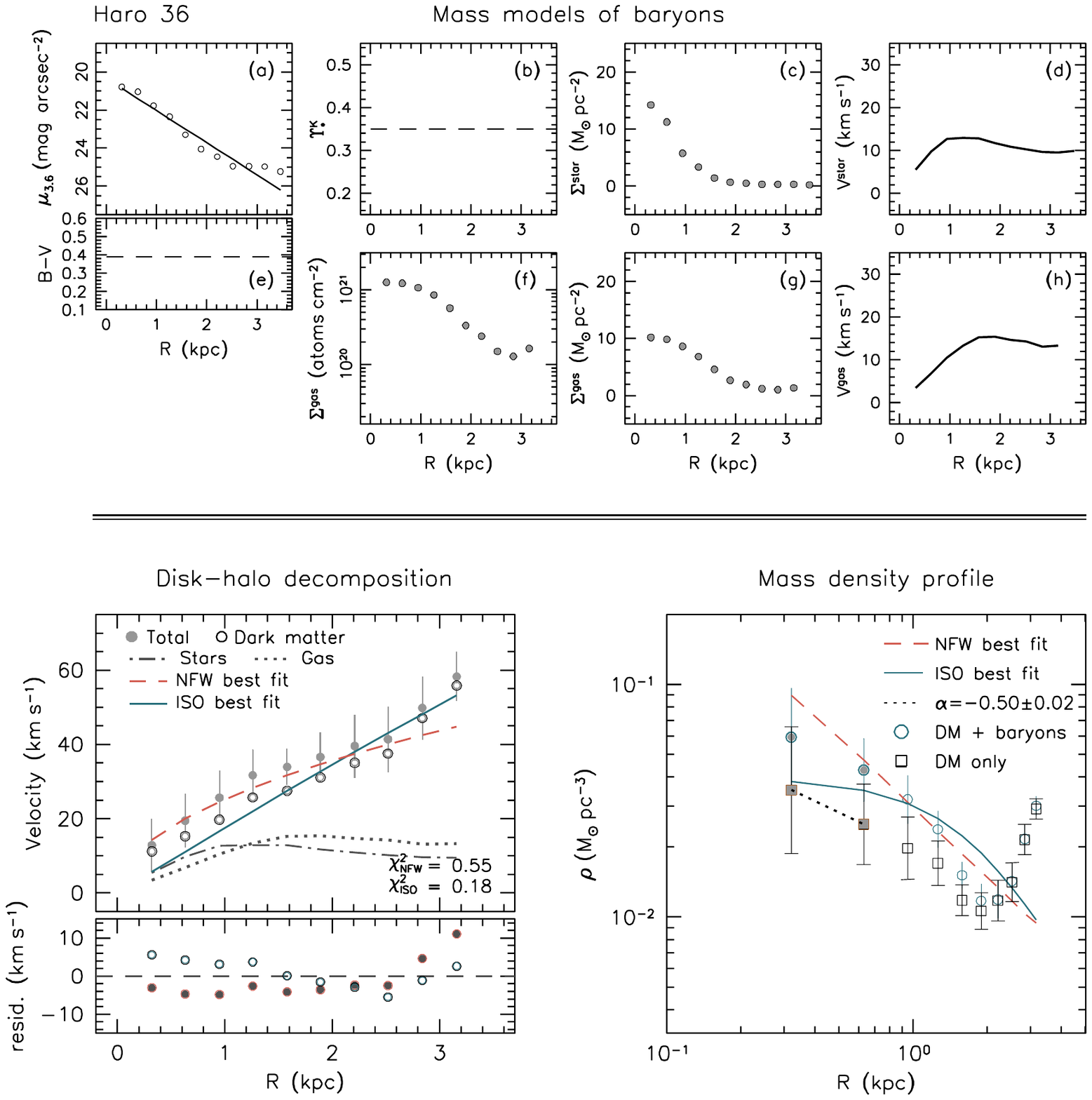}
\caption{The mass models of baryons, disk-halo decomposition and mass density
profile of Haro 36. Please refer to the text in Sections~\ref{MASS_MODELS} and
\ref{DARK_MATTER_DISTRIBUTION} for full information.
\label{MD_DH_DM_haro36}}
\end{figure}
{\clearpage}

\end{document}